\PassOptionsToPackage{unicode}{hyperref}
\PassOptionsToPackage{hyphens}{url}
\PassOptionsToPackage{dvipsnames,svgnames,x11names}{xcolor}
\documentclass[a4paper,11pt]{article}

\usepackage{amsmath,amssymb}
\usepackage{iftex}
\usepackage[T1]{fontenc}
\usepackage[utf8]{inputenc}
\usepackage{textcomp}
\usepackage{lmodern}
\IfFileExists{upquote.sty}{\usepackage{upquote}}{}
\IfFileExists{microtype.sty}{
  \usepackage[]{microtype}
  \UseMicrotypeSet[protrusion]{basicmath} 
}{}
\makeatletter
\@ifundefined{KOMAClassName}{
  \IfFileExists{parskip.sty}{%
    \usepackage{parskip}
  }{
    \setlength{\parindent}{0pt}
    \setlength{\parskip}{6pt plus 2pt minus 1pt}}
}{
  \KOMAoptions{parskip=half}}
\makeatother
\usepackage{xcolor}
\makeatletter
\ifx\paragraph\undefined\else
  \let\oldparagraph\paragraph
  \renewcommand{\paragraph}{
    \@ifstar
      \xxxParagraphStar
      \xxxParagraphNoStar
  }
  \newcommand{\xxxParagraphStar}[1]{\oldparagraph*{#1}\mbox{}}
  \newcommand{\xxxParagraphNoStar}[1]{\oldparagraph{#1}\mbox{}}
\fi
\ifx\subparagraph\undefined\else
  \let\oldsubparagraph\subparagraph
  \renewcommand{\subparagraph}{
    \@ifstar
      \xxxSubParagraphStar
      \xxxSubParagraphNoStar
  }
  \newcommand{\xxxSubParagraphStar}[1]{\oldsubparagraph*{#1}\mbox{}}
  \newcommand{\xxxSubParagraphNoStar}[1]{\oldsubparagraph{#1}\mbox{}}
\fi
\makeatother

\usepackage{longtable,booktabs,array}
\usepackage{calc} 
\usepackage{etoolbox}
\makeatletter
\patchcmd\longtable{\par}{\if@noskipsec\mbox{}\fi\par}{}{}
\makeatother
\IfFileExists{footnotehyper.sty}{\usepackage{footnotehyper}}{\usepackage{footnote}}
\makesavenoteenv{longtable}
\usepackage{graphicx}
\makeatletter
\def\maxwidth{\ifdim\Gin@nat@width>\linewidth\linewidth\else\Gin@nat@width\fi}
\def\maxheight{\ifdim\Gin@nat@height>\textheight\textheight\else\Gin@nat@height\fi}
\makeatother
\setkeys{Gin}{width=\maxwidth,height=\maxheight,keepaspectratio}
\makeatletter
\def\fps@figure{htbp}
\makeatother

\makeatletter
\@ifpackageloaded{caption}{}{\usepackage{caption}}
\AtBeginDocument{%
\ifdefined\contentsname
  \renewcommand*\contentsname{Table of contents}
\else
  \newcommand\contentsname{Table of contents}
\fi
\ifdefined\listfigurename
  \renewcommand*\listfigurename{List of Figures}
\else
  \newcommand\listfigurename{List of Figures}
\fi
\ifdefined\listtablename
  \renewcommand*\listtablename{List of Tables}
\else
  \newcommand\listtablename{List of Tables}
\fi
\ifdefined\figurename
  \renewcommand*\figurename{Figure}
\else
  \newcommand\figurename{Figure}
\fi
\ifdefined\tablename
  \renewcommand*\tablename{Table}
\else
  \newcommand\tablename{Table}
\fi
}
\@ifpackageloaded{float}{}{\usepackage{float}}
\floatstyle{ruled}
\@ifundefined{c@chapter}{\newfloat{codelisting}{h}{lop}}{\newfloat{codelisting}{h}{lop}[chapter]}
\floatname{codelisting}{Listing}

\makeatother
\makeatletter
\@ifpackageloaded{caption}{}{\usepackage{caption}}
\@ifpackageloaded{subcaption}{}{\usepackage{subcaption}}
\makeatother

\ifLuaTeX
  \usepackage{selnolig}  
\fi
\usepackage[]{natbib}
\usepackage{bookmark}

\IfFileExists{xurl.sty}{\usepackage{xurl}}{} 
\hypersetup{
  pdftitle={Double machine learning for causal inference in a multivariate sample selection model},
  pdfauthor={Sofiia Dolgikh; Bogdan Potanin},
  pdfkeywords={double machine learning, sample selection, causal inference},
  colorlinks=true,
  linkcolor={blue},
  filecolor={Maroon},
  citecolor={Blue},
  urlcolor={Blue},
  pdfcreator={LaTeX via pandoc}}

\newcommand{\anon}{1}

\usepackage{amsmath}
\usepackage{amssymb}
\usepackage{amsthm}

\theoremstyle{plain}
\newtheorem{theorem}{Theorem}
\newtheorem{lemma}{Lemma}

\theoremstyle{definition}
\newtheorem{assumption}{Assumption}

\usepackage[a4paper,margin=17mm,top=20mm,bottom=22mm]{geometry}
\allowdisplaybreaks[2]

\begin{document}

\def\spacingset#1{\renewcommand{\baselinestretch}%
{#1}\small\normalsize} \spacingset{1}


\if1\anon
{
  \title{\bf Double machine learning for causal inference in a multivariate sample selection model}
  \author{Sofiia Dolgikh\\
    Department of Theoretical Economics, HSE University\\
    and \\
    Bogdan Potanin \\
    Department of Applied Economics, HSE University}
  \maketitle
} \fi

\if0\anon
{
  \bigskip
  \bigskip
  \bigskip
  \begin{center}
    {\LARGE\bf Title}
\end{center}
  \medskip
} \fi

\bigskip
\begin{abstract}
We propose double machine learning (DML) estimators of the average treatment effect (ATE) and the average treatment effect on the treated (ATT) in a multivariate sample selection model with multiple ordinal selection equations and heterogeneous thresholds. We establish nonparametric identification of these causal parameters for both the full population and the selected subpopulation, derive Neyman-orthogonal scores accounting for the estimation of the control functions, and establish asymptotic normality of the proposed estimators. Using simulated data, we study their finite-sample properties and demonstrate the importance of accounting for multiple selection equations rather than replacing them with a single equation or ignoring sample selection altogether. Moreover, we apply the estimators to study the effect of the wife's education on the log-wage ratio of spouses in Russia. Finally, we discuss extensions of the proposed framework to various alternative scenarios, including conditional average treatment effects (CATE), nested sequential sample selection, multiple selection criteria, and treatment effect heterogeneity with respect to endogenous moderators.
\end{abstract}

\noindent%

\spacingset{1}


\section{Introduction}\label{sec-intro}

Machine learning methods are gaining popularity in causal inference \citep{Chernozhukov}. The reason is that these methods provide greater flexibility and robustness to functional form misspecification compared with conventional parametric estimators, which usually rely on restrictive assumptions about the relationships between variables. Nevertheless, some features of the data collection process may pose serious challenges to the accurate estimation of causal parameters, even when flexible estimators are used. Sample selection is one of these features.

Sample selection arises when an outcome is observable only under specific conditions. A classic example is wages, which are observable only for employed individuals \citep{Heckman1979}. If common or correlated factors influence both the outcome and the selection mechanism, estimators of causal parameters may be subject to selection bias. To avoid this bias, it is necessary to account for non-random sample selection. Methods for addressing sample selection have been extensively studied in classical econometrics \citep{Vella1998} and have recently been developed within the double machine learning framework for the estimation of the average treatment effect \citep{Bia} and the parameters of the partially linear model \citep{Potanin2026}.

Usually, researchers consider a single binary selection equation. However, in many applications, the observability of the outcome depends on multiple selection criteria. For example, the share of income that a household spends on food is observable only for households that have revealed information on both their income (first selection criterion) and food expenditure (second selection criterion) \citep{Potanin2026}. Selection in the estimation of Engel curves may also arise from both unit and item nonresponse \citep{DeLuca}. Another example is wages, which are observable only for employed individuals (first selection criterion) who are willing to reveal information on their wages (second selection criterion) \citep{Kossova2018}. Moreover, replacing multiple selection mechanisms with a single one may lead to model misspecification and biased estimates \citep{Mohanty2001, Potanin2026}.

The most commonly considered type of multivariate sample selection model has multiple binary selection equations. An early extension of the Heckman selection correction to multiple selection rules was proposed by \citet{Catsiapis1982} under the assumption of independence between the selection errors. Alternative parametric approaches to correcting selection bias under multiple selection rules were compared by \citet{Schmertmann1994}. Maximum-likelihood and two-step estimation methods for models with multiple binary selection equations under the assumption of joint normality of random errors have been considered by \citet{Ogundimu, Kossova2018, Rezaee}. The assumption of joint normality has been relaxed using Archimedean copulas \citep{Smith2003} and Edgeworth expansions \citep{Tunali2012}. Bayesian estimation methods have been proposed for an ordinal outcome model with two binary selection equations \citep{Zhang2015} and for regression models with multiple dependent selection rules \citep{Rezaee2024}.

Models with multiple selection mechanisms have been considered for panel data \citep{Cinzia, LeeVella2006, Cappellari2008, Lacroix2011}. Semiparametric estimators have been considered by \citet{DeLuca, LeeVella2006}, while nonparametric estimators have been proposed by \citet{Das, Potanin2026} and studied using simulated data by \citet{Kupriianova2020}. Furthermore, \citet{FernandezVal2023} consider nonseparable outcome models with multiple censored selection rules. Sequential selection, in which some selection variables are observable only for specific values of other selection variables, has been considered by \citet{Gatzlaff1997, Mohanty2001, Kossova2018}. Finally, \citet{Morozov2026} proposes two-step parametric estimators for the Poisson count data model with multiple binary selection equations.

Multivariate sample selection is closely related to systems of outcome equations with equation-specific selection mechanisms \citep{Tauchmann, Li}, bivariate probit models with partial observability \citep{Poirier}, and multinomial sample selection models based on the multinomial logit \citep{DMF, Bourguignon} and multinomial probit \citep{Kossova2022} models. Ordinal selection rules have also been considered in nonseparable sample selection models \citep{FernandezVal2023} and panel data models \citep{Been2023}. An alternative selection mechanism has been considered by \citet{Miao2025}, who establish nonparametric identification and propose doubly robust estimators for survey data in which the outcome becomes observable once at least one of several contact attempts is successful. Another closely related strand of the literature considers models combining sample selection and endogenous switching \citep{HH, Angela, Sheluntcova2020, Dolgikh2024, Dolgikh2026gender, Dolgikh2026, Potanin2026}.

In this article, we focus on the nonparametric identification and estimation of the average treatment effect (ATE) and the average treatment effect on the treated (ATT) in a multivariate sample selection model (MSSM) with multiple ordinal selection equations and heterogeneous thresholds. We consider these causal parameters for both the full population and the selected subpopulation. We concentrate on the case where selection is non-random (MNAR --- missing not at random), meaning that even conditional on the observed covariates, some unobserved factors affect both the outcome and the probability of selection.

Our primary focus on multiple ordinal selection equations is motivated by several considerations. First, ordinal equations nest binary ones. Thus, as a particular case, we cover the most commonly considered type of MSSM --- one with multiple binary selection equations. Second, ordinal selection covers cases in which the outcome is observable only if latent variables belong to bounded intervals, as may occur in multivariate sample selection arising in customer segmentation tasks \citep{Kim}. Moreover, allowing for heterogeneous thresholds makes it possible to account for the dependence of the distances between adjacent thresholds on observed characteristics, such as product prices in ordered choice models \citep{Aristodemou2022}. Nevertheless, we also consider extensions of the proposed framework to alternative selection mechanisms, including sequential selection and selection based on the union of multiple selection rules. Moreover, we demonstrate that the proposed framework is also applicable beyond conventional sample selection problems. In particular, we show how the established identification results can be used to study treatment effect heterogeneity with respect to multiple potentially endogenous ordinal moderators.

We establish the assumptions under which the causal parameters are nonparametrically identified in the MSSM. We then derive Neyman-orthogonal scores and propose double machine learning (DML) estimators of these parameters, for which we establish asymptotic normality. Specifically, we extend the DML framework proposed by \citet{Bia} for causal inference in a binary sample selection model to a model with multiple ordinal selection equations and heterogeneous thresholds. Moreover, even in the case of a single binary selection equation, the Neyman-orthogonal score we use for ATE differs from that of \citet{Bia}: it contains additional derivative terms that explicitly account for the estimation of the selection probabilities entering the control functions. The simulations in Section~\ref{sec:sim} show that these terms may substantially improve the finite-sample accuracy of the estimates. Moreover, we consider an alternative version of the proposed estimators that does not require estimating derivatives of the nuisance functions. To establish their asymptotic properties, we provide a modification of the general DML theorem of \citet{Chernozhukov} that replaces Neyman orthogonality and differentiability requirements with a direct restriction on the expected score. The established results reveal a trade-off between the smoothness requirements imposed on the nuisance functions and the restrictions imposed on the estimation errors in the generated control functions.

The paper is organized as follows. In Section~\ref{sec:model}, we introduce the nonparametric MSSM. In Section~\ref{sec:identification}, we establish and discuss the assumptions under which the causal parameters are identified in this model. In Sections~\ref{sec:neyman} and~\ref{sec:dml}, we derive Neyman-orthogonal scores and propose DML estimators of the causal parameters, respectively. In Section~\ref{sec:asymptotic}, we establish the asymptotic properties of the proposed estimators. In Section~\ref{sec:sim}, we study their finite-sample properties using simulated data. In Section~\ref{sec:extensions}, we extend the proposed framework to alternative selection mechanisms and causal parameters. In Section~\ref{sec:example}, we apply the estimators to study the effect of the wife's education on the log-wage ratio of spouses in Russia. Finally, Section~\ref{sec:conclusion} concludes. All proofs are provided in the appendix. In addition, we provide the R package \textbf{sdml}, which implements the proposed estimators.

\section{The model}\label{sec:model}

Consider a sample of $n$ i.i.d.\ observations indexed by $i = 1, \dots, n$. The main notation used throughout the paper is introduced below and follows that of \citet{Potanin2026}.

\begin{enumerate}
    \item $D_i$ --- a categorical treatment variable with support $\operatorname{supp}(D_i) = \{0, \dots, n_D - 1\}$, where $n_D$ is the number of possible treatment levels.

    \item $Y_{di}$ --- the potential outcome under treatment level $d$, where $d \in \{0, \dots, n_D - 1\}$. We assume that $\mathbb{E}|Y_{di}|<\infty$ for every $d\in\{0,\dots,n_D-1\}$.

    \item $Y_i = \sum\limits_{d=0}^{n_D - 1} Y_{di} \cdot \mathbb{I}(D_i = d)$ --- the realized outcome, where $\mathbb{I}(\text{condition})$ is an indicator function that equals $1$ if the condition is true and $0$ otherwise.

    \item $Z_{ji}$ --- the $j$-th ordinal selection variable, where $j \in \{1, \dots, n_Z\}$ and $\operatorname{supp}(Z_{ji}) = \{0, \dots, n_{Z_j} - 1\}$. Here, $n_Z$ denotes the number of selection equations and $n_{Z_j}$ denotes the number of possible values of the $j$-th selection variable. We denote by $Z_i = (Z_{1i}, \dots, Z_{n_Z i})$ the vector of selection variables.

    \item $X_i$ --- a row vector of covariates.

    \item $W_i^{(Z)}$ --- a row vector of variables serving as exclusion restrictions for the selection equations. The variables in $X_i$ and $W_i^{(Z)}$ are distinct.
\end{enumerate}

We assume that $D_i$, $Z_i$, $X_i$, and $W_i^{(Z)}$ are always observed, whereas $Y_i$ is observed if and only if $\tilde{Z}_i = 1$, where the selection indicator is defined as follows:
\begin{equation}
    \tilde{Z}_i = \mathbb{I}\left(Z_i = z^{*}\right) =
    \begin{cases}
        1, & \text{if } Z_i = z^{*}, \\
        0, & \text{otherwise},
    \end{cases}
\end{equation}
where $z^{*}\in\operatorname{supp}\left(Z_i\right)$ is the vector of selection variable values under which $Y_i$ is observable. For example, \citet{Potanin2026} considers a case where a researcher observes the share of income that a household spends on food $Y_i$ only if the respondent reveals information on both food expenditure and income. Therefore, $Z_{1i}$ and $Z_{2i}$ are binary selection variables corresponding to the answers to these questions (1 --- information revealed, 0 --- otherwise). The outcome is observed if and only if $Z_{1i}=Z_{2i}=1$, so $z^{*} = (1, 1)$. In this setting, a researcher may be interested in estimating the effect of the education level of the household head $D_i$ on the share of household income spent on food.

\section{Identification}\label{sec:identification}

To identify the causal parameters in the MSSM, we consider a generalization of the assumptions used by \citet{Bia} in the univariate case with binary selection.

\begin{assumption}[Multivariate ordinal selection]\label{assumption:selection}
Denote $Q_i = (D_i, X_i, W_i^{(Z)})$. For every $j \in \{1, \dots, n_Z\}$, the selection mechanism is as follows:
  \begin{equation}\label{eq:selection}
    Z_{ji} =
    \begin{cases}
      0, & \text{if } U_{ji}^{(Z)} \leq g_j(Q_i), \\
      1, & \text{if } g_j(Q_i) < U_{ji}^{(Z)}
             \leq g_j(Q_i) + c_{j1}(Q_i), \\
      \vdots & \\
      n_{Z_j}-2, & \text{if } g_j(Q_i) + c_{j(n_{Z_j}-3)}(Q_i)
             < U_{ji}^{(Z)}
             \leq g_j(Q_i) + c_{j(n_{Z_j}-2)}(Q_i), \\
      n_{Z_j}-1, & \text{if }
          g_j(Q_i) + c_{j(n_{Z_j}-2)}(Q_i)
          < U_{ji}^{(Z)},
    \end{cases}
  \end{equation}
where $U_{ji}^{(Z)}$ are random errors and $c_{jk}(Q_i)$ determine the potentially heterogeneous distances between the thresholds. For every $j \in \{1,\dots,n_Z\}$, the functions $g_j(q)$ and $c_{jk}(q)$, $k \in \{0,\dots,n_{Z_j}-2\}$, are measurable and defined on $\operatorname{supp}(Q_i)$. Without loss of generality, we impose the normalization $c_{j0}(q)=0$. Moreover, for every $q \in \operatorname{supp}(Q_i)$ and $k \in \{1,\dots,n_{Z_j}-2\}$, we have $c_{j(k-1)}(q) < c_{jk}(q)$. In the binary case ($n_{Z_j}=2$), only the first and last conditions in equation~(\ref{eq:selection}) apply.
\end{assumption}

\begin{assumption}[Conditional exogeneity of treatment and exclusion variables]
  \label{assumption:exogeneity}
  For every $d \in \{0, \dots, n_D - 1\}$, we have
  $(Y_{di}, U_i^{(Z)}) \perp (D_i, W_i^{(Z)}) \mid X_i$,
  where $U_i^{(Z)} = (U_{1i}^{(Z)}, \dots, U_{n_Z i}^{(Z)})$.
\end{assumption}

If there is only one binary selection equation, i.e., $n_Z = 1$ and $n_{Z_1} = 2$, then the selection mechanism specified in Assumption~\ref{assumption:selection} reduces to that considered in Assumption 4b of \citet{Bia}. If there are multiple binary selection equations, i.e., $n_{Z_j} = 2$ for every $j \in \{1, \dots, n_Z\}$, then the selection mechanism is similar to those considered in Section 2.2 of \citet{Das} and Section 2.1 of \citet{Potanin2026}. Note that Assumption~\ref{assumption:selection} allows for dependence among the selection errors $U_{ji}^{(Z)}$ across equations. In particular, the multivariate ordinal probit model is a special case of the selection mechanism specified in this assumption.

Assumption~\ref{assumption:exogeneity} has two crucial implications. First, it implies that the potential outcomes $Y_{di}$ and treatment $D_i$ are independent conditional on the covariates $X_i$. This is a fairly standard assumption in causal inference but may be violated in many empirical applications. For example, potential wages $Y_{di}$ and the level of education $D_i$ are likely to remain dependent even conditional on conventional covariates (age, health status, and so on), since individual abilities are usually unobserved and may confound the relationship between wages and education. To make the conditional independence assumption more plausible, some studies on the returns to education include proxies for these abilities, such as test scores \citep{Grogger, Rozhkova}. Nevertheless, finding adequate proxies for unobserved confounders may be challenging, making the reliance on conditional exogeneity questionable in some applications.

Second, Assumption~\ref{assumption:exogeneity} requires the exclusion variables $W_i^{(Z)}$ to be independent of the joint vector of potential outcome $Y_{di}$ and selection errors $U_i^{(Z)}$ conditional on the covariates $X_i$. For example, interviewer characteristics (age, education, and gender) may influence the probability that a respondent reveals information on food expenditure ($Z_{1i}=1$), while being plausibly independent of the unobserved determinants of household consumption and survey participation \citep{DeLuca}.

Note that Assumption~\ref{assumption:exogeneity} implies that $Y_{di} \perp D_i \mid X_i$. However, it does not imply $Y_{di} \perp D_i \mid (X_i, \tilde{Z}_i = 1)$. Hence, in the subsample of selected observations ($\tilde{Z}_i = 1$), the potential outcomes $Y_{di}$ may be associated with the treatment variable $D_i$ even conditional on the covariates $X_i$. To account for this selection-induced dependence, following \citet{Bia} and \citet{Potanin2026}, we use a control function approach based on the structural selection mechanism specified in Assumption~\ref{assumption:selection}. Specifically, this assumption motivates the following vector of nuisance functions representing conditional marginal cumulative probabilities associated with the selection variables\footnote{We follow the convention that for two vectors $a=\left(a_{1},\dots,a_{n_{a}}\right)$ and $b=\left(b_{1},\dots,b_{n_{b}}\right)$ with $n_{a}$ and $n_{b}$ elements, respectively, $\left(a,b\right)=\left(a_{1},\dots,a_{n_{a}},b_{1},\dots,b_{n_{b}}\right).$}:
\begin{align}
    p_{Z}^{0}(d,x,w) &= \left(p^{0,z^{*} - \left(1,\dots,1\right)}\left(d,x,w\right), p^{0,z^{*}}\left(d,x,w\right)\right),
\end{align}
where for $j \in \{1,\dots,n_Z\}$ and $z\in\prod_{j=1}^{n_Z}\{-1,\dots,n_{Z_j}-1\}$, we use the following notation:
\begin{align}
    p^{0,z}\left(d,x,w\right) &= \left(p_{1}^{0}(z,d,x,w),\dots,p_{n_Z}^{0}(z,d,x,w)\right), \\
    p_{j}^{0}(z,d,x,w) &= \mathbb{P}(Z_{ji} \leq z_j \mid D_i = d, X_i = x, W_i^{(Z)} = w).
\end{align}
The two cumulative probabilities associated with each selection equation characterize the lower and upper boundaries of the corresponding selection interval in the probability space. Thus, the vector of control functions is $P_{i}^{0}=\left(P_{0i}^{0},P_{1i}^{0}\right)=p_Z^0(D_i,X_i,W_i^{(Z)})$, where $P_{0i}^{0} = p^{0,z^{*}-\left(1,\dots,1\right)}\left(D_{i},X_{i},W_{i}^{(Z)}\right)$ and $P_{1i}^{0} = p^{0,z^{*}}\left(D_{i},X_{i},W_{i}^{(Z)}\right)$. The following lemma establishes that conditioning on the covariates and control functions ensures the independence of the potential outcomes from the treatment and exclusion variables in both the selected and non-selected subpopulations.

\begin{lemma}\label{lemma:independence}
  Assumptions~\ref{assumption:selection} and~\ref{assumption:exogeneity} imply that
  $Y_{di} \perp \left(D_i, W_i^{(Z)}\right) \mid (X_i, P_i^0, \tilde{Z}_i)$ and $(Y_{di}, \tilde{Z}_i) \perp \left(D_i, W_i^{(Z)}\right) \mid (X_i, P_i^0)$ for every $d \in \{0, \dots, n_D - 1\}$.
\end{lemma}

This lemma is central to the identification of ATE and ATT in the selected subpopulation ($\tilde{Z}_i=1$). We denote these causal parameters by ATES and ATTS, respectively, where the additional letter `S' stands for `selected'. For example, ATES may represent the average treatment effect of education on wages among employed individuals. To identify these causal parameters, we impose the following additional assumption.

\begin{assumption}[Common support of treatment]
    \label{assumption:support}
    Selection is non-deterministic, i.e., $0<\mathbb{P}(\tilde{Z}_i=1)<1$. Moreover, for every $d\in\operatorname{supp}(D_i)$, we have
    \begin{align*}
        \mathbb{P}(D_i=d\mid X_i,P_i^0,\tilde{Z}_i=1)>0
    \end{align*}
    almost surely with respect to the conditional distribution of $(X_i,P_i^0)$ given $\tilde{Z}_i=1$.
\end{assumption}

Assumption~\ref{assumption:support} ensures that the conditional expectations required for identification can be recovered from the observed data. For example, this assumption may be violated if a researcher studies the effect of university education $D_i$ on health $Y_i$ without imposing a minimum age restriction on the sample. Indeed, 2-year-old individuals cannot have completed university education, so $\mathbb{P}(D_i=1 \mid \text{Age}_i=2,\tilde{Z}_i=1)=0$, provided that such individuals are present in the selected subpopulation, where $\text{Age}_i=X_{ti}$ denotes age. Reasonable restrictions on the composition of the sample, along with sufficient variation in the exclusion variables $W_i^{(Z)}$, may help ensure common support with respect to both the covariates $X_i$ and the control functions $P_i^0$.

The following theorem establishes identification results for ATES and ATTS.

\begin{theorem}[Identification of ATES and ATTS]\label{theorem:ates}
  Assumptions~\ref{assumption:selection},~\ref{assumption:exogeneity}, and~\ref{assumption:support} imply that, for any distinct treatment levels $d,d^*\in\operatorname{supp}(D_i)$, we have:
      \begin{align}
        \mathrm{ATES}
        &= \mathbb{E}\!\left(Y_{di} - Y_{d^*i} \mid \tilde{Z}_i = 1\right) = \mathbb{E}\!\left(\mu_{Y}^{0}\!\left(d, X_i, P_i^0\right) - \mu_{Y}^{0}\!\left(d^*, X_i, P_i^0\right) \mid \tilde{Z}_i = 1\right), \\
        \mathrm{ATTS}
        &= \mathbb{E}\!\left(Y_{di} - Y_{d^*i} \mid D_i = d, \tilde{Z}_i = 1\right) = \nonumber \\
        &= \mathbb{E}\!\left(\mu_{Y}^{0}\!\left(d, X_i, P_i^0\right) - \mu_{Y}^{0}\!\left(d^*, X_i, P_i^0\right) \mid D_i = d, \tilde{Z}_i = 1\right),
      \end{align}
  where the nuisance function is defined as follows:
    \begin{align}
        \mu_Y^0(d,x,p) &= \mathbb{E}(Y_i \mid D_i = d, X_i = x, P_i^0 = p, \tilde{Z}_i = 1).
    \end{align}
\end{theorem}

The key difference from the identification formula for ATES obtained by \citet{Bia} is that the conditional outcome expectations depend on probabilities associated with multiple selection equations. Moreover, ordinal selection with heterogeneous thresholds generally requires conditioning on two vectors of probabilities $P_i^0=(P_{0i}^0,P_{1i}^0)$ rather than one. However, $z_j^*=n_{Z_j}-1$ implies that the $j$-th element of $P_{1i}^0$, denoted by $P_{1ji}^0$, equals $1$. Similarly, $z_j^*=0$ implies that the $j$-th element of $P_{0i}^0$, denoted by $P_{0ji}^0$, equals $0$. Such deterministic elements may therefore be removed from $P_i^0$ without affecting the identification results.

Following \citet{Bia}, we impose additional assumptions to identify ATE and ATT in the full population rather than only in the selected subpopulation.

\begin{assumption}[Conditional effect homogeneity]
  \label{assumption:homogeneity}
  For every $d,d^* \in \{0,\dots,n_D-1\}$, we have:
  \begin{equation}\label{eq:homogeneity}
      \mathbb{E}(Y_{di}-Y_{d^*i} \mid X_i,P_i^0,\tilde{Z}_i=1)
      = \mathbb{E}(Y_{di}-Y_{d^*i} \mid X_i,P_i^0).
  \end{equation}
\end{assumption}

\begin{assumption}[Common support of selection]
    \label{assumption:selection-support}
    The conditional probability of selection is positive almost surely:
    \begin{align*}
        \mathbb{P}(\tilde{Z}_i=1\mid D_i,X_i,W_i^{(Z)})>0.
    \end{align*}
\end{assumption}

Assumption~\ref{assumption:selection-support} imposes a common support requirement for selection rather than treatment. For example, suppose that a researcher studies the effect of health on wages among individuals aged 16--54 in a country where employment of individuals under 18 is prohibited. In this case, the probability of employment (selection) is zero for individuals aged 16--17, violating Assumption~\ref{assumption:selection-support}. An appropriate age restriction on the sample may help avoid such violations.

Assumption~\ref{assumption:homogeneity} is implied by the model $Y_{di} = g(X_i,d) + \varepsilon_i$ \citep{Bia}, under which treatment effect heterogeneity is entirely determined by the observed covariates $X_i$ \citep{Frolich}, since the contribution of unobserved factors $\varepsilon_i$ to the potential outcome is the same for all $d \in \{0,\dots,n_D-1\}$. Indeed, under this model Assumption~\ref{assumption:homogeneity} holds, because:
\begin{align}
    \mathbb{E}(Y_{di}-Y_{d^*i} \mid X_i,P_i^0,\tilde{Z}_i=1) = g(X_i,d)-g(X_i,d^*) = \mathbb{E}(Y_{di}-Y_{d^*i} \mid X_i,P_i^0).
\end{align}

The following theorem establishes identification of ATE and ATT in the full population under these additional assumptions.

\begin{theorem}[Identification of ATE and ATT]\label{theorem:ate}
Assumptions~\ref{assumption:selection},~\ref{assumption:exogeneity},~\ref{assumption:support},~\ref{assumption:homogeneity}, and~\ref{assumption:selection-support} imply that, for any distinct treatment levels $d,d^*\in\operatorname{supp}(D_i)$, we have:
    \begin{align}
        \mathrm{ATE} &= \mathbb{E}(Y_{di}-Y_{d^*i}) = \mathbb{E}\left(\mu_Y^0(d,X_i,P_i^0)-\mu_Y^0(d^*,X_i,P_i^0)\right), \label{eq:ATE} \\
        \mathrm{ATT} &= \mathbb{E}(Y_{di}-Y_{d^*i}\mid D_i=d) = \mathbb{E}\left(\mu_Y^0(d,X_i,P_i^0)-\mu_Y^0(d^*,X_i,P_i^0)\mid D_i=d\right).
    \end{align}
\end{theorem}

Notably, the identification formulas for all four causal parameters involve the same conditional outcome expectations and differ only in the subpopulations over which their differences are averaged. In the following sections, we use the established identification results to derive Neyman-orthogonal scores and construct DML estimators of ATES, ATTS, ATE, and ATT.

\section{Neyman-orthogonal scores}\label{sec:neyman}

To construct the scores, we introduce the following additional nuisance functions:
\begin{align}
    p_{D}^{0}\left(d,x,p\right)
    &= \mathbb{P}\left(D_i=d\mid X_i=x,P_i^0=p\right), \\
    \tilde{p}_{Z}^{0}\left(d,x,w\right)
    &= \mathbb{P}\left(\tilde{Z}_i=1\mid D_i=d,X_i=x,W_i^{(Z)}=w\right), \\
    \mu_{\partial Y}^{0}(d,x,p)
    &= \frac{\partial\mu_Y^0(d,x,p)}{\partial p},
\end{align}
where $\mu_{\partial Y}^{0}(d,x,p)$ is well defined under Assumption~\ref{assumption:differentiability} below. Here and throughout the paper, gradients are treated as column vectors, whereas all other vector-valued variables and nuisance function values are treated as row vectors. The following lemma establishes several useful identities that are used throughout the proofs.

\begin{lemma}\label{lemma:useful}
  Assumptions~\ref{assumption:selection},~\ref{assumption:exogeneity}, and~\ref{assumption:support} imply that, for every $d\in\operatorname{supp}(D_i)$,
  \begin{align}
      \mu_Y^0\left(d,X_i,p_Z^0\left(d,X_i,W_i^{(Z)}\right)\right) &= \mathbb{E}\left(Y_i\mid D_i=d,X_i,W_i^{(Z)},\tilde{Z}_i=1\right), \label{eq:useful:muY} \\
      p_D^0\left(d,X_i,P_i^0\right) &= \mathbb{P}\left(D_i=d\mid X_i,P_i^0,\tilde{Z}_i=1\right), \label{eq:useful:pD} \\
      \tilde{p}_Z^0\left(D_i,X_i,W_i^{(Z)}\right) &= \mathbb{P}\left(\tilde{Z}_i=1\mid D_i,X_i,P_i^0\right) = \mathbb{P}\left(\tilde{Z}_i=1\mid X_i,P_i^0\right), \label{eq:useful:pZ}
  \end{align}
  where the equalities hold almost surely with respect to the conditional distribution of $(X_i,W_i^{(Z)})$ given $D_i=d$ and $\tilde{Z}_i=1$ in~(\ref{eq:useful:muY}), the conditional distribution of $(X_i,P_i^0)$ given $\tilde{Z}_i=1$ in~(\ref{eq:useful:pD}), and the unconditional distribution of $(D_i,X_i,W_i^{(Z)})$ in~(\ref{eq:useful:pZ}).
\end{lemma}

For brevity, define $\tilde{P}_{i}^{0} = \tilde{p}_{Z}^{0}\left(D_i,X_i,W_i^{(Z)}\right)$, $P_i^{0(d)} = p_D^0\left(d,X_i,P_i^0\right)$, $\text{E}_i^{0(d)} = \mu_Y^0\left(d,X_i,P_i^0\right)$, $D_i^{(d)} = \mathbb{I}\left(D_i=d\right)$, $\text{E}_{\partial i}^{0(d)} = \mu_{\partial Y}^0\left(d,X_i,P_i^0\right)$, and
\begin{align}
    \overline{Z}_i = \left(\mathbb{I}\left(Z_{1i}\leq z_1^*-1\right),\dots,\mathbb{I}\left(Z_{n_Z i}\leq z_{n_Z}^*-1\right),\mathbb{I}\left(Z_{1i}\leq z_1^*\right),\dots,\mathbb{I}\left(Z_{n_Z i}\leq z_{n_Z}^*\right)\right).
\end{align}
By construction, $\mathbb{E}\left(\overline{Z}_i\mid D_i,X_i,W_i^{(Z)}\right)=P_i^0$. Hereafter, we omit the mechanically deterministic boundary components of $P_i^0$ and the corresponding components of $\overline{Z}_i$.

Denote by $\mu = \left(\mu_Y,\mu_{\partial Y},p_D,p_Z,\tilde{p}_Z\right)$ a generic vector of nuisance functions belonging to the nuisance realization set $\mathcal{T}$ and by $\mu^0 = \left(\mu_Y^0,\mu_{\partial Y}^0,p_D^0,p_Z^0,\tilde{p}_Z^0\right)$ the vector of true nuisance functions. For any $\mu\in\mathcal{T}$, define $P_i=p_Z\left(D_i,X_i,W_i^{(Z)}\right)$, $\tilde{P}_i=\tilde{p}_Z\left(D_i,X_i,W_i^{(Z)}\right)$, $P_i^{(d)}=p_D\left(d,X_i,P_i\right)$, $\text{E}_i^{(d)}=\mu_Y\left(d,X_i,P_i\right)$, and $\text{E}_{\partial i}^{(d)}=\mu_{\partial Y}\left(d,X_i,P_i\right)$. Moreover, define the vector of observed variables $O_i=\left(\tilde{Z}_iY_i,D_i,X_i,W_i^{(Z)},Z_i\right)$. Finally, assume that $\mu^0\in\mathcal{T}$ and that the nuisance functions representing probabilities take values in $[0,1]$.

To establish Neyman orthogonality of the scores, we impose the following additional regularity assumption:

\begin{assumption}[Differentiability and regularity]
  \label{assumption:differentiability}
  For every $\mu\in\mathcal{T}$, the functions $\mu_Y(d,x,p)$ and $p_D(d,x,p)$ are, respectively, twice and once continuously differentiable with respect to $p$ on an open neighborhood of $[0,1]^{\dim(P_i^0)}$, and
  \begin{align}
      \mu_{\partial Y}(d,x,p) &= \frac{\partial\mu_Y(d,x,p)}{\partial p}.
  \end{align}
  For almost every $x$ with respect to the distribution of $X_i$ conditional on $\tilde{Z}_i=1$, the support of $P_i^0$ conditional on $X_i=x$ and $\tilde{Z}_i=1$ is the closure of its interior in $\mathbb{R}^{\dim(P_i^0)}$. Moreover, along each perturbation $\mu^\epsilon=\mu^0+\epsilon(\mu-\mu^0)$, where $\mu\in\mathcal{T}$ and $\epsilon\in[0,1)$, the additive terms of each score are, under the respective assumptions of Lemma~\ref{lemma:neyman}, well defined, integrable, and almost surely continuously differentiable with respect to $\epsilon$. Each term arising from their differentiation by the chain rule is dominated by an integrable random variable uniformly in $\epsilon$.
\end{assumption}

We propose the following Neyman-orthogonal scores\footnote{Throughout, a ratio with a zero denominator is defined to be zero if its numerator is zero. Products involving ratios with zero denominators are also defined to be zero whenever a multiplicative factor outside the ratio is zero. Terms involving $Y_i$ are evaluated only for selected observations.}.

\begin{lemma}[Neyman-orthogonal scores]\label{lemma:neyman}
Under Assumptions~\ref{assumption:selection},~\ref{assumption:exogeneity},~\ref{assumption:support}, and~\ref{assumption:differentiability}, the following scores have zero expectation at the true values of the corresponding causal parameters and nuisance functions and are Neyman-orthogonal with respect to the nuisance realization set $\mathcal{T}$:
    \begin{align}
        \psi_\text{ATES}\left(O_i,\theta,\mu\right) &= \tilde{Z}_i\left(\text{E}_i^{(d)}-\text{E}_i^{(d^*)}\right)-\tilde{Z}_i\theta + \nonumber \\
        &+ \tilde{Z}_i\left(\frac{D_i^{(d)}}{P_i^{(d)}}\left(Y_i-\text{E}_i^{(d)}\right)-\frac{D_i^{(d^*)}}{P_i^{(d^*)}}\left(Y_i-\text{E}_i^{(d^*)}\right)\right) + \nonumber \\
        &+ \left(\overline{Z}_i-P_i\right)\tilde{P}_i\left(\text{E}_{\partial i}^{(d)}\left(1-\frac{D_i^{(d)}}{P_i^{(d)}}\right)-\text{E}_{\partial i}^{(d^*)}\left(1-\frac{D_i^{(d^*)}}{P_i^{(d^*)}}\right)\right), \label{eq:score_ATES} \\
        \psi_\text{ATTS}\left(O_i,\theta,\mu\right) &= D_i^{(d)}\tilde{Z}_i\left(Y_i-\text{E}_i^{(d^*)}\right)-D_i^{(d)}\tilde{Z}_i\theta + \nonumber \\
        &-D_i^{(d^*)}\tilde{Z}_i\frac{P_i^{(d)}}{P_i^{(d^*)}}\left(Y_i-\text{E}_i^{(d^*)}\right) + \nonumber \\
        &+\left(\overline{Z}_i-P_i\right)\tilde{P}_i\text{E}_{\partial i}^{(d^*)}\left(\frac{P_i^{(d)}}{P_i^{(d^*)}}D_i^{(d^*)}-D_i^{(d)}\right). \label{eq:score_ATTS}
    \end{align}
If, in addition, Assumptions~\ref{assumption:homogeneity} and~\ref{assumption:selection-support} hold, then the same properties hold for the following scores:
    \begin{align}
        \psi_\text{ATE}\left(O_i,\theta,\mu\right) &= \left(\text{E}_i^{(d)}-\text{E}_i^{(d^*)}\right)-\theta + \nonumber \\
        &+\frac{\tilde{Z}_i}{\tilde{P}_i}\left(\frac{D_i^{(d)}}{P_i^{(d)}}\left(Y_i-\text{E}_i^{(d)}\right)-\frac{D_i^{(d^*)}}{P_i^{(d^*)}}\left(Y_i-\text{E}_i^{(d^*)}\right)\right) + \nonumber \\
        &+\left(\overline{Z}_i-P_i\right)\left(\text{E}_{\partial i}^{(d)}\left(1-\frac{D_i^{(d)}}{P_i^{(d)}}\right)-\text{E}_{\partial i}^{(d^*)}\left(1-\frac{D_i^{(d^*)}}{P_i^{(d^*)}}\right)\right), \label{eq:score_ATE} \\
        \psi_\text{ATT}\left(O_i,\theta,\mu\right) &= D_i^{(d)}\left(\text{E}_i^{(d)}-\text{E}_i^{(d^*)}\right)-D_i^{(d)}\theta + \nonumber \\
        &+\frac{\tilde{Z}_i}{\tilde{P}_i}D_i^{(d)}\left(Y_i-\text{E}_i^{(d)}\right)-\frac{\tilde{Z}_i}{\tilde{P}_i}\frac{P_i^{(d)}}{P_i^{(d^*)}}D_i^{(d^*)}\left(Y_i-\text{E}_i^{(d^*)}\right) + \nonumber \\
        &+\left(\overline{Z}_i-P_i\right)\text{E}_{\partial i}^{(d^*)}\left(\frac{P_i^{(d)}}{P_i^{(d^*)}}D_i^{(d^*)}-D_i^{(d)}\right). \label{eq:score_ATT}
    \end{align}
\end{lemma}

The proposed scores are similar in form to those used by \citet{Chernozhukov} for ATE and ATT and by \citet{Bia} for ATE and ATES. However, our scores contain additional terms involving the derivatives $\text{E}_{\partial i}^{(d)}$ and $\text{E}_{\partial i}^{(d^{*})}$ that ensure Neyman orthogonality with respect to the nuisance functions generating the control probabilities $P_i^0$. These terms account for the first-order effects of estimating generated regressors entering other nuisance functions, an issue extensively studied in the semiparametric literature \citep{Hahn2013,Mammen2016,Chernozhukov2022}. In Section~\ref{sec:sim}, we show that including these terms may improve the finite-sample accuracy of the proposed DML estimators.

To illustrate the importance of the derivative terms, consider the score for ATE without the derivative correction terms, denoted by $\psi_{\mathrm{ATE}}^{*}$. Perturbing $p_Z^0$ according to $p_Z^\epsilon(Q_i)=p_Z^0(Q_i)+\epsilon h(Q_i)$, while keeping the other nuisance functions fixed, yields:
\begin{align}
    \left.\frac{\partial}{\partial\epsilon}\mathbb{E}\left[\psi_{\mathrm{ATE}}^{*}(O_i,\theta^0,\mu^\epsilon)\right]\right|_{\epsilon=0}
    &=\mathbb{E}\left[h(Q_i)\left(\text{E}_{\partial i}^{0(d)}\left(1-\frac{D_i^{(d)}}{P_i^{0(d)}}\right)
    -\text{E}_{\partial i}^{0(d^*)}\left(1-\frac{D_i^{(d^*)}}{P_i^{0(d^*)}}\right)\right)\right].
\end{align}
This expression equals zero if $h(Q_i)$ is a measurable function of $(X_i,P_i^0)$ but need not equal zero for unrestricted perturbations. Thus, the cancellation of the corresponding terms in the proof of Neyman orthogonality in \citet{Bia} (see Appendix B.3.1) is not generally guaranteed. The proposed derivative correction terms ensure Neyman orthogonality without imposing such additional restrictions on the perturbations of the generated control functions.

\section{Double machine learning estimators}\label{sec:dml}

Below, we construct DML estimators based on the scores from Lemma~\ref{lemma:neyman}.

\begin{enumerate}
    \item For a fixed integer $K\geq2$, independently of the observed values of the variables, partition the sample into $K$ (approximately) equal-sized, non-overlapping parts $A_k$, where $k\in\{1,\dots,K\}$. Denote by $\overline{A}_k$ the set of observations not in $A_k$. Following \citet{Bia} and \citet{Potanin2026}, to avoid evaluating estimated control functions on the observations used to estimate them when constructing other nuisance estimators, further partition $\overline{A}_k$ into two (approximately) equal-sized, non-overlapping parts $\overline{A}_{1k}$ and $\overline{A}_{2k}$ independently of the observed values of the variables.

    \item For each $k\in\{1,\dots,K\}$, obtain an estimator $\hat{\tilde{p}}_Z^{(k)}(\cdot)$ by regressing $\tilde{Z}_i$ on $(D_i,X_i,W_i^{(Z)})$ using observations from $\overline{A}_k$.

    \item For each $k\in\{1,\dots,K\}$ and $t\in\{1,2\}$, obtain an estimator $\hat{p}_Z^{(k,t)}(\cdot)$ by regressing $\mathbb{I}(Z_{ji}\leq z_j^*-1)$ and $\mathbb{I}(Z_{ji}\leq z_j^*)$ on $(D_i,X_i,W_i^{(Z)})$ for every $j\in\{1,\dots,n_Z\}$ using observations from $\overline{A}_{tk}$. The estimated probabilities are arranged in the same order as the corresponding indicators in $\overline{Z}_i$. Deterministic components of the control vector need not be estimated.

    \item For each $k\in\{1,\dots,K\}$ and $t\in\{1,2\}$, obtain an estimator $\hat{\mu}_Y^{(k,t)}(\cdot)$ by regressing $Y_i$ on $\left(D_i,X_i,\hat{p}_Z^{(k,3-t)}(D_i,X_i,W_i^{(Z)})\right)$ using observations from $\overline{A}_{tk}$ for which $\tilde{Z}_i=1$.

    \item For each $k\in\{1,\dots,K\}$ and $t\in\{1,2\}$, obtain an estimator $\hat{p}_D^{(k,t)}(\overline{d},\cdot)$ for $\overline{d}\in\{d,d^*\}$ by regressing the indicators $\mathbb{I}(D_i=\overline{d})$ on $\left(X_i,\hat{p}_Z^{(k,3-t)}(D_i,X_i,W_i^{(Z)})\right)$ using observations from $\overline{A}_{tk}$.

    \item For each $k\in\{1,\dots,K\}$, $t\in\{1,2\}$, and $\overline{d}\in\{d,d^*\}$, obtain an estimator $\hat{\mu}_{\partial Y}^{(k,t)}(\overline{d},x,p)$ by numerical or analytical differentiation of $\hat{\mu}_Y^{(k,t)}(\overline{d},x,p)$ with respect to $p$.

       \item Denote by $q_i$ the index of the fold containing the $i$-th observation, i.e., $q_i=k$ if and only if $i\in A_k$. For each $\overline{d}\in\{d,d^*\}$ and $t\in\{1,2\}$, calculate the following variables:
    \begin{align}
        \hat{E}_{it}^{(\overline{d})}
        &= \hat{\mu}_Y^{(q_i,t)}\left(\overline{d},X_i,\hat{p}_Z^{(q_i,3-t)}(D_i,X_i,W_i^{(Z)})\right), \\
        \hat{P}_{it}^{(\overline{d})}
        &= \hat{p}_D^{(q_i,t)}\left(\overline{d},X_i,\hat{p}_Z^{(q_i,3-t)}(D_i,X_i,W_i^{(Z)})\right), \\
        \hat{E}_{\partial it}^{(\overline{d})}
        &= \hat{\mu}_{\partial Y}^{(q_i,t)}\left(\overline{d},X_i,\hat{p}_Z^{(q_i,3-t)}(D_i,X_i,W_i^{(Z)})\right), \\
        \hat{P}_{it}
        &= \hat{p}_Z^{(q_i,3-t)}\left(D_i,X_i,W_i^{(Z)}\right), \\
        \hat{\tilde{P}}_i
        &= \hat{\tilde{p}}_Z^{(q_i)}\left(D_i,X_i,W_i^{(Z)}\right).
    \end{align}
    Denote the corresponding vector of nuisance function estimators by
    \begin{align}
        \hat{\mu}^{(k,t)} = \left(\hat{\mu}_Y^{(k,t)},\hat{\mu}_{\partial Y}^{(k,t)},\hat{p}_D^{(k,t)},\hat{p}_Z^{(k,3-t)},\hat{\tilde{p}}_Z^{(k)}\right).
    \end{align}

    \item For every $\theta^0\in\{\mathrm{ATES},\mathrm{ATTS},\mathrm{ATE},\mathrm{ATT}\}$, evaluate the corresponding score separately for $t=1$ and $t=2$ at $\theta=0$ and average the resulting values:
    \begin{align}
        \hat{r}_{it}^{\theta^0} &= \psi_{\theta^0}\left(O_i,0,\hat{\mu}^{(q_i,t)}\right), \\
        \hat{r}_i^{\theta^0} &= 0.5\hat{r}_{i1}^{\theta^0}+0.5\hat{r}_{i2}^{\theta^0}.
    \end{align}
    Since the scores are affine in the causal parameters and their derivatives with respect to these parameters do not depend on the nuisance functions, averaging the scores is numerically equivalent to obtaining two separate estimates of each causal parameter using $t=1$ and $t=2$ and then averaging these estimates.

    \item Obtain estimates of the causal parameters:
    \begin{align}
        \widehat{\mathrm{ATES}} &= \frac{1}{\sum\limits_{i=1}^n\tilde{Z}_i}\sum\limits_{i=1}^n\hat{r}_i^{\mathrm{ATES}}, \label{eq:DML_ATES}\\
        \widehat{\mathrm{ATTS}} &= \frac{1}{\sum\limits_{i=1}^n\tilde{Z}_iD_i^{(d)}}\sum\limits_{i=1}^n\hat{r}_i^{\mathrm{ATTS}}, \label{eq:DML_ATTS}\\
        \widehat{\mathrm{ATE}} &= \frac{1}{n}\sum\limits_{i=1}^n\hat{r}_i^{\mathrm{ATE}}, \label{eq:DML_ATE}\\
        \widehat{\mathrm{ATT}} &= \frac{1}{\sum\limits_{i=1}^nD_i^{(d)}}\sum\limits_{i=1}^n\hat{r}_i^{\mathrm{ATT}}. \label{eq:DML_ATT}
    \end{align}

   \item Estimate the variances of the proposed estimators using their asymptotic linear representations as follows:
    \begin{align}
        \widehat{\operatorname{As.Var}}\left(\widehat{\mathrm{ATES}}\right) &= \frac{1}{\left(\sum\limits_{i=1}^n\tilde{Z}_i\right)^2}\sum\limits_{i=1}^n\left(\hat{r}_i^{\mathrm{ATES}}-\tilde{Z}_i\widehat{\mathrm{ATES}}\right)^2, \label{eq:var_ates}\\
        \widehat{\operatorname{As.Var}}\left(\widehat{\mathrm{ATTS}}\right) &= \frac{1}{\left(\sum\limits_{i=1}^nD_i^{(d)}\tilde{Z}_i\right)^2}\sum\limits_{i=1}^n\left(\hat{r}_i^{\mathrm{ATTS}}-D_i^{(d)}\tilde{Z}_i\widehat{\mathrm{ATTS}}\right)^2, \label{eq:var_atts}\\
        \widehat{\operatorname{As.Var}}\left(\widehat{\mathrm{ATE}}\right) &= \frac{1}{n^2}\sum\limits_{i=1}^n\left(\hat{r}_i^{\mathrm{ATE}}-\widehat{\mathrm{ATE}}\right)^2, \label{eq:var_ate}\\
        \widehat{\operatorname{As.Var}}\left(\widehat{\mathrm{ATT}}\right) &= \frac{1}{\left(\sum\limits_{i=1}^nD_i^{(d)}\right)^2}\sum\limits_{i=1}^n\left(\hat{r}_i^{\mathrm{ATT}}-D_i^{(d)}\widehat{\mathrm{ATT}}\right)^2. \label{eq:var_att}
    \end{align}
\end{enumerate}

We use nested cross-fitting to account for the presence of generated regressors $P_i^0$. Although conventional cross-fitting is sufficient once the required nuisance convergence conditions are satisfied, nested cross-fitting may facilitate the construction of nuisance estimators satisfying these conditions. Despite its higher computational cost, theoretical arguments favoring the estimation of different nuisance functions on separate subsamples \citep{Newey2018} and simulation evidence \citep{Potanin2026} suggest that nested cross-fitting may substantially improve estimation accuracy. Moreover, the equal weights of $0.5$ assigned to the estimates obtained with $t=1$ and $t=2$ may be replaced with non-negative, possibly data-dependent weights summing to one and reflecting the out-of-sample predictive performance of the corresponding nuisance estimators. Since both estimators have the same asymptotic linear representation under the regularity conditions imposed below, such weighting does not affect their first-order asymptotic properties.

Typically, researchers use $K=5$ or $K=10$ folds for cross-fitting. Moreover, to reduce the sensitivity of the estimates to a particular sample split, the estimation procedure may be repeated $5$--$10$ times using different splits. Researchers may then report the mean or median estimate, with the corresponding variance estimated using the formulas in Definition 3.3 of \citet{Chernozhukov}.

The accuracy of DML estimates depends on the quality of the nuisance function estimators. A popular approach to improving their quality is hyperparameter tuning using cross-validation or other methods. To preserve the independence between the nuisance estimators and the observations in the corresponding test fold, hyperparameter tuning should be performed separately within each training fold. However, simulation evidence in \citet{Bach2024} suggests that tuning on the entire sample using independently generated splits for cross-validation and cross-fitting may perform similarly to fold-specific tuning, while substantially reducing the computational cost.

\begin{sloppypar}
It is natural to require the estimates $\hat{p}^{u}_{jit}$ and $\hat{p}^{l}_{jit}$ of $\mathbb{P}\left(Z_{ji}\leq z^{*}_{j}\mid D_i,X_i,W_i^{(Z)}\right)$ and $\mathbb{P}\left(Z_{ji}\leq z^{*}_{j}-1\mid D_i,X_i,W_i^{(Z)}\right)$, respectively, to satisfy the ordering $\hat{p}^{u}_{jit}\geq\hat{p}^{l}_{jit}$ for every $t\in\{1,2\}$. Some estimators, including the ordinal probit model, preserve this ordering by construction, while others, such as random forests fitted separately to the corresponding indicators, do not necessarily preserve it. In the latter case, one may apply an ordering-preserving projection by replacing both $\hat{p}^{l}_{jit}$ and $\hat{p}^{u}_{jit}$ with their arithmetic mean whenever $\hat{p}^{u}_{jit}<\hat{p}^{l}_{jit}$. This projection cannot increase the squared Euclidean distance between the estimated and true probabilities. Alternatively, when $n_Z$ and the number of categories of the selection variables are small, the control probabilities and the joint selection probability may be obtained from the same estimator of the conditional joint distribution of $Z_i$, thereby ensuring their mutual consistency by construction.
\end{sloppypar}

To reduce the variance of the proposed DML estimators, potentially at the cost of introducing bias, one may apply trimming by excluding observations with extremely small (e.g., below $0.01$ or $0.05$) estimated probabilities or their products, such as $\hat{\tilde{P}}_i\hat{P}_{it}^{(d)}$, appearing in the denominators of the corresponding score for either $t=1$ or $t=2$. However, trimming with a fixed threshold may affect the probability limit of the estimator.

\section{Asymptotic properties}\label{sec:asymptotic}

In this section, we establish consistency and asymptotic normality of the proposed estimators under additional regularity conditions. For brevity, we consider the case in which $0<z_j^*<n_{Z_j}-1$ for every $j\in\{1,\dots,n_Z\}$, so that none of the components of $P_i^0$ is mechanically equal to zero or one\footnote{Deterministic boundary components can be accommodated by omitting the corresponding components of $P_i^0$ and $\overline{Z}_i$.}. Moreover, we assume that analytical differentiation\footnote{The conclusions of Theorem~\ref{theorem:normality} also hold when numerical differentiation is used, provided that the differences between the numerical and analytical derivatives are $o_P(n^{-1/4})$ in the $L^2$ norm evaluated at the estimated control variables.} is used to obtain $\hat{\mu}_{\partial Y}^{(k,t)}(\overline{d},\cdot)$ from $\hat{\mu}_Y^{(k,t)}(\overline{d},\cdot)$.

\begin{assumption}[Regularity conditions]
\label{assumption:regularity}
Consider an observation indexed by $i$ from the $k$-th test fold $A_k$ and the estimators of the nuisance functions obtained using observations from $\overline{A}_k$. Let $\mathcal{T}_n\subseteq\mathcal{T}$ be a nuisance realization set whose elements satisfy Parts 3--6 below. There exist sequences of positive constants $\delta_n$ and $\Delta_n$ converging to zero as $n\to\infty$ and constants $q>4$, $C>0$, and $c\in(0,1)$ such that $\mu^0\in\mathcal{T}_n$ and, for every $k\in\{1,\dots,K\}$ and $t\in\{1,2\}$, the nuisance vector $\hat{\mu}^{(k,t)}=\left(\hat{\mu}_Y^{(k,t)},\hat{\mu}_{\partial Y}^{(k,t)},\hat{p}_D^{(k,t)},\hat{p}_Z^{(k,3-t)},\hat{\tilde{p}}_Z^{(k)}\right)$ belongs to $\mathcal{T}_n$ with probability at least $1-\Delta_n$. Moreover, the following conditions hold.\\
    \textbf{Part 1}. The following moment bound holds:
    \begin{align}
        \mathbb{E}\left(\left\|Y_i\right\|^q\right)^{1/q}\leq C.
    \end{align}
    \textbf{Part 2}. For every $\theta^0\in\{\mathrm{ATES},\mathrm{ATTS},\mathrm{ATE},\mathrm{ATT}\}$, the following lower bounds hold:
    \begin{align}
        \mathbb{P}\left(\tilde{Z}_i=1\right)\geq c&,\qquad \mathbb{P}\left(D_i=d\right)\geq c,\qquad \mathbb{P}\left(D_i=d^*\right)\geq c, \nonumber\\
        \mathbb{P}\left(\tilde{Z}_i=1,D_i=d\right)\geq c&,\qquad \mathbb{P}\left(\tilde{Z}_i=1,D_i=d^*\right)\geq c,\qquad \mathbb{E}\left(\psi_{\theta^0}\left(O_i,\theta^0,\mu^0\right)^2\right)\geq c.
    \end{align}
    \textbf{Part 3}. The following moment bound on the estimation errors holds:
    \begin{align}
        \mathbb{E}\left(\left\|\text{E}_i^{(d)}-\text{E}_i^{0(d)}\right\|^q\right)^{1/q}+\mathbb{E}\left(\left\|\text{E}_i^{(d^*)}-\text{E}_i^{0(d^*)}\right\|^q\right)^{1/q}\leq C.
    \end{align}
    \textbf{Part 4}. The following bounds on the probabilities hold almost surely:
    \begin{align}
        c\leq P_i^{(d)}\leq1,\qquad c\leq P_i^{(d^*)}\leq1,\qquad c\leq\tilde{P}_i\leq1,\qquad P_i\in[0,1]^{2n_Z}.
    \end{align}
    \textbf{Part 5}. The following bounds on the estimation errors hold:
    \begin{align}
        \mathbb{E}\left(\left\|\text{E}_i^{(d)}-\text{E}_i^{0(d)}\right\|^2\right)^{1/2}+\mathbb{E}\left(\left\|\text{E}_i^{(d^*)}-\text{E}_i^{0(d^*)}\right\|^2\right)^{1/2}&\leq\delta_nn^{-1/4},\\
        \mathbb{E}\left(\left\|\text{E}_{\partial i}^{(d)}-\text{E}_{\partial i}^{0(d)}\right\|^2\right)^{1/2}+\mathbb{E}\left(\left\|\text{E}_{\partial i}^{(d^*)}-\text{E}_{\partial i}^{0(d^*)}\right\|^2\right)^{1/2}&\leq\delta_n,\label{eq:derivative_consistency}\\
        \mathbb{E}\left(\left\|P_i^{(d)}-P_i^{0(d)}\right\|^2\right)^{1/2}+\mathbb{E}\left(\left\|P_i^{(d^*)}-P_i^{0(d^*)}\right\|^2\right)^{1/2}&\leq\delta_nn^{-1/4},\\
        \mathbb{E}\left(\left\|P_i-P_i^0\right\|^2\right)^{1/2}+\mathbb{E}\left(\left\|\tilde{P}_i-\tilde{P}_i^0\right\|^2\right)^{1/2}&\leq\delta_nn^{-1/4}.\label{eq:rates_selection}
    \end{align}
    \textbf{Part 6}. For every $\overline{d}\in\{d,d^*\}$, $j\in\{1,2,3\}$, and $t\in\{1,2\}$, the following bounds hold almost surely, uniformly in $p\in[0,1]^{2n_Z}$:
    \begin{align}
        c\leq p_D\left(\overline{d},X_i,p\right)\leq1,\qquad \left\|\frac{\partial^j\mu_Y\left(\overline{d},X_i,p\right)}{\partial p^j}\right\|_\infty\leq C,\qquad \left\|\frac{\partial^t p_D\left(\overline{d},X_i,p\right)}{\partial p^t}\right\|_\infty\leq C,
    \end{align}
    where the derivatives exist and are continuous on an open neighborhood of $[0,1]^{2n_Z}$. Moreover, $\mu_{\partial Y}(\overline{d},x,p)=\partial\mu_Y(\overline{d},x,p)/\partial p$.
\end{assumption}

Part 1 of Assumption~\ref{assumption:regularity} requires that the $q$-th absolute moment of the outcome $Y_i$ exists and is bounded for some $q>4$. This condition is satisfied whenever $Y_i$ has bounded support, as is the case for variables measured on a bounded scale, such as the share of income spent on food or a test score, and for categorical outcomes, such as marital status. However, sufficiently heavy-tailed outcome distributions may violate this condition. Transforming the outcome or winsorising extreme values may restore it, although such modifications may change the estimand and its interpretation.

Part 2 of Assumption~\ref{assumption:regularity} ensures that the score is non-degenerate, $\mathbb{E}(\psi_{\theta^0}^2)\geq c$, so that the asymptotic variance is bounded away from zero. It also requires that both treatment groups, $D_i=d$ and $D_i=d^*$, be sufficiently represented in both the full and selected populations, with the corresponding marginal and joint probabilities bounded below by $c$.

Part 3 of Assumption~\ref{assumption:regularity} requires uniformly bounded $q$-th moments of the estimation errors of the conditional outcome expectations. Given Part 1, this condition is satisfied when the estimator produces predictions bounded by a fixed constant, as is the case for probability models, regression trees and random forests when $Y_i$ is bounded, or estimators whose predictions are clipped to a fixed range. It is also satisfied for spline and polynomial estimators with a fixed number of basis functions, uniformly bounded coefficients, and basis functions with uniformly bounded $q$-th moments.

Part 4 of Assumption~\ref{assumption:regularity} requires that the estimated probabilities lie in $[0,1]$ and that those appearing in the denominators of the scores be bounded away from zero. Some estimators, such as the linear probability model, may produce estimates outside the required range. A common remedy is to clip the estimated probabilities to an interval such as $[\epsilon,1]$ for a small $\epsilon>0$. However, hard clipping of the conditional treatment probabilities may introduce kinks and thereby violate the differentiability requirement of Part 6 of Assumption~\ref{assumption:regularity}. A smooth alternative for estimators taking values in $[0,1]$ is to apply a smooth transformation that maps $[0,1]$ into $[\epsilon,1]$ and coincides with the identity function on $[2\epsilon,1]$, where $2\epsilon<c$. The chosen estimators of the conditional treatment probabilities must therefore satisfy both the range and smoothness restrictions.

Part 5 of Assumption~\ref{assumption:regularity} requires consistency of the derivative estimators $\hat{E}_{\partial i}^{(d)}$ and $\hat{E}_{\partial i}^{(d^*)}$ and $o(n^{-1/4})$ convergence rates of the estimators of the other nuisance functions in the $L^2$ norm. Such rates are attainable for many popular regression estimators under suitable conditions, including cross-validated LASSO under sparsity \citep{Chetverikov} and certain neural network estimators under appropriate restrictions on the structure and smoothness of the regression function \citep{Kohler}. In our setting, the estimation errors in the generated control functions must also be accounted for. Consistent estimation of the derivatives generally requires additional regularity conditions.

Part 6 rules out standard random forest and tree-based gradient boosting estimators of $\text{E}_i^{(d)}$, $\text{E}_i^{(d^*)}$, $P_i^{(d)}$, and $P_i^{(d^*)}$ when their predictions are discontinuous in $p$. However, it does not preclude tree-based estimation of the selection probabilities, since their derivatives do not enter the scores. Smooth tree-based alternatives, such as boosting with soft splits \citep{Giordani}, may satisfy the differentiability requirements of Part 6.

The following theorem establishes the key asymptotic properties of the proposed estimators.

\begin{theorem}[Consistency and asymptotic normality]\label{theorem:normality}
Assumptions~\ref{assumption:selection},~\ref{assumption:exogeneity},~\ref{assumption:support},~\ref{assumption:differentiability}, and~\ref{assumption:regularity} imply that the DML estimators $\widehat{\mathrm{ATES}}$ and $\widehat{\mathrm{ATTS}}$ are consistent and asymptotically normal. If, in addition, Assumptions~\ref{assumption:homogeneity} and~\ref{assumption:selection-support} hold, then the DML estimators $\widehat{\mathrm{ATE}}$ and $\widehat{\mathrm{ATT}}$ are also consistent and asymptotically normal. In each case, the corresponding estimator of the asymptotic variance provided in equations~(\ref{eq:var_ates})--(\ref{eq:var_att}) satisfies
\begin{align}
    \frac{n\widehat{\operatorname{As.Var}}(\hat{\theta})}{\sigma_{\theta^0}^{2}}\xrightarrow{p}1,
    \qquad
    \frac{\hat{\theta}-\theta^0}{\sqrt{\widehat{\operatorname{As.Var}}(\hat{\theta})}}\xrightarrow{d}\mathcal{N}(0,1),
\end{align}
where $\sigma_{\theta^0}^{2}$ is the asymptotic variance of $\sqrt{n}(\hat{\theta}-\theta^0)$.
\end{theorem}

Part 6 of Assumption~\ref{assumption:regularity} rules out many popular non-smooth machine learning estimators. A natural way to relax this restriction is to consider DML estimators $\widehat{\mathrm{ATES}}^{*}$, $\widehat{\mathrm{ATTS}}^{*}$, $\widehat{\mathrm{ATE}}^{*}$, and $\widehat{\mathrm{ATT}}^{*}$ obtained by omitting the derivative correction terms from the corresponding scores, i.e., by setting $\hat{E}_{\partial it}^{(d)}=\hat{E}_{\partial it}^{(d^*)}=0$. However, the resulting scores are not generally Neyman-orthogonal, so Theorem 3.1 of \citet{Chernozhukov} is not directly applicable. Following \citet[Lemma B.1]{Fan2026}, we consider a modification of this theorem that replaces the Neyman-orthogonality and differentiability requirements with a direct bound on the expected score. For completeness, we formulate this result in the original notation of \citet{Chernozhukov}, retaining all the conclusions of their theorem and providing an explicit bound on the remainder.

\begin{theorem}[Asymptotic normality without requiring Neyman orthogonality; cf. Lemma B.1 of \citealp{Fan2026}]\label{theorem:modified}
    Suppose that Assumptions 3.1 and 3.2 of \citet{Chernozhukov} hold with two modifications. First, drop the bound on $\lambda_{N}'$ from condition (c) of Assumption 3.2. Second, replace conditions (c) and (d) of Assumption 3.1 with the following bound, which holds for every $P\in\mathcal{P}_{N}$:
    \begin{align}\label{eq:NO_alternative}
        \lambda_{N}'':=\sup_{\eta\in \mathcal{T}_{N}}\left\|\mathbb{E}_{P}\left[\psi\left(W;\theta_{0},\eta\right)\right]\right\|\leq \delta_{N}/\sqrt{N},
    \end{align}
    where the notation is as in \citet{Chernozhukov}: $N$ is the total number of observations, $\delta_{N}$ is a deterministic sequence converging to zero such that $\delta_{N}\geq 1/\sqrt{N}$, $\eta$ is a vector of nuisance functions from the nuisance realization set $\mathcal{T}_{N}$ containing the vector of true nuisance functions $\eta_{0}$, $W$ is an observation, and $\psi\left(\cdot\right)$ is a score\footnote{In our notation, $N=n$, $\eta=\mu$, $\eta_0=\mu^0$, and $W=O_i$.}. Then all the conclusions of Theorem 3.1 of \citet{Chernozhukov} hold with the size of the remainder term given by:
    \begin{align}
        \rho_{N} = 1/\sqrt{N} + r_{N} + r_{N}' + \sqrt{N}\lambda_{N}''\lesssim \delta_{N},
    \end{align}
    where $r_{N}$ and $r_{N}'$ are specified in condition (c) of Assumption 3.2.
\end{theorem}

By replacing the Neyman-orthogonality and differentiability requirements with a direct bound on the expected score, Theorem~\ref{theorem:modified} allows us to dispense with the smoothness requirements imposed on nuisance functions with respect to generated regressors. In our setting, this makes it possible to establish asymptotic normality of the derivative-free estimators under alternative restrictions on the nuisance estimation errors, without requiring differentiability of the conditional outcome expectations or treatment probabilities with respect to the generated control variables.

\begin{assumption}[Alternative regularity conditions]
\label{assumption:regularity2}
For every $\theta^0\in\{\mathrm{ATES},\mathrm{ATTS},\mathrm{ATE},\mathrm{ATT}\}$, denote by $\psi_{\theta^0}^{*}(\cdot)$ the corresponding score from Lemma~\ref{lemma:neyman} with the derivative correction terms omitted, i.e., with $\text{E}_{\partial i}^{(d)}$ and $\text{E}_{\partial i}^{(d^*)}$ replaced by zero. Suppose that Parts 1--5 of Assumption~\ref{assumption:regularity} hold, except that equation~(\ref{eq:derivative_consistency}) in Part 5 is omitted and the bound on the score in Part 2 is replaced by
\begin{align}
    \mathbb{E}\left(\psi_{\theta^0}^{*}\left(O_i,\theta^0,\mu^0\right)^2\right)\geq c.
\end{align}
The nuisance vectors $\mu$ and $\mu^0$ and their estimators are defined without the derivative nuisance function, and every element of the nuisance realization set $\mathcal{T}_n$ satisfies Parts 3--5 of Assumption~\ref{assumption:regularity}, subject to the above modifications. For $\overline{d}\in\{d,d^*\}$, define
\begin{align}
    \Delta\text{E}_i^{(\overline{d})}=\text{E}_i^{(\overline{d})}-\text{E}_i^{0(\overline{d})},\qquad \Delta P_i^{(\overline{d})}=1-\frac{D_i^{(\overline{d})}}{P_i^{(\overline{d})}}.
\end{align}
Moreover, suppose that the bound corresponding to the causal parameter of interest holds:
\begin{align}
&\underbrace{\sup_{\mu\in\mathcal{T}_n}\left|\mathbb{E}\left(\tilde{P}_i^0\left(\Delta\text{E}_i^{(d)}\Delta P_i^{(d)}-\Delta\text{E}_i^{(d^*)}\Delta P_i^{(d^*)}\right)\right)\right|\leq\delta_n/\sqrt{n}}_{\text{Required for ATES only}},\label{eq:bound_atelrnative_ATES}\\
&\underbrace{\sup_{\mu\in\mathcal{T}_n}\left|\mathbb{E}\left(\tilde{P}_i^0\Delta\text{E}_i^{(d^*)}P_i^{(d)}\left(\Delta P_i^{(d)}-\Delta P_i^{(d^*)}\right)\right)\right|\leq\delta_n/\sqrt{n}}_{\text{Required for ATTS only}},\label{eq:bound_atelrnative_ATTS}\\
&\underbrace{\sup_{\mu\in\mathcal{T}_n}\left|\mathbb{E}\left(\Delta\text{E}_i^{(d)}\left(1-\frac{\tilde{P}_i^0}{\tilde{P}_i}\frac{D_i^{(d)}}{P_i^{(d)}}\right)-\Delta\text{E}_i^{(d^*)}\left(1-\frac{\tilde{P}_i^0}{\tilde{P}_i}\frac{D_i^{(d^*)}}{P_i^{(d^*)}}\right)\right)\right|\leq\delta_n/\sqrt{n}}_{\text{Required for ATE only}},\label{eq:bound_atelrnative_ATE}\\
&\underbrace{\sup_{\mu\in\mathcal{T}_n}\left|\mathbb{E}\left(\Delta\text{E}_i^{(d)}D_i^{(d)}\left(1-\frac{\tilde{P}_i^0}{\tilde{P}_i}\right)+\Delta\text{E}_i^{(d^*)}\left(\frac{\tilde{P}_i^0}{\tilde{P}_i}\frac{P_i^{(d)}}{P_i^{(d^*)}}D_i^{(d^*)}-D_i^{(d)}\right)\right)\right|\leq\delta_n/\sqrt{n}}_{\text{Required for ATT only}}.\label{eq:bound_atelrnative_ATT}
\end{align}
\end{assumption}

The following theorem establishes the asymptotic properties of the derivative-free estimators.

\begin{theorem}[Consistency and asymptotic normality without derivative correction terms]\label{theorem:normality2}
    Assumptions~\ref{assumption:selection},~\ref{assumption:exogeneity},~\ref{assumption:support}, and~\ref{assumption:regularity2} imply that the estimators $\widehat{\mathrm{ATES}}^{*}$ and $\widehat{\mathrm{ATTS}}^{*}$ are consistent, asymptotically linear, and asymptotically normal in the sense of Theorem~\ref{theorem:modified}. If, in addition, Assumptions~\ref{assumption:homogeneity} and~\ref{assumption:selection-support} hold, then the estimators $\widehat{\mathrm{ATE}}^{*}$ and $\widehat{\mathrm{ATT}}^{*}$ have the same properties. For each estimator, only the bound corresponding to the causal parameter of interest in Assumption~\ref{assumption:regularity2} is required. In each case, the corresponding asymptotic variance estimator, obtained from equations~(\ref{eq:var_ates})--(\ref{eq:var_att}) using the derivative-free scores, is consistent in the sense of Theorem~\ref{theorem:normality}.
\end{theorem}

Bounds~(\ref{eq:bound_atelrnative_ATES})--(\ref{eq:bound_atelrnative_ATT}) provide model-specific formulations of condition~(\ref{eq:NO_alternative}). At the end of the proof of Theorem~\ref{theorem:normality2}, we show that these bounds are implied by two stronger but more interpretable conditions. First, $\mu_Y(\overline{d},x,p)$ is uniformly Lipschitz continuous with respect to the generated control variables $p$ for $\overline{d}\in\{d,d^*\}$. Second, $\mathbb{E}\left(\|P_i-\mathbb{E}(P_i\mid X_i,P_i^0)\|^2\right)^{1/2}\leq\delta_n/\sqrt{n}$. Both conditions are imposed uniformly over $\mu\in\mathcal{T}_n$. Intuitively, the latter condition allows for relatively large estimation errors in the generated control variables, provided that their variation conditional on $X_i$ and $P_i^0$ is sufficiently small. In particular, it is satisfied whenever $P_i$ is a measurable function of $X_i$ and $P_i^0$ alone.

Therefore, the established theorems reveal a trade-off between the smoothness of the nuisance functions and the restrictions imposed on the estimation errors in the generated control variables. Including the derivative correction terms accommodates a wider class of estimation errors under stronger smoothness conditions. Omitting these terms relaxes the smoothness requirements but imposes additional restrictions on the structure of the estimation errors in the generated control variables. We conjecture that the sufficient conditions for asymptotic normality of both classes of estimators can be further relaxed, possibly using more sophisticated asymptotic arguments, which we leave for future research.

\section{Simulated data analysis}\label{sec:sim}

We consider $n_Z=2$ selection equations, one binary with $\operatorname{supp}(Z_{1i})=\{0,1\}$ and one ordinal with $\operatorname{supp}(Z_{2i})=\{0,1,2\}$. The outcome $Y_i$ is observable only if $Z_{1i}=1$ and $Z_{2i}=1$, so $z^*=(1,1)$. The treatment variable $D_i$ is binary, with $\operatorname{supp}(D_i)=\{0,1\}$. The data-generating process is as follows:
\begin{align}
    Y_{0i} &= 1 + X_i \beta + U_i^{(Y)}, \quad Y_{1i} = 2 - X_i \beta + U_i^{(Y)}, \\
    D_i &= \mathbb{I}\left(X_i \beta - 0.2 W_{3i}^{(Z)} + U_i^{(D)} \geq 0\right), \\
    Z_{1i}^* &= 2\mathbb{I}(X_i > 0)\beta + 0.5D_i + \mathbb{I}(W_{2i}^{(Z)} > 0) + 0.5W_{1i}^{(Z)} \nonumber \\
    &+ h(W_{1i}^{(Z)}) + U_{1i}^{(Z)} - 1, \\
    Z_{2i}^* &= \mathbb{I}(X_i > 0)\beta - 0.25D_i - 0.5\mathbb{I}(W_{1i}^{(Z)} > 0) + 0.5W_{3i}^{(Z)} \nonumber \\
    &+ h(W_{3i}^{(Z)}) - U_{2i}^{(Z)} - 0.5, \\
    Z_{1i} &= \mathbb{I}(Z_{1i}^* \geq 0), \\
    Z_{2i} &= \mathbb{I}(-1.2 \leq Z_{2i}^* \leq c_i) + 2\mathbb{I}(Z_{2i}^* > c_i), \\
    c_i &= 1.2 + 0.3\Phi(W_{2i}^{(Z)}), \quad h(w) = \min\{\max\{0.5 + 2w, 0\}, 1\}, \\
    X_i &\sim \mathcal{N}(0, \Sigma_X), \quad W_{1i}^{(Z)}, W_{2i}^{(Z)}, W_{3i}^{(Z)}, U_i^{(D)} \sim \mathcal{N}(0, 1), \\
    \left(U_{1i}^{(Z)}, U_{2i}^{(Z)}, U_i^{(Y)}\right) &\sim \mathcal{N}(0, \Sigma_U),
\end{align}
where $\Phi(\cdot)$ is the standard normal cumulative distribution function, $X_i$ is a row vector of length $n_X = 20$, and $\beta$ is a column vector of length $n_X$. The indicator function $\mathbb{I}(X_i > 0)$ is applied elementwise. Similarly to \citet{Bia} and \citet{Potanin2026}, we define the $j$-th element of $\beta$ as $\beta_j = 0.4 / j^2$. The covariate vector, the three exclusion variables, the treatment error, and the vector of selection and outcome errors are mutually independent. Covariates $X_i$ follow a multivariate normal distribution with covariance matrix $\Sigma_X$ with entries $(\Sigma_X)_{jt} = \operatorname{Cov}(X_{ji}, X_{ti}) = 0.5^{|j - t|}$. The joint distribution of random errors from the selection and outcome equations is multivariate normal with zero means and the following covariance matrix:
\begin{equation}
    \Sigma_U = \begin{bmatrix}
        1   & 0.3 & 0.7 \\
        0.3 & 1   & 0.8 \\
        0.7 & 0.8 & 1
    \end{bmatrix}.
\end{equation}

We use three exclusion variables $W_i^{(Z)} = (W_{1i}^{(Z)}, W_{2i}^{(Z)}, W_{3i}^{(Z)})$ since there is one probability $\mathbb{P}(Z_{1i} \leq 0 \mid D_i, X_i, W_i^{(Z)})$ induced by the first (binary) selection equation and there are two probabilities $\mathbb{P}(Z_{2i} \leq 0 \mid D_i, X_i, W_i^{(Z)})$, $\mathbb{P}(Z_{2i} \leq 1 \mid D_i, X_i, W_i^{(Z)})$, induced by the second (ordinal) selection equation. The upper threshold of the ordinal selection equation varies across observations to allow the distance between the thresholds, and hence the relationship between the two corresponding probabilities, to vary across observations. The nonlinear specifications of the selection equations motivate the use of flexible machine learning estimators of these probabilities.

In contrast to \citet{Bia}, the specifications of $Y_{0i}$ and $Y_{1i}$ differ not only in their constant terms but also in the signs of the coefficients of $X_i$. This specification allows the true values of ATE, ATES, ATT, and ATTS to differ from one another. Assumption~\ref{assumption:homogeneity} holds since the individual treatment effect $Y_{1i}-Y_{0i}=1-2X_i\beta$ depends only on $X_i$.

We conduct $m = 1000$ simulations for the small sample size $n = 1000$, $m = 200$ simulations for the medium sample size $n = 3000$, and $m = 100$ simulations for the large sample size $n = 10000$. We compare the accuracy of several estimators:
\begin{itemize}
    \item \textbf{DML} --- the proposed DML estimators for the MSSM.
    \item \textbf{DML-NG} --- the same as DML, but with the derivative correction terms omitted.
    \item \textbf{DML-NN} --- the same as DML, but with conventional (non-nested) cross-fitting.
    \item \textbf{DML-1S} --- the same as DML, but with two selection equations $Z_{1i}$, $Z_{2i}$ replaced by a single one $\tilde{Z}_{i} = \mathbb{I}\left(Z_{1i} = 1, Z_{2i} = 1\right)$.
    \item \textbf{DML-0S} --- conventional DML estimators\footnote{For DML-0S, the estimates of ATE and ATT coincide with those of ATES and ATTS, respectively.} that do not account for non-random sample selection: no selection control functions are used, and estimation is restricted to observations for which $\tilde{Z}_i = 1$.
\end{itemize}

We apply trimming by excluding observations for which any of the estimated treatment or joint selection probabilities entering the corresponding score is below $0.01$. We use five folds for cross-fitting, with the same partition employed by all estimators within each simulation.

We use ridge regression\footnote{We use the glmnet package in R for ridge regression with three-fold cross-validation to select the value of the regularization parameter from a grid of up to 10 values \citep{glmnet}.} to estimate $\text{E}_{i}^{0(d)}$ and $\text{E}_{i}^{0(d^{*})}$, random forest\footnote{We use the ranger package in R to fit random forests with 20 trees \citep{Wright}, a minimum node size parameter of 10, and other hyperparameters set to their default values. We use a small number of trees to speed up the calculation and test the robustness of the estimators to noisy estimates of the probabilities.} to estimate $P_i^0$ and $\tilde{P}_i^0$, and logistic regression without regularization to estimate $P_i^{0(d)}$ and $P_i^{0(d^{*})}$. We employ a two-learner estimation strategy for the outcome regressions, while the selection probabilities are estimated using pooled treatment and control groups with $D_i$ included as a regressor. Central finite differences are used to estimate $\text{E}_{\partial i}^{0(d)}$ and $\text{E}_{\partial i}^{0(d^{*})}$.

The true values of ATE and ATT can be obtained analytically:
\begin{align}
    \mathrm{ATE} &= 1, \\
    \mathrm{ATT} &= 1 - \sqrt{\frac{2}{\pi}}\frac{2\beta'\Sigma_X\beta}{\sqrt{\beta'\Sigma_X\beta + 1.04}} \approx 0.668.
\end{align}
The population values of ATES and ATTS are approximated by averaging their sample-specific values calculated from the simulated potential outcomes. The resulting reference values are approximately 0.838 and 0.540, respectively.

Using these reference values, we calculate the mean absolute error (MAE) and root mean squared error (RMSE) for each causal parameter and estimator. We also report the mean, standard deviation (SD), and empirical coverage probability (CP) of nominal $95\%$ confidence intervals across replications. Moreover, we compare the estimators pairwise using two-sided paired $t$-tests of equality of their mean squared errors (MSEs), without adjustment for multiple testing, and report the share of replications in which one estimator has a smaller absolute estimation error than the other. The approximated values of ATES and ATTS are treated as fixed in these comparisons.

\begin{table}[htbp]
\centering
\scriptsize
\caption{Simulation results for the four causal parameters. CP denotes the coverage probability of nominal 95\% confidence intervals. ATES and ATTS reference values are approximated from the simulated samples.}
\label{tbl:sim}
\setlength{\tabcolsep}{3pt}
\begin{tabular}{l ccccc ccccc ccccc}
\toprule
 & \multicolumn{5}{c}{Small sample ($n = 1000$)} & \multicolumn{5}{c}{Medium sample ($n = 3000$)} & \multicolumn{5}{c}{Large sample ($n = 10000$)} \\
\cmidrule(lr){2-6} \cmidrule(lr){7-11} \cmidrule(lr){12-16}
 & RMSE & MAE & Mean & SD & CP & RMSE & MAE & Mean & SD & CP & RMSE & MAE & Mean & SD & CP \\
\midrule
\multicolumn{16}{c}{\textbf{ATES} $\approx$ 0.838} \\
DML & 0.134 & 0.104 & 0.876 & 0.128 & 0.953 & 0.067 & 0.053 & 0.865 & 0.062 & 0.930 & 0.033 & 0.026 & 0.844 & 0.032 & 0.960 \\
DML-NG & 0.150 & 0.123 & 0.741 & 0.113 & 0.870 & 0.119 & 0.107 & 0.734 & 0.055 & 0.525 & 0.096 & 0.091 & 0.745 & 0.030 & 0.090 \\
DML-NN & 0.157 & 0.126 & 0.931 & 0.127 & 0.895 & 0.115 & 0.097 & 0.931 & 0.070 & 0.755 & 0.073 & 0.064 & 0.900 & 0.037 & 0.630 \\
DML-1S & 0.231 & 0.205 & 0.638 & 0.116 & 0.629 & 0.209 & 0.201 & 0.638 & 0.056 & 0.070 & 0.209 & 0.207 & 0.630 & 0.031 & 0.000 \\
DML-0S & 0.253 & 0.227 & 0.616 & 0.120 & 0.582 & 0.224 & 0.217 & 0.621 & 0.056 & 0.045 & 0.223 & 0.221 & 0.616 & 0.030 & 0.000 \\
\midrule
\multicolumn{16}{c}{\textbf{ATTS} $\approx$ 0.540} \\
DML & 0.144 & 0.112 & 0.525 & 0.143 & 0.953 & 0.068 & 0.055 & 0.558 & 0.066 & 0.950 & 0.036 & 0.030 & 0.543 & 0.036 & 0.960 \\
DML-NG & 0.189 & 0.156 & 0.403 & 0.130 & 0.832 & 0.122 & 0.108 & 0.436 & 0.061 & 0.555 & 0.094 & 0.088 & 0.451 & 0.032 & 0.210 \\
DML-NN & 0.164 & 0.129 & 0.617 & 0.146 & 0.916 & 0.117 & 0.097 & 0.631 & 0.077 & 0.795 & 0.075 & 0.064 & 0.602 & 0.041 & 0.700 \\
DML-1S & 0.274 & 0.242 & 0.303 & 0.136 & 0.569 & 0.208 & 0.198 & 0.344 & 0.064 & 0.140 & 0.202 & 0.199 & 0.340 & 0.034 & 0.000 \\
DML-0S & 0.297 & 0.263 & 0.284 & 0.148 & 0.596 & 0.227 & 0.217 & 0.325 & 0.065 & 0.115 & 0.222 & 0.219 & 0.319 & 0.033 & 0.000 \\
\midrule
\multicolumn{16}{c}{\textbf{ATE} $\approx$ 1.000} \\
DML & 0.149 & 0.117 & 1.033 & 0.146 & 0.965 & 0.072 & 0.057 & 1.024 & 0.068 & 0.930 & 0.039 & 0.031 & 1.005 & 0.039 & 0.970 \\
DML-NG & 0.169 & 0.139 & 0.892 & 0.131 & 0.839 & 0.130 & 0.116 & 0.886 & 0.063 & 0.520 & 0.108 & 0.101 & 0.899 & 0.037 & 0.180 \\
DML-NN & 0.156 & 0.126 & 1.084 & 0.132 & 0.923 & 0.113 & 0.096 & 1.088 & 0.070 & 0.795 & 0.075 & 0.066 & 1.065 & 0.038 & 0.630 \\
DML-1S & 0.277 & 0.248 & 0.756 & 0.132 & 0.542 & 0.249 & 0.241 & 0.759 & 0.066 & 0.075 & 0.265 & 0.262 & 0.738 & 0.041 & 0.000 \\
DML-0S & 0.402 & 0.384 & 0.616 & 0.120 & 0.132 & 0.383 & 0.379 & 0.621 & 0.056 & 0.000 & 0.385 & 0.384 & 0.616 & 0.030 & 0.000 \\
\midrule
\multicolumn{16}{c}{\textbf{ATT} $\approx$ 0.668} \\
DML & 0.154 & 0.121 & 0.636 & 0.151 & 0.956 & 0.070 & 0.056 & 0.684 & 0.068 & 0.955 & 0.039 & 0.031 & 0.673 & 0.039 & 0.930 \\
DML-NG & 0.214 & 0.180 & 0.502 & 0.135 & 0.803 & 0.134 & 0.120 & 0.549 & 0.063 & 0.515 & 0.104 & 0.098 & 0.570 & 0.035 & 0.150 \\
DML-NN & 0.164 & 0.130 & 0.741 & 0.147 & 0.921 & 0.122 & 0.104 & 0.764 & 0.075 & 0.785 & 0.083 & 0.074 & 0.741 & 0.041 & 0.530 \\
DML-1S & 0.328 & 0.298 & 0.371 & 0.141 & 0.458 & 0.247 & 0.238 & 0.430 & 0.066 & 0.065 & 0.250 & 0.247 & 0.421 & 0.038 & 0.000 \\
DML-0S & 0.411 & 0.384 & 0.284 & 0.148 & 0.235 & 0.349 & 0.343 & 0.325 & 0.065 & 0.000 & 0.350 & 0.349 & 0.319 & 0.033 & 0.000 \\
\bottomrule
\end{tabular}
\end{table}

The results of the analysis are provided in Table~\ref{tbl:sim}. The estimators accounting for both selection equations (DML, DML-NG, and DML-NN) outperform those accounting for only a single (DML-1S) or no (DML-0S) selection equation in terms of both RMSE and MAE for all four causal parameters and all considered sample sizes. This advantage becomes particularly pronounced in the large sample. For example, the RMSE of the DML estimator of ATES decreases from 0.134 in the small sample to 0.033 in the large sample, while the corresponding RMSEs of DML-1S and DML-0S decrease from 0.231 to 0.209 and from 0.253 to 0.223, respectively. The persistence of substantial estimation errors for the latter two estimators illustrates the importance of accounting for the multivariate nature of the selection mechanism.

The proposed DML estimators consistently outperform their counterparts without the derivative correction terms (DML-NG). Although omitting these terms reduces the standard deviations of the estimates in all considered cases, the resulting mean estimates exhibit substantial deviations from the reference values even in the large sample. For example, in the large sample, the mean estimates of ATES obtained using DML and DML-NG are 0.844 and 0.745, respectively, compared with the reference value of 0.838. The corresponding RMSEs are 0.033 and 0.096. Moreover, depending on the causal parameter, DML provides more accurate estimates than DML-NG in 61.7--76.6\% of replications in the small sample, 75.5--79.0\% in the medium sample, and 89--93\% in the large sample (Tables~\ref{tbl:sim-win-1000}, \ref{tbl:sim-win-3000}, and~\ref{tbl:sim-win-10000}). These findings illustrate the importance of the derivative correction terms in accounting for the first-order effects of estimation errors in the generated control functions.

The results also demonstrate the advantages of nested cross-fitting. DML outperforms DML-NN in terms of both RMSE and MAE for all four causal parameters and all considered sample sizes. For example, the RMSEs of the DML and DML-NN estimators of ATES are 0.134 and 0.157 in the small sample and 0.033 and 0.073 in the large sample, respectively. Depending on the causal parameter, DML provides more accurate estimates than DML-NN in 52.6--62.9\% of replications in the small sample, 77--82\% in the medium sample, and 79--85\% in the large sample. Thus, despite the smaller training samples used to estimate individual nuisance functions, nested cross-fitting improves estimation accuracy in the considered setting. This improvement is primarily associated with smaller absolute deviations of the mean estimates from the reference values rather than a uniform reduction in their standard deviations.

The advantages of the proposed DML estimators are further supported by the paired tests of equality of MSE reported in Tables~\ref{tbl:sim-pval-1000}, \ref{tbl:sim-pval-3000}, and~\ref{tbl:sim-pval-10000}. For every causal parameter and sample size, the MSE of DML is significantly lower than that of each competing estimator at the 5\% significance level. All these differences are significant at the 0.1\% significance level except for the comparisons of DML with DML-NN for ATE and ATT in the small sample, for which the $p$-values are 0.037 and 0.018, respectively. In the large sample, DML produces a smaller absolute estimation error than either DML-1S or DML-0S in every replication for all four causal parameters.

Finally, the proposed DML estimators exhibit empirical coverage probabilities of nominal 95\% confidence intervals ranging from 0.930 to 0.970 across all causal parameters and sample sizes. In contrast, the coverage probabilities of the competing estimators deteriorate as the sample size increases. In the large sample, they range from 0.090 to 0.210 for DML-NG and from 0.530 to 0.700 for DML-NN, while the confidence intervals constructed using DML-1S and DML-0S never cover the corresponding reference values. These findings are consistent with the persistence of bias in the competing estimators as their standard deviations decline. In the considered simulation design, accounting for both selection equations, including the derivative correction terms, and employing nested cross-fitting are therefore important not only for the accuracy of point estimates but also for the reliability of statistical inference. 

\begin{table}[H]
\centering
\footnotesize
\caption{P-values of two-sided paired tests of equality of MSE ($n = 1000$).}
\label{tbl:sim-pval-1000}
\begin{tabular}{lccccc}
\toprule
 & DML & DML-NG & DML-NN & DML-1S & DML-0S \\
\midrule
\multicolumn{6}{c}{\textbf{ATES}} \\
DML & -- & $<0.001$ & $<0.001$ & $<0.001$ & $<0.001$ \\
DML-NG & $<0.001$ & -- & 0.140 & $<0.001$ & $<0.001$ \\
DML-NN & $<0.001$ & 0.140 & -- & $<0.001$ & $<0.001$ \\
DML-1S & $<0.001$ & $<0.001$ & $<0.001$ & -- & $<0.001$ \\
DML-0S & $<0.001$ & $<0.001$ & $<0.001$ & $<0.001$ & -- \\
\midrule
\multicolumn{6}{c}{\textbf{ATTS}} \\
DML & -- & $<0.001$ & $<0.001$ & $<0.001$ & $<0.001$ \\
DML-NG & $<0.001$ & -- & $<0.001$ & $<0.001$ & $<0.001$ \\
DML-NN & $<0.001$ & $<0.001$ & -- & $<0.001$ & $<0.001$ \\
DML-1S & $<0.001$ & $<0.001$ & $<0.001$ & -- & $<0.001$ \\
DML-0S & $<0.001$ & $<0.001$ & $<0.001$ & $<0.001$ & -- \\
\midrule
\multicolumn{6}{c}{\textbf{ATE}} \\
DML & -- & $<0.001$ & 0.037 & $<0.001$ & $<0.001$ \\
DML-NG & $<0.001$ & -- & 0.008 & $<0.001$ & $<0.001$ \\
DML-NN & 0.037 & 0.008 & -- & $<0.001$ & $<0.001$ \\
DML-1S & $<0.001$ & $<0.001$ & $<0.001$ & -- & $<0.001$ \\
DML-0S & $<0.001$ & $<0.001$ & $<0.001$ & $<0.001$ & -- \\
\midrule
\multicolumn{6}{c}{\textbf{ATT}} \\
DML & -- & $<0.001$ & 0.018 & $<0.001$ & $<0.001$ \\
DML-NG & $<0.001$ & -- & $<0.001$ & $<0.001$ & $<0.001$ \\
DML-NN & 0.018 & $<0.001$ & -- & $<0.001$ & $<0.001$ \\
DML-1S & $<0.001$ & $<0.001$ & $<0.001$ & -- & $<0.001$ \\
DML-0S & $<0.001$ & $<0.001$ & $<0.001$ & $<0.001$ & -- \\
\bottomrule
\end{tabular}
\end{table}

\begin{table}[H]
\centering
\footnotesize
\caption{Shares of replications in which the row estimator is more accurate than the column estimator ($n = 1000$). Ties count as wins for neither estimator.}
\label{tbl:sim-win-1000}
\begin{tabular}{lccccc}
\toprule
 & DML & DML-NG & DML-NN & DML-1S & DML-0S \\
\midrule
\multicolumn{6}{c}{\textbf{ATES}} \\
DML & -- & 0.617 & 0.629 & 0.765 & 0.793 \\
DML-NG & 0.383 & -- & 0.484 & 0.903 & 0.926 \\
DML-NN & 0.371 & 0.516 & -- & 0.687 & 0.726 \\
DML-1S & 0.235 & 0.097 & 0.313 & -- & 0.673 \\
DML-0S & 0.207 & 0.074 & 0.274 & 0.327 & -- \\
\midrule
\multicolumn{6}{c}{\textbf{ATTS}} \\
DML & -- & 0.720 & 0.555 & 0.844 & 0.850 \\
DML-NG & 0.280 & -- & 0.403 & 0.890 & 0.886 \\
DML-NN & 0.445 & 0.597 & -- & 0.744 & 0.744 \\
DML-1S & 0.156 & 0.110 & 0.256 & -- & 0.596 \\
DML-0S & 0.150 & 0.114 & 0.256 & 0.404 & -- \\
\midrule
\multicolumn{6}{c}{\textbf{ATE}} \\
DML & -- & 0.627 & 0.553 & 0.802 & 0.935 \\
DML-NG & 0.373 & -- & 0.437 & 0.886 & 0.982 \\
DML-NN & 0.447 & 0.563 & -- & 0.754 & 0.912 \\
DML-1S & 0.198 & 0.114 & 0.246 & -- & 0.938 \\
DML-0S & 0.065 & 0.018 & 0.088 & 0.062 & -- \\
\midrule
\multicolumn{6}{c}{\textbf{ATT}} \\
DML & -- & 0.766 & 0.526 & 0.893 & 0.941 \\
DML-NG & 0.234 & -- & 0.370 & 0.907 & 0.947 \\
DML-NN & 0.474 & 0.630 & -- & 0.813 & 0.893 \\
DML-1S & 0.107 & 0.093 & 0.187 & -- & 0.804 \\
DML-0S & 0.059 & 0.053 & 0.107 & 0.196 & -- \\
\bottomrule
\end{tabular}
\end{table}

\begin{table}[H]
\centering
\footnotesize
\caption{P-values of two-sided paired tests of equality of MSE ($n = 3000$).}
\label{tbl:sim-pval-3000}
\begin{tabular}{lccccc}
\toprule
 & DML & DML-NG & DML-NN & DML-1S & DML-0S \\
\midrule
\multicolumn{6}{c}{\textbf{ATES}} \\
DML & -- & $<0.001$ & $<0.001$ & $<0.001$ & $<0.001$ \\
DML-NG & $<0.001$ & -- & 0.657 & $<0.001$ & $<0.001$ \\
DML-NN & $<0.001$ & 0.657 & -- & $<0.001$ & $<0.001$ \\
DML-1S & $<0.001$ & $<0.001$ & $<0.001$ & -- & $<0.001$ \\
DML-0S & $<0.001$ & $<0.001$ & $<0.001$ & $<0.001$ & -- \\
\midrule
\multicolumn{6}{c}{\textbf{ATTS}} \\
DML & -- & $<0.001$ & $<0.001$ & $<0.001$ & $<0.001$ \\
DML-NG & $<0.001$ & -- & 0.550 & $<0.001$ & $<0.001$ \\
DML-NN & $<0.001$ & 0.550 & -- & $<0.001$ & $<0.001$ \\
DML-1S & $<0.001$ & $<0.001$ & $<0.001$ & -- & $<0.001$ \\
DML-0S & $<0.001$ & $<0.001$ & $<0.001$ & $<0.001$ & -- \\
\midrule
\multicolumn{6}{c}{\textbf{ATE}} \\
DML & -- & $<0.001$ & $<0.001$ & $<0.001$ & $<0.001$ \\
DML-NG & $<0.001$ & -- & 0.017 & $<0.001$ & $<0.001$ \\
DML-NN & $<0.001$ & 0.017 & -- & $<0.001$ & $<0.001$ \\
DML-1S & $<0.001$ & $<0.001$ & $<0.001$ & -- & $<0.001$ \\
DML-0S & $<0.001$ & $<0.001$ & $<0.001$ & $<0.001$ & -- \\
\midrule
\multicolumn{6}{c}{\textbf{ATT}} \\
DML & -- & $<0.001$ & $<0.001$ & $<0.001$ & $<0.001$ \\
DML-NG & $<0.001$ & -- & 0.119 & $<0.001$ & $<0.001$ \\
DML-NN & $<0.001$ & 0.119 & -- & $<0.001$ & $<0.001$ \\
DML-1S & $<0.001$ & $<0.001$ & $<0.001$ & -- & $<0.001$ \\
DML-0S & $<0.001$ & $<0.001$ & $<0.001$ & $<0.001$ & -- \\
\bottomrule
\end{tabular}
\end{table}

\begin{table}[H]
\centering
\footnotesize
\caption{Shares of replications in which the row estimator is more accurate than the column estimator ($n = 3000$). Ties count as wins for neither estimator.}
\label{tbl:sim-win-3000}
\begin{tabular}{lccccc}
\toprule
 & DML & DML-NG & DML-NN & DML-1S & DML-0S \\
\midrule
\multicolumn{6}{c}{\textbf{ATES}} \\
DML & -- & 0.770 & 0.810 & 0.920 & 0.945 \\
DML-NG & 0.230 & -- & 0.435 & 1.000 & 1.000 \\
DML-NN & 0.190 & 0.565 & -- & 0.815 & 0.870 \\
DML-1S & 0.080 & 0.000 & 0.185 & -- & 0.850 \\
DML-0S & 0.055 & 0.000 & 0.130 & 0.150 & -- \\
\midrule
\multicolumn{6}{c}{\textbf{ATTS}} \\
DML & -- & 0.755 & 0.790 & 0.925 & 0.940 \\
DML-NG & 0.245 & -- & 0.415 & 0.995 & 1.000 \\
DML-NN & 0.210 & 0.585 & -- & 0.765 & 0.820 \\
DML-1S & 0.075 & 0.005 & 0.235 & -- & 0.865 \\
DML-0S & 0.060 & 0.000 & 0.180 & 0.135 & -- \\
\midrule
\multicolumn{6}{c}{\textbf{ATE}} \\
DML & -- & 0.775 & 0.770 & 0.965 & 1.000 \\
DML-NG & 0.225 & -- & 0.425 & 1.000 & 1.000 \\
DML-NN & 0.230 & 0.575 & -- & 0.885 & 0.995 \\
DML-1S & 0.035 & 0.000 & 0.115 & -- & 1.000 \\
DML-0S & 0.000 & 0.000 & 0.005 & 0.000 & -- \\
\midrule
\multicolumn{6}{c}{\textbf{ATT}} \\
DML & -- & 0.790 & 0.820 & 0.975 & 1.000 \\
DML-NG & 0.210 & -- & 0.440 & 1.000 & 1.000 \\
DML-NN & 0.180 & 0.560 & -- & 0.850 & 0.980 \\
DML-1S & 0.025 & 0.000 & 0.150 & -- & 0.990 \\
DML-0S & 0.000 & 0.000 & 0.020 & 0.010 & -- \\
\bottomrule
\end{tabular}
\end{table}

\begin{table}[H]
\centering
\footnotesize
\caption{P-values of two-sided paired tests of equality of MSE ($n = 10000$).}
\label{tbl:sim-pval-10000}
\begin{tabular}{lccccc}
\toprule
 & DML & DML-NG & DML-NN & DML-1S & DML-0S \\
\midrule
\multicolumn{6}{c}{\textbf{ATES}} \\
DML & -- & $<0.001$ & $<0.001$ & $<0.001$ & $<0.001$ \\
DML-NG & $<0.001$ & -- & $<0.001$ & $<0.001$ & $<0.001$ \\
DML-NN & $<0.001$ & $<0.001$ & -- & $<0.001$ & $<0.001$ \\
DML-1S & $<0.001$ & $<0.001$ & $<0.001$ & -- & $<0.001$ \\
DML-0S & $<0.001$ & $<0.001$ & $<0.001$ & $<0.001$ & -- \\
\midrule
\multicolumn{6}{c}{\textbf{ATTS}} \\
DML & -- & $<0.001$ & $<0.001$ & $<0.001$ & $<0.001$ \\
DML-NG & $<0.001$ & -- & 0.005 & $<0.001$ & $<0.001$ \\
DML-NN & $<0.001$ & 0.005 & -- & $<0.001$ & $<0.001$ \\
DML-1S & $<0.001$ & $<0.001$ & $<0.001$ & -- & $<0.001$ \\
DML-0S & $<0.001$ & $<0.001$ & $<0.001$ & $<0.001$ & -- \\
\midrule
\multicolumn{6}{c}{\textbf{ATE}} \\
DML & -- & $<0.001$ & $<0.001$ & $<0.001$ & $<0.001$ \\
DML-NG & $<0.001$ & -- & $<0.001$ & $<0.001$ & $<0.001$ \\
DML-NN & $<0.001$ & $<0.001$ & -- & $<0.001$ & $<0.001$ \\
DML-1S & $<0.001$ & $<0.001$ & $<0.001$ & -- & $<0.001$ \\
DML-0S & $<0.001$ & $<0.001$ & $<0.001$ & $<0.001$ & -- \\
\midrule
\multicolumn{6}{c}{\textbf{ATT}} \\
DML & -- & $<0.001$ & $<0.001$ & $<0.001$ & $<0.001$ \\
DML-NG & $<0.001$ & -- & 0.002 & $<0.001$ & $<0.001$ \\
DML-NN & $<0.001$ & 0.002 & -- & $<0.001$ & $<0.001$ \\
DML-1S & $<0.001$ & $<0.001$ & $<0.001$ & -- & $<0.001$ \\
DML-0S & $<0.001$ & $<0.001$ & $<0.001$ & $<0.001$ & -- \\
\bottomrule
\end{tabular}
\end{table}

\begin{table}[H]
\centering
\footnotesize
\caption{Shares of replications in which the row estimator is more accurate than the column estimator ($n = 10000$). Ties count as wins for neither estimator.}
\label{tbl:sim-win-10000}
\begin{tabular}{lccccc}
\toprule
 & DML & DML-NG & DML-NN & DML-1S & DML-0S \\
\midrule
\multicolumn{6}{c}{\textbf{ATES}} \\
DML & -- & 0.930 & 0.850 & 1.000 & 1.000 \\
DML-NG & 0.070 & -- & 0.330 & 1.000 & 1.000 \\
DML-NN & 0.150 & 0.670 & -- & 0.980 & 0.990 \\
DML-1S & 0.000 & 0.000 & 0.020 & -- & 0.990 \\
DML-0S & 0.000 & 0.000 & 0.010 & 0.010 & -- \\
\midrule
\multicolumn{6}{c}{\textbf{ATTS}} \\
DML & -- & 0.890 & 0.790 & 1.000 & 1.000 \\
DML-NG & 0.110 & -- & 0.360 & 1.000 & 1.000 \\
DML-NN & 0.210 & 0.640 & -- & 0.970 & 0.980 \\
DML-1S & 0.000 & 0.000 & 0.030 & -- & 0.970 \\
DML-0S & 0.000 & 0.000 & 0.020 & 0.030 & -- \\
\midrule
\multicolumn{6}{c}{\textbf{ATE}} \\
DML & -- & 0.890 & 0.830 & 1.000 & 1.000 \\
DML-NG & 0.110 & -- & 0.250 & 1.000 & 1.000 \\
DML-NN & 0.170 & 0.750 & -- & 1.000 & 1.000 \\
DML-1S & 0.000 & 0.000 & 0.000 & -- & 1.000 \\
DML-0S & 0.000 & 0.000 & 0.000 & 0.000 & -- \\
\midrule
\multicolumn{6}{c}{\textbf{ATT}} \\
DML & -- & 0.900 & 0.840 & 1.000 & 1.000 \\
DML-NG & 0.100 & -- & 0.370 & 1.000 & 1.000 \\
DML-NN & 0.160 & 0.630 & -- & 1.000 & 1.000 \\
DML-1S & 0.000 & 0.000 & 0.000 & -- & 1.000 \\
DML-0S & 0.000 & 0.000 & 0.000 & 0.000 & -- \\
\bottomrule
\end{tabular}
\end{table}

\section{Extensions}\label{sec:extensions}

In this section, we discuss extensions of the proposed framework to more general selection mechanisms, alternative causal parameters, and settings in which the treatment variable itself is subject to non-random sample selection.

\subsection{Nested sequential sample selection}\label{sec:nested}

In many empirical settings, sample selection occurs sequentially, with subsequent selection variables being observed only for individuals who have passed the preceding selection stages. Sequential sample selection mechanisms have been considered by, among others, \citet{Mohanty2001}, \citet{Zhang2015}, and \cite{Kossova2018}. We show how the proposed framework can accommodate such mechanisms. Specifically, suppose that the selection equations are binary\footnote{The framework can also be extended to ordinal selection equations, but this requires considerably more cumbersome notation and is therefore omitted for brevity.}, while the selection mechanism is nested and sequential.

\begin{assumption}[Nested sequential selection]\label{assumption:selection_nested}
For every $j\in\{1,\ldots,n_Z\}$, the selection mechanism is given by
\begin{equation}\label{eq:selection_nested}
    Z_{ji} =
    \begin{cases}
      0, & \text{if } U_{ji}^{(Z)} \leq g_j(D_i,X_i,W_i^{(Z)},Z_{(j-1)i},\ldots,Z_{1i}), \\
      1, & \text{if } U_{ji}^{(Z)} > g_j(D_i,X_i,W_i^{(Z)},Z_{(j-1)i},\ldots,Z_{1i}).
    \end{cases}
\end{equation}
For $j=1$, the function $g_1(\cdot)$ depends only on $(D_i,X_i,W_i^{(Z)})$.
\end{assumption}

Suppose that the outcome $Y_i$ is observed if and only if $\tilde{Z}_i=1$, where $\tilde{Z}_i=\prod\limits_{t=1}^{n_Z}Z_{ti}$. The selection is nested in the sense that the function $g_j(\cdot)$ may depend on the preceding selection variables $Z_{(j-1)i},\ldots,Z_{1i}$. The selection is sequential in the sense that, for every $j\in\{2,\ldots,n_Z\}$, the selection variable $Z_{ji}$ is observed if and only if $\prod\limits_{t=1}^{j-1}Z_{ti}=1$, while $Z_{1i}$ is always observed. For example, wages $Y_i$ may be observed only for employed individuals ($Z_{1i}=1$) who reveal their wage information ($Z_{2i}=1$). The selection is sequential because the researcher observes the willingness to reveal wage information only for employed individuals, and nested because employment status may itself affect this willingness.

Define $Z_{ji}^{*}=\prod\limits_{t=1}^{j}Z_{ti}$ for $j\in\{1,\ldots,n_Z\}$, where the product is defined as zero whenever at least one of the observed selection variables entering the product equals zero. Notice that all $Z_{ji}^{*}$ are observable, even when some of the original selection variables are unobserved. This motivates the following redefinition of the control functions:
\begin{align*}
    p_Z^0(d,x,w) &= \left(p_1^0(d,x,w),p_2^0(d,x,w),\ldots,p_{n_Z}^0(d,x,w)\right), \\
    p_j^0(d,x,w) &= \mathbb{P}\left(Z_{ji}^{*}=1\mid D_i=d,X_i=x,W_i^{(Z)}=w\right).
\end{align*}

As before, we define $P_i^0=p_Z^0(D_i,X_i,W_i^{(Z)})$. Under this definition, we establish the following analogue of Lemma~\ref{lemma:independence}.

\begin{lemma}\label{lemma:independence_nested}
    Assumptions~\ref{assumption:selection_nested} and~\ref{assumption:exogeneity} imply that $Y_{di}\perp(D_i,W_i^{(Z)})\mid(X_i,P_i^0,\tilde{Z}_i)$ and $(Y_{di},\tilde{Z}_i)\perp(D_i,W_i^{(Z)})\mid(X_i,P_i^0)$ for every $d\in\{0,\dots,n_D-1\}$.
\end{lemma}

Moreover, redefining $\overline{Z}_i=(Z_{1i}^{*},\ldots,Z_{n_Zi}^{*})$ preserves the conditional moment restriction $\mathbb{E}(\overline{Z}_i-P_i^0\mid D_i,X_i,W_i^{(Z)})=0$. Consequently, the previously established identification results, Neyman-orthogonal scores, and asymptotic results remain valid if Assumption~\ref{assumption:selection} is replaced with Assumption~\ref{assumption:selection_nested}, the control variables and corresponding nuisance functions are redefined as described above, and the remaining assumptions hold under these modifications. In particular, the functional form of the proposed estimators remains unchanged.

\subsection{Multiple selection criteria}\label{sec:msc}

In some empirical settings, the outcome may be observable under several alternative combinations of selection variables rather than a single one. Such selection mechanisms have been considered in the context of parametric multivariate sample selection models by \citet{Kossova2018}. The proposed framework can also accommodate such mechanisms. Specifically, suppose that $\tilde{Z}_i=\mathbb{I}(Z_i\in\mathcal{Z}^{*})$, where $\mathcal{Z}^{*}$ is a subset of $\mathrm{supp}(Z_i)$. For example, wages may be observable if an individual reveals wage information or the corresponding administrative records are available.

For every $z\in\mathcal{Z}^{*}$, define the corresponding vector of control variables and the conditional expectation of the outcome as follows:
\begin{align}
    P_i^0(z) &= \left(p^{0,z-(1,\ldots,1)}\left(D_i,X_i,W_i^{(Z)}\right),p^{0,z}\left(D_i,X_i,W_i^{(Z)}\right)\right), \nonumber \\
    \mu_Y^0(d,x,p,z) &= \mathbb{E}\left(Y_i\mid D_i=d,X_i=x,P_i^0(z)=p,Z_i=z\right).
\end{align}

Suppose that the assumptions of Theorem~\ref{theorem:ates} hold separately for every $z\in\mathcal{Z}^{*}$ with the selection criterion $\mathbb{I}(Z_i=z)$. Then, by this theorem and the law of iterated expectations, we obtain
\begin{align}
    \text{ATES} &= \sum\limits_{z\in\mathcal{Z}^{*}}\mathbb{P}\left(Z_i=z\mid\tilde{Z}_i=1\right)\times \nonumber \\
    &\times \mathbb{E}\left(\mu_Y^0(d,X_i,P_i^0(z),z)-\mu_Y^0(d^*,X_i,P_i^0(z),z)\mid Z_i=z\right), \\
    \text{ATTS} &= \sum\limits_{z\in\mathcal{Z}^{*}}\mathbb{P}\left(Z_i=z\mid D_i=d,\tilde{Z}_i=1\right)\times \nonumber \\
    &\times \mathbb{E}\left(\mu_Y^0(d,X_i,P_i^0(z),z)-\mu_Y^0(d^*,X_i,P_i^0(z),z)\mid D_i=d,Z_i=z\right).
\end{align}

If, in addition, the assumptions of Theorem~\ref{theorem:ate} hold separately for every $z\in\mathcal{Z}^{*}$, then
\begin{align}
    \text{ATE} &= \sum\limits_{z\in\mathcal{Z}^{*}}\mathbb{P}\left(Z_i=z\mid\tilde{Z}_i=1\right)\times \nonumber \\
    &\times \mathbb{E}\left(\mu_Y^0(d,X_i,P_i^0(z),z)-\mu_Y^0(d^*,X_i,P_i^0(z),z)\right), \\
    \text{ATT} &= \sum\limits_{z\in\mathcal{Z}^{*}}\mathbb{P}\left(Z_i=z\mid D_i=d,\tilde{Z}_i=1\right)\times \nonumber \\
    &\times \mathbb{E}\left(\mu_Y^0(d,X_i,P_i^0(z),z)-\mu_Y^0(d^*,X_i,P_i^0(z),z)\mid D_i=d\right).
\end{align}

Therefore, these causal parameters can be estimated as weighted averages of the corresponding estimates obtained separately for every $z\in\mathcal{Z}^{*}$, with the weights estimated using the corresponding sample frequencies. Note that, unlike for ATES and ATTS, the weights for ATE and ATT are not uniquely determined by the identification results, since each component of the weighted average identifies the same causal parameter. Consequently, alternative weighting schemes may be employed to improve estimation efficiency.

\subsection{Conditional average treatment effects}\label{sec:cate}

Under the assumptions of Theorem~\ref{theorem:ate}, the proposed framework also identifies the conditional average treatment effect (CATE), which is useful for studying treatment effect heterogeneity with respect to the covariates $X_i$. Since $P_i^0$ is a function of $(D_i,X_i,W_i^{(Z)})$, Assumption~\ref{assumption:exogeneity} implies that $Y_{di}\perp P_i^0\mid X_i$ for every $d\in\{0,\dots,n_D-1\}$. Consequently, we have:
\begin{align}
    \text{CATE}(X_i) &= \mathbb{E}(Y_{di}\mid X_i)-\mathbb{E}(Y_{d^*i}\mid X_i) \nonumber \\
    &= \underbrace{\mathbb{E}(Y_{di}\mid X_i,P_i^0)-\mathbb{E}(Y_{d^*i}\mid X_i,P_i^0)}_{\text{Assumption~\ref{assumption:exogeneity}}} \nonumber \\
    &= \underbrace{\mathbb{E}(Y_{di}\mid X_i,P_i^0,\tilde{Z}_i=1)-\mathbb{E}(Y_{d^*i}\mid X_i,P_i^0,\tilde{Z}_i=1)}_{\text{Assumption~\ref{assumption:homogeneity}}} \nonumber \\
    &= \underbrace{\mu_Y^0(d,X_i,P_i^0)-\mu_Y^0(d^*,X_i,P_i^0)}_{\text{By equation~(\ref{eq:T1:conditional_mean})}}.
\end{align}

Therefore, plug-in estimates of the CATE can be obtained using the same nuisance function estimates as those used to estimate the ATE, without estimating any additional nuisance functions.

\subsection{Treatment effect heterogeneity with respect to endogenous moderators}\label{sec:moderation}

The analysis of treatment effect heterogeneity with respect to potentially endogenous post-treatment variables has been considered, among others, by \citet{Joffe2007} and \citet{Ertefaie2018}. The former consider, in particular, treatment effects conditional on the realized values of post-treatment variables and treatment status, while the latter develop instrumental variable and control function approaches to estimating treatment effects conditional on the potential values of post-treatment variables. Our framework can also be used to study treatment effect heterogeneity with respect to potentially endogenous ordinal moderators.

Suppose that $Y_i$ is observable for every realization of $Z_i$ and that the researcher is interested in the conditional average treatment effect\footnote{The framework can also accommodate simultaneous multivariate sample selection and endogenous moderation by allowing some elements of $Z_i$ to serve as selection variables and others as moderators.}:
\begin{align}
    \text{CATE}(z^*) = \mathbb{E}\left(Y_{di}-Y_{d^*i}\mid Z_i=z^*\right).
\end{align}
By introducing an artificial selection indicator $\tilde{Z}_i=\mathbb{I}(Z_i=z^*)$, this causal parameter can be interpreted as ATES and, therefore, identified under the assumptions of Theorem~\ref{theorem:ates} and estimated using the proposed estimators under the corresponding regularity conditions. Similarly, ATTS can be interpreted as the average treatment effect among treated individuals with particular realized values of the moderators, as in the observed auxiliary stratification approach of \citet{Joffe2007}. By varying $z^*$, the researcher can investigate treatment effect heterogeneity with respect to multiple potentially endogenous ordinal moderators.

Conditioning on endogenous moderators $Z_i$ may induce dependence between the treatment variable and the potential outcomes, even if they are independent conditional on $X_i$. For example, suppose that a researcher studies the effect of education on fertility among individuals with different marital statuses. Education may affect marital status, while unobservable individual preferences may affect both marital status and fertility. In this case, conventional CATE estimators treating marital status as an exogenous covariate may yield inconsistent estimates of the causal parameter of interest.

Note that the proposed approach identifies treatment effects conditional on the realized values of the moderators rather than the causal effect of changing the moderators on the magnitude of the treatment effect\footnote{See \citealp{Bansak2021} for the distinction between treatment effect heterogeneity and causal moderation}. Moreover, when the moderators are affected by treatment, membership in the corresponding subpopulations may not be known prior to treatment assignment. Nevertheless, such an analysis may be useful for explaining treatment effect heterogeneity across subpopulations.

\subsection{Sample selection in the treatment variable}\label{sec:st}

The established identification results can also be adapted to settings in which both the outcome and treatment variables are subject to non-random sample selection, as in \citet{Potanin2026}. To accommodate such settings, one may exclude $D_i$ from the selection functions and thresholds in Assumption~\ref{assumption:selection}. Under this modification, $P_i^0$ becomes a function of $(X_i,W_i^{(Z)})$ only, and the previously established identification results remain valid under the corresponding assumptions. In particular, the conditional treatment probabilities $p_D^0(d,x,p)$ can be identified from the selected sample, since $D_i\perp\tilde{Z}_i\mid(X_i,P_i^0)$. However, the proposed estimators require additional modifications whenever their construction involves the values of $D_i$ for non-selected individuals ($\tilde{Z}_i=0$). In particular, under the assumptions of Theorem~\ref{theorem:ate}, ATT remains identified by:
\begin{align}
    \mathrm{ATT}=\frac{\mathbb{E}\left[P_i^{0(d)}\left(\text{E}_i^{0(d)}-\text{E}_i^{0(d^*)}\right)\right]}{\mathbb{E}\left(P_i^{0(d)}\right)}.
\end{align}

Excluding $D_i$ from the selection functions and thresholds may, however, be restrictive in some applications. For example, suppose that a researcher studies the effect of household income (categorized into income groups) on expenditure using survey data in which both variables are observed only for households completing a questionnaire on their financial circumstances. In this case, the probability of completing the questionnaire may depend directly on household income, i.e., on the treatment variable itself.

\section{The effect of education on the wage gap between spouses}\label{sec:example}

We use data from the Russia Longitudinal Monitoring Survey (RLMS)\footnote{``Russia Longitudinal Monitoring Survey, RLMS-HSE'', conducted by National Research University ``Higher School of Economics'' and OOO ``Demoscope'' together with Carolina Population Center, University of North Carolina at Chapel Hill and the Institute of Sociology of the Federal Center of Theoretical and Applied Sociology of the Russian Academy of Sciences. (RLMS-HSE web sites: https://rlms-hse.cpc.unc.edu, https://www.hse.ru/org/hse/rlms). We use both individual and household datasets available at https://www.hse.ru/en/rlms/downloads.} for the years 2023--2024 (waves 32 and 33) to estimate the effect of the wife's education level $D_i$ (2 --- university, 1 --- vocational, 0 --- lower) on the difference between the log hourly wage rates of spouses $Y_i$, i.e., the log of the ratio of the wife's wage rate to that of her husband. Our initial sample comprises 4410 couple-year observations with complete covariates, with both spouses aged $25$--$60$. The outcome is observable only if both spouses are employed and both wage rates are available. Let $Z_{1i}$ and $Z_{2i}$ denote binary indicators of the employment status (1 --- employed, 0 --- not employed) of the wife and husband, respectively. The relationship between education and the wage gap between spouses is of particular interest in the Russian context, given persistent positive educational assortative mating and the growing share of unions in which the wife is more educated than the husband \citep{Zinchenko2021}.

We use a set of control variables including age, squared age, self-rated health, and personality traits\footnote{Following \citet{Rozhkova2025}, we construct the Big Five measures of openness, conscientiousness, extraversion, agreeableness, and neuroticism.} of both spouses. Moreover, we include the number of children younger than 18 years old, a binary indicator for the year 2024, and dummy variables for the place of residence (regional center, city, urban-type settlement, or rural area) and the education level of the husband (university, vocational, or lower). Similarly to \citet{Mroz1987}, we use the number of children aged 6 years or younger in the household as an exclusion restriction. Following \citet{Dolgikh2024}, we additionally use gender-specific regional employment rates as exclusion restrictions. Descriptive statistics are provided in Table~\ref{tbl:descriptive}, where the reference categories of the dummy variables are omitted for brevity. The causal interpretation of the estimates relies on the identification assumptions in Section~\ref{sec:identification} and treats the included covariates as fixed with respect to the education contrast.

As in Section~\ref{sec:sim}, we estimate $\text{E}_i^{0(d)}$ and $\text{E}_i^{0(d^*)}$ using ridge regression and the treatment probabilities $P_i^{0(d)}$ and $P_i^{0(d^*)}$ using logistic regression. We estimate the selection probabilities $P_i^0$ and $\tilde{P}_i^0$ using random forests in the first specification and logistic regression in the second, pooling the treatment groups and including treatment as a regressor. We use five folds for cross-fitting and repeat the estimation procedure ten times, reporting the median estimate and its associated standard error. Since the same woman may be observed in both waves, we assign all her observations to the same fold at each stage of cross-fitting, including the nested splits, and cluster the standard errors by her individual identifier.

In addition to the estimators considered in Section~\ref{sec:sim}, we consider DML-4S, which accounts for the fact that employment does not guarantee observability of the wage rate: among dual-earner couples, the wife's wage rate is missing in approximately 20\% of cases and the husband's in approximately 27\%. For each spouse, employment and wage-rate observability constitute a separate sequential selection mechanism. Let $Z_{3i}$ and $Z_{4i}$ denote binary indicators of the observability of the wife's and husband's wage rates conditional on their respective employment. We construct four observable indicators $Z_{1i}$, $Z_{2i}$, $Z_{1i}Z_{3i}$, and $Z_{2i}Z_{4i}$, setting the last two products to zero when the corresponding spouse is not employed. Identification follows by extending the arguments for nested sequential selection in Section~\ref{sec:nested} to these two parallel sequences of selection equations, allowing for dependence between the random errors of all four equations. We additionally include two binary exclusion restrictions indicating whether another household member or a person outside the household is present during the wife's interview, as their presence may affect reporting behavior.

All estimators use the same initial sample with complete covariates and selection indicators. However, for the estimators other than DML-4S, we exclude dual-earner couples for whom at least one wage rate is missing, so that the observability of the outcome is determined by the employment statuses of the spouses alone. DML-4S explicitly accounts for these additional selection mechanisms.

The estimates are provided in Table~\ref{tbl:example}. Higher education of the wife is associated with a higher wife-to-husband wage ratio. Across all estimators and both specifications, the estimated effects of moving from vocational to university education are larger than those of moving from lower to vocational education. For example, the DML estimates of the ATE for the upper education contrast are 0.182 and 0.181 when the selection probabilities are estimated using random forests and logistic regression, respectively, corresponding to an increase in the geometric mean of the wife-to-husband wage ratio of approximately 20\%. All estimates for this education contrast are statistically significant at the 5\% level or better, whereas those for the lower education contrast are less precise.

Accounting for the selection mechanisms separately appears particularly relevant for the upper education contrast. When logistic regression is used to estimate the selection probabilities, DML-1S yields an ATE estimate of 0.230, which is approximately 27\% larger than the corresponding DML estimate of 0.181. The DML-4S estimate is 0.139. When random forests are used instead, the ATE estimates of DML and DML-4S are close (0.182 and 0.175, respectively), while DML-4S yields a larger estimate for the lower education contrast (0.131 compared with 0.107). Thus, accounting for wage-rate observability may affect both the magnitude and precision of the estimates, although differences in the estimation samples and the sets of exclusion restrictions may also contribute to these differences.

The DML-NG and DML-NN estimates are generally close to those of DML, suggesting that the derivative correction terms and nested cross-fitting have a relatively modest effect on the point estimates in this particular empirical application. Overall, the qualitative conclusions are robust to the choice of the selection-probability estimator, but the differences across selection specifications illustrate the potential empirical importance of modeling the individual selection mechanisms separately.

\begin{table}[H]
\centering
\scriptsize
\setlength{\tabcolsep}{3pt}
\renewcommand{\arraystretch}{0.88}
\setlength{\aboverulesep}{2pt}
\setlength{\belowrulesep}{2pt}
\caption{Descriptive statistics by wife's education level}
\label{tbl:descriptive}
\begin{tabular}{l cc cc cc l}
\toprule
 & \multicolumn{2}{c}{$D_{i} = 0$} & \multicolumn{2}{c}{$D_{i} = 1$} & \multicolumn{2}{c}{$D_{i} = 2$} & \\
\cmidrule(lr){2-3} \cmidrule(lr){4-5} \cmidrule(lr){6-7}
 & Wife & Husband & Wife & Husband & Wife & Husband & Role \\
\midrule
Log-wage ratio & \multicolumn{2}{c}{-0.360} & \multicolumn{2}{c}{-0.279} & \multicolumn{2}{c}{-0.190} & $Y_{i}$ \\
 & \multicolumn{2}{c}{(0.023)} & \multicolumn{2}{c}{(0.022)} & \multicolumn{2}{c}{(0.017)} &  \\
Employment & 0.612 & 0.815 & 0.739 & 0.884 & 0.830 & 0.922 & $Z_{i}$ \\
(1 --- employed, 0 --- unemployed) & (0.015) & (0.012) & (0.015) & (0.011) & (0.010) & (0.007) &  \\
Age & 43.373 & 45.315 & 43.254 & 44.788 & 41.232 & 42.975 & $X_{i}$ \\
 & (0.264) & (0.265) & (0.312) & (0.306) & (0.212) & (0.221) &  \\
Age$^{2}$ / 100 & 19.515 & 21.240 & 19.524 & 20.843 & 17.625 & 19.144 & $X_{i}$ \\
 & (0.230) & (0.239) & (0.268) & (0.272) & (0.179) & (0.193) &  \\
Self-rated health & 2.401 & 2.463 & 2.440 & 2.501 & 2.502 & 2.568 & $X_{i}$ \\
(0 --- worst, 4 --- best) & (0.019) & (0.019) & (0.021) & (0.020) & (0.015) & (0.016) &  \\
Openness & 1.844 & 1.792 & 1.945 & 1.877 & 2.118 & 1.990 & $X_{i}$ \\
 & (0.016) & (0.017) & (0.018) & (0.019) & (0.012) & (0.014) &  \\
Conscientiousness & 1.764 & 1.781 & 1.848 & 1.777 & 1.924 & 1.913 & $X_{i}$ \\
 & (0.013) & (0.014) & (0.014) & (0.014) & (0.011) & (0.011) &  \\
Extraversion & 1.780 & 1.584 & 1.798 & 1.545 & 1.828 & 1.627 & $X_{i}$ \\
 & (0.016) & (0.018) & (0.018) & (0.019) & (0.015) & (0.016) &  \\
Agreeableness & 1.807 & 1.688 & 1.824 & 1.694 & 1.890 & 1.722 & $X_{i}$ \\
 & (0.014) & (0.016) & (0.016) & (0.017) & (0.012) & (0.013) &  \\
Neuroticism & 1.176 & 1.096 & 1.084 & 1.031 & 1.027 & 0.961 & $X_{i}$ \\
 & (0.012) & (0.012) & (0.013) & (0.012) & (0.010) & (0.010) &  \\
Year 2024 & \multicolumn{2}{c}{0.491} & \multicolumn{2}{c}{0.489} & \multicolumn{2}{c}{0.483} & $X_{i}$ \\
 & \multicolumn{2}{c}{(0.016)} & \multicolumn{2}{c}{(0.017)} & \multicolumn{2}{c}{(0.013)} &  \\
Regional center & \multicolumn{2}{c}{0.353} & \multicolumn{2}{c}{0.340} & \multicolumn{2}{c}{0.539} & $X_{i}$ \\
 & \multicolumn{2}{c}{(0.015)} & \multicolumn{2}{c}{(0.016)} & \multicolumn{2}{c}{(0.013)} &  \\
City & \multicolumn{2}{c}{0.266} & \multicolumn{2}{c}{0.328} & \multicolumn{2}{c}{0.284} & $X_{i}$ \\
 & \multicolumn{2}{c}{(0.014)} & \multicolumn{2}{c}{(0.016)} & \multicolumn{2}{c}{(0.012)} &  \\
Urban-type settlement & \multicolumn{2}{c}{0.057} & \multicolumn{2}{c}{0.082} & \multicolumn{2}{c}{0.040} & $X_{i}$ \\
 & \multicolumn{2}{c}{(0.007)} & \multicolumn{2}{c}{(0.010)} & \multicolumn{2}{c}{(0.005)} &  \\
Number of minor children & \multicolumn{2}{c}{1.087} & \multicolumn{2}{c}{0.879} & \multicolumn{2}{c}{1.106} & $X_{i}$ \\
 & \multicolumn{2}{c}{(0.037)} & \multicolumn{2}{c}{(0.033)} & \multicolumn{2}{c}{(0.026)} &  \\
Vocational education &  & 0.196 &  & 0.301 &  & 0.236 & $X_{i}$ \\
 &  & (0.013) &  & (0.016) &  & (0.011) &  \\
University education &  & 0.104 &  & 0.230 &  & 0.474 & $X_{i}$ \\
 &  & (0.010) &  & (0.015) &  & (0.013) &  \\
Number of children aged 6 or younger & \multicolumn{2}{c}{0.320} & \multicolumn{2}{c}{0.272} & \multicolumn{2}{c}{0.299} & $W_{i}^{(Z)}$ \\
 & \multicolumn{2}{c}{(0.020)} & \multicolumn{2}{c}{(0.019)} & \multicolumn{2}{c}{(0.014)} &  \\
Regional employment rate & 0.618 & 0.708 & 0.627 & 0.714 & 0.651 & 0.741 & $W_{i}^{(Z)}$ \\
 & (0.003) & (0.003) & (0.003) & (0.003) & (0.002) & (0.002) &  \\
\bottomrule
\end{tabular}
\par\medskip
\parbox{\textwidth}{\scriptsize \textit{Note.} The table reports 
sample means and standard errors. The latter are shown in 
parentheses. Dual-earner couples with missing wage rates are excluded.}
\end{table}

\begin{table}[H]
\centering
\footnotesize
\caption{The estimates of the effect of wife's education on the log wage ratio}
\label{tbl:example}
\begin{tabular}{l cc cc cc cc}
\toprule
 & \multicolumn{4}{c}{$d^{*} = 0$, $d = 1$} & \multicolumn{4}{c}{$d^{*} = 1$, $d = 2$} \\
\cmidrule(lr){2-5} \cmidrule(lr){6-9}
 & ATES & ATTS & ATE & ATT & ATES & ATTS & ATE & ATT \\
\midrule
\multicolumn{9}{l}{\textit{Selection probabilities: random forest}} \\
DML & 0.086$^{**}$ & 0.110$^{**}$ & 0.107$^{**}$ & 0.126$^{**}$ & 0.175$^{***}$ & 0.172$^{***}$ & 0.182$^{***}$ & 0.179$^{***}$ \\
 & (0.043) & (0.045) & (0.047) & (0.054) & (0.042) & (0.038) & (0.048) & (0.044) \\
DML-NG & 0.084$^{*}$ & 0.110$^{**}$ & 0.103$^{**}$ & 0.126$^{***}$ & 0.170$^{***}$ & 0.166$^{***}$ & 0.178$^{***}$ & 0.177$^{***}$ \\
 & (0.046) & (0.045) & (0.046) & (0.047) & (0.040) & (0.044) & (0.043) & (0.049) \\
DML-NN & 0.096$^{**}$ & 0.119$^{***}$ & 0.119$^{***}$ & 0.140$^{***}$ & 0.172$^{***}$ & 0.175$^{***}$ & 0.179$^{***}$ & 0.179$^{***}$ \\
 & (0.039) & (0.037) & (0.045) & (0.049) & (0.038) & (0.037) & (0.042) & (0.043) \\
DML-1S & 0.094$^{**}$ & 0.112$^{*}$ & 0.109$^{**}$ & 0.126$^{**}$ & 0.185$^{***}$ & 0.176$^{***}$ & 0.192$^{***}$ & 0.188$^{***}$ \\
 & (0.046) & (0.061) & (0.054) & (0.053) & (0.042) & (0.037) & (0.051) & (0.041) \\
DML-0S & 0.093$^{*}$ & 0.111$^{*}$ & 0.093$^{*}$ & 0.111$^{*}$ & 0.158$^{***}$ & 0.161$^{***}$ & 0.158$^{***}$ & 0.161$^{***}$ \\
 & (0.053) & (0.058) & (0.053) & (0.058) & (0.037) & (0.041) & (0.037) & (0.041) \\
DML-4S & 0.107$^{**}$ & 0.126$^{***}$ & 0.131$^{***}$ & 0.139$^{***}$ & 0.168$^{***}$ & 0.168$^{***}$ & 0.175$^{***}$ & 0.168$^{***}$ \\
 & (0.043) & (0.042) & (0.048) & (0.050) & (0.041) & (0.039) & (0.047) & (0.045) \\
\midrule
\multicolumn{9}{l}{\textit{Selection probabilities: logistic regression}} \\
DML & 0.086 & 0.114$^{**}$ & 0.088$^{*}$ & 0.110$^{**}$ & 0.178$^{***}$ & 0.179$^{***}$ & 0.181$^{***}$ & 0.183$^{***}$ \\
 & (0.055) & (0.044) & (0.051) & (0.044) & (0.049) & (0.039) & (0.060) & (0.040) \\
DML-NG & 0.085 & 0.113$^{**}$ & 0.086$^{*}$ & 0.111$^{**}$ & 0.177$^{***}$ & 0.171$^{***}$ & 0.180$^{***}$ & 0.177$^{***}$ \\
 & (0.054) & (0.045) & (0.051) & (0.044) & (0.053) & (0.037) & (0.058) & (0.038) \\
DML-NN & 0.085$^{*}$ & 0.113$^{***}$ & 0.090$^{*}$ & 0.111$^{**}$ & 0.183$^{***}$ & 0.182$^{***}$ & 0.178$^{***}$ & 0.187$^{***}$ \\
 & (0.048) & (0.044) & (0.047) & (0.044) & (0.045) & (0.038) & (0.060) & (0.039) \\
DML-1S & 0.096$^{*}$ & 0.119$^{**}$ & 0.093$^{*}$ & 0.119$^{**}$ & 0.208$^{***}$ & 0.183$^{***}$ & 0.230$^{***}$ & 0.191$^{***}$ \\
 & (0.053) & (0.048) & (0.050) & (0.048) & (0.054) & (0.037) & (0.069) & (0.038) \\
DML-0S & 0.093$^{*}$ & 0.111$^{*}$ & 0.093$^{*}$ & 0.111$^{*}$ & 0.158$^{***}$ & 0.161$^{***}$ & 0.158$^{***}$ & 0.161$^{***}$ \\
 & (0.053) & (0.058) & (0.053) & (0.058) & (0.037) & (0.041) & (0.037) & (0.041) \\
DML-4S & 0.084 & 0.109$^{*}$ & 0.081 & 0.097 & 0.119$^{**}$ & 0.153$^{***}$ & 0.139$^{**}$ & 0.162$^{***}$ \\
 & (0.074) & (0.058) & (0.071) & (0.061) & (0.060) & (0.031) & (0.068) & (0.033) \\
\bottomrule
\end{tabular}
\par\medskip
\parbox{\textwidth}{\footnotesize \textit{Note.} $^{***}$ --- $p < 0.01$, 
$^{**}$ --- $p < 0.05$, $^{*}$ --- $p < 0.10$, standard errors 
clustered by wife are provided in parentheses.}
\end{table}

\section{Conclusion}\label{sec:conclusion}

We have proposed DML estimators of ATES, ATTS, ATE, and ATT in the MSSM with multiple ordinal selection equations. The proposed estimation procedure incorporates correction terms accounting for the estimation of generated regressors and employs nested cross-fitting. Under the established regularity conditions, we have proved the asymptotic normality of the proposed estimators. We have also established alternative conditions under which asymptotically normal estimators can be obtained without estimating derivatives of the nuisance functions. The simulation results demonstrate the importance of accounting for multiple selection mechanisms separately and illustrate the advantages of the proposed estimation procedure. These results contribute to the literature on MSSMs \citep{Poirier, DMF, Bourguignon, Cinzia, Tauchmann, DeLuca, Li, Ogundimu, Kim, Kossova2018, Kupriianova2020, Rezaee, Kossova2022, Dolgikh2026gender, Morozov2026} and the recent literature on DML estimation of sample selection models \citep{Bia, Potanin2026, Bjelac}.

The proposed estimators are useful in settings in which an outcome is observable only when several selection conditions hold simultaneously. Examples include wages observed only when a worker is both employed and willing to report pay \citep{Kossova2018}, or food expenditure observed only when a household discloses both income and spending \citep{DeLuca, Potanin2026}. The analysis of the spousal wage gap in this study illustrates a class of applications in which the outcome is constructed from several variables subject to distinct selection mechanisms. Another example is the analysis of the effect of a policy on the gap between chief executive officer (CEO) pay and median worker pay within a firm when the two figures are subject to separate disclosure decisions. Moreover, the proposed framework accommodates more general selection mechanisms and allows researchers to investigate treatment effect heterogeneity with respect to potentially endogenous ordinal moderators.

Promising directions for future research include developing DML estimators of other causal parameters, such as the local average treatment effect (LATE) in the MSSM, and adapting the proposed estimation procedure to other types of selection mechanisms, such as those arising in multinomial sample selection models \citep{Bourguignon, Kossova2022}. Another direction is to investigate estimation strategies that relax the smoothness requirements for nuisance functions without compromising the asymptotic properties of the proposed estimators.

\section*{Disclosure Statement}

The authors have no competing interests to declare.

\section*{Use of generative artificial intelligence}

During the preparation of this manuscript, the authors used AI to improve the clarity and readability of the text and to assist in checking mathematical derivations and programming code. The authors reviewed and verified the AI-assisted content and take full responsibility for the accuracy and integrity of the manuscript.

\section*{Funding}

This work was supported by the Russian Science Foundation [25-78-00095].

\clearpage
\appendix
\numberwithin{equation}{section}
\small

\section{Proof of Lemma 1}

Recall that $Q_i = (D_i, X_i, W_i^{(Z)})$. According to Assumption~\ref{assumption:exogeneity}, we have $(Y_{di}, U_i^{(Z)}) \perp (D_i, W_i^{(Z)}) \mid X_i$. By the decomposition property, this implies that $U_{ji}^{(Z)} \perp (W_i^{(Z)}, D_i) \mid X_i$, so almost surely $F_{U_{ji}^{(Z)} \mid D_i, X_i, W_i^{(Z)}}(u) = F_{U_{ji}^{(Z)} \mid X_i}(u)$ for all $u \in \mathbb{R}$. By combining this fact with Assumption~\ref{assumption:selection}, we obtain:
\begin{align}
\label{eq:L1-1}
        P_{1ji}^{0} &= \underbrace{\mathbb{P}\left(Z_{ji} \le z_{j}^{*} \mid D_i, X_i, W_i^{(Z)}\right)}_{P_{1ji}^{0} \text{ is the $j$-th element of } P_{1i}^{0}} = \nonumber \\
        &= \underbrace{\mathbb{P}\left(U_{ji}^{(Z)} \le g_j\left(Q_i\right) + c_{jz_{j}^{*}}\left(Q_i\right) \mid Q_{i}\right)}_{\text{Assumption~\ref{assumption:selection}}} = \nonumber \\
        &= \underbrace{F_{U_{ji}^{(Z)} \mid Q_{i}}\left(g_j\left(Q_i\right) + c_{jz_{j}^{*}}\left(Q_i\right)\right)}_{\text{By definition of CDF}} = \nonumber \\
        &= \underbrace{F_{U_{ji}^{(Z)} \mid X_i}\left(g_j\left(Q_i\right) + c_{jz_{j}^{*}}\left(Q_i\right)\right)}_{\text{Since } U_{ji}^{(Z)} \perp (W_i^{(Z)}, D_i) \mid X_i},
\end{align}
where $c_{j0}\left(Q_i\right) = 0$, $c_{j(-1)}\left(Q_i\right) = -\infty$, $c_{j(n_{Z_j} - 1)}\left(Q_i\right) = \infty$ and $z_{j}^{*}$ is the $j$-th element of $z^{*}$. Throughout the proof, we extend cumulative distribution functions to $[-\infty,\infty]$ by setting $F(-\infty)=0$ and $F(\infty)=1$.

Define the generalized inverse of the conditional cumulative distribution function as follows:
\begin{align}
F_{U_{ji}^{(Z)} \mid X_i}^{-1}(p) =
\begin{cases}
    -\infty, & \text{if } p = 0,\\
    \inf\left\{u:
        F_{U_{ji}^{(Z)} \mid X_i}(u)\geq p
    \right\}, & \text{if } p\in(0,1),\\
    +\infty, & \text{if } p = 1.
\end{cases}
\end{align}
Notice that for any random variable $U$ with cumulative distribution function $F\left(\cdot\right)$ and any constant $a\in\mathbb{R}$ such that $F(a)\in(0,1)$, by the monotonicity and right-continuity of $F\left(\cdot\right)$, we have
$F^{-1}(F(a))\leq a$ and $F(F^{-1}(F(a)))=F(a)$. Hence, $\mathbb{P}(F^{-1}(F(a))<U\leq a)=0$, implying that the following equality holds almost surely:
\begin{align}
\label{eq:L1-1B}
    \mathbb{I}(U\leq a)
    = \mathbb{I}(U\leq F^{-1}(F(a))).
\end{align}
Moreover, under the adopted definition of the generalized inverse, the same equality holds whenever $F(a)\in\{0,1\}$, including the cases $a=-\infty$ and $a=\infty$.

Therefore, applying this result conditionally on $Q_i$ and using Assumption~\ref{assumption:exogeneity}, equality~(\ref{eq:L1-1}), and equality~(\ref{eq:L1-1B}), we obtain that the following equality holds almost surely:
\begin{align}
\label{eq:L1-2}
    \mathbb{I}\left(
        U_{ji}^{(Z)} \leq g_j(Q_i)+c_{jz_j^*}(Q_i)
    \right)
    =
    \mathbb{I}\left(
        U_{ji}^{(Z)} \leq
        F_{U_{ji}^{(Z)}\mid X_i}^{-1}(P_{1ji}^{0})
    \right).
\end{align}
By the same steps, we obtain that the following equality also holds almost surely:
\begin{align}
\label{eq:L1-2B}
    \mathbb{I}\left(
        U_{ji}^{(Z)} \leq
        g_j(Q_i)+c_{j(z_j^*-1)}(Q_i)
    \right)
    =
    \mathbb{I}\left(
        U_{ji}^{(Z)} \leq
        F_{U_{ji}^{(Z)}\mid X_i}^{-1}(P_{0ji}^{0})
    \right),
\end{align}
where $P_{0ji}^{0} = \mathbb{P}(Z_{ji}\leq z_j^*-1\mid Q_i)$ is the $j$-th element of $P_{0i}^{0}$.

Equalities~(\ref{eq:L1-2}) and~(\ref{eq:L1-2B}) imply that the random variable $\mathbb{I}(Z_{ji}=z_j^*)$ is almost surely equal to a measurable function of $(X_i,P_{0ji}^{0},P_{1ji}^{0},U_{ji}^{(Z)})$, because the following equality holds almost surely:
\begin{align}
        \mathbb{I}\left(Z_{ji} = z_{j}^{*}\right) & = \mathbb{I}\left(g_j\left(Q_i\right) + c_{j(z_{j}^{*}-1)}\left(Q_i\right) < U_{ji}^{(Z)} \le g_j\left(Q_i\right) + c_{jz_{j}^{*}}\left(Q_i\right)\right) = \nonumber \\
        &= \underbrace{\mathbb{I}\left(F_{U_{ji}^{(Z)} \mid X_i}^{-1}(P_{0ji}^{0}) < U_{ji}^{(Z)} \le F_{U_{ji}^{(Z)} \mid X_i}^{-1}(P_{1ji}^{0})\right)}_{\text{Depends only on }X_i, {P}_{0ji}^{0}, {P}_{1ji}^{0}\text{, and }U_{ji}^{(Z)}}.
\end{align}

Hence, the random variable $\tilde{Z}_i$ is almost surely equal to a measurable function of $(X_i,P_{0i}^{0},P_{1i}^{0},U_i^{(Z)})$, since:
\begin{align}
\tilde{Z}_{i}=\mathbb{I}\left(Z_{i} = z^{*}\right) = \underbrace{\mathbb{I}(Z_{1i} = z_1^{*}) \times \mathbb{I}(Z_{2i} = z_2^{*}) \times \cdots \times \mathbb{I}(Z_{n_Z i} = z_{n_Z}^{*})}_{\text{The product of the functions of }X_i, {P}_{0ji}^{0}, {P}_{1ji}^{0}\text{, and }U_{ji}^{(Z)}}.
\end{align}

Because $P_i^0=(P_{0i}^0,P_{1i}^0)$ is a measurable function of $(X_i,D_i,W_i^{(Z)})$ and $\tilde{Z}_i$ is almost surely equal to a measurable function of $(X_i,P_i^0,U_i^{(Z)})$, we have:
    \begin{align}
        \underbrace{(Y_{di}, U_i^{(Z)}) \perp (D_i, W_i^{(Z)}) \mid X_i}_{\text{Assumption~\ref{assumption:exogeneity}}} &\implies \underbrace{(Y_{di}, U_i^{(Z)}) \perp (D_i, W_i^{(Z)}, P_{i}^{0}) \mid X_{i}}_{\text{Since $P_i^0$ is a function of $(D_{i},W_i^{(Z)},X_{i})$}} \implies \nonumber \\ %
         &\implies \underbrace{(Y_{di}, U_i^{(Z)}) \perp (D_i, W_i^{(Z)}) \mid (X_i, P_{i}^{0})}_{\text{Weak union}} \implies \nonumber \\ %
         &\implies \underbrace{(Y_{di}, \tilde{Z}_{i}) \perp \left(D_i, W_{i}^{(Z)}\right) \mid (X_i, P_{i}^{0})}_{\text{Since }\tilde{Z}_{i}\text{ is a function of }(X_{i}, P_{i}^{0}, U_{i}^{(Z)})} \implies \nonumber \\ &\implies \underbrace{Y_{di} \perp \left(D_i, W_{i}^{(Z)}\right) \mid (X_i, P_{i}^{0}, \tilde{Z}_{i})}_{\text{Symmetry and weak union}}.
\end{align}

\section{Proof of Lemma 2}

Consider $d \in \operatorname{supp}(D_i)$. Below we prove the equalities of the lemma.

\subsection{Proof of the first equation}

Recall that $Y_i=Y_{di}$ and $P_i^0=p_Z^0(d,X_i,W_i^{(Z)})$ whenever $D_i=d$. The formula follows from the following steps, where all equalities hold almost surely with respect to the conditional distribution of $(X_i,W_i^{(Z)})$ given $D_i=d$ and $\tilde{Z}_i=1$:
\begin{align}
    \mu_{Y}^{0}\left(d, X_{i}, p_{Z}^{0}\left(d, X_{i}, W_{i}^{(Z)}\right)\right)
    &= \underbrace{\mathbb{E}\left(Y_{i}\mid D_{i} = d,X_{i}, p_{Z}^{0}\left(d, X_{i}, W_{i}^{(Z)}\right), \tilde{Z}_{i} = 1\right)}_{\text{By definition}} = \nonumber \\ 
    &= \underbrace{\mathbb{E}\left(Y_{i}\mid D_{i} = d,X_{i}, p_{Z}^{0}\left(d, X_{i}, W_{i}^{(Z)}\right), W_{i}^{(Z)}, \tilde{Z}_{i} = 1\right)}_{\substack{\text{By Lemma~\ref{lemma:independence} and weak union:}\\Y_{di}\perp W_{i}^{(Z)}\mid D_{i},X_{i},P_{i}^{0},\tilde{Z}_{i}}} = \nonumber \\
    &= \underbrace{\mathbb{E}\left(Y_{i}\mid D_{i} = d, X_{i}, W_{i}^{(Z)}, \tilde{Z}_{i} = 1\right)}_{\text{Since }P_{i}^{0}\text{ is a function of }D_{i}, X_{i}, W_{i}^{(Z)}}.
\end{align}

\subsection{Proof of the second equation}

The formula follows from the following steps, where all equalities hold almost surely with respect to the conditional distribution of $(X_i,P_i^0)$ given $\tilde{Z}_i=1$:
\begin{align}
    p_{D}^{0}\left(d, X_{i}, P_{i}^{0}\right) = \underbrace{\mathbb{P}\left(D_{i} = d\mid X_{i}, P_{i}^{0}\right)}_{\text{By definition}} = \underbrace{\mathbb{P}\left(D_{i} = d\mid X_{i}, P_{i}^{0}, \tilde{Z}_{i} = 1\right)}_{\text{Lemma~\ref{lemma:independence} implies }\tilde{Z}_{i}\perp D_{i}\mid X_{i}, P_{i}^{0}}.
\end{align}

\subsection{Proof of the third equation}

The formula follows from the following steps, where all equalities hold almost surely:
\begin{align}
    \tilde{p}_{Z}^{0}\left(D_{i}, X_{i}, W_{i}^{(Z)}\right) &= \underbrace{\mathbb{P}\left(\tilde{Z}_{i} = 1\mid D_{i}, X_{i}, W_{i}^{(Z)}\right)}_{\text{By definition}} = \underbrace{\mathbb{P}\left(\tilde{Z}_{i} = 1\mid D_{i}, X_{i}, W_{i}^{(Z)}, P_{i}^{0}\right)}_{\text{Since }P_{i}^{0}\text{ is a function of }D_{i}, X_{i}, W_{i}^{(Z)}} =  \nonumber \\ 
    &= \underbrace{\mathbb{P}\left(\tilde{Z}_{i} = 1\mid D_{i}, X_{i}, P_{i}^{0}\right)}_{\substack{\text{By Lemma~\ref{lemma:independence} and weak union:}\\\tilde{Z}_{i}\perp W_{i}^{(Z)}\mid D_{i},X_{i},P_{i}^{0}}} = \underbrace{\mathbb{P}\left(\tilde{Z}_{i} = 1\mid X_{i}, P_{i}^{0}\right)}_{\substack{\text{Lemma~\ref{lemma:independence} implies }\\\tilde{Z}_{i}\perp D_{i}\mid X_{i}, P_{i}^{0}}}.
\end{align}

\section{Proof of Lemma 3}

Here and throughout the subsequent proofs, we use the following abbreviations:
\begin{enumerate}
    \item LTP --- the law of total probability.
    \item LIE --- the law of iterated expectations.
    \item LTE --- the law of total expectation. 
\end{enumerate}

For every $\overline{d}\in \{0,\cdots,n_{D} - 1\}$, define for brevity $D_{i}^{(\overline{d})} = \mathbb{I}\left(D_{i} = \overline{d}\right)$ and recall that:
\begin{align}
    \overline{Z}_{i} = \left(\mathbb{I}\left(Z_{1i}\leq z_{1}^{*} - 1\right),\cdots,\mathbb{I}\left(Z_{n_{Z}i}\leq z_{n_{Z}}^{*} - 1\right),\mathbb{I}\left(Z_{1i}\leq z_{1}^{*}\right),\cdots,\mathbb{I}\left(Z_{n_{Z}i}\leq z_{n_Z}^{*}\right)\right).
\end{align}

For $\epsilon\in[0,1)$, define the perturbations of the nuisance functions in the direction $\mu-\mu^0$:
\begin{align}
    \mu_{Y}^{\epsilon}\left(d, x, p\right) &= \mu_{Y}^{0}\left(d, x, p\right) + \epsilon \left(\mu_{Y}\left(d, x, p\right) - \mu_{Y}^{0}\left(d, x, p\right)\right), \\ 
    p_{Z}^{\epsilon}(d,x,w)   &= p_{Z}^{0}(d,x,w) + \epsilon \left(p_{Z}(d,x,w) - p_{Z}^{0}(d,x,w)\right),  \\ 
    p_{D}^{\epsilon}(d,x,p)   &= p_{D}^{0}(d,x,p) + \epsilon \left(p_{D}(d,x,p) - p_{D}^{0}(d,x,p)\right), \\ 
    \tilde{p}_{Z}^{\epsilon}(d, x, w) &= \tilde{p}_{Z}^{0}(d, x, w) + \epsilon\left(\tilde{p}_{Z}(d, x, w) - \tilde{p}_{Z}^{0}(d, x, w)\right). 
\end{align}
Throughout the proof, derivatives with respect to $\epsilon$ evaluated at $\epsilon=0$ are understood as right-hand derivatives.

Notice that $P_i^\epsilon=p_Z^\epsilon(D_i,X_i,W_i^{(Z)})$ takes values in $[0,1]^{\dim(P_i^0)}$ since it is a convex combination of $P_i^0$ and $P_i$. Therefore, the differentiability properties imposed by Assumption~\ref{assumption:differentiability} apply at the perturbed control probabilities. Moreover, by the same assumption and the linearity of differentiation:
\begin{align}
    \frac{\partial\mu_Y^\epsilon(d,x,p)}{\partial p}
    &= \mu_{\partial Y}^0(d,x,p)
    + \epsilon\left(\mu_{\partial Y}(d,x,p)
    - \mu_{\partial Y}^0(d,x,p)\right)
    = \mu_{\partial Y}^{\epsilon}(d,x,p).
\end{align}

For each $\overline{d}\in\{d,d^{*}\}$, denote for brevity \footnote{The definitions hold wherever the corresponding quantities are well-defined. For example, $R_{\partial^2 D}^{\epsilon,\overline{d}}$ is defined under the additional assumptions we impose later in Theorem~\ref{theorem:normality} to establish asymptotic normality of the proposed estimators.}:
\begin{align}
    R_{D}^{\epsilon, \overline{d}} &= \frac{D_{i}^{(\overline{d})}}{p_{D}^{\epsilon}\left(\overline{d}, X_{i}, p_{Z}^{\epsilon}\left(D_{i}, X_{i}, W_{i}^{(Z)}\right)\right)}, \\
    R_{PD}^{\epsilon} &= \frac{p_{D}^{\epsilon}\left(d,X_{i},p_{Z}^{\epsilon}\left(D_{i},X_{i},W_{i}^{(Z)}\right)\right)} {p_{D}^{\epsilon}\left(d^{*},X_{i},p_{Z}^{\epsilon}\left(D_{i},X_{i},W_{i}^{(Z)}\right)\right)}, \\
    R_{\partial Y}^{\epsilon, \overline{d}} &= \left.\frac{\partial \mu_{Y}^{\epsilon}\left(\overline{d}, X_{i}, p\right)}{\partial p}\right|_{p = p_{Z}^{\epsilon}\left(D_{i},X_{i},W_{i}^{(Z)}\right)}, \\
    R_{\partial D}^{\epsilon, \overline{d}} &= \left.\frac{\partial p_{D}^{\epsilon}\left(\overline{d}, X_{i}, p\right)}{\partial p}\right|_{p = p_{Z}^{\epsilon}\left(D_{i},X_{i},W_{i}^{(Z)}\right)}, \\
    \triangle R_{\partial^2 Y}^{\epsilon, \overline{d}} &= \left(\frac{\partial^2 \mu_Y\left(\overline{d}, X_i, p\right)}{\partial p^2}\bigg|_{p=P_i^\epsilon}\right) - \left(\frac{\partial^2 \mu_Y^0\left(\overline{d}, X_i, p\right)}{\partial p^2}\bigg|_{p=P_i^\epsilon}\right),\\
    R_{\partial^2 D}^{\epsilon, \overline{d}} &= \left.\frac{\partial^2 p_D^\epsilon\left(\overline{d}, X_i, p\right)}{\partial p^2}\right|_{p = P_{i}^{\epsilon}}, \\ \triangle R_{\partial D}^{\epsilon, \overline{d}} &= \left(\left.\frac{\partial p_D\left(\overline{d}, X_i, p\right)}{\partial p}\right|_{p = P_{i}^{\epsilon}}\right) - \left(\left.\frac{\partial p_D^0\left(\overline{d}, X_i, p\right)}{\partial p}\right|_{p = P_{i}^{\epsilon}}\right),\\
    \triangle \mu_{Y}^{\overline{d}} &= \mu_{Y}\left(\overline{d}, X_{i}, p_{Z}^{0}\left(D_{i}, X_{i}, W_{i}^{(Z)}\right)\right) - \mu_{Y}^{0}\left(\overline{d}, X_{i}, p_{Z}^{0}\left(D_{i}, X_{i}, W_{i}^{(Z)}\right)\right), \\
    \triangle \mu_{Y}^{\overline{d}}\!\left(P_{i}^{\epsilon}\right) &= \mu_{Y}\left(\overline{d}, X_{i}, P_{i}^{\epsilon}\right) - \mu_{Y}^{0}\left(\overline{d}, X_{i}, P_{i}^{\epsilon}\right),\\
    \triangle p_{D}^{\overline{d}} &= p_{D}\left(\overline{d}, X_{i}, p_{Z}^{0}\left(D_{i}, X_{i}, W_{i}^{(Z)}\right)\right) - p_{D}^{0}\left(\overline{d}, X_{i}, p_{Z}^{0}\left(D_{i}, X_{i}, W_{i}^{(Z)}\right)\right), \\
    \triangle p_{Z} &= p_{Z}\left(D_{i}, X_{i}, W_{i}^{(Z)}\right) - p_{Z}^{0}\left(D_{i}, X_{i}, W_{i}^{(Z)}\right), \\
    \triangle \tilde{p}_{Z} &= \tilde{p}_{Z}\left(D_{i}, X_{i}, W_{i}^{(Z)}\right) - \tilde{p}_{Z}^{0}\left(D_{i}, X_{i}, W_{i}^{(Z)}\right),\\
    \triangle \mu_{Y}^{\epsilon} &= \mu_{Y}^{\epsilon}\left(d, X_{i}, p_{Z}^{\epsilon}\left(D_{i}, X_{i}, W_{i}^{(Z)}\right)\right) - \mu_{Y}^{\epsilon}\left(d^{*}, X_{i}, p_{Z}^{\epsilon}\left(D_{i}, X_{i}, W_{i}^{(Z)}\right)\right), \\
    \triangle Y^{\epsilon, \overline{d}} &= Y_{i} - \mu_{Y}^{\epsilon}\left(\overline{d}, X_{i}, p_{Z}^{\epsilon}\left(D_{i},X_{i},W_{i}^{(Z)}\right)\right), \\
    \triangle \overline{Z}^{\epsilon} &= \overline{Z}_{i} - P_{i}^{\epsilon}, \\
    P_{i}^{\epsilon} &= p_{Z}^{\epsilon}\left(D_{i}, X_{i}, W_{i}^{(Z)}\right), \\
    \tilde{P}_{i}^{\epsilon} &= \tilde{p}_{Z}^{\epsilon}\left(D_{i}, X_{i}, W_{i}^{(Z)}\right).
\end{align}

Moreover, denote by $\text{E}_{i}^{\epsilon(\overline{d})}$, $\text{E}_{\partial i}^{\epsilon(\overline{d})}$, and $P_{i}^{\epsilon(\overline{d})}$ the values of $\text{E}_{i}^{0(\overline{d})}$, $\text{E}_{\partial i}^{0(\overline{d})}$, and $P_{i}^{0(\overline{d})}$ calculated with $\mu^{0}$ replaced with $\mu^{\epsilon}$.

Throughout the proof, terms involving undefined ratios of probabilities are defined to be zero whenever they are multiplied by a zero selection probability, selection indicator, or treatment indicator. This convention is needed because Assumption~\ref{assumption:support} ensures positivity of the treatment probabilities among selected observations, whereas the derivative correction terms are not multiplied by $\tilde{Z}_i$. For example, the expression $\tilde{P}_{i}^{0}\text{E}_{\partial i}^{0(d^{*})}\left(1-\frac{D_{i}^{(d^{*})}}{P_{i}^{0(d^{*})}}\right)$ is defined to be zero whenever $\tilde{P}_{i}^{0}=0$. The same convention applies to expressions obtained by grouping such terms when both the corresponding selection probability and selection indicator are zero. Notice also that, by definition of the treatment probabilities, $P_i^{0(\overline{d})}>0$ almost surely whenever $D_i^{(\overline{d})}=1$. When differentiating products involving such ratios, we apply the same convention. In particular, whenever $D_i^{(d^*)}=0$, the product $D_i^{(d^*)}R_{PD}^{\epsilon}$ is identically zero along the perturbation path, and its derivative is therefore zero. Whenever $D_i^{(d^*)}=1$, the positivity of $P_i^{0(d^*)}$ and continuity imply that, almost surely, there exists a neighborhood of $\epsilon=0$ in which the denominator of $R_{PD}^{\epsilon}$ is positive.

\subsection{Neyman-orthogonality of the score of ATES}

\subsubsection{Mean zero of ATES score}

Consider the score at true values of the parameter and nuisance functions:
\begin{align}
    \psi_{\mathrm{ATES}}^{0} &= \underbrace{\tilde{Z}_{i}\left(\text{E}_{i}^{0(d)} - \text{E}_{i}^{0(d^{*})}\right) -\tilde{Z}_{i} \mathrm{ATES}}_{\psi^{0}_{1}} + \nonumber \\
        &+ \underbrace{\tilde{Z}_{i}\frac{D_{i}^{(d)}}{P_{i}^{0(d)}}\left(Y_{i} - \text{E}_{i}^{0(d)}\right)}_{\psi^{0}_{2}} - \underbrace{\tilde{Z}_{i}\frac{D_{i}^{(d^{*})}}{P_{i}^{0(d^{*})}}\left(Y_{i} - \text{E}_{i}^{0(d^{*})}\right)}_{\psi^{0}_{3}} + \nonumber \\ 
    &+\underbrace{\left(\overline{Z}_{i} - P_{i}^{0}\right)\tilde{P}_{i}^{0}\left(\text{E}_{\partial i}^{0(d)}\left( 1 - \frac{D_{i}^{(d)}}{P_{i}^{0(d)}}\right) - \text{E}_{\partial i}^{0(d^{*})}\left(1 - \frac{D_{i}^{(d^{*})}}{P_{i}^{0(d^{*})}}\right)\right)}_{\psi^{0}_{4}}.
\end{align}

By Assumption~\ref{assumption:differentiability}, all these terms are well-defined and integrable. Next, we show that the expectations of $\psi_{1}^{0}$, $\psi_{2}^{0}$, $\psi_{3}^{0}$, and $\psi_{4}^{0}$ are $0$. Specifically, notice that Theorem~\ref{theorem:ates} implies $\mathbb{E}\left(\psi_{1}^{0}\right) = 0$. Further, notice that:
\begin{align}
    \mathbb{E}\left(\psi_{2}^{0}\right) &= \mathbb{E}\left(\tilde{Z}_{i}\frac{D_{i}^{(d)}}{P_{i}^{0(d)}}\left(Y_{i} - \text{E}_{i}^{0(d)}\right)\right) = \nonumber \\
    &= \underbrace{\mathbb{E}\left(\frac{1}{P_{i}^{0(d)}}\left(Y_{i} - \text{E}_{i}^{0(d)}\right)\mid D_{i} = d, \tilde{Z}_{i} = 1\right)\mathbb{P}\left(D_{i} = d, \tilde{Z}_{i} = 1\right)}_{\substack{\text{LTE with the events }\{D_{i} = d, \tilde{Z}_{i} = 1\}\\ \text{ and }\{D_{i} \ne d\text{ or }\tilde{Z}_{i} \ne 1\}}} \propto \nonumber \\
    &\propto\underbrace{\mathbb{E}\left(\mathbb{E}\left(\frac{1}{P_{i}^{0(d)}}\left(Y_{i} - \text{E}_{i}^{0(d)}\right)\mid X_{i}, P_{i}^{0}, D_{i} = d, \tilde{Z}_{i} = 1\right)\mid D_{i} = d, \tilde{Z}_{i} = 1\right)}_{\text{LIE}} = \nonumber \\
    &=\underbrace{\mathbb{E}\left(\frac{1}{P_{i}^{0(d)}}\left(\mathbb{E}\left(Y_{i}\mid X_{i}, P_{i}^{0}, D_{i} = d, \tilde{Z}_{i} = 1\right) - \text{E}_{i}^{0(d)}\right)\mid D_{i} = d, \tilde{Z}_{i} = 1\right)}_{\text{Since }P_{i}^{0(d)}\text{ and }\text{E}_{i}^{0(d)}\text{ are functions of }X_{i}, P_{i}^{0}} = \nonumber \\
    &=\underbrace{\mathbb{E}\left(\frac{1}{P_{i}^{0(d)}}\left(\text{E}_{i}^{0(d)} - \text{E}_{i}^{0(d)}\right)\mid D_{i} = d, \tilde{Z}_{i} = 1\right)}_{\text{By the definition of }\text{E}_{i}^{0(d)}} = 0,
\end{align}
where here and throughout the proof we use the proportionality symbol $\propto$ for brevity of notation to omit positive constant factors, such as $\mathbb{P}(D_i=d,\tilde{Z}_i=1)$. By the same steps, we obtain $\mathbb{E}\left(\psi_{3}^{0}\right) = 0$. Finally, notice that:
\begin{align}
    \mathbb{E}\left(\psi_{4}^{0}\right) &= \mathbb{E}\left(\left(\overline{Z}_{i} - P_{i}^{0}\right)\underbrace{\tilde{P}_{i}^{0}\left(\text{E}_{\partial i}^{0(d)}\left( 1 - \frac{D_{i}^{(d)}}{P_{i}^{0(d)}}\right) - \text{E}_{\partial i}^{0(d^{*})}\left(1 - \frac{D_{i}^{(d^{*})}}{P_{i}^{0(d^{*})}}\right)\right)}_{A_{i}}\right) = \nonumber \\
    &= \underbrace{\mathbb{E}\left(\mathbb{E}\left(\left(\overline{Z}_{i} - P_{i}^{0}\right)A_{i}\mid D_{i}, X_{i}, W_{i}^{(Z)}\right)\right)}_{\text{LIE}} = \nonumber \\
    &= \underbrace{\mathbb{E}\left(\mathbb{E}\left(\overline{Z}_{i} - P_{i}^{0}\mid D_{i}, X_{i}, W_{i}^{(Z)}\right)A_{i}\right)}_{\text{Since }A_{i}\text{ is a function of }D_{i}, X_{i}, W_{i}^{(Z)}} = \nonumber \\
    &=\underbrace{\mathbb{E}\left(\mathbb{E}\left(P_{i}^{0} - P_{i}^{0}\mid D_{i}, X_{i}, W_{i}^{(Z)}\right)A_{i}\right)}_{\text{Since }\mathbb{E}\left(\overline{Z}_{i}\mid D_{i}, X_{i}, W_{i}^{(Z)}\right)=P_{i}^{0}} = 0.
\end{align}
Therefore, we have:
\begin{align}
    \mathbb{E}\left(\psi_{\mathrm{ATES}}^{0}\right)
    = \mathbb{E}\left(\psi_{1}^{0}\right)
    + \mathbb{E}\left(\psi_{2}^{0}\right)
    - \mathbb{E}\left(\psi_{3}^{0}\right)
    + \mathbb{E}\left(\psi_{4}^{0}\right)
    = 0 + 0 - 0 + 0 = 0.
\end{align}

\subsubsection{Zero Gateaux derivative of ATES score}

The perturbed score is as follows:
\begin{align}
    \psi_{\mathrm{ATES}}^{\epsilon} &= \underbrace{\tilde{Z}_{i}\triangle \mu_{Y}^{\epsilon} - \tilde{Z}_{i}\mathrm{ATES}}_{\psi_{1}^{\epsilon}} + \underbrace{\tilde{Z}_{i}R_{D}^{\epsilon, d}\triangle Y^{\epsilon, d}}_{\psi_{2}^{\epsilon}} - \underbrace{\tilde{Z}_{i}R_{D}^{\epsilon, d^{*}}\triangle Y^{\epsilon, d^{*}}}_{\psi_{3}^{\epsilon}} + \nonumber \\
    &+ \underbrace{\triangle \overline{Z}^{\epsilon}\tilde{P}_{i}^{\epsilon}\left(R_{\partial Y}^{\epsilon, d}\left(1 - R_{D}^{\epsilon, d}\right) - R_{\partial Y}^{\epsilon, d^{*}}\left(1 - R_{D}^{\epsilon, d^{*}}\right)\right)}_{\psi_{4}^{\epsilon}}.
\end{align}

By Assumption~\ref{assumption:differentiability}, we can differentiate each term by the chain rule. Specifically, using continuous differentiability of $\mu_Y$ with respect to $p$, we obtain the derivative associated with the first term:
\begin{align}
    \frac{\partial}{\partial \epsilon}\psi_{1}^{\epsilon}\mid_{\epsilon = 0} &= \underbrace{\tilde{Z}_{i}\left(\triangle \mu_{Y}^{d} - \triangle \mu_{Y}^{d^{*}}\right)}_{\psi_{1A}} + \underbrace{\tilde{Z}_{i}\triangle p_{Z}\left(R_{\partial Y}^{0, d} - R_{\partial Y}^{0, d^{*}}\right)}_{\psi_{1B}}. 
\end{align}

By using also the continuous differentiability of $p_D$ with respect to $p$, imposed by Assumption~\ref{assumption:differentiability}, we obtain the derivative associated with the second term:
\begin{align}
    \frac{\partial}{\partial \epsilon}\psi_{2}^{\epsilon}\mid_{\epsilon = 0} &= -\underbrace{\frac{D_{i}^{(d)}\tilde{Z}_{i}\triangle p_{D}^{d}\triangle Y^{0, d}}{\left(P_{i}^{0(d)}\right)^{2}}}_{\psi_{2A}} - \underbrace{\frac{D_{i}^{(d)}\tilde{Z}_{i}\triangle p_{Z}R_{\partial D}^{0, d}\triangle Y^{0, d}}{\left(P_{i}^{0(d)}\right)^{2}}}_{\psi_{2B}} - \nonumber \\ 
    &-\underbrace{\frac{D_{i}^{(d)}\tilde{Z}_{i}\triangle \mu_{Y}^{d}}{P_{i}^{0(d)}}}_{\psi_{2C}} - \underbrace{\frac{D_{i}^{(d)}\tilde{Z}_{i}\triangle p_{Z}R_{\partial Y}^{0, d}}{P_{i}^{0(d)}}}_{\psi_{2D}}. 
\end{align}

The derivative associated with the third term is as follows:
\begin{align}
    \frac{\partial}{\partial \epsilon}\psi_{3}^{\epsilon}\mid_{\epsilon = 0} &= -\underbrace{\frac{D_{i}^{(d^{*})}\tilde{Z}_{i}\triangle p_{D}^{d^{*}}\triangle Y^{0, d^{*}}}{\left(P_{i}^{0(d^{*})}\right)^{2}}}_{\psi_{3A}} - \underbrace{\frac{D_{i}^{(d^{*})}\tilde{Z}_{i}\triangle p_{Z}R_{\partial D}^{0, d^{*}}\triangle Y^{0, d^{*}}}{\left(P_{i}^{0(d^{*})}\right)^{2}}}_{\psi_{3B}} - \nonumber \\ 
    &-\underbrace{\frac{D_{i}^{(d^{*})}\tilde{Z}_{i}\triangle \mu_{Y}^{d^{*}}}{P_{i}^{0(d^{*})}}}_{\psi_{3C}} - \underbrace{\frac{D_{i}^{(d^{*})}\tilde{Z}_{i}\triangle p_{Z}R_{\partial Y}^{0, d^{*}}}{P_{i}^{0(d^{*})}}}_{\psi_{3D}}. 
\end{align}

Differentiation of the fourth term involves derivatives of $\text{E}_{\partial i}^{\epsilon(d)}$ and $\text{E}_{\partial i}^{\epsilon(d^{*})}$ with respect to $\epsilon$, which are well-defined due to twice continuous differentiability of $\mu_Y$ imposed by Assumption~\ref{assumption:differentiability}. Specifically, the derivative associated with the fourth term is as follows:
\begin{align}
    \frac{\partial}{\partial \epsilon}\psi_{4}^{\epsilon}\mid_{\epsilon = 0} &= -\underbrace{\triangle p_{Z}\tilde{P}_{i}^{0}\left(\text{E}_{\partial i}^{0(d)}\left(1 - \frac{D_{i}^{(d)}}{P_{i}^{0(d)}}\right) - \text{E}_{\partial i}^{0(d^{*})}\left(1 - \frac{D_{i}^{(d^{*})}}{P_{i}^{0(d^{*})}}\right)\right)}_{\psi_{4A}} + \nonumber \\
    &+\underbrace{\left(\overline{Z}_{i} - P_{i}^{0}\right)\left(\frac{\partial}{\partial \epsilon}\left[\tilde{P}_{i}^{\epsilon}\left(\text{E}_{\partial i}^{\epsilon(d)}\left(1 - \frac{D_{i}^{(d)}}{P_{i}^{\epsilon(d)}}\right) - \text{E}_{\partial i}^{\epsilon(d^{*})}\left(1 - \frac{D_{i}^{(d^{*})}}{P_{i}^{\epsilon(d^{*})}}\right)\right)\right]\mid_{\epsilon = 0}\right)}_{\psi_{4B}}. 
\end{align}

By Assumption~\ref{assumption:differentiability}, all the derivative terms are integrable. Therefore, by summing the derivatives, taking the expectation, and grouping the terms, we obtain:
\begin{align}
    \mathbb{E}\left(\frac{\partial}{\partial \epsilon}\psi^{\epsilon}_{\mathrm{ATES}}\mid_{\epsilon = 0}\right) &=\underbrace{\mathbb{E}\left(\psi_{3A} - \psi_{2A} + \psi_{3B} - \psi_{2B}\right)}_{E_{1}} + \underbrace{\mathbb{E}\left(\psi_{1A} - \psi_{2C} + \psi_{3C}\right)}_{E_{2}} + \nonumber \\
    &+\underbrace{\mathbb{E}\left(\psi_{1B} - \psi_{2D} + \psi_{3D} - \psi_{4A}\right)}_{E_{3}} + \underbrace{\mathbb{E}\left(\psi_{4B}\right)}_{E_{4}}.
\end{align}

Next, we show that $E_{1} = 0$, $E_{2} = 0$, $E_{3} = 0$, and $E_{4} = 0$. Specifically, notice that:
\begin{align}
    \mathbb{E}\left(\psi_{2A}\right) &= \mathbb{E}\left(\frac{D_{i}^{(d)}\tilde{Z}_{i}\triangle p_{D}^{d}\triangle Y^{0, d}}{\left(P_{i}^{0(d)}\right)^{2}}\right) \propto \underbrace{\mathbb{E}\left(\frac{\triangle p_{D}^{d}\triangle Y^{0, d}}{\left(P_{i}^{0(d)}\right)^{2}}\mid D_{i} = d, \tilde{Z}_{i} = 1\right)}_{\text{LTE}} = \nonumber \\
    &= \underbrace{\mathbb{E}\left(\frac{\triangle p_{D}^{d}}{\left(P_{i}^{0(d)}\right)^{2}}\mathbb{E}\left(\triangle Y^{0, d}\mid X_{i}, P_{i}^{0}, D_{i} = d, \tilde{Z}_{i} = 1\right)\mid D_{i} = d, \tilde{Z}_{i} = 1\right)}_{\text{LIE}} = \nonumber \\
    &= \underbrace{\mathbb{E}\left(\frac{\triangle p_{D}^{d}}{\left(P_{i}^{0(d)}\right)^{2}}\times 0 \mid D_{i} = d, \tilde{Z}_{i} = 1\right)}_{\text{By the definition of }\mu_Y^0\left(\cdot\right)} = 0.
\end{align}

By the same steps, we obtain $\mathbb{E}\left(\psi_{3A}\right) = 0$. Further, notice that:
\begin{align}
    \mathbb{E}\left(\psi_{2B}\right)
    &= \mathbb{E}\left(
        \frac{D_i^{(d)}\tilde{Z}_i\triangle p_Z
        R_{\partial D}^{0,d}\triangle Y^{0,d}}
        {(P_i^{0(d)})^2}
    \right) \propto \nonumber \\
    &\propto
    \underbrace{\mathbb{E}\left(
        \frac{\triangle p_Z R_{\partial D}^{0,d}
        \triangle Y^{0,d}}{(P_i^{0(d)})^2}
        \mid D_i=d,\tilde{Z}_i=1
    \right)}_{\text{LTE}} = \nonumber \\
    &= \underbrace{\mathbb{E}\left(
        \frac{\triangle p_Z R_{\partial D}^{0,d}}
        {(P_i^{0(d)})^2}
        \mathbb{E}\left(
            \triangle Y^{0,d}
            \mid D_i=d,X_i,W_i^{(Z)},\tilde{Z}_i=1
        \right)
        \mid D_i=d,\tilde{Z}_i=1
    \right)}_{\text{LIE}} = \nonumber \\
    &= \underbrace{\mathbb{E}\left(
        \frac{\triangle p_Z R_{\partial D}^{0,d}}
        {(P_i^{0(d)})^2}\times 0
        \mid D_i=d,\tilde{Z}_i=1
    \right)}_{\text{Equation~(\ref{eq:useful:muY})
    of Lemma~\ref{lemma:useful}}}=0.
\end{align}
By the same steps, we obtain
$\mathbb{E}(\psi_{3B})=0$.
These equalities imply $E_1=0$.

Further, notice that the terms grouped in $E_{2}$ can be rewritten as the sum of the following terms:
\begin{align}
    E_{2} &= \underbrace{\mathbb{E}\left(\tilde{Z}_{i}\triangle \mu_{Y}^{d} - \frac{D_{i}^{(d)}\tilde{Z}_{i}\triangle \mu_{Y}^{d}}{P_{i}^{0(d)}}\right)}_{E_{2A}} - \underbrace{\mathbb{E}\left(\tilde{Z}_{i}\triangle \mu_{Y}^{d^{*}} - \frac{D_{i}^{(d^{*})}\tilde{Z}_{i}\triangle \mu_{Y}^{d^{*}}}{P_{i}^{0(d^{*})}}\right)}_{E_{2B}}.
\end{align}

For the first term, we obtain:
\begin{align}
    E_{2A} &= \mathbb{E}\left(\tilde{Z}_{i}\triangle \mu_{Y}^{d} - \frac{D_{i}^{(d)}\tilde{Z}_{i}\triangle \mu_{Y}^{d}}{P_{i}^{0(d)}}\right) \propto \underbrace{\mathbb{E}\left(\triangle \mu_{Y}^{d} - \frac{\triangle \mu_{Y}^{d}}{P_{i}^{0(d)}}D_{i}^{(d)}\mid \tilde{Z}_{i} = 1\right)}_{\text{LTE}} = \nonumber \\
    &= \underbrace{\mathbb{E}\left(\triangle \mu_{Y}^{d} - \frac{\triangle \mu_{Y}^{d}}{P_{i}^{0(d)}}\mathbb{E}\left(D_{i}^{(d)}\mid X_{i}, P_{i}^{0}, \tilde{Z}_{i} = 1\right)\mid \tilde{Z}_{i} = 1\right)}_{\text{LIE}} = \nonumber \\
    &= \underbrace{\mathbb{E}\left(\triangle \mu_{Y}^{d} - \frac{\triangle \mu_{Y}^{d}}{P_{i}^{0(d)}}P_{i}^{0(d)}\mid \tilde{Z}_{i} = 1\right)}_{\text{Equation~(\ref{eq:useful:pD}) of Lemma~\ref{lemma:useful}}} = \mathbb{E}\left(\triangle \mu_{Y}^{d} - \triangle \mu_{Y}^{d}\mid \tilde{Z}_{i} = 1\right) = 0.
\end{align}
By the same steps, for the second term, we obtain $E_{2B} = 0$. These equalities imply $E_{2} = 0$.

Further, notice that the terms grouped in $E_{3}$ can be rewritten as follows:
\begin{align}
    E_{3} &= \mathbb{E}\left(\psi_{1B} - \psi_{2D} + \psi_{3D} - \psi_{4A}\right) = \nonumber \\
    &= \mathbb{E}\left(\triangle p_{Z}\left(R_{\partial Y}^{0, d}\left(1 - \frac{D_{i}^{(d)}}{P_{i}^{0(d)}}\right) - R_{\partial Y}^{0, d^{*}}\left(1 - \frac{D_{i}^{(d^{*})}}{P_{i}^{0(d^{*})}}\right)\right)\left(\tilde{Z}_{i} - \tilde{P}_{i}^{0}\right)\right) = \nonumber \\
    &= \underbrace{\mathbb{E}\left(\triangle p_{Z}\left(R_{\partial Y}^{0, d}\left(1 - \frac{D_{i}^{(d)}}{P_{i}^{0(d)}}\right) - R_{\partial Y}^{0, d^{*}}\left(1 - \frac{D_{i}^{(d^{*})}}{P_{i}^{0(d^{*})}}\right)\right)\mathbb{E}\left(\tilde{Z}_{i} - \tilde{P}_{i}^{0}\mid D_{i}, X_{i}, W_{i}^{(Z)}\right)\right)}_{\text{LIE}} = \nonumber \\
    &= \underbrace{\mathbb{E}\left(\triangle p_{Z}\left(R_{\partial Y}^{0, d}\left(1 - \frac{D_{i}^{(d)}}{P_{i}^{0(d)}}\right) - R_{\partial Y}^{0, d^{*}}\left(1 - \frac{D_{i}^{(d^{*})}}{P_{i}^{0(d^{*})}}\right)\right)\times 0\right)}_{\text{Since }\tilde{P}_{i}^{0} = \mathbb{E}\left(\tilde{Z}_{i}\mid D_{i}, X_{i}, W_{i}^{(Z)}\right)} = 0.
\end{align}

Finally, notice that:
\begin{align}
    E_{4} &= \mathbb{E}\left(\left(\overline{Z}_{i} - P_{i}^{0}\right)\left(\frac{\partial}{\partial \epsilon}\underbrace{\left[\tilde{P}_{i}^{\epsilon}\left(\text{E}_{\partial i}^{\epsilon(d)}\left(1 - \frac{D_{i}^{(d)}}{P_{i}^{\epsilon(d)}}\right) - \text{E}_{\partial i}^{\epsilon(d^{*})}\left(1 - \frac{D_{i}^{(d^{*})}}{P_{i}^{\epsilon(d^{*})}}\right)\right)\right]}_{\text{Denote by }A_{i}^{\epsilon}}\mid_{\epsilon = 0}\right)\right) = \nonumber \\
    &= \underbrace{\mathbb{E}\left(\mathbb{E}\left(\left(\overline{Z}_{i} - P_{i}^{0}\right)\left(\frac{\partial}{\partial \epsilon}A_{i}^{\epsilon}\mid_{\epsilon = 0}\right)\mid D_{i}, X_{i}, W_{i}^{(Z)}\right)\right)}_{\text{LIE}} = \nonumber \\
    &= \underbrace{\mathbb{E}\left(\mathbb{E}\left(\overline{Z}_{i} - P_{i}^{0}\mid D_{i}, X_{i}, W_{i}^{(Z)}\right)\left(\frac{\partial}{\partial \epsilon}A_{i}^{\epsilon}\mid_{\epsilon = 0}\right)\right)}_{\substack{\text{Since }\frac{\partial}{\partial \epsilon}A_{i}^{\epsilon}\mid_{\epsilon = 0}\text{ is a function of }D_{i}, X_{i}, W_{i}^{(Z)}}} = \nonumber \\
    &= \underbrace{\mathbb{E}\left(\mathbb{E}\left(P_{i}^{0} - P_{i}^{0}\mid D_{i}, X_{i}, W_{i}^{(Z)}\right)\left(\frac{\partial}{\partial \epsilon}A_{i}^{\epsilon}\mid_{\epsilon = 0}\right)\right)}_{\text{Since }\mathbb{E}\left(\overline{Z}_{i}\mid D_{i}, X_{i}, W_{i}^{(Z)}\right)=P_{i}^{0}} = 0.
\end{align}

By the mean value theorem, Assumption~\ref{assumption:differentiability} ensures that the difference quotients of the score are dominated by an integrable random variable. Therefore, by the dominated convergence theorem, we obtain:
\begin{align}
    \left.\frac{\partial}{\partial\epsilon}
    \mathbb{E}\left(\psi_{\mathrm{ATES}}^\epsilon\right)
    \right|_{\epsilon=0}
    &=
    \mathbb{E}\left(
        \left.\frac{\partial}{\partial\epsilon}
        \psi_{\mathrm{ATES}}^\epsilon
        \right|_{\epsilon=0}
    \right) = E_1+E_2+E_3+E_4=0.
\end{align}

\subsection{Neyman-orthogonality of the score of ATTS}

\subsubsection{Mean zero of ATTS score}

Consider the score at true values:
\begin{align}
    \psi_{\mathrm{ATTS}}^{0} &= \underbrace{D_{i}^{(d)}\tilde{Z}_{i}\left(Y_{i} - \text{E}_{i}^{0(d^{*})}\right) - D_{i}^{(d)}\tilde{Z}_{i}\mathrm{ATTS}}_{\psi^{0}_{1}} - \underbrace{D_{i}^{(d^{*})}\tilde{Z}_{i}R_{PD}^{0}\left(Y_{i} - \text{E}_{i}^{0(d^{*})}\right)}_{\psi^{0}_{2}} - \nonumber \\ &-\underbrace{\left(\overline{Z}_{i} - P_{i}^{0}\right)\tilde{P}_{i}^{0}\text{E}_{\partial i}^{0(d^{*})}\left(D_{i}^{(d)} - D_{i}^{(d^{*})}R_{PD}^{0}\right)}_{\psi^{0}_{3}}.
\end{align}

Next, we show that the expectations of $\psi_{1}^{0}$, $\psi_{2}^{0}$, and $\psi_{3}^{0}$ are $0$. Specifically, notice that:
\begin{align}
    \mathbb{E}\left(\psi_{1}^{0}\right) &= \mathbb{E}\left(D_{i}^{(d)}\tilde{Z}_{i}\left(Y_{i} - \text{E}_{i}^{0(d^{*})}\right)\right) - \mathbb{E}\left(D_{i}^{(d)}\tilde{Z}_{i}\mathrm{ATTS}\right) = \nonumber \\
    &= \underbrace{\left[\mathbb{E}\left(Y_{i} - \text{E}_{i}^{0(d^{*})}\mid D_{i} = d, \tilde{Z}_{i} = 1\right)- \mathrm{ATTS}\right]\mathbb{P}\left(D_{i} = d, \tilde{Z}_{i} = 1\right)}_{\substack{\text{LTE with the events }\{D_{i} = d, \tilde{Z}_{i} = 1\}\text{ and }\{D_{i} \ne d\text{ or }\tilde{Z}_{i} \ne 1\}}} \propto \nonumber \\
    &\propto \underbrace{\mathbb{E}\left(\mathbb{E}\left(Y_{i} - \text{E}_{i}^{0(d^{*})}\mid X_{i}, W_{i}^{(Z)}, D_{i} = d, \tilde{Z}_{i} = 1\right)\mid D_{i} = d, \tilde{Z}_{i} = 1\right)}_{\text{LIE}} - \mathrm{ATTS} = \nonumber \\
    &= \underbrace{\mathbb{E}\left(
    \text{E}_{i}^{0(d)} - \text{E}_{i}^{0(d^{*})}
    \mid D_{i} = d, \tilde{Z}_{i} = 1\right)}
    _{\text{Equation~(\ref{eq:useful:muY}) of Lemma~\ref{lemma:useful}}}
    - \mathrm{ATTS}
    = \underbrace{\mathrm{ATTS} - \mathrm{ATTS}}
    _{\text{Theorem~\ref{theorem:ates}}} = 0.
\end{align}

Further, notice that:
\begin{align}
    \mathbb{E}\left(\psi_{2}^{0}\right) &= \mathbb{E}\left(D_{i}^{(d^{*})}\tilde{Z}_{i}R_{PD}^{0}\left(Y_{i} - \text{E}_{i}^{0(d^{*})}\right)\right) = \nonumber \\
    &= \underbrace{\mathbb{E}\left(R_{PD}^{0}\left(Y_{i} - \text{E}_{i}^{0(d^{*})}\right)\mid D_{i} = d^{*}, \tilde{Z}_{i} = 1\right)\mathbb{P}\left(D_{i} = d^{*}, \tilde{Z}_{i} = 1\right)}_{\text{LTE}} \propto \nonumber \\
    &\propto \underbrace{\mathbb{E}\left(R_{PD}^{0}\mathbb{E}\left(Y_{i} - \text{E}_{i}^{0(d^{*})}\mid X_{i}, P_{i}^{0}, D_{i} = d^{*}, \tilde{Z}_{i} = 1\right)\mid D_{i} = d^{*}, \tilde{Z}_{i} = 1\right)}_{\text{LIE}} = \nonumber \\
    &= \underbrace{\mathbb{E}\left(R_{PD}^{0}\left(\text{E}_{i}^{0(d^{*})} - \text{E}_{i}^{0(d^{*})}\right)\mid D_{i} = d^{*}, \tilde{Z}_{i} = 1\right)}_{\text{Since }\mathbb{E}\left(Y_{i}\mid X_{i}, P_{i}^{0}, D_{i} = d^{*}, \tilde{Z}_{i} = 1\right) = \text{E}_{i}^{0(d^{*})}} = 0.
\end{align}

Finally, notice that:
\begin{align}
    \mathbb{E}\left(\psi_{3}^{0}\right) &= \mathbb{E}\left(\left(\overline{Z}_{i} - P_{i}^{0}\right)\underbrace{\tilde{P}_{i}^{0}\text{E}_{\partial i}^{0(d^{*})}\left(D_{i}^{(d)} - D_{i}^{(d^{*})}R_{PD}^{0}\right)}_{B_{i}}\right) = \nonumber \\
    &= \underbrace{\mathbb{E}\left(\mathbb{E}\left(\left(\overline{Z}_{i} - P_{i}^{0}\right)B_{i}\mid D_{i}, X_{i}, W_{i}^{(Z)}\right)\right)}_{\text{LIE}} = \nonumber \\
    &= \underbrace{\mathbb{E}\left(\mathbb{E}\left(\overline{Z}_{i} - P_{i}^{0}\mid D_{i}, X_{i}, W_{i}^{(Z)}\right)B_{i}\right)}_{\text{Since }B_{i}\text{ is a function of }D_{i}, X_{i}, W_{i}^{(Z)}} = \nonumber \\
    &= \underbrace{\mathbb{E}\left(\mathbb{E}\left(P_{i}^{0} - P_{i}^{0}\mid D_{i}, X_{i}, W_{i}^{(Z)}\right)B_{i}\right)}_{\text{Since }\mathbb{E}\left(\overline{Z}_{i}\mid D_{i}, X_{i}, W_{i}^{(Z)}\right)=P_{i}^{0}} = 0.
\end{align}
Therefore, we have:
\begin{align}
    \mathbb{E}\left(\psi_{\mathrm{ATTS}}^{0}\right)
    = \mathbb{E}\left(\psi_{1}^{0}\right)
    - \mathbb{E}\left(\psi_{2}^{0}\right)
    - \mathbb{E}\left(\psi_{3}^{0}\right)
    = 0 - 0 - 0 = 0.
\end{align}

\subsubsection{Zero Gateaux derivative of ATTS score}

The perturbed score is as follows:
\begin{align}
    \psi_{\mathrm{ATTS}}^{\epsilon} &= \underbrace{D_{i}^{(d)}\tilde{Z}_{i}\left(Y_{i} - \mu_{Y}^{\epsilon}\left(d^{*}, X_{i}, P_{i}^{\epsilon}\right)\right) - D_{i}^{(d)}\tilde{Z}_{i}\mathrm{ATTS}}_{\psi_{1}^{\epsilon}} - \nonumber \\
    &- \underbrace{D_{i}^{(d^{*})}\tilde{Z}_{i}R_{PD}^{\epsilon}\left(Y_{i} - \mu_{Y}^{\epsilon}\left(d^{*}, X_{i}, P_{i}^{\epsilon}\right)\right)}_{\psi_{2}^{\epsilon}} - \underbrace{\triangle \overline{Z}^{\epsilon}\tilde{P}_{i}^{\epsilon}R_{\partial Y}^{\epsilon, d^{*}}\left(D_{i}^{(d)} - D_{i}^{(d^{*})}R_{PD}^{\epsilon}\right)}_{\psi_{3}^{\epsilon}}.
\end{align}

As in the proof for ATES, by Assumption~\ref{assumption:differentiability}, we can differentiate each term by the chain rule. Specifically, the derivative associated with the first term is as follows:
\begin{align}
    \frac{\partial}{\partial \epsilon}\psi_{1}^{\epsilon}\mid_{\epsilon = 0} &= -\underbrace{D_{i}^{(d)}\tilde{Z}_{i}\triangle \mu_{Y}^{d^{*}}}_{\psi_{1A}} - \underbrace{D_{i}^{(d)}\tilde{Z}_{i}\triangle p_{Z}R_{\partial Y}^{0, d^{*}}}_{\psi_{1B}}.
\end{align}

The derivative associated with the second term is as follows:
\begin{align}
    \frac{\partial}{\partial \epsilon}\psi_{2}^{\epsilon}\mid_{\epsilon = 0} &= \underbrace{D_{i}^{(d^{*})}\tilde{Z}_{i}\left(Y_{i} - \text{E}_{i}^{0(d^{*})}\right)\left(\frac{\partial}{\partial \epsilon}R_{PD}^{\epsilon}\mid_{\epsilon = 0}\right)}_{\psi_{2A}} - \nonumber \\
    &- \underbrace{D_{i}^{(d^{*})}\tilde{Z}_{i}\triangle \mu_{Y}^{d^{*}}R_{PD}^{0}}_{\psi_{2B}} - \underbrace{D_{i}^{(d^{*})}\tilde{Z}_{i}\triangle p_{Z}R_{\partial Y}^{0, d^{*}}R_{PD}^{0}}_{\psi_{2C}}.
\end{align}

The derivative associated with the third term is as follows:
\begin{align}
    \frac{\partial}{\partial \epsilon}\psi_{3}^{\epsilon}\mid_{\epsilon = 0} &= -\underbrace{\triangle p_{Z}\tilde{P}_{i}^{0}R_{\partial Y}^{0, d^{*}}\left(D_{i}^{(d)} - D_{i}^{(d^{*})}R_{PD}^{0}\right)}_{\psi_{3A}} + \nonumber \\
    &+ \underbrace{\triangle \overline{Z}^{0}\left(\frac{\partial}{\partial \epsilon}\left[\tilde{P}_{i}^{\epsilon}R_{\partial Y}^{\epsilon, d^{*}}\left(D_{i}^{(d)} - D_{i}^{(d^{*})}R_{PD}^{\epsilon}\right)\right]\mid_{\epsilon = 0}\right)}_{\psi_{3B}}.
\end{align}

Since all the derivative terms are integrable by Assumption~\ref{assumption:differentiability}, we obtain by summing the derivatives, taking the expectation, and grouping the terms:
\begin{align}
    \mathbb{E}\left(\frac{\partial}{\partial \epsilon}\psi^{\epsilon}_{\mathrm{ATTS}}\mid_{\epsilon = 0}\right) &=\underbrace{\mathbb{E}\left(-\psi_{2A}\right)}_{E_{1}} + \underbrace{\mathbb{E}\left(\psi_{2B} - \psi_{1A}\right)}_{E_{2}} + \underbrace{\mathbb{E}\left(\psi_{2C} - \psi_{1B} + \psi_{3A}\right)}_{E_{3}} + \underbrace{\mathbb{E}\left(-\psi_{3B}\right)}_{E_{4}}.
\end{align}

Next, we show that $E_{1} = 0$, $E_{2} = 0$, $E_{3} = 0$, and $E_{4} = 0$. Specifically, notice that:
\begin{align}
    \mathbb{E}\left(\psi_{2A}\right) &= \mathbb{E}\left(D_{i}^{(d^{*})}\tilde{Z}_{i}\left(Y_{i} - \text{E}_{i}^{0(d^{*})}\right)\left(\frac{\partial}{\partial \epsilon}R_{PD}^{\epsilon}\mid_{\epsilon = 0}\right)\right) \propto \nonumber \\
    &\propto \underbrace{\mathbb{E}\left(\left(Y_{i} - \text{E}_{i}^{0(d^{*})}\right)\left(\frac{\partial}{\partial \epsilon}R_{PD}^{\epsilon}\mid_{\epsilon = 0}\right)\mid D_{i} = d^{*}, \tilde{Z}_{i} = 1\right)}_{\text{LTE}} = \nonumber \\
    &= \underbrace{\mathbb{E}\left(\left(\frac{\partial}{\partial \epsilon}R_{PD}^{\epsilon}\mid_{\epsilon = 0}\right)\mathbb{E}\left(Y_{i} - \text{E}_{i}^{0(d^{*})}\mid X_{i}, W_{i}^{(Z)}, D_{i} = d^{*}, \tilde{Z}_{i} = 1\right)\mid D_{i} = d^{*}, \tilde{Z}_{i} = 1\right)}_{\text{LIE}} = \nonumber \\
    &= \underbrace{\mathbb{E}\left(\left(\frac{\partial}{\partial \epsilon}R_{PD}^{\epsilon}\mid_{\epsilon = 0}\right)\times 0\mid D_{i} = d^{*}, \tilde{Z}_{i} = 1\right)}_{\text{Equation~(\ref{eq:useful:muY}) of Lemma~\ref{lemma:useful}}} = 0.
\end{align}
This equality implies $E_{1} = 0$.

Further, notice that the terms grouped in $E_{2}$ can be rewritten as follows:
\begin{align}
    E_{2} &= \mathbb{E}\left(\psi_{2B} - \psi_{1A}\right) = \mathbb{E}\left(\tilde{Z}_{i}\triangle \mu_{Y}^{d^{*}}\left(D_{i}^{(d^{*})}R_{PD}^{0} - D_{i}^{(d)}\right)\right) \propto \nonumber \\ 
    &\propto \underbrace{\mathbb{E}\left(\triangle \mu_{Y}^{d^{*}}\left(D_{i}^{(d^{*})}R_{PD}^{0} - D_{i}^{(d)}\right)\mid \tilde{Z}_{i} = 1\right)}_{\text{LTE}} = \nonumber \\ 
    &= \underbrace{\mathbb{E}\left(\mathbb{E}\left(\triangle \mu_{Y}^{d^{*}}\left(D_{i}^{(d^{*})}R_{PD}^{0} - D_{i}^{(d)}\right)\mid X_{i}, P_{i}^{0}, \tilde{Z}_{i} = 1\right)\mid \tilde{Z}_{i} = 1\right)}_{\text{LIE}} = \nonumber \\ 
    &=\mathbb{E}\left(\triangle \mu_{Y}^{d^{*}}\mathbb{E}\left(D_{i}^{(d^{*})}\mid X_{i}, P_{i}^{0}, \tilde{Z}_{i} = 1\right)R_{PD}^{0}\mid \tilde{Z}_{i} = 1\right) - \nonumber \\
    &-\mathbb{E}\left(\triangle \mu_{Y}^{d^{*}}\mathbb{E}\left(D_{i}^{(d)}\mid X_{i}, P_{i}^{0}, \tilde{Z}_{i} = 1\right)\mid \tilde{Z}_{i} = 1\right) = \nonumber \\ 
    &= \underbrace{\mathbb{E}\left(\triangle \mu_{Y}^{d^{*}}P_{i}^{0(d^{*})}R_{PD}^{0}\mid \tilde{Z}_{i} = 1\right) - \mathbb{E}\left(\triangle\mu_{Y}^{d^{*}}P_{i}^{0(d)}\mid \tilde{Z}_{i} = 1\right)}_{\text{Equation~(\ref{eq:useful:pD}) of Lemma~\ref{lemma:useful}}} = \nonumber \\
    &= \underbrace{\mathbb{E}\left(\triangle \mu_{Y}^{d^{*}}P_{i}^{0(d)}\mid \tilde{Z}_{i} = 1\right) - \mathbb{E}\left(\triangle \mu_{Y}^{d^{*}}P_{i}^{0(d)}\mid \tilde{Z}_{i} = 1\right)}_{\text{Since }R_{PD}^{0} = P_{i}^{0(d)}/P_{i}^{0(d^{*})}} = 0.
\end{align}

Further, notice that the terms grouped in $E_{3}$ can be rewritten as follows:
\begin{align}
    E_{3} &= \mathbb{E}\left(\psi_{2C} - \psi_{1B} + \psi_{3A}\right) = \nonumber \\
    &= \mathbb{E}\left(\triangle p_{Z}R_{\partial Y}^{0, d^{*}}\left(D_{i}^{(d^{*})}\tilde{Z}_{i}R_{PD}^{0} - D_{i}^{(d)}\tilde{Z}_{i} + \tilde{P}_{i}^{0}\left(D_{i}^{(d)} - D_{i}^{(d^{*})}R_{PD}^{0}\right)\right)\right) = \nonumber \\
    &= \mathbb{E}\left(\triangle p_{Z}R_{\partial Y}^{0, d^{*}}\left(\tilde{Z}_{i} - \tilde{P}_{i}^{0}\right)\left(D_{i}^{(d^{*})}R_{PD}^{0} - D_{i}^{(d)}\right)\right) = \nonumber \\
    &= \underbrace{\mathbb{E}\left(\mathbb{E}\left(\triangle p_{Z}R_{\partial Y}^{0, d^{*}}\left(\tilde{Z}_{i} - \tilde{P}_{i}^{0}\right)\left(D_{i}^{(d^{*})}R_{PD}^{0} - D_{i}^{(d)}\right)\mid D_{i}, X_{i}, W_{i}^{(Z)}\right)\right)}_{\text{LIE}} = \nonumber \\
    &= \underbrace{\mathbb{E}\left(\triangle p_{Z}R_{\partial Y}^{0, d^{*}}\left(D_{i}^{(d^{*})}R_{PD}^{0} - D_{i}^{(d)}\right)\mathbb{E}\left(\tilde{Z}_{i} - \tilde{P}_{i}^{0}\mid D_{i}, X_{i}, W_{i}^{(Z)}\right)\right)}_{\substack{\text{Since }\triangle p_{Z}, R_{\partial Y}^{0,d^{*}},R_{PD}^{0}, D_{i}^{(d)}\text{, and }D_{i}^{(d^{*})}\\ \text{are functions of }D_{i}, X_{i}, W_{i}^{(Z)}}} = \nonumber \\
    &= \underbrace{\mathbb{E}\left(\triangle p_{Z}R_{\partial Y}^{0, d^{*}}\left(D_{i}^{(d^{*})}R_{PD}^{0} - D_{i}^{(d)}\right)\times 0\right)}_{\text{Since }\tilde{P}_{i}^{0} = \mathbb{E}\left(\tilde{Z}_{i}\mid D_{i}, X_{i}, W_{i}^{(Z)}\right)} = 0.
\end{align}
These equalities imply $E_{3} = 0$.

Finally, notice that:
\begin{align}
    E_{4} &= \mathbb{E}\left(-\psi_{3B}\right) = -\mathbb{E}\left(\triangle \overline{Z}^{0}\left(\frac{\partial}{\partial \epsilon}\left[\tilde{P}_{i}^{\epsilon}R_{\partial Y}^{\epsilon, d^{*}}\left(D_{i}^{(d)} - D_{i}^{(d^{*})}R_{PD}^{\epsilon}\right)\right]\mid_{\epsilon = 0}\right)\right) = \nonumber \\
    &= -\underbrace{\mathbb{E}\left(\mathbb{E}\left(\left(\overline{Z}_{i} - P_{i}^{0}\right)\left(\frac{\partial}{\partial \epsilon}\left[\tilde{P}_{i}^{\epsilon}R_{\partial Y}^{\epsilon, d^{*}}\left(D_{i}^{(d)} - D_{i}^{(d^{*})}R_{PD}^{\epsilon}\right)\right]\mid_{\epsilon = 0}\right)\mid D_{i}, X_{i}, W_{i}^{(Z)}\right)\right)}_{\text{LIE}} = \nonumber \\
    &= -\underbrace{\mathbb{E}\left(\mathbb{E}\left(\overline{Z}_{i} - P_{i}^{0}\mid D_{i}, X_{i}, W_{i}^{(Z)}\right)\left(\frac{\partial}{\partial \epsilon}\left[\tilde{P}_{i}^{\epsilon}R_{\partial Y}^{\epsilon, d^{*}}\left(D_{i}^{(d)} - D_{i}^{(d^{*})}R_{PD}^{\epsilon}\right)\right]\mid_{\epsilon = 0}\right)\right)}_{\substack{\text{Since the derivative }\frac{\partial}{\partial \epsilon}\left[\ldots\right]\mid_{\epsilon = 0}\text{ is a function of }D_{i}, X_{i}, W_{i}^{(Z)}}} = \nonumber \\
    &= -\underbrace{\mathbb{E}\left(\mathbb{E}\left(P_{i}^{0} - P_{i}^{0}\mid D_{i}, X_{i}, W_{i}^{(Z)}\right)\left(\frac{\partial}{\partial \epsilon}\left[\tilde{P}_{i}^{\epsilon}R_{\partial Y}^{\epsilon, d^{*}}\left(D_{i}^{(d)} - D_{i}^{(d^{*})}R_{PD}^{\epsilon}\right)\right]\mid_{\epsilon = 0}\right)\right)}_{\text{Since }\mathbb{E}\left(\overline{Z}_{i}\mid D_{i}, X_{i}, W_{i}^{(Z)}\right)=P_{i}^{0}} = 0.
\end{align}
These equalities imply $E_{4} = 0$.

Therefore, by Assumption~\ref{assumption:differentiability} and the dominated convergence theorem, as in the proof for ATES, we obtain:
\begin{align}
    \left.\frac{\partial}{\partial\epsilon}
    \mathbb{E}\left(\psi_{\mathrm{ATTS}}^\epsilon\right)
    \right|_{\epsilon=0}
    &=
    \mathbb{E}\left(
        \left.\frac{\partial}{\partial\epsilon}
        \psi_{\mathrm{ATTS}}^\epsilon
        \right|_{\epsilon=0}
    \right) = E_1+E_2+E_3+E_4=0.
\end{align}

\subsection{Neyman-orthogonality of the score of ATE}

\subsubsection{Mean zero of ATE score}

Consider the score at true values of the parameter and nuisance functions\footnote{To justify divisions by $\tilde{P}_{i}^{0}$ and $P_{i}^{0(\overline{d})}$, where $\overline{d}\in\{d,d^{*}\}$, notice that Assumption~\ref{assumption:selection-support} implies $\tilde{P}_{i}^{0}>0$ almost surely. Moreover, by Assumption~\ref{assumption:support}, Equation~(\ref{eq:useful:pD}) of Lemma~\ref{lemma:useful}, and the law of iterated expectations, we obtain:
\begin{align*}
    0 &= \mathbb{P}\left(P_{i}^{0(\overline{d})}=0,\tilde{Z}_{i}=1\right)
    = \mathbb{E}\left(\mathbb{I}\left(P_{i}^{0(\overline{d})}=0\right)\tilde{P}_{i}^{0}\right).
\end{align*}
Since $\tilde{P}_{i}^{0}>0$ almost surely, this equality implies $P_{i}^{0(\overline{d})}>0$ almost surely.}:
\begin{align}
    \psi_{\mathrm{ATE}}^{0} &= \underbrace{\left(\text{E}_{i}^{0(d)} - \text{E}_{i}^{0(d^{*})}\right) - \mathrm{ATE}}_{\psi^{0}_{1}} + \underbrace{\tilde{Z}_{i}\frac{D_{i}^{(d)}}{P_{i}^{0(d)}\tilde{P}_{i}^{0}}\left(Y_{i} - \text{E}_{i}^{0(d)}\right)}_{\psi^{0}_{2}} - \nonumber \\
    &- \underbrace{\tilde{Z}_{i}\frac{D_{i}^{(d^{*})}}{P_{i}^{0(d^{*})}\tilde{P}_{i}^{0}}\left(Y_{i} - \text{E}_{i}^{0(d^{*})}\right)}_{\psi^{0}_{3}} - \underbrace{\left(\overline{Z}_{i} - P_{i}^{0}\right)\left(\text{E}_{\partial i}^{0(d)}\left(\frac{D_{i}^{(d)}}{P_{i}^{0(d)}}-1\right) - \text{E}_{\partial i}^{0(d^{*})} \left(\frac{D_{i}^{(d^{*})}}{P_{i}^{0(d^{*})}}-1\right)\right)}_{\psi^{0}_{4}}.
\end{align}

By Assumption~\ref{assumption:differentiability}, all these terms are well-defined and integrable. Next, we show that the expectations of $\psi_{1}^{0}$, $\psi_{2}^{0}$, $\psi_{3}^{0}$, and $\psi_{4}^{0}$ are $0$. Specifically, notice that Theorem~\ref{theorem:ate} implies $\mathbb{E}\left(\psi_{1}^{0}\right) = 0$. Further, notice that:
\begin{align}
    \mathbb{E}\left(\psi_{2}^{0}\right) &= \mathbb{E}\left(\tilde{Z}_{i}\frac{D_{i}^{(d)}}{P_{i}^{0(d)}\tilde{P}_{i}^{0}}\left(Y_{i} - \text{E}_{i}^{0(d)}\right)\right) = \nonumber \\
    &= \underbrace{\mathbb{E}\left(\frac{1}{P_{i}^{0(d)}\tilde{P}_{i}^{0}}\left(Y_{i} - \text{E}_{i}^{0(d)}\right)\mid D_{i} = d, \tilde{Z}_{i} = 1\right)\mathbb{P}\left(D_{i} = d, \tilde{Z}_{i} = 1\right)}_{\substack{\text{LTE with the events }\{D_{i} = d, \tilde{Z}_{i} = 1\}\text{ and }\{D_{i} \ne d\text{ or }\tilde{Z}_{i} \ne 1\}}} \propto \nonumber \\
    &\propto \underbrace{\mathbb{E}\left(\mathbb{E}\left(\frac{1}{P_{i}^{0(d)}\tilde{P}_{i}^{0}}\left(Y_{i} - \text{E}_{i}^{0(d)}\right)\mid X_{i}, W_{i}^{(Z)}, D_{i} = d, \tilde{Z}_{i} = 1\right)\mid D_{i} = d, \tilde{Z}_{i} = 1\right)}_{\text{LIE}} = \nonumber \\
    &= \underbrace{\mathbb{E}\left(\frac{1}{P_{i}^{0(d)}\tilde{P}_{i}^{0}}\left(\mathbb{E}\left(Y_{i}\mid X_{i}, W_{i}^{(Z)}, D_{i} = d, \tilde{Z}_{i} = 1\right) - \text{E}_{i}^{0(d)}\right)\mid D_{i} = d, \tilde{Z}_{i} = 1\right)}_{\substack{\text{Since }P_{i}^{0(d)},\tilde{P}_{i}^{0}\text{, and }\text{E}_{i}^{0(d)}\\ \text{are functions of }X_{i}, W_{i}^{(Z)}\text{ on }\{D_{i} = d\}}} = \nonumber \\
    &= \underbrace{\mathbb{E}\left(\frac{1}{P_{i}^{0(d)}\tilde{P}_{i}^{0}}\left(\text{E}_{i}^{0(d)} - \text{E}_{i}^{0(d)}\right)\mid D_{i} = d, \tilde{Z}_{i} = 1\right)}_{\text{Equation~(\ref{eq:useful:muY}) of Lemma~\ref{lemma:useful}}} = 0.
\end{align}

By the same steps, we obtain $\mathbb{E}\left(\psi_{3}^{0}\right) = 0$. Finally, notice that:
\begin{align}
    \mathbb{E}\left(\psi_{4}^{0}\right) &= \mathbb{E}\left(
    \left(\overline{Z}_{i} - P_{i}^{0}\right)
    \underbrace{
        \left(
            \text{E}_{\partial i}^{0(d)}
            \left(\frac{D_{i}^{(d)}}{P_{i}^{0(d)}}-1\right)
            -
            \text{E}_{\partial i}^{0(d^{*})}
            \left(\frac{D_{i}^{(d^{*})}}{P_{i}^{0(d^{*})}}-1\right)
        \right)
    }_{A_{i}}\right) = \nonumber \\
    &= \underbrace{\mathbb{E}\left(\mathbb{E}\left(\left(\overline{Z}_{i} - P_{i}^{0}\right)A_{i}\mid D_{i}, X_{i}, W_{i}^{(Z)}\right)\right)}_{\text{LIE}} = \nonumber \\
    &= \underbrace{\mathbb{E}\left(\mathbb{E}\left(\overline{Z}_{i} - P_{i}^{0}\mid D_{i}, X_{i}, W_{i}^{(Z)}\right)A_{i}\right)}_{\text{Since }A_{i}\text{ is a function of }D_{i}, X_{i}, W_{i}^{(Z)}} = \nonumber \\
    &= \underbrace{\mathbb{E}\left(\mathbb{E}\left(P_{i}^{0} - P_{i}^{0}\mid D_{i}, X_{i}, W_{i}^{(Z)}\right)A_{i}\right)}_{\text{Since }\mathbb{E}\left(\overline{Z}_{i}\mid D_{i}, X_{i}, W_{i}^{(Z)}\right)=P_{i}^{0}} = 0.
\end{align}
Therefore, we have:
\begin{align}
    \mathbb{E}\left(\psi_{\mathrm{ATE}}^{0}\right)
    = \mathbb{E}\left(\psi_{1}^{0}\right)
    + \mathbb{E}\left(\psi_{2}^{0}\right)
    - \mathbb{E}\left(\psi_{3}^{0}\right)
    - \mathbb{E}\left(\psi_{4}^{0}\right)
    = 0 + 0 - 0 - 0 = 0.
\end{align}

\subsubsection{Zero Gateaux derivative of ATE score}

The perturbed score is as follows:
\begin{align}
    \psi_{\mathrm{ATE}}^{\epsilon} &= \underbrace{\triangle \mu_{Y}^{\epsilon} - \mathrm{ATE}}_{\psi_{1}^{\epsilon}} + \underbrace{\tilde{Z}_{i}\frac{D_{i}^{(d)}\triangle Y^{\epsilon, d}}{p_{D}^{\epsilon}\left(d, X_{i}, P_{i}^{\epsilon}\right)\tilde{p}_{Z}^{\epsilon}\left(D_{i}, X_{i}, W_{i}^{(Z)}\right)}}_{\psi_{2}^{\epsilon}} - \nonumber \\
    &- \underbrace{\tilde{Z}_{i}\frac{D_{i}^{(d^{*})}\triangle Y^{\epsilon, d^{*}}}{p_{D}^{\epsilon}\left(d^{*}, X_{i}, P_{i}^{\epsilon}\right)\tilde{p}_{Z}^{\epsilon}\left(D_{i}, X_{i}, W_{i}^{(Z)}\right)}}_{\psi_{3}^{\epsilon}} - \nonumber \\
    &-\underbrace{
    \triangle \overline{Z}^{\epsilon}
    \left(
        R_{\partial Y}^{\epsilon,d}
        \left(R_{D}^{\epsilon,d}-1\right)
        -
        R_{\partial Y}^{\epsilon,d^{*}}
        \left(R_{D}^{\epsilon,d^{*}}-1\right)
    \right)
}_{\psi_{4}^{\epsilon}}.
\end{align}

As in the proof for ATES, by Assumption~\ref{assumption:differentiability}, we can differentiate each term by the chain rule. Specifically, the derivative associated with the first term is as follows:
\begin{align}
    \frac{\partial}{\partial \epsilon}
    \psi_{1}^{\epsilon}\mid_{\epsilon = 0}
    &=
    \underbrace{
        \left(\triangle \mu_{Y}^{d}
        - \triangle \mu_{Y}^{d^{*}}\right)
    }_{\psi_{1A}}
    +
    \underbrace{
        \triangle p_{Z}
        \left(R_{\partial Y}^{0,d}
        - R_{\partial Y}^{0,d^{*}}\right)
    }_{\psi_{1B}}.
\end{align}

The derivative associated with the second term is as follows:
\begin{align}
    \frac{\partial}{\partial \epsilon}\psi_{2}^{\epsilon}\mid_{\epsilon = 0} &= -\underbrace{\frac{D_{i}^{(d)}\tilde{Z}_{i}\triangle Y^{0, d}\tilde{P}_{i}^{0}\triangle p_{D}^{d}}{\left(P_{i}^{0(d)}\tilde{P}_{i}^{0}\right)^{2}}}_{\psi_{2A}} - \underbrace{\frac{D_{i}^{(d)}\tilde{Z}_{i}\triangle Y^{0, d}\tilde{P}_{i}^{0}\triangle p_{Z}R_{\partial D}^{0, d}}{\left(P_{i}^{0(d)}\tilde{P}_{i}^{0}\right)^{2}}}_{\psi_{2B}} - \nonumber \\ 
    &- \underbrace{\frac{D_{i}^{(d)}\tilde{Z}_{i}\triangle Y^{0, d}P_{i}^{0(d)}\triangle \tilde{p}_{Z}}{\left(P_{i}^{0(d)}\tilde{P}_{i}^{0}\right)^{2}}}_{\psi_{2C}} - \underbrace{\frac{D_{i}^{(d)}\tilde{Z}_{i}\triangle \mu_{Y}^{d}}{P_{i}^{0(d)}\tilde{P}_{i}^{0}}}_{\psi_{2D}} - \underbrace{\frac{D_{i}^{(d)}\tilde{Z}_{i}\triangle p_{Z}R_{\partial Y}^{0, d}}{P_{i}^{0(d)}\tilde{P}_{i}^{0}}}_{\psi_{2E}}. 
\end{align}

The derivative associated with the third term is as follows:
\begin{align}
    \frac{\partial}{\partial \epsilon}\psi_{3}^{\epsilon}\mid_{\epsilon = 0} &= -\underbrace{\frac{D_{i}^{(d^{*})}\tilde{Z}_{i}\triangle Y^{0, d^{*}}\tilde{P}_{i}^{0}\triangle p_{D}^{d^{*}}}{\left(P_{i}^{0(d^{*})}\tilde{P}_{i}^{0}\right)^{2}}}_{\psi_{3A}} - \underbrace{\frac{D_{i}^{(d^{*})}\tilde{Z}_{i}\triangle Y^{0, d^{*}}\tilde{P}_{i}^{0}\triangle p_{Z}R_{\partial D}^{0, d^{*}}}{\left(P_{i}^{0(d^{*})}\tilde{P}_{i}^{0}\right)^{2}}}_{\psi_{3B}} - \nonumber \\ 
    &- \underbrace{\frac{D_{i}^{(d^{*})}\tilde{Z}_{i}\triangle Y^{0, d^{*}}P_{i}^{0(d^{*})}\triangle \tilde{p}_{Z}}{\left(P_{i}^{0(d^{*})}\tilde{P}_{i}^{0}\right)^{2}}}_{\psi_{3C}} - \underbrace{\frac{D_{i}^{(d^{*})}\tilde{Z}_{i}\triangle \mu_{Y}^{d^{*}}}{P_{i}^{0(d^{*})}\tilde{P}_{i}^{0}}}_{\psi_{3D}} - \underbrace{\frac{D_{i}^{(d^{*})}\tilde{Z}_{i}\triangle p_{Z}R_{\partial Y}^{0, d^{*}}}{P_{i}^{0(d^{*})}\tilde{P}_{i}^{0}}}_{\psi_{3E}}. 
\end{align}

The derivative associated with the fourth term is as follows:
\begin{align}
    \frac{\partial}{\partial \epsilon}
    \psi_{4}^{\epsilon}\mid_{\epsilon = 0}
    &=
    -\underbrace{
        \triangle p_{Z}
        \left(
            R_{\partial Y}^{0,d}
            \left(R_{D}^{0,d}-1\right)
            -
            R_{\partial Y}^{0,d^{*}}
            \left(R_{D}^{0,d^{*}}-1\right)
        \right)
    }_{\psi_{4A}} + \nonumber \\
    &+
    \underbrace{
        \triangle \overline{Z}^{0}
        \left(
            \frac{\partial}{\partial \epsilon}
            \left[
                R_{\partial Y}^{\epsilon,d}
                \left(R_{D}^{\epsilon,d}-1\right)
                -
                R_{\partial Y}^{\epsilon,d^{*}}
                \left(R_{D}^{\epsilon,d^{*}}-1\right)
            \right]\mid_{\epsilon = 0}
        \right)
    }_{\psi_{4B}}.
\end{align}

By summing the derivatives, taking the expectation, and grouping the terms, we obtain:
\begin{align}
    \mathbb{E}\left(\frac{\partial}{\partial \epsilon}\psi^{\epsilon}_{\mathrm{ATE}}\mid_{\epsilon = 0}\right) &=\underbrace{\mathbb{E}\left(\psi_{3A} - \psi_{2A} + \psi_{3B} - \psi_{2B} + \psi_{3C} - \psi_{2C}\right)}_{E_{1}} + \underbrace{\mathbb{E}\left(\psi_{1A} - \psi_{2D} + \psi_{3D}\right)}_{E_{2}} + \nonumber \\
    &+ \underbrace{\mathbb{E}\left( \psi_{1B}-\psi_{2E}+\psi_{3E}+\psi_{4A}\right)}_{E_{3}} + \underbrace{\mathbb{E}\left(-\psi_{4B}\right)}_{E_{4}}.
\end{align}

Next, we show that $E_{1} = 0$, $E_{2} = 0$, $E_{3} = 0$, and $E_{4} = 0$. Specifically, notice that:
\begin{align}
    \mathbb{E}\left(\psi_{2A}\right) &= \mathbb{E}\left(\frac{D_{i}^{(d)}\tilde{Z}_{i}\triangle Y^{0, d}\tilde{P}_{i}^{0}\triangle p_{D}^{d}}{\left(P_{i}^{0(d)}\tilde{P}_{i}^{0}\right)^{2}}\right) \propto \underbrace{\mathbb{E}\left(\frac{\triangle Y^{0, d}\tilde{P}_{i}^{0}\triangle p_{D}^{d}}{\left(P_{i}^{0(d)}\tilde{P}_{i}^{0}\right)^{2}}\mid D_{i} = d, \tilde{Z}_{i} = 1\right)}_{\text{LTE}} = \nonumber \\
    &= \underbrace{\mathbb{E}\left(\frac{\tilde{P}_{i}^{0}\triangle p_{D}^{d}}{\left(P_{i}^{0(d)}\tilde{P}_{i}^{0}\right)^{2}}\mathbb{E}\left(\triangle Y^{0, d}\mid X_{i}, W_{i}^{(Z)}, D_{i} = d, \tilde{Z}_{i} = 1\right)\mid D_{i} = d, \tilde{Z}_{i} = 1\right)}_{\text{LIE}} = \nonumber \\
    &= \underbrace{\mathbb{E}\left(\frac{\tilde{P}_{i}^{0}\triangle p_{D}^{d}}{\left(P_{i}^{0(d)}\tilde{P}_{i}^{0}\right)^{2}}\times 0 \mid D_{i} = d, \tilde{Z}_{i} = 1\right)}_{\text{Equation~(\ref{eq:useful:muY}) of Lemma~\ref{lemma:useful}}} = 0.
\end{align}
By the same steps, we obtain $\mathbb{E}\left(\psi_{2B}\right) = 0$, $\mathbb{E}\left(\psi_{2C}\right) = 0$, $\mathbb{E}\left(\psi_{3A}\right) = 0$, $\mathbb{E}\left(\psi_{3B}\right) = 0$, and $\mathbb{E}\left(\psi_{3C}\right) = 0$. These equalities imply $E_{1} = 0$.

Further, notice that the terms grouped in $E_{2}$ can be rewritten as the sum of the following terms:
\begin{align}
    E_{2} &= \underbrace{\mathbb{E}\left(\triangle \mu_{Y}^{d} - \frac{D_{i}^{(d)}\tilde{Z}_{i}\triangle \mu_{Y}^{d}}{P_{i}^{0(d)}\tilde{P}_{i}^{0}}\right)}_{E_{2A}} - \underbrace{\mathbb{E}\left(\triangle \mu_{Y}^{d^{*}} - \frac{D_{i}^{(d^{*})}\tilde{Z}_{i}\triangle \mu_{Y}^{d^{*}}}{P_{i}^{0(d^{*})}\tilde{P}_{i}^{0}}\right)}_{E_{2B}}.
\end{align}

For the first term, we obtain:
\begin{align}
    E_{2A} &= \mathbb{E}\left(\triangle \mu_{Y}^{d}\left(1 - \frac{D_{i}^{(d)}\tilde{Z}_{i}}{P_{i}^{0(d)}\tilde{P}_{i}^{0}}\right)\right) = \nonumber \\
    &= \underbrace{\mathbb{E}\left(\mathbb{E}\left[\triangle \mu_{Y}^{d}\left(1 - \frac{D_{i}^{(d)}\tilde{Z}_{i}}{P_{i}^{0(d)}\tilde{P}_{i}^{0}}\right)\mid X_{i}, W_{i}^{(Z)}, D_{i}\right]\right)}_{\text{LIE}} = \nonumber \\
    &= \underbrace{\mathbb{E}\left(\triangle \mu_{Y}^{d}\left(1 - \frac{D_{i}^{(d)}\tilde{P}_{i}^{0}}{P_{i}^{0(d)}\tilde{P}_{i}^{0}}\right)\right)}_{\text{Since }\mathbb{E}\left(\tilde{Z}_{i}\mid X_{i}, W_{i}^{(Z)}, D_{i}\right)=\tilde{P}_{i}^{0}} = \mathbb{E}\left(\triangle \mu_{Y}^{d}\left(1 - \frac{D_{i}^{(d)}}{P_{i}^{0(d)}}\right)\right) = \nonumber \\
    &= \underbrace{\mathbb{E}\left(\mathbb{E}\left[\triangle \mu_{Y}^{d}\left(1 - \frac{D_{i}^{(d)}}{P_{i}^{0(d)}}\right)\mid X_{i}, P_{i}^{0}\right]\right)}_{\text{LIE}} = \nonumber \\
    &= \mathbb{E}\left(\triangle \mu_{Y}^{d}\left(1 - \frac{\mathbb{E}\left[D_{i}^{(d)}\mid X_{i}, P_{i}^{0}\right]}{P_{i}^{0(d)}}\right)\right) = \nonumber \\
    &= \underbrace{\mathbb{E}\left(\triangle \mu_{Y}^{d}\left(1 - \frac{P_{i}^{0(d)}}{P_{i}^{0(d)}}\right)\right)}_{\text{By the definition of }P_{i}^{0(d)}} = \mathbb{E}\left(\triangle \mu_{Y}^{d}\left(1 - 1\right)\right) = 0.
\end{align}
By the same steps, for the second term, we obtain $E_{2B} = 0$. These equalities imply $E_{2} = 0$.

Further, notice that the terms grouped in $E_{3}$ can be rewritten as follows:
\begin{align}
    E_{3}
&= \mathbb{E}\left(
    \psi_{1B}-\psi_{2E}+\psi_{3E}+\psi_{4A}
\right) = \nonumber \\
&= \mathbb{E}\left(
    -\psi_{2E}+\psi_{3E}
    +
    \underbrace{
        \triangle p_{Z}
        \left(
            R_{\partial Y}^{0,d}R_{D}^{0,d}
            -
            R_{\partial Y}^{0,d^{*}}R_{D}^{0,d^{*}}
        \right)
    }_{\psi_{1B}+\psi_{4A}}
\right) = \nonumber \\
    &= \mathbb{E}\left(\triangle p_{Z}\left(-\frac{D_{i}^{(d)}\tilde{Z}_{i}R_{\partial Y}^{0, d}}{P_{i}^{0(d)}\tilde{P}_{i}^{0}} + \frac{D_{i}^{(d^{*})}\tilde{Z}_{i}R_{\partial Y}^{0, d^{*}}}{P_{i}^{0(d^{*})}\tilde{P}_{i}^{0}} + R_{\partial Y}^{0, d}\frac{D_{i}^{(d)}}{P_{i}^{0(d)}} - R_{\partial Y}^{0, d^{*}}\frac{D_{i}^{(d^{*})}}{P_{i}^{0(d^{*})}}\right)\right) = \nonumber \\
    &= \mathbb{E}\left(\triangle p_{Z}R_{\partial Y}^{0, d}\frac{D_{i}^{(d)}}{P_{i}^{0(d)}}\left(1 - \frac{\tilde{Z}_{i}}{\tilde{P}_{i}^{0}}\right)\right) - \mathbb{E}\left(\triangle p_{Z}R_{\partial Y}^{0, d^{*}}\frac{D_{i}^{(d^{*})}}{P_{i}^{0(d^{*})}}\left(1 - \frac{\tilde{Z}_{i}}{\tilde{P}_{i}^{0}}\right)\right) = \nonumber \\
    &= \underbrace{\mathbb{E}\left(\mathbb{E}\left[\triangle p_{Z}\left(R_{\partial Y}^{0, d}\frac{D_{i}^{(d)}}{P_{i}^{0(d)}} - R_{\partial Y}^{0, d^{*}}\frac{D_{i}^{(d^{*})}}{P_{i}^{0(d^{*})}}\right)\left(1 - \frac{\tilde{Z}_{i}}{\tilde{P}_{i}^{0}}\right)\mid D_{i}, X_{i}, W_{i}^{(Z)}\right]\right)}_{\text{LIE}} = \nonumber \\
    &= \underbrace{\mathbb{E}\left(\triangle p_{Z}\left(R_{\partial Y}^{0, d}\frac{D_{i}^{(d)}}{P_{i}^{0(d)}} - R_{\partial Y}^{0, d^{*}}\frac{D_{i}^{(d^{*})}}{P_{i}^{0(d^{*})}}\right)\frac{1}{\tilde{P}_{i}^{0}}\left(\tilde{P}_{i}^{0} - \mathbb{E}\left[\tilde{Z}_{i}\mid D_{i}, X_{i}, W_{i}^{(Z)}\right]\right)\right)}_{\substack{\text{Since }\triangle p_{Z}, R_{\partial Y}^{0, d}, R_{\partial Y}^{0, d^{*}}, D_{i}^{(d)}, D_{i}^{(d^{*})}, P_{i}^{0(d)}, P_{i}^{0(d^{*})}\text{, and }\tilde{P}_{i}^{0}\\ \text{are functions of }D_{i}, X_{i}, W_{i}^{(Z)}}} = \nonumber \\
    &= \underbrace{\mathbb{E}\left(\triangle p_{Z}\left(R_{\partial Y}^{0, d}\frac{D_{i}^{(d)}}{P_{i}^{0(d)}} - R_{\partial Y}^{0, d^{*}}\frac{D_{i}^{(d^{*})}}{P_{i}^{0(d^{*})}}\right)\frac{1}{\tilde{P}_{i}^{0}}\times 0\right)}_{\text{Since }\tilde{P}_{i}^{0}=\mathbb{E}\left(\tilde{Z}_{i}\mid D_{i}, X_{i}, W_{i}^{(Z)}\right)} = 0.
\end{align}

Finally, notice that:
\begin{align}
    E_{4} &= \mathbb{E}\left(-\psi_{4B}\right) = -\mathbb{E}\left(\triangle \overline{Z}^{0}\left(\frac{\partial}{\partial \epsilon}\left[R_{\partial Y}^{\epsilon, d}\left(R_{D}^{\epsilon, d}-1\right) - R_{\partial Y}^{\epsilon, d^{*}}\left(R_{D}^{\epsilon, d^{*}}-1\right)\right]\mid_{\epsilon = 0}\right)\right) = \nonumber \\
    &= -\underbrace{\mathbb{E}\left(\mathbb{E}\left[\left(\overline{Z}_{i} - P_{i}^{0}\right)\left(\frac{\partial}{\partial \epsilon}\left[R_{\partial Y}^{\epsilon, d}\left(R_{D}^{\epsilon, d}-1\right) - R_{\partial Y}^{\epsilon, d^{*}}\left(R_{D}^{\epsilon, d^{*}}-1\right)\right]\mid_{\epsilon = 0}\right)\mid D_{i}, X_{i}, W_{i}^{(Z)}\right]\right)}_{\text{LIE}} = \nonumber \\
    &= -\underbrace{\mathbb{E}\left(\mathbb{E}\left[\overline{Z}_{i} - P_{i}^{0}\mid D_{i}, X_{i}, W_{i}^{(Z)}\right]\left(\frac{\partial}{\partial \epsilon}\left[R_{\partial Y}^{\epsilon, d}\left(R_{D}^{\epsilon, d}-1\right) - R_{\partial Y}^{\epsilon, d^{*}}\left(R_{D}^{\epsilon, d^{*}}-1\right)\right]\mid_{\epsilon = 0}\right)\right)}_{\substack{\text{Since the derivative }\frac{\partial}{\partial \epsilon}\left[\ldots\right]\mid_{\epsilon = 0}\text{ is a function of }D_{i}, X_{i}, W_{i}^{(Z)}}} = \nonumber \\
    &= -\underbrace{\mathbb{E}\left(\mathbb{E}\left[P_{i}^{0} - P_{i}^{0}\mid D_{i}, X_{i}, W_{i}^{(Z)}\right]\left(\frac{\partial}{\partial \epsilon}\left[R_{\partial Y}^{\epsilon, d}\left(R_{D}^{\epsilon, d}-1\right) - R_{\partial Y}^{\epsilon, d^{*}}\left(R_{D}^{\epsilon, d^{*}}-1\right)\right]\mid_{\epsilon = 0}\right)\right)}_{\text{Since }\mathbb{E}\left[\overline{Z}_{i}\mid D_{i}, X_{i}, W_{i}^{(Z)}\right]=P_{i}^{0}} = 0.
\end{align}

Therefore, by Assumption~\ref{assumption:differentiability} and the dominated convergence theorem, as in the proof for ATES, we obtain:
\begin{align}
    \left.\frac{\partial}{\partial\epsilon}
    \mathbb{E}\left(\psi_{\mathrm{ATE}}^\epsilon\right)
    \right|_{\epsilon=0}
    &=
    \mathbb{E}\left(
        \left.\frac{\partial}{\partial\epsilon}
        \psi_{\mathrm{ATE}}^\epsilon
        \right|_{\epsilon=0}
    \right) = E_1+E_2+E_3+E_4=0.
\end{align}

\subsection{Neyman-orthogonality of the score of ATT}

\subsubsection{Mean zero of ATT score}

Consider the score at true values of the parameter and nuisance functions\footnote{As in the proof for ATE, Assumptions~\ref{assumption:support} and~\ref{assumption:selection-support} imply that $\tilde{P}_{i}^{0}>0$ and $P_{i}^{0(\overline{d})}>0$ almost surely for every $\overline{d}\in\{d,d^{*}\}$.}:
\begin{align}
    \psi_{\mathrm{ATT}}^{0} &= \underbrace{D_{i}^{(d)}\left(\text{E}_{i}^{0(d)} - \text{E}_{i}^{0(d^{*})}\right) - D_{i}^{(d)}\mathrm{ATT}}_{\psi^{0}_{1}} + \underbrace{\frac{D_{i}^{(d)}\tilde{Z}_{i}}{\tilde{P}_{i}^{0}}\left(Y_{i} - \text{E}_{i}^{0(d)}\right)}_{\psi^{0}_{2}} - \nonumber \\
    &- \underbrace{\frac{D_{i}^{(d^{*})}\tilde{Z}_{i}R_{PD}^{0}}{\tilde{P}_{i}^{0}}\left(Y_{i} - \text{E}_{i}^{0(d^{*})}\right)}_{\psi^{0}_{3}} - \underbrace{\left(\overline{Z}_{i} - P_{i}^{0}\right)\text{E}_{\partial i}^{0(d^{*})}D_{i}^{(d)}}_{\psi^{0}_{4}} + \underbrace{\left(\overline{Z}_{i} - P_{i}^{0}\right)\text{E}_{\partial i}^{0(d^{*})}D_{i}^{(d^{*})}R_{PD}^{0}}_{\psi^{0}_{5}}.
\end{align}
By Assumption~\ref{assumption:differentiability}, all these terms are well-defined and integrable.

Next, we show that the expectations of $\psi_{1}^{0}$, $\psi_{2}^{0}$, $\psi_{3}^{0}$, $\psi_{4}^{0}$, and $\psi_{5}^{0}$ are $0$. Specifically, notice that Theorem~\ref{theorem:ate} implies $\mathbb{E}\left(\psi_{1}^{0}\right) = 0$. Further, notice that:
\begin{align}
    \mathbb{E}\left(\psi_{2}^{0}\right) &= \mathbb{E}\left(\frac{D_{i}^{(d)}\tilde{Z}_{i}}{\tilde{P}_{i}^{0}}\left(Y_{i} - \text{E}_{i}^{0(d)}\right)\right) = \nonumber \\
    &= \underbrace{\mathbb{E}\left(\frac{1}{\tilde{P}_{i}^{0}}\left(Y_{i} - \text{E}_{i}^{0(d)}\right)\mid D_{i} = d, \tilde{Z}_{i} = 1\right)\mathbb{P}\left(D_{i} = d, \tilde{Z}_{i} = 1\right)}_{\substack{\text{LTE with the events }\{D_{i} = d, \tilde{Z}_{i} = 1\}\text{ and }\{D_{i} \ne d\text{ or }\tilde{Z}_{i} \ne 1\}}} \propto \nonumber \\
    &\propto \underbrace{\mathbb{E}\left(\mathbb{E}\left(\frac{1}{\tilde{P}_{i}^{0}}\left(Y_{i} - \text{E}_{i}^{0(d)}\right)\mid X_{i}, W_{i}^{(Z)}, D_{i} = d, \tilde{Z}_{i} = 1\right)\mid D_{i} = d, \tilde{Z}_{i} = 1\right)}_{\text{LIE}} = \nonumber \\
    &= \underbrace{\mathbb{E}\left(\frac{1}{\tilde{P}_{i}^{0}}\left(\mathbb{E}\left(Y_{i}\mid X_{i}, W_{i}^{(Z)}, D_{i} = d, \tilde{Z}_{i} = 1\right) - \text{E}_{i}^{0(d)}\right)\mid D_{i} = d, \tilde{Z}_{i} = 1\right)}_{\text{Since }\tilde{P}_{i}^{0}\text{ and }\text{E}_{i}^{0(d)}\text{ are functions of }X_{i}, W_{i}^{(Z)}\text{ on }\{D_{i} = d\}} = \nonumber \\
    &= \underbrace{\mathbb{E}\left(\frac{1}{\tilde{P}_{i}^{0}}\left(\text{E}_{i}^{0(d)} - \text{E}_{i}^{0(d)}\right)\mid D_{i} = d, \tilde{Z}_{i} = 1\right)}_{\text{Equation~(\ref{eq:useful:muY}) of Lemma~\ref{lemma:useful}}} = 0.
\end{align}
By the same steps, we obtain $\mathbb{E}\left(\psi_{3}^{0}\right) = 0$. 

Finally, notice that:
\begin{align}
    \mathbb{E}\left(\psi_{4}^{0}\right) &= \mathbb{E}\left(\left(\overline{Z}_{i} - P_{i}^{0}\right)\underbrace{\text{E}_{\partial i}^{0(d^{*})}D_{i}^{(d)}}_{A_{i}}\right) = \nonumber \\
    &= \underbrace{\mathbb{E}\left(\mathbb{E}\left(\left(\overline{Z}_{i} - P_{i}^{0}\right)A_{i}\mid D_{i}, X_{i}, W_{i}^{(Z)}\right)\right)}_{\text{LIE}} = \nonumber \\
    &= \underbrace{\mathbb{E}\left(\mathbb{E}\left(\overline{Z}_{i} - P_{i}^{0}\mid D_{i}, X_{i}, W_{i}^{(Z)}\right)A_{i}\right)}_{\text{Since }A_{i}\text{ is a function of }D_{i}, X_{i}, W_{i}^{(Z)}} = \nonumber \\
    &= \underbrace{\mathbb{E}\left(\mathbb{E}\left(P_{i}^{0} - P_{i}^{0}\mid D_{i}, X_{i}, W_{i}^{(Z)}\right)A_{i}\right)}_{\text{Since }\mathbb{E}\left(\overline{Z}_{i}\mid D_{i}, X_{i}, W_{i}^{(Z)}\right)=P_{i}^{0}} = 0.
\end{align}
By the same steps, we obtain $\mathbb{E}\left(\psi_{5}^{0}\right) = 0$. Therefore, we have:
\begin{align}
    \mathbb{E}\left(\psi_{\mathrm{ATT}}^{0}\right) = \mathbb{E}\left(\psi_{1}^{0}\right) + \mathbb{E}\left(\psi_{2}^{0}\right) - \mathbb{E}\left(\psi_{3}^{0}\right) - \mathbb{E}\left(\psi_{4}^{0}\right) + \mathbb{E}\left(\psi_{5}^{0}\right) = 0 + 0 - 0 - 0 + 0 = 0.
\end{align}

\subsubsection{Zero Gateaux derivative of ATT score}

The perturbed score is as follows:
\begin{align}
    \psi_{\mathrm{ATT}}^{\epsilon} &= \underbrace{D_{i}^{(d)}\triangle \mu_{Y}^{\epsilon} - D_{i}^{(d)}\mathrm{ATT}}_{\psi_{1}^{\epsilon}} + \underbrace{\frac{D_{i}^{(d)}\tilde{Z}_{i}\triangle Y^{\epsilon, d}}{\tilde{P}_{i}^{\epsilon}}}_{\psi_{2}^{\epsilon}} - \underbrace{\frac{D_{i}^{(d^{*})}\tilde{Z}_{i}R_{PD}^{\epsilon}\triangle Y^{\epsilon, d^{*}}}{\tilde{P}_{i}^{\epsilon}}}_{\psi_{3}^{\epsilon}} - \nonumber \\
    &- \underbrace{\triangle \overline{Z}^{\epsilon}R_{\partial Y}^{\epsilon, d^{*}}D_{i}^{(d)}}_{\psi_{4}^{\epsilon}} + \underbrace{\triangle \overline{Z}^{\epsilon}R_{\partial Y}^{\epsilon, d^{*}}D_{i}^{(d^{*})}R_{PD}^{\epsilon}}_{\psi_{5}^{\epsilon}}.
\end{align}

As in the proof for ATES, by Assumption~\ref{assumption:differentiability}, we can differentiate each term by the chain rule. Specifically, the derivative associated with the first term is as follows:
\begin{align}
    \frac{\partial}{\partial \epsilon}\psi_{1}^{\epsilon}\mid_{\epsilon = 0} &= \underbrace{D_{i}^{(d)}\left(\triangle \mu_{Y}^{d} - \triangle \mu_{Y}^{d^{*}}\right)}_{\psi_{1A}} + \underbrace{D_{i}^{(d)}\triangle p_{Z}\left(R_{\partial Y}^{0, d} - R_{\partial Y}^{0, d^{*}}\right)}_{\psi_{1B}}. 
\end{align}

The derivative associated with the second term is as follows:
\begin{align}
    \frac{\partial}{\partial \epsilon}\psi_{2}^{\epsilon}\mid_{\epsilon = 0} &= -\underbrace{\frac{D_{i}^{(d)}\tilde{Z}_{i}\triangle Y^{0, d}\triangle \tilde{p}_{Z}}{\left(\tilde{P}_{i}^{0}\right)^{2}}}_{\psi_{2A}} - \underbrace{\frac{D_{i}^{(d)}\tilde{Z}_{i}\triangle \mu_{Y}^{d}}{\tilde{P}_{i}^{0}}}_{\psi_{2B}} - \underbrace{\frac{D_{i}^{(d)}\tilde{Z}_{i}\triangle p_{Z}R_{\partial Y}^{0, d}}{\tilde{P}_{i}^{0}}}_{\psi_{2C}}. 
\end{align}

The derivative associated with the third term is as follows:
\begin{align}
    \frac{\partial}{\partial \epsilon}\psi_{3}^{\epsilon}\mid_{\epsilon = 0} &= \underbrace{\frac{D_{i}^{(d^{*})}\tilde{Z}_{i}\triangle Y^{0, d^{*}}}{\tilde{P}_{i}^{0}}\left[\frac{\partial}{\partial \epsilon}R_{PD}^{\epsilon}\mid_{\epsilon = 0}\right]}_{\psi_{3A}} - \underbrace{\frac{D_{i}^{(d^{*})}\tilde{Z}_{i}\triangle Y^{0, d^{*}}R_{PD}^{0}\triangle \tilde{p}_{Z}}{\left(\tilde{P}_{i}^{0}\right)^{2}}}_{\psi_{3B}} - \nonumber \\ 
    &- \underbrace{\frac{D_{i}^{(d^{*})}\tilde{Z}_{i}R_{PD}^{0}\triangle \mu_{Y}^{d^{*}}}{\tilde{P}_{i}^{0}}}_{\psi_{3C}} - \underbrace{\frac{D_{i}^{(d^{*})}\tilde{Z}_{i}R_{PD}^{0}\triangle p_{Z}R_{\partial Y}^{0, d^{*}}}{\tilde{P}_{i}^{0}}}_{\psi_{3D}}. 
\end{align}

The derivative associated with the fourth term is as follows:
\begin{align}
    \frac{\partial}{\partial \epsilon}\psi_{4}^{\epsilon}\mid_{\epsilon = 0} &= -\underbrace{\triangle p_{Z}R_{\partial Y}^{0, d^{*}}D_{i}^{(d)}}_{\psi_{4A}} + \underbrace{\left(\overline{Z}_{i} - P_{i}^{0}\right)\left[\frac{\partial}{\partial \epsilon}R_{\partial Y}^{\epsilon, d^{*}}\mid_{\epsilon = 0}\right]D_{i}^{(d)}}_{\psi_{4B}}. 
\end{align}

The derivative associated with the fifth term is as follows:
\begin{align}
    \frac{\partial}{\partial \epsilon}\psi_{5}^{\epsilon}\mid_{\epsilon = 0} &= -\underbrace{\triangle p_{Z}R_{\partial Y}^{0, d^{*}}R_{PD}^{0}D_{i}^{(d^{*})}}_{\psi_{5A}} + \underbrace{\left(\overline{Z}_{i} - P_{i}^{0}\right)\left[\frac{\partial}{\partial \epsilon}\left[R_{\partial Y}^{\epsilon, d^{*}}R_{PD}^{\epsilon}\right]\mid_{\epsilon = 0}\right]D_{i}^{(d^{*})}}_{\psi_{5B}}. 
\end{align}

By summing the derivatives, taking the expectation, and grouping the terms, we obtain:
\begin{align}
    \mathbb{E}\left(\frac{\partial}{\partial \epsilon}\psi^{\epsilon}_{\mathrm{ATT}}\mid_{\epsilon = 0}\right) &=\underbrace{\mathbb{E}\left(-\psi_{3A} - \psi_{2A} + \psi_{3B}\right)}_{E_{1}} + \underbrace{\mathbb{E}\left(\psi_{1A} - \psi_{2B} + \psi_{3C}\right)}_{E_{2}} + \nonumber \\
    &+ \underbrace{\mathbb{E}\left(\psi_{1B} + \psi_{4A} - \psi_{2C} + \psi_{3D} - \psi_{5A}\right)}_{E_{3}} + \underbrace{\mathbb{E}\left(\psi_{5B} - \psi_{4B}\right)}_{E_{4}}.
\end{align}

Next, we show that $E_{1} = 0$, $E_{2} = 0$, $E_{3} = 0$, and $E_{4} = 0$. Specifically, notice that:
\begin{align}
    \mathbb{E}\left(\psi_{2A}\right) &= \mathbb{E}\left(\frac{D_{i}^{(d)}\tilde{Z}_{i}\triangle Y^{0, d}\triangle \tilde{p}_{Z}}{\left(\tilde{P}_{i}^{0}\right)^{2}}\right) \propto \underbrace{\mathbb{E}\left(\frac{\triangle Y^{0, d}\triangle \tilde{p}_{Z}}{\left(\tilde{P}_{i}^{0}\right)^{2}}\mid D_{i} = d, \tilde{Z}_{i} = 1\right)}_{\text{LTE}} = \nonumber \\
    &= \underbrace{\mathbb{E}\left(\frac{\triangle \tilde{p}_{Z}}{\left(\tilde{P}_{i}^{0}\right)^{2}}\mathbb{E}\left(\triangle Y^{0, d}\mid X_{i}, W_{i}^{(Z)}, D_{i} = d, \tilde{Z}_{i} = 1\right)\mid D_{i} = d, \tilde{Z}_{i} = 1\right)}_{\text{LIE}} = \nonumber \\
    &= \underbrace{\mathbb{E}\left(\frac{\triangle \tilde{p}_{Z}}{\left(\tilde{P}_{i}^{0}\right)^{2}}\times 0 \mid D_{i} = d, \tilde{Z}_{i} = 1\right)}_{\text{Equation~(\ref{eq:useful:muY}) of Lemma~\ref{lemma:useful}}} = 0.
\end{align}
By the same steps, we obtain $\mathbb{E}\left(\psi_{3A}\right) = 0$ and $\mathbb{E}\left(\psi_{3B}\right) = 0$. These equalities imply $E_{1} = 0$.

Further, notice that the terms grouped in $E_{2}$ can be rewritten as the sum of the following terms:
\begin{align}
    E_{2} &= \underbrace{\mathbb{E}\left(D_{i}^{(d)}\triangle \mu_{Y}^{d} - \frac{D_{i}^{(d)}\tilde{Z}_{i}\triangle \mu_{Y}^{d}}{\tilde{P}_{i}^{0}}\right)}_{E_{2A}} - \underbrace{\mathbb{E}\left(D_{i}^{(d)}\triangle \mu_{Y}^{d^{*}} - \frac{D_{i}^{(d^{*})}\tilde{Z}_{i}R_{PD}^{0}\triangle \mu_{Y}^{d^{*}}}{\tilde{P}_{i}^{0}}\right)}_{E_{2B}}.
\end{align}

For the first term, we obtain:
\begin{align}
    E_{2A} &= \mathbb{E}\left(D_{i}^{(d)}\triangle \mu_{Y}^{d}\left(1 - \frac{\tilde{Z}_{i}}{\tilde{P}_{i}^{0}}\right)\right) \propto \underbrace{\mathbb{E}\left(\triangle \mu_{Y}^{d}\left(1 - \frac{\tilde{Z}_{i}}{\tilde{P}_{i}^{0}}\right)\mid D_{i} = d\right)}_{\text{LTE}} = \nonumber \\
    &= \underbrace{\mathbb{E}\left(\mathbb{E}\left[\triangle \mu_{Y}^{d}\left(1 - \frac{\tilde{Z}_{i}}{\tilde{P}_{i}^{0}}\right)\mid X_{i}, W_{i}^{(Z)}, D_{i} = d\right]\mid D_{i} = d\right)}_{\text{LIE}} = \nonumber \\
    &= \mathbb{E}\left(\triangle \mu_{Y}^{d}\left(1 - \frac{\mathbb{E}\left(\tilde{Z}_{i}\mid X_{i}, W_{i}^{(Z)}, D_{i} = d\right)}{\tilde{P}_{i}^{0}}\right)\mid D_{i} = d\right) = \nonumber \\
    &= \underbrace{\mathbb{E}\left(\triangle \mu_{Y}^{d}\left(1 - \frac{\tilde{P}_{i}^{0}}{\tilde{P}_{i}^{0}}\right)\mid D_{i} = d\right)}_{\text{Since }\tilde{P}_{i}^{0}=\mathbb{E}\left[\tilde{Z}_{i}\mid X_{i}, W_{i}^{(Z)}, D_{i}\right]} = \mathbb{E}\left(\triangle \mu_{Y}^{d}\times 0\mid D_{i} = d\right) = 0.
\end{align}

For the second term, we obtain:
\begin{align}
    E_{2B} &= \mathbb{E}\left(D_{i}^{(d)}\triangle \mu_{Y}^{d^{*}} - \frac{D_{i}^{(d^{*})}\tilde{Z}_{i}R_{PD}^{0}\triangle \mu_{Y}^{d^{*}}}{\tilde{P}_{i}^{0}}\right) = \nonumber \\
    &= \underbrace{\mathbb{E}\left(\mathbb{E}\left[D_{i}^{(d)}\triangle \mu_{Y}^{d^{*}} - \frac{D_{i}^{(d^{*})}\tilde{Z}_{i}R_{PD}^{0}\triangle \mu_{Y}^{d^{*}}}{\tilde{P}_{i}^{0}}\mid D_{i}, X_{i}, W_{i}^{(Z)}\right]\right)}_{\text{LIE}} = \nonumber \\
    &= \underbrace{\mathbb{E}\left(\triangle \mu_{Y}^{d^{*}}\left(D_{i}^{(d)} - D_{i}^{(d^{*})}R_{PD}^{0}\right)\right)}_{\text{Since }\mathbb{E}\left[\tilde{Z}_{i}\mid D_{i}, X_{i}, W_{i}^{(Z)}\right]=\tilde{P}_{i}^{0}} = \nonumber \\
    &= \underbrace{\mathbb{E}\left(\mathbb{E}\left[\triangle \mu_{Y}^{d^{*}}\left(D_{i}^{(d)} - D_{i}^{(d^{*})}R_{PD}^{0}\right)\mid X_{i}, P_{i}^{0}\right]\right)}_{\text{LIE}} = \nonumber \\
    &= \underbrace{\mathbb{E}\left(\triangle \mu_{Y}^{d^{*}}\left(P_{i}^{0(d)} - P_{i}^{0(d)}\right)\right)}_{\substack{\text{Since }\mathbb{E}\left[D_{i}^{(d)}\mid X_{i}, P_{i}^{0}\right] = P_{i}^{0(d)}\text{, and}\\ \mathbb{E}\left[D_{i}^{(d^{*})}R_{PD}^{0}\mid X_{i}, P_{i}^{0}\right] = P_{i}^{0(d^{*})}R_{PD}^{0} = P_{i}^{0(d)}}} = 0.
\end{align}
These equalities imply $E_{2} = 0$.

Further, notice that the terms grouped in $E_{3}$ can be rewritten as follows:
\begin{align}
    E_{3} &= \mathbb{E}\left(\psi_{1B} + \psi_{4A} - \psi_{2C} + \psi_{3D} - \psi_{5A}\right) = \nonumber \\
    &= \mathbb{E}\left(\underbrace{\triangle p_{Z}R_{\partial Y}^{0, d}D_{i}^{(d)}}_{\psi_{1B}+\psi_{4A}} - \psi_{2C} + \psi_{3D} - \psi_{5A}\right) = \nonumber \\
    &= \mathbb{E}\left(\triangle p_{Z}\left(R_{\partial Y}^{0, d}D_{i}^{(d)} - \frac{D_{i}^{(d)}\tilde{Z}_{i}R_{\partial Y}^{0, d}}{\tilde{P}_{i}^{0}} + \frac{D_{i}^{(d^{*})}\tilde{Z}_{i}R_{PD}^{0}R_{\partial Y}^{0, d^{*}}}{\tilde{P}_{i}^{0}} - R_{\partial Y}^{0, d^{*}}R_{PD}^{0}D_{i}^{(d^{*})}\right)\right) = \nonumber \\
    &= \mathbb{E}\left(\triangle p_{Z}\left(R_{\partial Y}^{0, d}D_{i}^{(d)} - R_{\partial Y}^{0, d^{*}}R_{PD}^{0}D_{i}^{(d^{*})}\right)\left(1 - \frac{\tilde{Z}_{i}}{\tilde{P}_{i}^{0}}\right)\right) = \nonumber \\
    &= \underbrace{\mathbb{E}\left(\mathbb{E}\left[\triangle p_{Z}\left(R_{\partial Y}^{0, d}D_{i}^{(d)} - R_{\partial Y}^{0, d^{*}}R_{PD}^{0}D_{i}^{(d^{*})}\right)\left(1 - \frac{\tilde{Z}_{i}}{\tilde{P}_{i}^{0}}\right)\mid D_{i}, X_{i}, W_{i}^{(Z)}\right]\right)}_{\text{LIE}} = \nonumber \\
    &= \underbrace{\mathbb{E}\left(\triangle p_{Z}\left(R_{\partial Y}^{0, d}D_{i}^{(d)} - R_{\partial Y}^{0, d^{*}}R_{PD}^{0}D_{i}^{(d^{*})}\right)\frac{1}{\tilde{P}_{i}^{0}}\left(\tilde{P}_{i}^{0} - \mathbb{E}\left[\tilde{Z}_{i}\mid D_{i}, X_{i}, W_{i}^{(Z)}\right]\right)\right)}_{\substack{\text{Since }\triangle p_{Z}, R_{\partial Y}^{0, d}, R_{\partial Y}^{0, d^{*}}, R_{PD}^{0}, D_{i}^{(d)}, D_{i}^{(d^{*})}\text{, and }\tilde{P}_{i}^{0}\text{ are functions of }D_{i}, X_{i}, W_{i}^{(Z)}}} = \nonumber \\
    &= \underbrace{\mathbb{E}\left(\triangle p_{Z}\left(R_{\partial Y}^{0, d}D_{i}^{(d)} - R_{\partial Y}^{0, d^{*}}R_{PD}^{0}D_{i}^{(d^{*})}\right)\frac{1}{\tilde{P}_{i}^{0}}\times 0\right)}_{\text{Since }\tilde{P}_{i}^{0}=\mathbb{E}\left(\tilde{Z}_{i}\mid D_{i}, X_{i}, W_{i}^{(Z)}\right)} = 0.
\end{align}

Finally, notice that:
\begin{align}
    E_{4} &= \mathbb{E}\left(\psi_{5B} - \psi_{4B}\right) = \nonumber \\
    &= \mathbb{E}\left(\left(\overline{Z}_{i} - P_{i}^{0}\right)\left(\frac{\partial}{\partial \epsilon}\left[R_{\partial Y}^{\epsilon, d^{*}}R_{PD}^{\epsilon}\right]\mid_{\epsilon = 0}D_{i}^{(d^{*})} - \frac{\partial}{\partial \epsilon}R_{\partial Y}^{\epsilon, d^{*}}\mid_{\epsilon = 0}D_{i}^{(d)}\right)\right) = \nonumber \\
    &= \underbrace{\mathbb{E}\left(\mathbb{E}\left[\left(\overline{Z}_{i} - P_{i}^{0}\right)\left(\frac{\partial}{\partial \epsilon}\left[R_{\partial Y}^{\epsilon, d^{*}}R_{PD}^{\epsilon}\right]\mid_{\epsilon = 0}D_{i}^{(d^{*})} - \frac{\partial}{\partial \epsilon}R_{\partial Y}^{\epsilon, d^{*}}\mid_{\epsilon = 0}D_{i}^{(d)}\right)\mid D_{i}, X_{i}, W_{i}^{(Z)}\right]\right)}_{\text{LIE}} = \nonumber \\
    &= \underbrace{\mathbb{E}\left(\mathbb{E}\left[\overline{Z}_{i} - P_{i}^{0}\mid D_{i}, X_{i}, W_{i}^{(Z)}\right]\left(\frac{\partial}{\partial \epsilon}\left[R_{\partial Y}^{\epsilon, d^{*}}R_{PD}^{\epsilon}\right]\mid_{\epsilon = 0}D_{i}^{(d^{*})} - \frac{\partial}{\partial \epsilon}R_{\partial Y}^{\epsilon, d^{*}}\mid_{\epsilon = 0}D_{i}^{(d)}\right)\right)}_{\substack{\text{Since the derivatives }\frac{\partial}{\partial \epsilon}\left[\ldots\right]\mid_{\epsilon = 0}\text{ are functions of }D_{i}, X_{i}, W_{i}^{(Z)}}} = \nonumber \\
    &= \underbrace{\mathbb{E}\left(\mathbb{E}\left[P_{i}^{0} - P_{i}^{0}\mid D_{i}, X_{i}, W_{i}^{(Z)}\right]\left(\frac{\partial}{\partial \epsilon}\left[R_{\partial Y}^{\epsilon, d^{*}}R_{PD}^{\epsilon}\right]\mid_{\epsilon = 0}D_{i}^{(d^{*})} - \frac{\partial}{\partial \epsilon}R_{\partial Y}^{\epsilon, d^{*}}\mid_{\epsilon = 0}D_{i}^{(d)}\right)\right)}_{\text{Since }\mathbb{E}\left[\overline{Z}_{i}\mid D_{i}, X_{i}, W_{i}^{(Z)}\right]=P_{i}^{0}} = 0.
\end{align}

Therefore, by Assumption~\ref{assumption:differentiability} and the dominated convergence theorem, as in the proof for ATES, we obtain:
\begin{align}
    \left.\frac{\partial}{\partial\epsilon}
    \mathbb{E}\left(\psi_{\mathrm{ATT}}^\epsilon\right)
    \right|_{\epsilon=0}
    &=
    \mathbb{E}\left(
        \left.\frac{\partial}{\partial\epsilon}
        \psi_{\mathrm{ATT}}^\epsilon
        \right|_{\epsilon=0}
    \right) = E_1+E_2+E_3+E_4=0.
\end{align}

\section{Proof of Lemma 4}

Recall that $Q_i = (D_i, X_i, W_i^{(Z)})$. According to Assumption~\ref{assumption:exogeneity}, we have $(Y_{di}, U_i^{(Z)}) \perp (D_i, W_i^{(Z)}) \mid X_i$. By the decomposition property, this implies that $U_i^{(Z)} \perp (D_i,W_i^{(Z)})\mid X_i$.

Denote by $P_{ji}^{0}$ the $j$-th element of $P_i^0$. First, we show by induction that $Z_{ji}^{*}$ is almost surely equal to a measurable function of $(X_i,P_{1i}^{0},\ldots,P_{ji}^{0},U_{1i}^{(Z)},\ldots,U_{ji}^{(Z)})$.

For $j=1$, by the same arguments as in the proof of Lemma~\ref{lemma:independence}, we obtain that the following equality holds almost surely:
\begin{align}
    Z_{1i}^{*} = Z_{1i}
    = \mathbb{I}\left(U_{1i}^{(Z)}>F_{U_{1i}^{(Z)}\mid X_i}^{-1}\left(1-P_{1i}^{0}\right)\right).
\end{align}
Therefore, the result holds for $j=1$. Suppose that this result holds for $Z_{(j-1)i}^{*}$, where $j\in\{2,\ldots,n_Z\}$. Since $U_i^{(Z)}\perp(D_i,W_i^{(Z)})\mid X_i$ and $Z_{(j-1)i}^{*}$ is almost surely equal to a measurable function of $(X_i,P_{1i}^{0},\ldots,P_{(j-1)i}^{0},U_{1i}^{(Z)},\ldots,U_{(j-1)i}^{(Z)})$, the following equality holds almost surely for all $u\in\mathbb{R}$:
\begin{align}\label{eq:L1nested-1}
    &\mathbb{P}\left(U_{ji}^{(Z)}\leq u,Z_{(j-1)i}^{*}=1\mid Q_i\right) = \mathbb{P}\left(U_{ji}^{(Z)}\leq u,Z_{(j-1)i}^{*}=1\mid X_i,P_{1i}^{0},\ldots,P_{(j-1)i}^{0}\right).
\end{align}
This equality follows because, conditional on $Q_i$, the induction hypothesis allows us to express the event $\{Z_{(j-1)i}^{*}=1\}$ in terms of the random errors $U_{i}^{(Z)}$ and $(X_i,P_{1i}^{0},\ldots,P_{(j-1)i}^{0})$, while the conditional distribution of the selection errors depends only on $X_i$.

Moreover, by the law of iterated expectations, we obtain:
\begin{align}\label{eq:L1nested-2}
    &\mathbb{P}\left(Z_{(j-1)i}^{*}=1\mid X_i,P_{1i}^{0},\ldots,P_{(j-1)i}^{0}\right) = \nonumber \\
    =&\mathbb{E}\left(\underbrace{\mathbb{P}\left(Z_{(j-1)i}^{*}=1\mid Q_i\right)}_{P_{(j-1)i}^{0}}\mid X_i,P_{1i}^{0},\ldots,P_{(j-1)i}^{0}\right) = P_{(j-1)i}^{0}.
\end{align}

Therefore, by using equalities (\ref{eq:L1nested-1}) and (\ref{eq:L1nested-2}), we obtain that whenever $P_{(j-1)i}^{0}>0$:
\begin{align}\label{eq:L1nested-3}
    F_{U_{ji}^{(Z)}\mid Q_i,Z_{(j-1)i}^{*}=1}(u) &= \underbrace{\frac{\mathbb{P}\left(U_{ji}^{(Z)}\leq u,Z_{(j-1)i}^{*}=1\mid Q_i\right)}{P_{(j-1)i}^{0}}}_{\text{Conditional probability formula}} = \nonumber \\
    &=\underbrace{\frac{\mathbb{P}\left(U_{ji}^{(Z)}\leq u,Z_{(j-1)i}^{*}=1\mid X_i,P_{1i}^{0},\ldots,P_{(j-1)i}^{0}\right)}{\mathbb{P}\left(Z_{(j-1)i}^{*}=1\mid X_i,P_{1i}^{0},\ldots,P_{(j-1)i}^{0}\right)}}_{\text{By equations (\ref{eq:L1nested-1}) and (\ref{eq:L1nested-2})}} = \nonumber \\
    &= F_{U_{ji}^{(Z)}\mid X_i,P_{1i}^{0},\ldots,P_{(j-1)i}^{0},Z_{(j-1)i}^{*}=1}(u).
\end{align}
For brevity, denote the conditional cumulative distribution function on the right-hand side of equality~(\ref{eq:L1nested-3}) by $F_j(u)$, setting it equal to an arbitrary fixed cumulative distribution function whenever $P_{(j-1)i}^{0}=0$.

By combining equality (\ref{eq:L1nested-3}) with Assumption~\ref{assumption:selection_nested}, we obtain the following equalities, which hold almost surely on the event $\{P_{(j-1)i}^{0}>0\}$:
\begin{align}\label{eq:L1nested-4}
    P_{ji}^{0} &= \underbrace{\mathbb{P}\left(Z_{ji}^{*}=1\mid Q_i\right)}_{P_{ji}^{0}\text{ is the $j$-th element of }P_i^0} = \nonumber \\
    &= \mathbb{P}\left(Z_{(j-1)i}^{*}=1,Z_{ji}=1\mid Q_i\right) = \nonumber \\
    &= P_{(j-1)i}^{0}\mathbb{P}\left(Z_{ji}=1\mid Q_i,Z_{(j-1)i}^{*}=1\right) = \nonumber \\
    &= P_{(j-1)i}^{0}\underbrace{\mathbb{P}\left(U_{ji}^{(Z)}>g_j(Q_i,1,\ldots,1)\mid Q_i,Z_{(j-1)i}^{*}=1\right)}_{\text{Assumption~\ref{assumption:selection_nested}}} = \nonumber \\
    &= P_{(j-1)i}^{0}\underbrace{\left[1-F_j\left(g_j(Q_i,1,\ldots,1)\right)\right]}_{\text{By equality (\ref{eq:L1nested-3})}} \implies \nonumber \\
    &\implies F_j\left(g_j(Q_i,1,\ldots,1)\right)=1-\frac{P_{ji}^{0}}{P_{(j-1)i}^{0}}.
\end{align}

Define the generalized inverse $F_j^{-1}(\cdot)$ in the same way as in the proof of Lemma~\ref{lemma:independence}.

Notice that $P_{(j-1)i}^{0}>0$ almost surely on the event $\{Z_{(j-1)i}^{*}=1\}$, since
\begin{align}
    \mathbb{P}\left(P_{(j-1)i}^{0}=0,Z_{(j-1)i}^{*}=1\right)
    &=\mathbb{E}\left(\mathbb{I}\left(P_{(j-1)i}^{0}=0\right)P_{(j-1)i}^{0}\right)=0.
\end{align}

Therefore, applying the result from equality (\ref{eq:L1-1B}) conditionally on $Q_i$ and $Z_{(j-1)i}^{*}=1$, and using equalities (\ref{eq:L1nested-3}) and (\ref{eq:L1nested-4}), we obtain that the following equality holds almost surely on the event $\{Z_{(j-1)i}^{*}=1\}$:
\begin{align}\label{eq:L1nested-5}
    \mathbb{I}\left(U_{ji}^{(Z)}>g_j(Q_i,1,\ldots,1)\right)
    =
    \mathbb{I}\left(U_{ji}^{(Z)}>F_j^{-1}\left(1-\frac{P_{ji}^{0}}{P_{(j-1)i}^{0}}\right)\right).
\end{align}

Hence, the following equality holds almost surely:
\begin{align}\label{eq:L1nested-6}
    Z_{ji}^{*}
    &=
    \begin{cases}
        Z_{(j-1)i}^{*}\mathbb{I}\left(U_{ji}^{(Z)}>g_j(Q_i,1,\ldots,1)\right), & \text{if }P_{(j-1)i}^{0}>0,\\
        0, & \text{if }P_{(j-1)i}^{0}=0
    \end{cases} = \nonumber \\
    &=
    \begin{cases}
        Z_{(j-1)i}^{*}\mathbb{I}\left(U_{ji}^{(Z)}>F_j^{-1}\left(1-\frac{P_{ji}^{0}}{P_{(j-1)i}^{0}}\right)\right), & \text{if }P_{(j-1)i}^{0}>0,\\
        0, & \text{if }P_{(j-1)i}^{0}=0,
    \end{cases}
\end{align}
where the second equality follows from (\ref{eq:L1nested-5}), and $Z_{ji}^{*}=0$ almost surely whenever $P_{(j-1)i}^{0}=0$, since $P_{(j-1)i}^{0}=\mathbb{P}(Z_{(j-1)i}^{*}=1\mid Q_i)$.

Since $F_j(\cdot)$ depends only on $(X_i,P_{1i}^{0},\ldots,P_{(j-1)i}^{0})$, equality (\ref{eq:L1nested-6}) implies that $Z_{ji}^{*}$ is almost surely equal to a measurable function of $(X_i,P_{1i}^{0},\ldots,P_{ji}^{0},U_{1i}^{(Z)},\ldots,U_{ji}^{(Z)})$. Therefore, by induction, this result holds for every $j\in\{1,\ldots,n_Z\}$. In particular, $\tilde{Z}_i=Z_{n_Zi}^{*}$ is almost surely equal to a measurable function of $(X_i,P_i^0,U_i^{(Z)})$.

Because $P_i^0$ is a measurable function of $(X_i,D_i,W_i^{(Z)})$ and $\tilde{Z}_i$ is almost surely equal to a measurable function of $(X_i,P_i^0,U_i^{(Z)})$, we have:
\begin{align}
    \underbrace{(Y_{di}, U_i^{(Z)}) \perp (D_i, W_i^{(Z)}) \mid X_i}_{\text{Assumption~\ref{assumption:exogeneity}}} &\implies \underbrace{(Y_{di}, U_i^{(Z)}) \perp (D_i, W_i^{(Z)}, P_{i}^{0}) \mid X_{i}}_{\text{Since $P_i^0$ is a function of $(D_{i},W_i^{(Z)},X_{i})$}} \implies \nonumber \\
    &\implies \underbrace{(Y_{di}, U_i^{(Z)}) \perp (D_i, W_i^{(Z)}) \mid (X_i, P_{i}^{0})}_{\text{Weak union}} \implies \nonumber \\
    &\implies \underbrace{(Y_{di}, \tilde{Z}_{i}) \perp \left(D_i, W_{i}^{(Z)}\right) \mid (X_i, P_{i}^{0})}_{\text{Since }\tilde{Z}_{i}\text{ is a function of }(X_{i}, P_{i}^{0}, U_i^{(Z)})} \implies \nonumber \\
    &\implies \underbrace{Y_{di} \perp \left(D_i, W_{i}^{(Z)}\right) \mid (X_i, P_{i}^{0}, \tilde{Z}_{i})}_{\text{Symmetry and weak union}}.
\end{align}

\section{Proof of Theorem 1}

Consider $(x,p)\in\mathrm{supp}(X_i,P_i^0\mid\tilde{Z}_i=1)$ and $\overline{d},\breve{d}\in\{0,\cdots,n_D-1\}$. Lemma~\ref{lemma:independence} implies the following equality for almost every $(x,p)$ with respect to the conditional distribution of $(X_i,P_i^0)$ given $\tilde{Z}_i=1$:
\begin{align}\label{eq:T1-1}
    \mathbb{E}\left(Y_{\overline{d}i}\mid D_i=\breve{d},X_i=x,P_i^0=p,\tilde{Z}_i=1\right)
    = \mathbb{E}\left(Y_{\overline{d}i}\mid X_i=x,P_i^0=p,\tilde{Z}_i=1\right),
\end{align}
where conditioning on an arbitrary treatment level $\breve{d}\in\{0,\cdots,n_D-1\}$ on the left-hand side is possible by Assumption~\ref{assumption:support}.

Therefore, by the definition of $\mu_Y^0(\cdot)$, we obtain:
\begin{align}\label{eq:T1:conditional_mean}
    \mu_Y^0\left(\overline{d},X_i,P_i^0\right)
    &= \mathbb{E}\left(Y_i\mid D_i=\overline{d},X_i,P_i^0,\tilde{Z}_i=1\right) = \nonumber \\
    &= \underbrace{\mathbb{E}\left(Y_{\overline{d}i}\mid D_i=\overline{d},X_i,P_i^0,\tilde{Z}_i=1\right)}_{\text{Since }Y_i=Y_{\overline{d}i}\text{ when }D_i=\overline{d}\text{ and }\tilde{Z}_i=1} = \nonumber \\
    &= \underbrace{\mathbb{E}\left(Y_{\overline{d}i}\mid X_i,P_i^0,\tilde{Z}_i=1\right)}_{\text{By equation (\ref{eq:T1-1})}}.
\end{align}

By using equation (\ref{eq:T1:conditional_mean}) and the law of iterated expectations, we obtain:
\begin{align}\label{eq:T1-2}
    \mathbb{E}\left(\mu_Y^0\left(\overline{d},X_i,P_i^0\right)\mid\tilde{Z}_i=1\right)
    &= \underbrace{\mathbb{E}\left(\mathbb{E}\left(Y_{\overline{d}i}\mid X_i,P_i^0,\tilde{Z}_i=1\right)\mid\tilde{Z}_i=1\right)}_{\text{By equation (\ref{eq:T1:conditional_mean})}} = \nonumber \\
    &= \underbrace{\mathbb{E}\left(Y_{\overline{d}i}\mid\tilde{Z}_i=1\right)}_{\text{LIE}}.
\end{align}

Therefore, for any $d,d^*\in\mathrm{supp}(D_i)$, we have:
\begin{align}
    \mathrm{ATES} &= \mathbb{E}\left(Y_{di}-Y_{d^*i}\mid\tilde{Z}_i=1\right) = \mathbb{E}\left(Y_{di}\mid\tilde{Z}_i=1\right)-\mathbb{E}\left(Y_{d^*i}\mid\tilde{Z}_i=1\right) = \nonumber \\
    &= \underbrace{\mathbb{E}\left(\mu_Y^0\left(d,X_i,P_i^0\right)\mid\tilde{Z}_i=1\right)-\mathbb{E}\left(\mu_Y^0\left(d^*,X_i,P_i^0\right)\mid\tilde{Z}_i=1\right)}_{\text{By equation (\ref{eq:T1-2})}}.
\end{align}

Thus, we have established the result for ATES and proceed with ATTS. Specifically, notice that:
\begin{align}\label{eq:T1-4}
    &\mathbb{E}\left(\mu_Y^0\left(d,X_i,P_i^0\right)\mid\tilde{Z}_i=1,D_i=d\right) = \nonumber \\
    = &\underbrace{\mathbb{E}\left(\mathbb{E}\left(Y_{di}\mid X_i,P_i^0,\tilde{Z}_i=1\right)\mid\tilde{Z}_i=1,D_i=d\right)}_{\text{By equation (\ref{eq:T1:conditional_mean})}} = \nonumber \\
    = &\underbrace{\mathbb{E}\left(\mathbb{E}\left(Y_{di}\mid X_i,P_i^0,\tilde{Z}_i=1,D_i=d\right)\mid\tilde{Z}_i=1,D_i=d\right)}_{\text{By equation (\ref{eq:T1-1})}} = \nonumber \\
    = &\underbrace{\mathbb{E}\left(Y_{di}\mid\tilde{Z}_i=1,D_i=d\right)}_{\text{LIE}},
\end{align}
and:
\begin{align}\label{eq:T1-5}
    &\mathbb{E}\left(\mu_Y^0\left(d^*,X_i,P_i^0\right)\mid\tilde{Z}_i=1,D_i=d\right) = \nonumber \\
    = &\underbrace{\mathbb{E}\left(\mathbb{E}\left(Y_{d^*i}\mid X_i,P_i^0,\tilde{Z}_i=1\right)\mid\tilde{Z}_i=1,D_i=d\right)}_{\text{By equation (\ref{eq:T1:conditional_mean})}} = \nonumber \\
    = &\underbrace{\mathbb{E}\left(\mathbb{E}\left(Y_{d^*i}\mid X_i,P_i^0,\tilde{Z}_i=1,D_i=d\right)\mid\tilde{Z}_i=1,D_i=d\right)}_{\text{By equation (\ref{eq:T1-1})}} = \nonumber \\
    = &\underbrace{\mathbb{E}\left(Y_{d^*i}\mid\tilde{Z}_i=1,D_i=d\right)}_{\text{LIE}}.
\end{align}

By using the established equalities, we finally obtain:
\begin{align}
    \mathrm{ATTS} &= \mathbb{E}\left(Y_{di}-Y_{d^*i}\mid D_i=d,\tilde{Z}_i=1\right) = \nonumber \\
    &= \mathbb{E}\left(\left[\begin{aligned}
        &\underbrace{\mu_Y^0\left(d,X_i,P_i^0\right)}_{\text{By equation (\ref{eq:T1-4})}} - \underbrace{\mu_Y^0\left(d^*,X_i,P_i^0\right)}_{\text{By equation (\ref{eq:T1-5})}}
    \end{aligned}\right]\mid D_i=d,\tilde{Z}_i=1\right).
\end{align}

\section{Proof of Theorem 2}

First, notice that:
\begin{align}\label{eq:T2-1}
    \mathbb{P}\left(\tilde{Z}_{i}=1\mid X_i,P_i^0\right)
    &= \underbrace{\mathbb{E}\left(\underbrace{\mathbb{P}\left(\tilde{Z}_{i}=1\mid D_i,X_i,W_i^{(Z)}\right)}_{\substack{>0\\ \text{Assumption~\ref{assumption:selection-support}}}}\mid X_i,P_i^0\right)}_{\substack{\text{LIE, since }P_i^0\text{ is a function}\\ \text{of }D_i,X_i,W_i^{(Z)}}} > 0.
\end{align}
Consequently, any event $A$ determined by $(X_i,P_i^0)$ satisfying $\mathbb{P}(A\mid\tilde{Z}_i=1)=0$ also satisfies $\mathbb{P}(A)=0$, since:
\begin{align}
    0 = \mathbb{P}\left(A,\tilde{Z}_i=1\right)
    = \mathbb{E}\left(\mathbb{I}(A)\underbrace{\mathbb{P}\left(\tilde{Z}_i=1\mid X_i,P_i^0\right)}_{\substack{>0\\ \text{By (\ref{eq:T2-1})}}}\right),
\end{align}
where $\mathbb{I}\left(A\right)$ is a Bernoulli random variable taking the value $1$ if event $A$ occurs and $0$ otherwise. Hence, equation~(\ref{eq:T1:conditional_mean}), established almost surely with respect to the conditional distribution of $(X_i,P_i^0)$ given $\tilde{Z}_i=1$, also holds almost surely with respect to their unconditional distribution.

Using equation (\ref{eq:T1:conditional_mean}), which now holds almost surely in the entire population, and Assumption~\ref{assumption:homogeneity}, we obtain:
\begin{align}
    &\mathbb{E}\left(\mu_Y^0\left(d,X_i,P_i^0\right)-\mu_Y^0\left(d^*,X_i,P_i^0\right)\right) = \nonumber \\
    = &\underbrace{\mathbb{E}\left(\mathbb{E}\left(Y_{di}-Y_{d^*i}\mid X_i,P_i^0,\tilde{Z}_i=1\right)\right)}_{\text{By equation (\ref{eq:T1:conditional_mean})}} = \nonumber \\
    = &\underbrace{\mathbb{E}\left(\mathbb{E}\left(Y_{di}-Y_{d^*i}\mid X_i,P_i^0\right)\right)}_{\text{Assumption~\ref{assumption:homogeneity}}} = \underbrace{\mathbb{E}\left(Y_{di}-Y_{d^*i}\right)}_{\text{LIE}} = \mathrm{ATE}. \label{eq:T2-2}
\end{align}

To obtain the formula for ATT, we follow similar steps:
\begin{align}
    &\mathbb{E}\left(\mu_Y^0\left(d,X_i,P_i^0\right)-\mu_Y^0\left(d^*,X_i,P_i^0\right)\mid D_i=d\right) = \nonumber \\
    = &\underbrace{\mathbb{E}\left(\mathbb{E}\left(Y_{di}-Y_{d^*i}\mid X_i,P_i^0,\tilde{Z}_i=1\right)\mid D_i=d\right)}_{\text{By equation (\ref{eq:T1:conditional_mean})}} = \nonumber \\
    = &\underbrace{\mathbb{E}\left(\mathbb{E}\left(Y_{di}-Y_{d^*i}\mid X_i,P_i^0\right)\mid D_i=d\right)}_{\text{Assumption~\ref{assumption:homogeneity}}} = \nonumber \\
    = &\underbrace{\mathbb{E}\left(\mathbb{E}\left(Y_{di}-Y_{d^*i}\mid X_i,P_i^0,D_i=d\right)\mid D_i=d\right)}_{\text{Lemma~\ref{lemma:independence}}} = \nonumber \\
    = &\underbrace{\mathbb{E}\left(Y_{di}-Y_{d^*i}\mid D_i=d\right)}_{\text{LIE}} = \mathrm{ATT}.
\end{align}

\section{Proof of Theorem 3}

We provide a detailed proof for the DML estimator of ATES and subsequently explain how the established bounds extend to the other estimators. First, we establish the asymptotic normality of the estimator obtained for a fixed $t\in\{1,2\}$. Since the vector of nuisance estimators $\hat{\mu}^{(k,t)}$ is obtained using observations from $\overline{A}_{k}$, it is independent of observations from $A_{k}$, so the results for conventional cross-fitting apply. For brevity, we omit the index $t$ throughout this part of the proof. Below, we verify the conditions of Assumptions 3.1 and 3.2 of \citet{Chernozhukov} required by their Theorem 3.1. Most properties used throughout the proof may be found in the appendix of \citet{Hansen}.

\subsection{Assumption 3.1}\label{a31}

Conditions (a) and (d) are implied by Lemma~\ref{lemma:neyman}. Condition (b) holds since the score $\psi(\cdot)$ is affine in $\theta$. Specifically, in terms of notation from Section 3.2 of \citet{Chernozhukov}, we have $\psi^{a}(\cdot)=-\tilde{Z}_{i}$ and $\psi^{b}(\cdot)$ is the remaining part of the score. For the purpose of applying Theorem 3.1 of \citet{Chernozhukov}, we take the nuisance parameter space to be the convex hull of $\bigcup_{n\geq1}\mathcal{T}_n$. Condition (c) is implied by the continuous differentiability of $\mu_Y$ and $p_D$ with respect to the control probabilities up to the third and second orders, respectively (Part 6 of Assumption~\ref{assumption:regularity}), the fact that the probabilities are bounded away from zero (Part 4 of Assumption~\ref{assumption:regularity}), and the moment and derivative bounds imposed by Parts 1, 3, and 6 of Assumption~\ref{assumption:regularity}, which allow us to interchange differentiation and expectation twice. Finally, condition (e) holds, because the fact that $\tilde{Z}_{i}$ is a Bernoulli random variable and Part 2 of Assumption~\ref{assumption:regularity} imply $c\leq \left|\mathbb{E}\left(-\tilde{Z}_{i}\right)\right|\leq 1$, since $\left|\mathbb{E}\left(-\tilde{Z}_{i}\right)\right| = \mathbb{P}\left(\tilde{Z}_{i} = 1\right)$.

\subsection{Assumption 3.2}\label{a32}

Condition (a) holds by Assumption~\ref{assumption:regularity} and the construction of the realization set $\mathcal{T}_{n}$. The fact that $\psi^{a}(\cdot)=-\tilde{Z}_{i}$ implies $m_{N}^{'} \leq 1$ and $r_{N} = 0$ in conditions (b) and (c), respectively. Condition (d) is implied by Part 2 of Assumption~\ref{assumption:regularity}. Below, we verify the bound $m_{N}$ from condition (b) and the bounds $r_{N}^{'}$ and $\lambda_{N}^{\prime}$ from condition (c). For brevity, throughout the remainder of the proof, the parts of Assumption~\ref{assumption:regularity} are referred to without explicitly mentioning this assumption. That is, Part~$A$ of Assumption~\ref{assumption:regularity} is referred to simply as Part~$A$.

\subsubsection{Bound \texorpdfstring{$m_{N}$}{mN} from condition (b)}

We are going to show that the following holds:
\begin{align}
    m_{N} := \sup_{\mu\in\mathcal{T}_{n}}\mathbb{E}\left(\left|\psi_{\text{ATES}}\left(O_{i}, \text{ATES}, \mu\right)\right|^{q}\right)^{1/q}&\leq 10C/c+ 4n_{Z}\left(1 + \frac{1}{c}\right)C + \frac{2C}{c^{2/q}}.
\end{align}

First, notice that:
\begin{align}
    \mathbb{E}\left(\tilde{Z}_{i}D_{i}^{(d)}\left\|\text{E}_{i}^{0(d)}\right\|^{q}\right)^{1/q}
    &\leq \underbrace{\mathbb{E}\left(\tilde{Z}_{i}D_{i}^{(d)}\left\|Y_{i}\right\|^{q}\right)^{1/q}}_{\text{Conditional expectation inequality and LIE}} \leq \nonumber \\
    &\leq \underbrace{\mathbb{E}\left(\left\|Y_{i}\right\|^{q}\right)^{1/q}}_{\text{Since }\tilde{Z}_{i},D_{i}^{(d)}\in\{0,1\}} \leq \underbrace{C}_{\text{Part 1}}.
\end{align}

Hence, we obtain:
\begin{align}
    \mathbb{E}\left(\left\|\tilde{Z}_{i}\text{E}_{i}^{0(d)}\right\|^{q}\right)^{1/q}
    &\leq \underbrace{\mathbb{E}\left(\tilde{Z}_{i}\frac{P_{i}^{0(d)}}{c}\left\|\text{E}_{i}^{0(d)}\right\|^{q}\right)^{1/q}}_{\text{Part 4}}\leq \nonumber \\
    &\leq \underbrace{\frac{1}{c^{1/q}}\mathbb{E}\left(\tilde{Z}_{i}D_{i}^{(d)}\left\|\text{E}_{i}^{0(d)}\right\|^{q}\right)^{1/q}}_{\text{LIE and Equation~(\ref{eq:useful:pD}) of Lemma~\ref{lemma:useful}}} \leq C/c^{1/q}.
\end{align}

Therefore, for every $\mu\in\mathcal{T}_{n}$, we obtain:
\begin{align}\label{eq:mN:Ed}
    \mathbb{E}\left(\left\|\tilde{Z}_{i}\text{E}_{i}^{(d)}\right\|^{q}\right)^{1/q}
    &\leq \underbrace{\mathbb{E}\left(\left\|\tilde{Z}_{i}\text{E}_{i}^{0(d)}\right\|^{q}\right)^{1/q} + \mathbb{E}\left(\left\|\tilde{Z}_{i}\left(\text{E}_{i}^{(d)}-\text{E}_{i}^{0(d)}\right)\right\|^{q}\right)^{1/q}}_{\text{Minkowski's inequality}} \leq \nonumber \\
    &\leq \underbrace{C/c^{1/q}+C}_{\text{Part 3 and the previous inequality}} \leq \underbrace{2C/c}_{\text{Since }q>4\text{ and }c\in(0,1)}.
\end{align}

Second, we obtain:
\begin{align}
    \mathbb{E}\left(\left\|\tilde{Z}_{i}\frac{D_{i}^{(d)}}{P_{i}^{(d)}}\left(Y_{i} - \text{E}_{i}^{(d)}\right)\right\|^{q}\right)^{1/q}
    &\leq \underbrace{\frac{1}{c}\mathbb{E}\left(\left\|\tilde{Z}_{i}D_{i}^{(d)}\left(Y_{i} - \text{E}_{i}^{(d)}\right)\right\|^{q}\right)^{1/q}}_{\text{Part 4}} \leq \nonumber \\
    &\leq \frac{1}{c}\Bigg(\mathbb{E}\left(\left\|\tilde{Z}_{i}D_{i}^{(d)}Y_{i}\right\|^{q}\right)^{1/q} + \mathbb{E}\left(\left\|\tilde{Z}_{i}D_{i}^{(d)}\text{E}_{i}^{0(d)}\right\|^{q}\right)^{1/q} + \nonumber \\
    &+\underbrace{\mathbb{E}\left(\left\|\tilde{Z}_{i}D_{i}^{(d)}\left(\text{E}_{i}^{(d)}-\text{E}_{i}^{0(d)}\right)\right\|^{q}\right)^{1/q}\Bigg)}_{\text{Minkowski's inequality}} \leq \nonumber \\
    &\leq \underbrace{\left(C+C+C\right)/c=3C/c}_{\text{Parts 1 and 3 and the conditional expectation inequality}}.
\end{align}

Third, we obtain:
\begin{align}
    \mathbb{E}\left(\left\|\left(\overline{Z}_{i} - P_{i}\right)\tilde{P}_{i}\left(\text{E}_{\partial i}^{(d)}\left(1 - \frac{D_{i}^{(d)}}{P_{i}^{(d)}}\right)\right)\right\|^{q}\right)^{1/q} &\leq \mathbb{E}\left(\left\|\left(\overline{Z}_{i} - P_{i}\right)\tilde{P}_{i}\left(\text{E}_{\partial i}^{(d)}\left(1 + \frac{1}{P_{i}^{(d)}}\right)\right)\right\|^{q}\right)^{1/q}\leq \nonumber \\ 
    &\leq  \underbrace{\sqrt{2n_{Z}}\left(1 + \frac{1}{c}\right)\mathbb{E}\left(\left\|\left(\text{E}_{\partial i}^{(d)}\right)\right\|^{q}\right)^{1/q}}_{\substack{\text{Since }\overline{Z}_{i}\in\{0, 1\}^{2n_{Z}}\text{ and Part 4}\\ \text{imply }P_i\in[0,1]^{2n_Z},\\ c\leq\tilde{P}_i\leq1\text{ and }c\leq P_i^{(d)}\leq1}} \leq \nonumber \\
    &\leq \underbrace{2n_{Z}\left(1 + \frac{1}{c}\right)\mathbb{E}\left(\left\|\text{E}_{\partial i}^{(d)}\right\|_{\infty}^{q}\right)^{1/q}}_{\text{Since }\|v\|\leq\sqrt{2n_{Z}}\|v\|_{\infty}\text{ for }v\in\mathbb{R}^{2n_{Z}}} \leq \nonumber \\
    &\leq \underbrace{2n_{Z}\left(1 + \frac{1}{c}\right)C}_{\text{Part 6}}.
\end{align}

The bounds associated with the $d^{*}$-parts of the score are the same by symmetry. Thus, by aggregating the results and using Minkowski's inequality, we finally obtain:
\begin{align}
    m_{N} &:= \sup_{\mu\in\mathcal{T}_{n}}\mathbb{E}\left(\left|\psi_{\text{ATES}}\left(O_{i}, \text{ATES}, \mu\right)\right|^{q}\right)^{1/q} \leq \nonumber \\
    &\leq \underbrace{2\times\left(\frac{2C}{c} + 3C/c+ 2n_{Z}\left(1 + \frac{1}{c}\right)C\right)}_{\substack{\text{Multiplication by 2 addresses the fact that the}\\ \text{same bounds apply for }d\text{ and }d^{*}\text{ parts of the score}}} + \underbrace{\left|\text{ATES}\right|}_{\text{Since }\tilde{Z}_{i}\in\{0, 1\}} = \nonumber \\
    &= 10C/c+ 4n_{Z}\left(1 + \frac{1}{c}\right)C + \nonumber \\
    &+ \underbrace{\left|\mathbb{E}\left(\text{E}_{i}^{0(d)}-\text{E}_{i}^{0(d^{*})}\mid \tilde{Z}_{i}=1\right)\right|}_{\text{Theorem~\ref{theorem:ates}}} \leq \nonumber \\
    &\leq 10C/c+ 4n_{Z}\left(1 + \frac{1}{c}\right)C + \nonumber \\
    &+ \underbrace{\left[\begin{aligned}
        &\mathbb{E}\left(\left|\text{E}_{i}^{0(d)}\right|^{q}\mid\tilde{Z}_{i}=1\right)^{1/q} + \\
        &+ \mathbb{E}\left(\left|\text{E}_{i}^{0(d^{*})}\right|^{q}\mid\tilde{Z}_{i}=1\right)^{1/q}
    \end{aligned}\right]}_{\text{Minkowski's and Lyapunov's inequalities}} \leq \nonumber \\
    &\leq 10C/c+ 4n_{Z}\left(1 + \frac{1}{c}\right)C + \underbrace{\frac{2C}{c^{2/q}}}_{\substack{\text{Part 2 and the previously}\\ \text{established bounds}}}.
\end{align}
\subsubsection{Bound \texorpdfstring{$r_{N}^{'}$}{rN'} from condition (c)}

Denote for brevity $\kappa_{n} = \max\left(\delta_{n}n^{-1/4},\delta_{n}\right)$. We are going to show that the following holds:
\begin{align}
    r_{N}^{'} &:= \sup_{\mu\in\mathcal{T}_{n}}\mathbb{E}\left(\left|\psi_{\text{ATES}}\left(O_{i}, \text{ATES}, \mu\right) - \psi_{\text{ATES}}\left(O_{i}, \text{ATES}, \mu^{0}\right)\right|^{2}\right)^{1/2} \leq \nonumber \\
    &\leq 2\left(1 + \frac{1}{c}\right)\left(1 + 2n_{Z}\right)\kappa_{n} + \frac{6C}{c^{2}} \kappa_{n}^{(q-2)/q} + 2\left(4n_{Z}\left(1+\frac{1}{c}\right) C + \frac{2n_{Z} C}{c^{2}}\right)\kappa_{n}.
\end{align}

First, we obtain:
\begin{align}\label{eq:rN:Ed}
    \mathbb{E}\left(\left\|\tilde{Z}_{i}\text{E}_{i}^{(d)} - \tilde{Z}_{i}\text{E}_{i}^{0(d)}\right\|^{2}\right)^{1/2} \leq \underbrace{\mathbb{E}\left(\left\|\text{E}_{i}^{(d)} - \text{E}_{i}^{0(d)}\right\|^{2}\right)^{1/2} \leq \kappa_{n}}_{\text{Part 5}}.
\end{align}

Second, we obtain:
\begin{align}
    &\mathbb{E}\left(\left\|\tilde{Z}_{i}\frac{D_{i}^{(d)}}{P_{i}^{(d)}}\left(Y_{i} - \text{E}_{i}^{(d)}\right) - \tilde{Z}_{i}\frac{D_{i}^{(d)}}{P_{i}^{0(d)}}\left(Y_{i} - \text{E}_{i}^{0(d)}\right)\right\|^{2}\right)^{1/2} = \nonumber \\
    = &\mathbb{E}\left(\left\|\tilde{Z}_{i}D_{i}^{(d)}\left(\frac{1}{P_{i}^{(d)}}\left(Y_{i} - \text{E}_{i}^{(d)}\right) - \frac{1}{P_{i}^{0(d)}}\left(Y_{i} - \text{E}_{i}^{0(d)}\right)\right)\right\|^{2}\right)^{1/2} = \nonumber \\
    =&\mathbb{E}\left(\left\|\tilde{Z}_{i}D_{i}^{(d)}\left(\frac{\left(Y_{i} - \text{E}_{i}^{(d)}\right)\left(P_{i}^{0(d)} - P_{i}^{(d)}\right)}{P_{i}^{(d)}P_{i}^{0(d)}} - \frac{\left(\text{E}_{i}^{(d)} - \text{E}_{i}^{0(d)}\right)}{P_{i}^{0(d)}}\right)\right\|^{2}\right)^{1/2} \leq \nonumber \\
\leq&\frac{1}{c^{2}}\mathbb{E}\left(\left\|\tilde{Z}_{i}D_{i}^{(d)}\left(Y_{i} - \text{E}_{i}^{(d)}\right)\right\|^{q}\right)^{1/q}\times \nonumber \\
&\times \underbrace{\mathbb{E}\left(\left\|P_{i}^{0(d)} - P_{i}^{(d)}\right\|^{2q/(q-2)}\right)^{(q-2)/(2q)}}_{\text{Hölder's inequality with }q/2\text{ and }q/(q-2)} + \underbrace{\frac{1}{c}\kappa_{n}}_{\text{Part 5}} \leq \nonumber \\
\leq &\frac{1}{c^{2}} \mathbb{E}\left(\left\|\tilde{Z}_{i}D_{i}^{(d)}\left(Y_{i} - \text{E}_{i}^{(d)}\right)\right\|^{q}\right)^{1/q}\times \nonumber \\
&\times \mathbb{E}\left(\underbrace{\left\|P_{i}^{0(d)} - P_{i}^{(d)}\right\|^{2q/(q-2) - 2}}_{\text{Bounded above by 1 by Part 4}} \left\|P_{i}^{0(d)} - P_{i}^{(d)}\right\|^{2}\right)^{(q-2)/(2q)} + \frac{1}{c}\kappa_{n} \leq \nonumber \\
\leq &\frac{1}{c^{2}} \mathbb{E}\left(\left\|\tilde{Z}_{i}D_{i}^{(d)}\left(Y_{i} - \text{E}_{i}^{(d)}\right)\right\|^{q}\right)^{1/q}\underbrace{\mathbb{E}\left(\left\|P_{i}^{0(d)} - P_{i}^{(d)}\right\|^{2}\right)^{(q-2)/(2q)}}_{\text{By the aforementioned bound}} + \frac{1}{c}\kappa_{n} \leq \nonumber \\
\leq &\frac{1}{c^{2}} \mathbb{E}\left(\left\|\tilde{Z}_{i}D_{i}^{(d)}\left(Y_{i} - \text{E}_{i}^{(d)}\right)\right\|^{q}\right)^{1/q} \left(\mathbb{E}\left(\left\|P_{i}^{0(d)} - P_{i}^{(d)}\right\|^{2}\right)^{1/2}\right)^{(q-2)/q} + \frac{1}{c}\kappa_{n} \leq \nonumber \\
\leq &\frac{1}{c^{2}} \underbrace{\mathbb{E}\left(\left\|\tilde{Z}_{i}D_{i}^{(d)}\left(Y_{i} - \text{E}_{i}^{(d)}\right)\right\|^{q}\right)^{1/q}}_{\text{Bounded above by }3C\text{ as established above}} \underbrace{\kappa_{n}^{(q-2)/q}}_{\text{Part 5}} + \frac{1}{c}\kappa_{n} \leq \nonumber \\
\leq &\frac{3}{c^{2}}C \kappa_{n}^{(q-2)/q} + \frac{1}{c}\kappa_{n}.
\end{align}

Third, we obtain:
\begin{align}
    &\underbrace{\mathbb{E}\left(\left\|\left(\overline{Z}_{i} - P_{i}\right)\tilde{P}_{i}\left(E_{\partial i}^{(d)}\left(1 - \frac{D_{i}^{(d)}}{P_{i}^{(d)}}\right)\right) - \left(\overline{Z}_{i} - P_{i}^{0}\right)\tilde{P}_{i}^{0}\left(E_{\partial i}^{0(d)}\left(1 - \frac{D_{i}^{(d)}}{P_{i}^{0(d)}}\right)\right)\right\|^{2}\right)^{1/2}}_{\text{Next, use Minkowski's inequality}} \leq
    \nonumber \\
    \leq& \underbrace{\mathbb{E}\left(\left\|\left(P_{i}^{0} - P_{i}\right)\tilde{P}_{i}E_{\partial i}^{(d)}\left(1 - \frac{D_{i}^{(d)}}{P_{i}^{(d)}}\right)\right\|^{2}\right)^{1/2}}_{E_{1}} + \nonumber \\
    + &\underbrace{\mathbb{E}\left(\left\|\left(\overline{Z}_{i} - P_{i}^{0}\right)\left(\tilde{P}_{i} - \tilde{P}_{i}^{0}\right)E_{\partial i}^{(d)}\left(1 - \frac{D_{i}^{(d)}}{P_{i}^{(d)}}\right)\right\|^{2}\right)^{1/2}}_{E_{2}} + \nonumber \\
    + &\underbrace{\mathbb{E}\left(\left\|\left(\overline{Z}_{i} - P_{i}^{0}\right)\tilde{P}_{i}^{0}\left(E_{\partial i}^{(d)} - E_{\partial i}^{0(d)}\right)\left(1 - \frac{D_{i}^{(d)}}{P_{i}^{(d)}}\right)\right\|^{2}\right)^{1/2}}_{E_{3}} + \nonumber \\
    + &\underbrace{\mathbb{E}\left(\left\|\left(\overline{Z}_{i} - P_{i}^{0}\right)\tilde{P}_{i}^{0}E_{\partial i}^{0(d)}\left(\frac{D_{i}^{(d)}}{P_{i}^{0(d)}} - \frac{D_{i}^{(d)}}{P_{i}^{(d)}}\right)\right\|^{2}\right)^{1/2}}_{E_{4}}.
\end{align}

Further, we bound each of these terms. For the first term, we obtain:
\begin{align}
    E_{1} &\leq \underbrace{\left(1+\frac{1}{c}\right)\mathbb{E}\left(\left\|P_{i}^{0} - P_{i}\right\|^{2}\left\|E_{\partial i}^{(d)}\right\|^{2}\right)^{1/2}}_{\text{Part 4 and Cauchy--Schwarz inequality}} \leq \nonumber \\
    &\leq \underbrace{\sqrt{2n_{Z}}\left(1+\frac{1}{c}\right)C\mathbb{E}\left(\left\|P_{i}^{0} - P_{i}\right\|^{2}\right)^{1/2}}_{\text{Since Part 6 implies }\left\|E_{\partial i}^{(d)}\right\|\leq\sqrt{2n_{Z}}\left\|E_{\partial i}^{(d)}\right\|_{\infty}\leq\sqrt{2n_{Z}}C} \leq \nonumber \\
    &\leq \underbrace{2n_{Z}\left(1+\frac{1}{c}\right)\kappa_{n}C}_{\text{Part 5 and }\sqrt{2n_{Z}}\leq 2n_{Z}}.
\end{align}

For the second term, we obtain:
\begin{align}
E_{2} &\leq \underbrace{\left(1+\frac{1}{c}\right) \mathbb{E}\left(\left\|\overline{Z}_{i} - P_{i}^{0}\right\|^{2} \left\|\tilde{P}_{i} - \tilde{P}_{i}^{0}\right\|^{2} \left\|E_{\partial i}^{(d)}\right\|^{2}\right)^{1/2}}_{\text{Since Part 4 implies }\left|1 - D_{i}^{(d)}/P_{i}^{(d)}\right|\leq 1 + 1/c} \leq \nonumber \\
&\leq \underbrace{2n_{Z}\left(1+\frac{1}{c}\right) C\mathbb{E}\left(\left\|\tilde{P}_{i} - \tilde{P}_{i}^{0}\right\|^{2}\right)^{1/2}}_{\substack{\text{Since }\left\|\overline{Z}_{i} - P_{i}^{0}\right\|\leq\sqrt{2n_{Z}}\\ \text{and Part 6 implies }\left\|E_{\partial i}^{(d)}\right\|\leq\sqrt{2n_{Z}}C}} \leq \nonumber \\
&\leq 2n_{Z}\left(1+\frac{1}{c}\right) \underbrace{\kappa_{n}}_{\text{Part 5}} C.
\end{align}

For the third term, we obtain:
\begin{align}
E_{3} &\leq \underbrace{\left(1+\frac{1}{c}\right)\mathbb{E}\left(\left\|\left(\overline{Z}_{i} - P_{i}^{0}\right)\left(E_{\partial i}^{(d)} - E_{\partial i}^{0(d)}\right)\right\|^{2}\right)^{1/2}}_{\text{Since Part 4 implies }\left|\tilde{P}_{i}^{0}\right|\leq 1\text{ and }\left|1 - D_{i}^{(d)}/P_{i}^{(d)}\right|\leq 1 + 1/c} \leq \nonumber \\
&\leq \underbrace{\left(1+\frac{1}{c}\right)\mathbb{E}\left(\left\|\overline{Z}_{i} - P_{i}^{0}\right\|^{2}\left\|E_{\partial i}^{(d)} - E_{\partial i}^{0(d)}\right\|^{2}\right)^{1/2}}_{\text{Cauchy--Schwarz inequality}} \leq \nonumber \\
&\leq \underbrace{2n_{Z}\left(1+\frac{1}{c}\right)\mathbb{E}\left(\left\|E_{\partial i}^{(d)} - E_{\partial i}^{0(d)}\right\|^{2}\right)^{1/2}}_{\text{Since }\left\|\overline{Z}_{i} - P_{i}^{0}\right\|_{2}\leq\sqrt{2n_{Z}}\leq 2n_{Z}} \leq \nonumber \\
&\leq \underbrace{2n_{Z}\left(1+\frac{1}{c}\right)\kappa_{n}}_{\text{Part 5}}.
\end{align}

For the fourth term, we obtain:
\begin{align}
E_{4} &\leq \underbrace{\frac{1}{c^{2}} \mathbb{E}\left(\left\|\left(\overline{Z}_{i} - P_{i}^{0}\right) E_{\partial i}^{0(d)} \left(P_{i}^{(d)} - P_{i}^{0(d)}\right)\right\|^{2}\right)^{1/2}}_{\substack{\text{Since Part 4 implies }\left|\tilde{P}_{i}^{0}\right|\leq 1\text{, }\left|D_{i}^{(d)}\right|\leq 1\text{,}\\ \text{and }1/\left(P_{i}^{0(d)}P_{i}^{(d)}\right)\leq 1/c^{2}}} \leq \nonumber \\
&\leq \underbrace{\frac{1}{c^{2}} \mathbb{E}\left(\left\|\overline{Z}_{i} - P_{i}^{0}\right\|^{2} \left\|E_{\partial i}^{0(d)}\right\|^{2} \left(P_{i}^{(d)} - P_{i}^{0(d)}\right)^{2}\right)^{1/2}}_{\text{Cauchy--Schwarz inequality}} \leq \nonumber \\
&\leq \underbrace{\frac{2n_{Z}C}{c^{2}} \mathbb{E}\left(\left(P_{i}^{(d)} - P_{i}^{0(d)}\right)^{2}\right)^{1/2}}_{\substack{\text{Since }\left\|\overline{Z}_{i} - P_{i}^{0}\right\|\leq\sqrt{2n_{Z}}\\ \text{and Part 6 implies }\left\|E_{\partial i}^{0(d)}\right\|\leq\sqrt{2n_{Z}}C}} \leq \nonumber \\
&\leq \frac{2n_{Z}C}{c^{2}} \underbrace{\kappa_{n}}_{\text{Part 5}}.
\end{align}

The bounds associated with the $d^{*}$-parts of the score are the same by symmetry. By combining the established inequalities and using Minkowski's inequality, we obtain:
\begin{align}
r_{N}^{'} &:= \sup_{\mu\in\mathcal{T}_{n}}\mathbb{E}\left(\|\psi_{\text{ATES}}\left(O_{i}, \text{ATES}, \mu\right) - \psi_{\text{ATES}}\left(O_{i}, \text{ATES}, \mu^{0}\right)\|^{2}\right)^{1/2} \leq \nonumber \\
&\leq 2\left(\kappa_{n} + \frac{3}{c^{2}}C \kappa_{n}^{(q-2)/q} + \frac{1}{c}\kappa_{n}\right) + \nonumber \\
&+ 2\left(4n_{Z}\left(1+\frac{1}{c}\right) \kappa_{n} C + 2n_{Z}\left(1+\frac{1}{c}\right)\kappa_{n} + \frac{2n_{Z}C}{c^{2}} \kappa_{n}\right) = \nonumber \\
&= 2\left(1 + \frac{1}{c}\right)\left(1 + 2n_{Z}\right)\kappa_{n} + \frac{6C}{c^{2}} \kappa_{n}^{(q-2)/q} + 2\left(4n_{Z}\left(1+\frac{1}{c}\right) C + \frac{2n_{Z} C}{c^{2}}\right)\kappa_{n}.
\end{align}

\subsubsection{Bound \texorpdfstring{$\lambda_{N}^{\prime}$}{lambdaN'} from condition (c)}

We are going to show that the following holds:
\begin{align}
    \lambda_N' &= \sup_{\epsilon\in(0,1)}\sup_{\mu\in\mathcal T_n}\left|\frac{\partial^2}{\partial\epsilon^2}\mathbb{E}\left[\psi\left(O_i,\theta^0,\mu^\epsilon\right)\right]\right| \leq \delta_{n}^{*}n^{-1/2},
\end{align}
where:
\begin{align}\label{eq:lambda:deltastar}
    \delta_n^{*} &= \left[\begin{aligned}
        &2(3C\sqrt{2n_Z}+1)\frac{C\sqrt{2n_Z}(2n_Z+4)}{c^2} + \\
        &+ \frac{4(2n_Z C+1)(3C\sqrt{2n_Z}+1)^2}{c^3} + \\
        &+ \frac{4(3C\sqrt{2n_Z}+1)}{c^2}\left(6C\sqrt{2n_Z}+1+C(2n_Z)^{3/2}\right) + \\
        &+ \frac{2(10C\sqrt{2n_Z}+16n_Z C+C(2n_Z)^2)}{c}
    \end{aligned}\right]\delta_n^2,
\end{align}

Below, we use the same notations and shorthands as in the proof of Lemma~\ref{lemma:neyman}. In addition, for $\overline{d}\in \{d, d^{*}\}$, we denote:
\begin{align}
    \overline{R}_{D}^{\epsilon, \overline{d}} &= 1 - R_{D}^{\epsilon, \overline{d}}, \\
    \triangle R_{\partial Y}^{\epsilon, \overline{d}} &= \left(\frac{\partial \mu_Y\left( \overline{d}, X_i, p\right)}{\partial p}\mid_{p=P_i^{\epsilon}}\right) - \left(\frac{\partial \mu_Y^{0}\left( \overline{d}, X_i, p\right)}{\partial p}\mid_{p=P_i^{\epsilon}}\right), \\
    \triangle R_{\partial^{2} Y}^{\epsilon, \overline{d}} &= \left(\frac{\partial^{2} \mu_{Y}\left( \overline{d}, X_{i}, p\right)}{\partial p^{2}}\mid_{p=P_{i}^{\epsilon}}\right) - \left(\frac{\partial^{2} \mu_{Y}^{0}\left( \overline{d}, X_{i}, p\right)}{\partial p^{2}}\mid_{p=P_{i}^{\epsilon}}\right), \\
    R_{\partial^2 Y}^{\epsilon, \overline{d}} &= \frac{\partial^2 \mu_Y^\epsilon\!\left( \overline{d}, X_i, p \right)}{\partial p^2} \mid_{p = P_{i}^{\epsilon}}, \\
    R_{\partial^{3} Y}^{\epsilon, \overline{d}} &= \frac{\partial^{3} \mu_{Y}^{\epsilon}\!\left( \overline{d}, X_{i}, p\right)}{\partial p^{3}}\mid_{p=P_i^\epsilon}, \\
    B_{i}^{\epsilon, \overline{d}} &= \tilde{P}_{i}^{0}\left(\mu_{Y}^{\epsilon}\left(\overline{d}, X_{i}, P_{i}^{\epsilon}\right) - \mu_{Y}^{0}\left(\overline{d}, X_{i}, P_{i}^{0}\right)\right) - \epsilon\triangle p_{Z} \tilde{P}_{i}^{\epsilon}R_{\partial Y}^{\epsilon, \overline{d}}, \nonumber \\
    R_{\partial^2D}^{\epsilon,\overline{d}} &=
    \frac{\partial^2 p_D^\epsilon\left(\overline{d},X_i,p\right)}{\partial p^2}\mid_{p=P_i^\epsilon}, \\
    \triangle R_{\partial D}^{\epsilon,\overline{d}} &=
    \left(\frac{\partial p_D\left(\overline{d},X_i,p\right)}{\partial p}\mid_{p=P_i^\epsilon}\right) - \left(\frac{\partial p_D^0\left(\overline{d},X_i,p\right)}{\partial p}\mid_{p=P_i^\epsilon}\right).
\end{align}

Here and further, for a third-order derivative tensor $R$, we use $R[h,h,h]$ to denote its contraction with a row vector $h$ in all three arguments and $R[\cdot,h,h]$ to denote the column vector obtained by contracting it in the last two arguments. Notice that
\begin{align*}
    |R[h,h,h]|\leq\|R\|_\infty\left(\sum_{j=1}^{2n_Z}|h_j|\right)^3\leq (2n_Z)^{3/2}\|R\|_\infty\|h\|^3.
\end{align*}

The perturbed score is as follows:
\begin{align}
    \psi_{\text{ATES}}^{\epsilon} &= \underbrace{\tilde{Z}_{i}\triangle \mu_{Y}^{\epsilon}}_{\psi_{1}^{\epsilon}} - \tilde{Z}_{i}\text{ATES} + \underbrace{\tilde{Z}_{i}R_{D}^{\epsilon, d}\triangle Y^{\epsilon, d}}_{\psi_{2}^{\epsilon}} - \underbrace{\tilde{Z}_{i}R_{D}^{\epsilon, d^{*}}\triangle Y^{\epsilon, d^{*}}}_{\psi_{3}^{\epsilon}} + \nonumber \\
    &+ \underbrace{\triangle \overline{Z}^{\epsilon}\tilde{P}_{i}^{\epsilon}\left(R_{\partial Y}^{\epsilon, d}\left(1 - R_{D}^{\epsilon, d}\right) - R_{\partial Y}^{\epsilon, d^{*}}\left(1 - R_{D}^{\epsilon, d^{*}}\right)\right)}_{\psi_{4}^{\epsilon}}.
\end{align}

By applying the law of iterated expectations with the conditioning on $\left(D_{i}, X_{i}, W_{i}^{(Z)}\right)$, and equations (\ref{eq:useful:muY}) and (\ref{eq:useful:pZ}) from Lemma \ref{lemma:useful}, we obtain:
\begin{align}
    \mathbb{E}\left(\psi_{1}^{\epsilon}\right) &= \mathbb{E}\left(\tilde{P}_{i}^{0}\triangle \mu_{Y}^{\epsilon}\right), \\
    \mathbb{E}\left(\psi_{2}^{\epsilon}\right) &= \mathbb{E}\left(R_{D}^{\epsilon,d}\tilde{P}_{i}^{0}\left(\mu_{Y}^{0}\left(d, X_{i}, P_{i}^{0}\right) - \mu_{Y}^{\epsilon}\left(d, X_{i}, P_{i}^{\epsilon}\right)\right)\right), \\
    \mathbb{E}\left(\psi_{3}^{\epsilon}\right) &= \mathbb{E}\left(R_{D}^{\epsilon,d^{*}}\tilde{P}_{i}^{0}\left(\mu_{Y}^{0}\left(d^{*}, X_{i}, P_{i}^{0}\right) - \mu_{Y}^{\epsilon}\left(d^{*}, X_{i}, P_{i}^{\epsilon}\right)\right)\right), \\
    \mathbb{E}\left(\psi_{4}^{\epsilon}\right) &= -\epsilon\mathbb{E}\left(\triangle p_{Z}\tilde{P}_{i}^{\epsilon}\left(R_{\partial Y}^{\epsilon, d}\overline{R}_{D}^{\epsilon, d} - R_{\partial Y}^{\epsilon, d^{*}}\overline{R}_{D}^{\epsilon, d^{*}}\right)\right).
\end{align}

Therefore, we obtain:
\begin{align}
    &\mathbb{E}\left( \psi_{\text{ATES}}^{\epsilon}\right) = \underbrace{\mathbb{E}\left(\tilde{P}_{i}^{0}\triangle \mu_{Y}^{0}\right) - \text{ATES}\mathbb{E}\left(\tilde{Z}_{i}\right)}_{\text{Does not depend on }\epsilon} + \underbrace{\mathbb{E}\left(\overline{R}_{D}^{\epsilon, d}B_{i}^{\epsilon, d}-\overline{R}_{D}^{\epsilon, d^{*}}B_{i}^{\epsilon, d^{*}}\right)}_{\text{Depends on }\epsilon}.
\end{align}

Notice that, by Part 6, $p_D^\epsilon(\overline d,X_i,p)\geq c$ for every $p\in[0,1]^{2n_Z}$ since $p_D^\epsilon$ is a convex combination of $p_D^0$ and $p_D$. Therefore, by Parts 4 and 6, $\overline R_D^{\epsilon,\overline d}$ and its first two derivatives with respect to $\epsilon$ are bounded uniformly in $\epsilon$. Moreover, $B_i^{\epsilon,\overline d}$ and its first two derivatives with respect to $\epsilon$ are bounded in absolute value, uniformly in $\epsilon$, by a constant times $1+|\triangle\mu_Y^{\overline d}|$. The latter random variable is integrable since Part 5, Part 6, and the mean value theorem imply that $\triangle\mu_Y^{\overline d}$ is square-integrable. Therefore, by the mean value theorem and the dominated convergence theorem, we can interchange differentiation and expectation twice. By taking the second derivative, we obtain:
\begin{align}
    &\frac{\partial^{2} }{\partial \epsilon^{2}}\mathbb{E}\left( \psi_{\text{ATES}}^{\epsilon}\right) = \nonumber \\
    &=  \underbrace{\frac{\partial^{2} }{\partial \epsilon^{2}}\mathbb{E}\left(\overline{R}_{D}^{\epsilon, d}B_{i}^{\epsilon, d}-\overline{R}_{D}^{\epsilon, d^{*}}B_{i}^{\epsilon, d^{*}}\right)}_{\text{Next, apply the Leibniz rule}} = \nonumber \\
    &= \underbrace{\mathbb{E}\left(\left[\frac{\partial^{2} }{\partial \epsilon^{2}}\overline{R}_{D}^{\epsilon, d}\right]B_{i}^{\epsilon, d}\right)}_{T_{1}^{d}} + \nonumber \\
    &+ 2\underbrace{\mathbb{E}\left(\left[\frac{\partial }{\partial \epsilon}\overline{R}_{D}^{\epsilon, d}\right]\times\left[\frac{\partial }{\partial \epsilon}B_{i}^{\epsilon, d}\right]\right)}_{T_{2}^{d}} + \nonumber \\
    &+ \underbrace{\mathbb{E}\left(\overline{R}_{D}^{\epsilon, d}\left[\frac{\partial^{2} }{\partial \epsilon^{2}}B_{i}^{\epsilon, d}\right]\right)}_{T_{3}^{d}} - \nonumber \\
    &-\underbrace{\mathbb{E}\left(\left[\frac{\partial^{2} }{\partial \epsilon^{2}}\overline{R}_{D}^{\epsilon, d^{*}}\right]B_{i}^{\epsilon, d^{*}}\right)}_{T_{1}^{d^{*}}} - \nonumber \\
    &- 2\underbrace{\mathbb{E}\left(\left[\frac{\partial }{\partial \epsilon}\overline{R}_{D}^{\epsilon, d^{*}}\right]\times\left[\frac{\partial }{\partial \epsilon}B_{i}^{\epsilon, d^{*}}\right]\right)}_{T_{2}^{d^{*}}} - \nonumber \\
    &- \underbrace{\mathbb{E}\left(\overline{R}_{D}^{\epsilon, d^{*}}\left[\frac{\partial^{2} }{\partial \epsilon^{2}}B_{i}^{\epsilon, d^{*}}\right]\right)}_{T_{3}^{d^{*}}}.
\end{align}

Next, we bound each of the six terms above. Before proceeding, we denote $\overline{d}\in \{d, d^{*}\}$ and derive some preliminary bounds. Notice that the bounds imposed by Part 6 also hold for the derivatives of the perturbed nuisance functions since these derivatives are convex combinations of the corresponding derivatives of $\mu$ and $\mu^0$. First, we obtain:
\begin{align}\label{eq:lambda:au1}
    &\mathbb{E}\left[\left(\mu_Y^\epsilon(\overline d,X_i,P_i^\epsilon)-\mu_Y^\epsilon(\overline d,X_i,P_i^0)\right)^2\right]^{1/2} = \nonumber \\
    &= \underbrace{\mathbb{E}\left[\left(\left(P_i^\epsilon-P_i^0\right)\frac{\partial\mu_Y^\epsilon}{\partial p}(\overline d,X_i,\overline p)\right)^2\right]^{1/2}}_{\text{Mean value theorem in }p\text{ between }P_i^0\text{ and }P_i^\epsilon}\leq \nonumber \\
    &\leq \underbrace{\mathbb{E}\left[2n_Z C^2\left\|P_i^\epsilon-P_i^0\right\|^2\right]^{1/2}}_{\text{Part 6 and }\|v\|\leq\sqrt{2n_Z}\|v\|_\infty} \leq \underbrace{C\sqrt{2n_Z}\mathbb{E}\left[\left\|\epsilon\triangle p_Z\right\|^2\right]^{1/2}}_{\text{Since }P_i^\epsilon-P_i^0=\epsilon\triangle p_Z}\leq \nonumber \\
    &\leq \underbrace{C\sqrt{2n_Z}|\epsilon|\mathbb{E}\left[\left\|\triangle p_Z\right\|^2\right]^{1/2}}_{\text{Since }\epsilon\text{ is a constant}} \leq \underbrace{C\sqrt{2n_Z}\delta_n n^{-1/4}}_{\text{Part 5 and }|\epsilon|\leq 1}.
\end{align}

Second, by the same steps, we obtain:
\begin{align}\label{eq:lambda:au2}
    &\mathbb{E}\left[\left(\mu_Y(\overline d,X_i,P_i^{0})-\mu_Y(\overline d,X_i,P_i)\right)^2\right]^{1/2} \leq C\sqrt{2n_Z}\delta_n n^{-1/4}.
\end{align}

Third, we obtain:
\begin{align}\label{eq:lambda:au3}
    &\mathbb{E}\left[\left(\mu_Y^\epsilon(\overline d,X_i,P_i^\epsilon)-\mu_Y^0(\overline d,X_i,P_i^0)\right)^2\right]^{1/2} \leq \nonumber \\
    &\leq \underbrace{\left[\begin{aligned}
        &\mathbb{E}\left[\left(\mu_Y^\epsilon(\overline d,X_i,P_i^\epsilon)-\mu_Y^\epsilon(\overline d,X_i,P_i^0)\right)^2\right]^{1/2} \\
        &+ \mathbb{E}\left[\left(\mu_Y^\epsilon(\overline d,X_i,P_i^0)-\mu_Y^0(\overline d,X_i,P_i^0)\right)^2\right]^{1/2}
    \end{aligned}\right]}_{\text{Minkowski's inequality}}\leq \nonumber \\
    &\leq \underbrace{C\sqrt{2n_Z}\delta_n n^{-1/4}}_{\text{Inequality (\ref{eq:lambda:au1})}} + \underbrace{|\epsilon|\mathbb{E}\left[\left(\mu_Y(\overline d,X_i,P_i^0)-\mu_Y^0(\overline d,X_i,P_i^0)\right)^2\right]^{1/2}}_{\text{Since }\mu_Y^\epsilon-\mu_Y^0=\epsilon(\mu_Y-\mu_Y^0)}\leq \nonumber \\
    &\leq C\sqrt{2n_Z}\delta_n n^{-1/4} + \nonumber \\
    &+ |\epsilon|\underbrace{\left[\begin{aligned}
        &\mathbb{E}\left[\left(\mu_Y(\overline d,X_i,P_i^0)-\mu_Y(\overline d,X_i,P_i)\right)^2\right]^{1/2} + \\
        &+ \mathbb{E}\left[\left(\mu_Y(\overline d,X_i,P_i)-\mu_Y^0(\overline d,X_i,P_i^0)\right)^2\right]^{1/2}
    \end{aligned}\right]}_{\text{Minkowski's inequality}}\leq \nonumber \\
    &\leq C\sqrt{2n_Z}\delta_n n^{-1/4} + |\epsilon|\left(\underbrace{C\sqrt{2n_Z}\delta_n n^{-1/4}}_{\text{Inequality (\ref{eq:lambda:au2})}} + \underbrace{\delta_n n^{-1/4}}_{\text{Part 5}}\right) \leq \underbrace{(2C\sqrt{2n_Z}+1)\delta_n n^{-1/4}}_{\text{Since }|\epsilon|\leq 1}.
\end{align}

Fourth, we obtain:
\begin{align}\label{eq:lambda:au4}
    &\mathbb{E}\left[\left(p_D(\overline d,X_i,P_i^\epsilon)-p_D(\overline d,X_i,P_i)\right)^2\right]^{1/2} = \nonumber \\
    &= \underbrace{\mathbb{E}\left[\left(\left(P_i^\epsilon-P_i\right)\frac{\partial p_D}{\partial p}(\overline d,X_i,\overline p)\right)^2\right]^{1/2}}_{\text{Mean value theorem in }p\text{ between }P_i\text{ and }P_i^\epsilon}\leq \nonumber \\
    &\leq \underbrace{\mathbb{E}\left[2n_Z C^2\left\|P_i^\epsilon-P_i\right\|^2\right]^{1/2}}_{\text{Part 6 and }\|v\|\leq\sqrt{2n_Z}\|v\|_\infty} \leq \underbrace{C\sqrt{2n_Z}\mathbb{E}\left[\left\|(\epsilon-1)\triangle p_Z\right\|^2\right]^{1/2}}_{\text{Since }P_i^\epsilon-P_i=(\epsilon-1)\triangle p_Z}\leq \nonumber \\
    &\leq \underbrace{C\sqrt{2n_Z}|\epsilon-1|\mathbb{E}\left[\left\|\triangle p_Z\right\|^2\right]^{1/2}}_{\text{Since }\epsilon\text{ is a constant}} \leq \underbrace{C\sqrt{2n_Z}\delta_n n^{-1/4}}_{\text{Part 5 and }|\epsilon-1|\leq 1}.
\end{align}

Fifth, by the same steps, we obtain:
\begin{align}\label{eq:lambda:au5}
    &\mathbb{E}\left[\left(p_D^0(\overline d,X_i,P_i^{0})-p_D^0(\overline d,X_i,P_i^{\epsilon})\right)^2\right]^{1/2} \leq C\sqrt{2n_Z}\delta_n n^{-1/4}.
\end{align}

Sixth, we obtain:
\begin{align}\label{eq:lambda:au6}
    &\mathbb{E}\left[\left(p_D(\overline d,X_i,P_i^\epsilon)-p_D^0(\overline d,X_i,P_i^\epsilon)\right)^2\right]^{1/2} \leq \nonumber \\
    &\leq \underbrace{\left[\begin{aligned}
        &\mathbb{E}\left[\left(p_D(\overline d,X_i,P_i^\epsilon)-p_D(\overline d,X_i,P_i)\right)^2\right]^{1/2} \\
        &+ \mathbb{E}\left[\left(p_D(\overline d,X_i,P_i)-p_D^0(\overline d,X_i,P_i^0)\right)^2\right]^{1/2}
    \end{aligned}\right]}_{\text{Minkowski's inequality}} + \nonumber \\
    &+ \mathbb{E}\left[\left(p_D^0(\overline d,X_i,P_i^0)-p_D^0(\overline d,X_i,P_i^\epsilon)\right)^2\right]^{1/2} \leq \underbrace{C\sqrt{2n_Z}\delta_n n^{-1/4}}_{\text{Inequality (\ref{eq:lambda:au4})}} + \nonumber \\
    &+ \underbrace{\delta_n n^{-1/4}}_{\text{Part 5}} + \underbrace{C\sqrt{2n_Z}\delta_n n^{-1/4}}_{\text{Inequality (\ref{eq:lambda:au5})}} \leq \nonumber \\
    &\leq (2C\sqrt{2n_Z}+1)\delta_n n^{-1/4}.
\end{align}

Seventh, we obtain:
\begin{align}\label{eq:lambda:au7}
    &\mathbb{E}\left[\left(B_i^{\epsilon,\overline d}\right)^2\right]^{1/2} \leq \nonumber \\
    &\leq \underbrace{\left[\begin{aligned}
        &\mathbb{E}\left[\left(\tilde P_i^0\left(\mu_Y^\epsilon(\overline d,X_i,P_i^\epsilon)-\mu_Y^0(\overline d,X_i,P_i^0)\right)\right)^2\right]^{1/2} \\
        &+ \mathbb{E}\left[\left(\epsilon\triangle p_Z\tilde P_i^\epsilon R_{\partial Y}^{\epsilon,\overline d}\right)^2\right]^{1/2}
    \end{aligned}\right]}_{\text{Minkowski's inequality}}\leq \nonumber \\
    &\leq \underbrace{(2C\sqrt{2n_Z}+1)\delta_n n^{-1/4}}_{\text{Inequality (\ref{eq:lambda:au3}) and }|\tilde P_i^0|\leq 1} + \underbrace{C\sqrt{2n_Z}\mathbb{E}\left[\left\|\triangle p_Z\right\|^2\right]^{1/2}}_{\text{Part 6, } |\epsilon|\leq 1\text{, and }|\tilde P_i^\epsilon|\leq 1}\leq \nonumber \\
    &\leq (2C\sqrt{2n_Z}+1)\delta_n n^{-1/4} + \underbrace{C\sqrt{2n_Z}\delta_n n^{-1/4}}_{\text{Part 5}} = (3C\sqrt{2n_Z}+1)\delta_n n^{-1/4}.
\end{align}

Eighth, we obtain:
\begin{align}\label{eq:lambda:au8}
    &\mathbb{E}\left[\left(\frac{\partial}{\partial\epsilon}p_D^\epsilon(\overline d,X_i,P_i^\epsilon)\right)^2\right]^{1/2} \leq \nonumber \\
    &\leq \underbrace{\left[\begin{aligned}
        &\mathbb{E}\left[\left(\triangle p_Z R_{\partial D}^{\epsilon,\overline d}\right)^2\right]^{1/2} \\
        &+ \mathbb{E}\left[\left(p_D(\overline d,X_i,P_i^\epsilon)-p_D^0(\overline d,X_i,P_i^\epsilon)\right)^2\right]^{1/2}
    \end{aligned}\right]}_{\text{Minkowski's inequality}}\leq \nonumber \\
    &\leq C\sqrt{2n_Z}\underbrace{\mathbb{E}\left[\|\triangle p_Z\|^2\right]^{1/2}}_{\text{Parts 5 and 6}} + \underbrace{(2C\sqrt{2n_Z}+1)\delta_n n^{-1/4}}_{\text{Inequality (\ref{eq:lambda:au6})}}\leq \nonumber \\
    &\leq C\sqrt{2n_Z}\delta_n n^{-1/4} + \underbrace{(2C\sqrt{2n_Z}+1)\delta_n n^{-1/4}}_{\text{Inequality (\ref{eq:lambda:au6})}} \leq (3C\sqrt{2n_Z}+1)\delta_n n^{-1/4}.
\end{align}

Ninth, we obtain:
\begin{align}\label{eq:lambda:au9}
    &\mathbb{E}\left[\left(\frac{\partial}{\partial\epsilon}p_D^\epsilon(\overline d,X_i,P_i^\epsilon)\right)^4\right]^{1/2} \leq \nonumber \\
    &\leq \underbrace{\mathbb{E}\left[\left(2n_Z C+1\right)^2\left(\frac{\partial}{\partial\epsilon}p_D^\epsilon(\overline d,X_i,P_i^\epsilon)\right)^2\right]^{1/2}}_{\substack{\text{Since Parts 4 and 6 imply }\\\left(\frac{\partial}{\partial\epsilon}p_D^\epsilon(\overline d,X_i,P_i^\epsilon)\right)^2\leq (2n_Z C+1)^2}}\leq \nonumber \\
    &\leq \underbrace{(2n_Z C+1)\mathbb{E}\left[\left(\frac{\partial}{\partial\epsilon}p_D^\epsilon(\overline d,X_i,P_i^\epsilon)\right)^2\right]^{1/2}}_{\text{Since }2n_Z C+1\text{ is deterministic}} \leq (2n_Z C+1)\underbrace{(3C\sqrt{2n_Z}+1)\delta_n n^{-1/4}}_{\text{Inequality (\ref{eq:lambda:au8})}}.
\end{align}

Tenth, we obtain:
\begin{align}\label{eq:lambda:au10}
    &\mathbb{E}\left[\left(\frac{\partial^2}{\partial\epsilon^2}p_D^\epsilon(\overline d,X_i,P_i^\epsilon)\right)^2\right]^{1/2} \leq \nonumber \\
    &\leq \underbrace{\left[\begin{aligned}
        &\mathbb{E}\left[\left(\triangle p_Z R_{\partial^2D}^{\epsilon,\overline d}\triangle p_Z^\top\right)^2\right]^{1/2} \\
        &+ 2\mathbb{E}\left[\left(\triangle p_Z\triangle R_{\partial D}^{\epsilon,\overline d}\right)^2\right]^{1/2}
    \end{aligned}\right]}_{\text{Minkowski's inequality}}\leq \nonumber \\
    &\leq \underbrace{2n_Z C\mathbb{E}\left[\left\|\triangle p_Z\right\|^4\right]^{1/2}}_{\text{Part 6 and }\|H\|_{\mathrm{op}}\leq 2n_Z\max_{j,l}|H_{jl}|\leq 2n_Z\|H\|_\infty} + \underbrace{4C\sqrt{2n_Z}\mathbb{E}\left[\left\|\triangle p_Z\right\|^2\right]^{1/2}}_{\text{Part 6}}\leq \nonumber \\
    &\leq \underbrace{C(2n_Z)^{3/2}\mathbb{E}\left[\left\|\triangle p_Z\right\|^2\right]^{1/2}}_{\text{Parts 4 and 5}} + \underbrace{4C\sqrt{2n_Z}\mathbb{E}\left[\left\|\triangle p_Z\right\|^2\right]^{1/2}}_{\text{Part 5}}\leq \nonumber \\
    &\leq C(2n_Z)^{3/2}\delta_n n^{-1/4} + 4C\sqrt{2n_Z}\delta_n n^{-1/4} \leq C\sqrt{2n_Z}(2n_Z+4)\delta_n n^{-1/4}.
\end{align}

Finally, we obtain:
\begin{align}\label{eq:lambda:au11}
    &\mathbb{E}\left[\left(\tilde P_i^0\left(\mu_Y(\overline d,X_i,P_i^\epsilon)-\mu_Y^0(\overline d,X_i,P_i^\epsilon)\right)\right)^2\right]^{1/2} \leq \nonumber \\
    &\leq \underbrace{\mathbb{E}\left[\left(\mu_Y(\overline d,X_i,P_i^\epsilon)-\mu_Y^0(\overline d,X_i,P_i^\epsilon)\right)^2\right]^{1/2}}_{\text{Since }|\tilde P_i^0|\leq 1} \leq \nonumber \\
    &\leq \underbrace{\left[\begin{aligned}
        &\mathbb{E}\left[\left(\mu_Y(\overline d,X_i,P_i^\epsilon)-\mu_Y(\overline d,X_i,P_i)\right)^2\right]^{1/2} \\
        &+ \mathbb{E}\left[\left(\mu_Y(\overline d,X_i,P_i)-\mu_Y^0(\overline d,X_i,P_i^0)\right)^2\right]^{1/2}
    \end{aligned}\right]}_{\text{Minkowski's inequality}} + \nonumber \\
    &+ \mathbb{E}\left[\left(\mu_Y^0(\overline d,X_i,P_i^0)-\mu_Y^0(\overline d,X_i,P_i^\epsilon)\right)^2\right]^{1/2} \leq \underbrace{C\sqrt{2n_Z}\delta_n n^{-1/4}}_{\text{Similar to Inequality (\ref{eq:lambda:au1})}} + \nonumber \\
    &+ \underbrace{\delta_n n^{-1/4}}_{\text{Part 5}} + \underbrace{C\sqrt{2n_Z}\delta_n n^{-1/4}}_{\text{Similar to Inequality (\ref{eq:lambda:au1})}} \leq \nonumber \\
    &\leq (2C\sqrt{2n_Z}+1)\delta_n n^{-1/4}.
\end{align}

Next, we bound the terms. For $T_{1}^{\overline{d}}$, we obtain:
\begin{align}\label{eq:lambda:T1}
    &|T_1^{\overline d}| \leq \underbrace{\mathbb{E}\left[\left(\frac{\partial^2}{\partial\epsilon^2}\overline R_D^{\epsilon,\overline d}\right)^2\right]^{1/2} \mathbb{E}\left[\left(B_i^{\epsilon,\overline d}\right)^2\right]^{1/2}}_{\text{Cauchy--Schwarz inequality}}\leq \nonumber \\
    &\leq \left[\begin{aligned}
        &\frac{1}{c^2}\mathbb{E}\left[\left(\frac{\partial^2}{\partial\epsilon^2}p_D^\epsilon(\overline d,X_i,P_i^\epsilon)\right)^2\right]^{1/2} + \\
        &+ \frac{2}{c^3}\mathbb{E}\left[\left(\frac{\partial}{\partial\epsilon}p_D^\epsilon(\overline d,X_i,P_i^\epsilon)\right)^4\right]^{1/2}
    \end{aligned}\right]\times \nonumber \\
    &\times \mathbb{E}\left[\left(B_i^{\epsilon,\overline d}\right)^2\right]^{1/2}\leq \nonumber \\
    &\leq \left[\begin{aligned}
        &\frac{1}{c^2}\underbrace{C\sqrt{2n_Z}(2n_Z+4)\delta_n n^{-1/4}}_{\text{Inequality (\ref{eq:lambda:au10})}} + \\
        &+ \frac{2}{c^3}\underbrace{(2n_Z C+1)(3C\sqrt{2n_Z}+1)\delta_n n^{-1/4}}_{\text{Inequality (\ref{eq:lambda:au9})}}
    \end{aligned}\right]\times \nonumber \\
    &\times \underbrace{(3C\sqrt{2n_Z}+1)\delta_n n^{-1/4}}_{\text{Inequality (\ref{eq:lambda:au7})}}\leq \nonumber \\
    &\leq \left[\frac{C\sqrt{2n_Z}(2n_Z+4)}{c^2} + \frac{2(2n_Z C+1)(3C\sqrt{2n_Z}+1)}{c^3}\right]\times \nonumber \\
    &\times(3C\sqrt{2n_Z}+1)\delta_n^2 n^{-1/2}.
\end{align}

For $T_{2}^{\overline{d}}$, we obtain:
\begin{align}\label{eq:lambda:T2}
    &|T_2^{\overline d}| \leq \underbrace{\mathbb{E}\left[\left(\frac{\partial}{\partial\epsilon}\overline R_D^{\epsilon,\overline d}\right)^2\right]^{1/2} \mathbb{E}\left[\left(\frac{\partial}{\partial\epsilon}B_i^{\epsilon,\overline d}\right)^2\right]^{1/2}}_{\text{Cauchy--Schwarz inequality}}\leq \nonumber \\
    &\leq \frac{1}{c^2}\underbrace{\mathbb{E}\left[\left(\frac{\partial}{\partial\epsilon}p_D^\epsilon(\overline d,X_i,P_i^\epsilon)\right)^2\right]^{1/2}}_{\text{Inequality (\ref{eq:lambda:au8})}}\times \nonumber \\
    &\times\left[\begin{aligned}
        &\underbrace{\mathbb{E}\left[\left(\tilde P_i^0\triangle\mu_Y^{\overline d}(P_i^\epsilon)\right)^2\right]^{1/2}}_{\text{Inequality (\ref{eq:lambda:au11})}} + \\
        &+\underbrace{2\mathbb{E}\left[\left(\triangle\tilde p_Z\triangle p_Z R_{\partial Y}^{\epsilon,\overline d}\right)^2\right]^{1/2}}_{\text{Parts 4-6}} + \\
        &+ \underbrace{\mathbb{E}\left[\left(\tilde P_i^\epsilon\triangle p_Z\left(R_{\partial^2Y}^{\epsilon,\overline d}\triangle p_Z^\top + \triangle R_{\partial Y}^{\epsilon,\overline d}\right)\right)^2\right]^{1/2}}_{\text{Parts 4-6}}
    \end{aligned}\right]\leq \nonumber \\
    &\leq \frac{(3C\sqrt{2n_Z}+1)\delta_n n^{-1/4}}{c^2}\left[\begin{aligned}
        &\underbrace{(2C\sqrt{2n_Z}+1)\delta_n n^{-1/4}}_{\text{Inequality (\ref{eq:lambda:au11})}} + \\
        &+\underbrace{2C\sqrt{2n_Z}\delta_n n^{-1/4}}_{\text{Parts 4-6}} + \\
        &+ \underbrace{C(2n_Z)^{3/2}\delta_n n^{-1/4}}_{\text{Parts 4-6}} + \\
        &+\underbrace{2C\sqrt{2n_Z}\delta_n n^{-1/4}}_{\text{Parts 5-6}}
    \end{aligned}\right]\leq \nonumber \\
    &\leq \frac{(3C\sqrt{2n_Z}+1)(6C\sqrt{2n_Z}+1+C(2n_Z)^{3/2})}{c^2}\delta_n^2 n^{-1/2}.
\end{align}

Finally, we bound $T_3^{\overline d}$. Before proceeding, notice that:
\begin{align}\label{eq:lambda:d2B}
    &\frac{\partial^{2} }{\partial \epsilon^{2}}B_{i}^{\epsilon, \overline{d}} = \frac{\partial}{\partial \epsilon}\left(\frac{\partial}{\partial \epsilon}B_{i}^{\epsilon, \overline{d}}\right) = \nonumber \\
    &= \frac{\partial }{\partial \epsilon}\left[\begin{aligned}
        &\tilde{P}_{i}^{0} \triangle \mu_{Y}^{\overline{d}}\!\left( P_{i}^{\epsilon} \right) - \\
        &- \underbrace{2 \epsilon  \triangle \tilde{p}_{Z}  \triangle p_{Z} R_{\partial Y}^{\epsilon, \overline{d}}}_{\text{Since } \tilde{P}_{i}^{0} - \tilde{P}_{i}^{\epsilon} = -\epsilon \triangle \tilde{p}_{Z}} - \\
        &- \underbrace{\epsilon  \tilde{P}_{i}^{\epsilon} \triangle p_{Z} \frac{\partial R_{\partial Y}^{\epsilon, \overline{d}}}{\partial \epsilon}}_{\text{Differentiating } \tilde{P}_{i}^{\epsilon} R_{\partial Y}^{\epsilon, \overline{d}}}
    \end{aligned}\right] = \nonumber\displaybreak[1] \\
    &= \underbrace{\tilde{P}_{i}^{0} \triangle p_{Z}\triangle R_{\partial Y}^{\epsilon, \overline{d}}}_{\substack{\text{Differentiating } \tilde{P}_{i}^{0} \triangle \mu_{Y}^{\overline{d}}\!\left( P_{i}^{\epsilon} \right) \\\text{ in } \epsilon}} - \nonumber \\
    &- \underbrace{2 \triangle \tilde{p}_{Z} \triangle p_{Z} R_{\partial Y}^{\epsilon, \overline{d}}}_{\text{Differentiating } 2\epsilon} - \nonumber \\
    &- \underbrace{2 \epsilon \triangle \tilde{p}_{Z} \triangle p_{Z}\left( R_{\partial^{2} Y}^{\epsilon, \overline{d}} \triangle p_{Z}^\top + \triangle R_{\partial Y}^{\epsilon, \overline{d}} \right)}_{\substack{\text{Since }\\ \partial R_{\partial Y}^{\epsilon, \overline{d}}/\partial \epsilon = R_{\partial^{2} Y}^{\epsilon, \overline{d}} \triangle p_{Z}^\top + \triangle R_{\partial Y}^{\epsilon, \overline{d}}}} - \nonumber \\
    &- \underbrace{\left( \tilde{P}_{i}^{\epsilon} + \epsilon \triangle \tilde{p}_{Z} \right) \triangle p_{Z}\left( R_{\partial^{2} Y}^{\epsilon, \overline{d}} \triangle p_{Z}^\top + \triangle R_{\partial Y}^{\epsilon, \overline{d}} \right)}_{\substack{\text{Product rule on } \epsilon \tilde{P}_{i}^{\epsilon}, \\\text{ with } \partial \tilde{P}_{i}^{\epsilon}/\partial \epsilon = \triangle \tilde{p}_{Z}}} - \nonumber \\
    &- \underbrace{\epsilon \tilde{P}_{i}^{\epsilon} \triangle p_{Z} \frac{\partial^{2} R_{\partial Y}^{\epsilon, \overline{d}}}{\partial \epsilon^{2}}}_{\text{Differentiating } \partial R_{\partial Y}^{\epsilon, \overline{d}}/\partial \epsilon} = \nonumber\displaybreak[1] \\
    &= \underbrace{-2 \triangle \tilde{p}_{Z} \triangle p_{Z} R_{\partial Y}^{\epsilon, \overline{d}}}_{\text{Since } \partial (-2\epsilon)/\partial \epsilon = -2} - \nonumber \\
    &- \underbrace{4\epsilon \triangle \tilde{p}_{Z} \triangle p_{Z}\triangle R_{\partial Y}^{\epsilon, \overline{d}}}_{\substack{\text{Since }\\ \tilde{P}_{i}^{0} - \tilde{P}_{i}^{\epsilon} - 3\epsilon \triangle \tilde{p}_{Z} = -4\epsilon \triangle \tilde{p}_{Z}}} - \nonumber \\
    &- \underbrace{\left( \tilde{P}_{i}^{\epsilon} + 3\epsilon \triangle \tilde{p}_{Z} \right) \triangle p_{Z} R_{\partial^{2} Y}^{\epsilon, \overline{d}} \triangle p_{Z}^{\top}}_{\substack{\text{Since }\\ -2\epsilon \triangle \tilde{p}_{Z} - \tilde{P}_{i}^{\epsilon} - \epsilon \triangle \tilde{p}_{Z} = -\left( \tilde{P}_{i}^{\epsilon} + 3\epsilon \triangle \tilde{p}_{Z} \right)}} - \nonumber \\
    &- \underbrace{\left[\begin{aligned}
        &2\epsilon \tilde{P}_{i}^{\epsilon} \triangle p_{Z}\triangle R_{\partial^{2} Y}^{\epsilon, \overline{d}} \triangle p_{Z}^{\top} \\
        &+ \epsilon \tilde{P}_{i}^{\epsilon} R_{\partial^{3} Y}^{\epsilon, \overline{d}}[\triangle p_{Z},\triangle p_{Z},\triangle p_{Z}]
    \end{aligned}\right]}_{\substack{\text{Expanding }\partial^{2} R_{\partial Y}^{\epsilon, \overline{d}}/\partial \epsilon^{2} = \\ = 2 \triangle R_{\partial^{2} Y}^{\epsilon, \overline{d}} \triangle p_{Z}^{\top} + R_{\partial^{3} Y}^{\epsilon, \overline{d}}[\cdot,\triangle p_{Z},\triangle p_{Z}]}}.
\end{align}

By replacing $\partial^2 B_i^{\epsilon,\overline d}/\partial\epsilon^2$ in $T_3^{\overline{d}}$ with (\ref{eq:lambda:d2B}) and noticing that $|\overline R_D^{\epsilon,\overline d}|\leq 1/c$, we obtain:
\begin{align}\label{eq:lambda:T3}
    &|T_3^{\overline d}| \leq \nonumber \\
    &\leq \frac{1}{c}\left[\begin{aligned}
        &\underbrace{\mathbb{E}\left[\left|2\triangle\tilde p_Z\triangle p_Z R_{\partial Y}^{\epsilon,\overline d}\right|\right]}_{\text{First summand of }\partial^2 B_i^{\epsilon,\overline d}/\partial\epsilon^2} + \\
        &+ \underbrace{\mathbb{E}\left[\left|4\epsilon\triangle\tilde p_Z\triangle p_Z\triangle R_{\partial Y}^{\epsilon,\overline d}\right|\right]}_{\text{Second summand}} + \\
        &+ \underbrace{\mathbb{E}\left[\left|\left(\tilde P_i^\epsilon+3\epsilon\triangle\tilde p_Z\right)\triangle p_Z R_{\partial^2Y}^{\epsilon,\overline d}\triangle p_Z^\top\right|\right]}_{\text{Third summand}} + \\
        &+ \underbrace{\mathbb{E}\left[\left|2\epsilon\tilde P_i^\epsilon\triangle p_Z\triangle R_{\partial^2Y}^{\epsilon,\overline d}\triangle p_Z^\top\right|\right]}_{\text{Fourth summand}} + \\
        &+ \underbrace{\mathbb{E}\left[\left|\epsilon\tilde P_i^\epsilon R_{\partial^3Y}^{\epsilon,\overline d}[\triangle p_Z,\triangle p_Z,\triangle p_Z]\right|\right]}_{\text{Fifth summand}}
    \end{aligned}\right] \leq \nonumber\displaybreak[1] \\
    &\leq \frac{1}{c}\left[\begin{aligned}
        &\underbrace{2C\sqrt{2n_Z}\mathbb{E}\left[|\triangle\tilde p_Z|\|\triangle p_Z\|\right]}_{\text{Part 4, Part 6, and }|\epsilon|\leq 1} + \\
        &+ \underbrace{8C\sqrt{2n_Z}\mathbb{E}\left[|\triangle\tilde p_Z|\|\triangle p_Z\|\right]}_{\text{Part 4, Part 6, and }|\epsilon|\leq 1} + \\
        &+ \underbrace{8n_Z C\mathbb{E}\left[\|\triangle p_Z\|^2\right]}_{\text{Parts 4 and 6 and the operator norm bound}} + \\
        &+ \underbrace{8n_Z C\mathbb{E}\left[\|\triangle p_Z\|^2\right]}_{\text{Parts 4 and 6 and the operator norm bound}} + \\
        &+ \underbrace{C(2n_Z)^{3/2}\mathbb{E}\left[\|\triangle p_Z\|^3\right]}_{\text{Parts 4 and 6}} 
    \end{aligned}\right] \leq \nonumber\displaybreak[1] \\
    &\leq \frac{1}{c}\left[\begin{aligned}
        &\underbrace{2C\sqrt{2n_Z}\delta_n^2 n^{-1/2}}_{\substack{\text{Cauchy--Schwarz}\\ \text{inequality and Part 5}}} + \\
        &+ \underbrace{8C\sqrt{2n_Z}\delta_n^2 n^{-1/2}}_{\substack{\text{Cauchy--Schwarz}\\ \text{inequality and Part 5}}} + \\
        &+ \underbrace{8n_Z C\delta_n^2 n^{-1/2}}_{\text{Part 5}} + \\
        &+ \underbrace{8n_Z C\delta_n^2 n^{-1/2}}_{\text{Part 5}} + \\
        &+ C(2n_Z)^{3/2}\underbrace{\sqrt{2n_Z}\delta_n^2 n^{-1/2}}_{\text{Parts 4-5}} 
    \end{aligned}\right] \leq \nonumber\displaybreak[1] \\
    &\leq \frac{1}{c}(10C\sqrt{2n_Z}+16n_Z C+C(2n_Z)^2)\delta_n^2 n^{-1/2}.
\end{align}

By aggregating the results, we obtain:
\begin{align}\label{eq:lambda:assembly}
    &\left|\frac{\partial^2}{\partial\epsilon^2}\mathbb{E}\left(\psi_{\mathrm{ATES}}^\epsilon\right)\right| \leq \nonumber \\
    &\leq |T_1^d|+2|T_2^d|+|T_3^d|+|T_1^{d^*}|+2|T_2^{d^*}|+|T_3^{d^*}| \leq \nonumber \\
    \leq &2(3C\sqrt{2n_Z}+1)\frac{C\sqrt{2n_Z}(2n_Z+4)}{c^2}\delta_n^2n^{-1/2} + \nonumber \\
    + &\frac{4(2n_Z C+1)(3C\sqrt{2n_Z}+1)^2}{c^3}\delta_n^2n^{-1/2} + \nonumber \\
    + &\frac{4(3C\sqrt{2n_Z}+1)(6C\sqrt{2n_Z}+1+C(2n_Z)^{3/2})}{c^2}\delta_n^2n^{-1/2} + \nonumber \\
    + &\frac{2(10C\sqrt{2n_Z}+16n_Z C+C(2n_Z)^2)}{c}\delta_n^2n^{-1/2} = \nonumber \\
    = &\delta_n^*n^{-1/2}.
\end{align}

Therefore, we finally obtain:
\begin{align}
    \lambda_N' &= \sup_{\epsilon\in(0,1)}\sup_{\mu\in\mathcal T_n}\left|\frac{\partial^2}{\partial\epsilon^2}\mathbb{E}\left[\psi\left(O_i,\theta^0,\mu^\epsilon\right)\right]\right| \leq \delta_{n}^{*}n^{-1/2},
\end{align}
where $\delta_{n}^{*}$ is defined in the equation (\ref{eq:lambda:deltastar}).

\subsection{Asymptotic normality}

By aggregating the results obtained above, we conclude that $m_N$ and $m_N'$ are bounded above by constants, $r_N=0$, $r_N'=O\left(\delta_n^{(q-2)/q}\right)$, and $\lambda_N'=O\left(\delta_n^2n^{-1/2}\right)$. Hence, there exists a sufficiently large constant $C_1\geq1$ such that $\overline{\delta}_{n} = C_1\max\left(\delta_n^{(q-2)/q},\delta_n^2,n^{-1/2}\right)$ converges to zero and satisfies the following inequalities:
\begin{align}\label{eq:normality:rates}
    r_N\leq\overline{\delta}_{n},\qquad
    r_N'\leq\overline{\delta}_{n},\qquad
    \lambda_N'\leq\overline{\delta}_{n}n^{-1/2},\qquad
    \overline{\delta}_{n}\geq n^{-1/2}.
\end{align}
Therefore, given (\ref{eq:normality:rates}) and the other conditions verified above, Assumptions 3.1 and 3.2 of \citet{Chernozhukov} are satisfied for the estimator obtained for every fixed $t\in\{1,2\}$.

Denote by $\hat{\theta}^{(t)}$ the estimator of the causal parameter $\theta^0$ obtained using $t\in\{1,2\}$. By Theorem 3.1 of \citet{Chernozhukov}, for every $t\in\{1,2\}$, we have:
\begin{align}\label{eq:normality:branch}
    \sqrt{n}\left(\hat{\theta}^{(t)}-\theta^0\right)
    &= -J_0^{-1}\frac{1}{\sqrt{n}}\sum\limits_{i=1}^{n}\psi\left(O_i,\theta^0,\mu^0\right) + o_p(1),
\end{align}
where $J_0=\mathbb{E}\left(\psi^a(O_i,\mu^0)\right)$. Therefore, the proposed estimator $\hat{\theta}=\frac{1}{2}\sum_{t=1}^{2}\hat{\theta}^{(t)}$ admits the following asymptotic linear representation:
\begin{align}\label{eq:normality:averaged}
    \sqrt{n}\left(\hat{\theta}-\theta^0\right)
    &= \frac{1}{2}\sum\limits_{t=1}^{2}\sqrt{n}\left(\hat{\theta}^{(t)}-\theta^0\right) = \nonumber \\
    &= \underbrace{\frac{1}{2}\sum\limits_{t=1}^{2}\left(-J_0^{-1}\frac{1}{\sqrt{n}}\sum\limits_{i=1}^{n}\psi\left(O_i,\theta^0,\mu^0\right) + o_p(1)\right)}_{\text{Equation (\ref{eq:normality:branch})}} = \nonumber \\
    &= -J_0^{-1}\frac{1}{\sqrt{n}}\sum\limits_{i=1}^{n}\psi\left(O_i,\theta^0,\mu^0\right) + \underbrace{o_p(1)}_{\substack{\text{Since the sum of two}\\ o_p(1)\text{ terms is }o_p(1)}}.
\end{align}
Hence, we obtain\footnote{Notice that asymptotic normality holds if the equal weights of $0.5$ are replaced with any potentially random non-negative weights summing to one, since the two estimators have the same asymptotic linear representation (\ref{eq:normality:branch}).}:
\begin{align}\label{eq:normality:distribution}
    \sqrt{n}\left(\hat{\theta}-\theta^0\right)
    &= \underbrace{-J_0^{-1}\frac{1}{\sqrt{n}}\sum\limits_{i=1}^{n}\psi\left(O_i,\theta^0,\mu^0\right)}_{\substack{\xrightarrow{d}\mathcal{N}\left(0,\sigma^2\right)\\ \text{Central limit theorem}}} + \underbrace{o_p(1)}_{\substack{\text{Equation}\\ (\ref{eq:normality:averaged})}} \xrightarrow[\text{Slutsky's theorem}]{d}\mathcal{N}\left(0,\sigma^{2}\right),
\end{align}
where:
\begin{align}\label{eq:normality:sigma}
    \sigma^{2} &= \underbrace{J_0^{-2}}_{\substack{\geq 1\\ \text{Since }c\leq |J_0|\leq 1}}\underbrace{\mathbb{E}\left(\psi\left(O_i,\theta^0,\mu^0\right)^2\right)}_{\substack{\geq c\\ \text{Part 2}}} \geq c > 0.
\end{align}

\subsection{Asymptotic variance}

Denote for brevity:
\begin{align}
    \hat{\psi}_{i}(\theta) &= \frac{1}{2}\sum\limits_{t=1}^{2}\psi\left(O_{i},\theta,\hat{\mu}^{(q_i,t)}\right), \qquad
    \psi_{i}^{0} = \psi\left(O_{i},\theta^{0},\mu^{0}\right).
\end{align}
Here and further, conditioning on $\overline{A}_k$ denotes conditioning on the training observations from $\overline{A}_k$ and the sample partition. For every $k\in\{1,\ldots,K\}$ and $i\in A_k$, on the event that $\hat{\mu}^{(k,1)},\hat{\mu}^{(k,2)}\in\mathcal{T}_n$, whose probability is at least $1-2\Delta_n$ by the union bound, we obtain:
\begin{align}\label{eq:variance:conditional-score}
    &\mathbb{E}\left(\left|\hat{\psi}_{i}(\theta^0)-\psi_i^0\right|^2\mid\overline{A}_k\right)^{1/2} = \nonumber \\
    = &\mathbb{E}\left(\left|\frac{1}{2}\sum\limits_{t=1}^{2}\psi\left(O_i,\theta^0,\hat{\mu}^{(k,t)}\right)-\psi\left(O_i,\theta^0,\mu^0\right)\right|^2\mid\overline{A}_k\right)^{1/2} \leq \nonumber \\
    \leq &\underbrace{\frac{1}{2}\sum\limits_{t=1}^{2}\mathbb{E}\left(\left|\psi\left(O_i,\theta^0,\hat{\mu}^{(k,t)}\right)-\psi\left(O_i,\theta^0,\mu^0\right)\right|^2\mid\overline{A}_k\right)^{1/2}}_{\text{Minkowski's inequality}} \leq \nonumber \\
    \leq &\underbrace{\frac{1}{2}\sum\limits_{t=1}^{2}r_N'}_{\substack{\text{By the definition of }r_N'\\ \text{since }\hat{\mu}^{(k,t)}\in\mathcal{T}_n}} = r_N' \leq \underbrace{\overline{\delta}_n}_{\substack{\text{Inequality}\\ (\ref{eq:normality:rates})}} \xrightarrow{n\to\infty} 0.
\end{align}

Then, for every $k\in\{1,\ldots,K\}$ and $\varepsilon>0$, we obtain:
\begin{align}\label{eq:variance:fold-score}
    &\mathbb{P}\left(\frac{1}{|A_k|}\sum\limits_{i\in A_k}\left|\hat{\psi}_{i}(\theta^{0})-\psi_{i}^{0}\right|^{2}>\varepsilon\right) \leq \nonumber \\
    \leq &\underbrace{\frac{(r_N')^{2}}{\varepsilon}}_{\substack{\text{Conditional Markov's inequality}\\ \text{and Inequality (\ref{eq:variance:conditional-score})}}} + \underbrace{2\Delta_n}_{\substack{\text{Probability bound on}\\ \text{at least one }\hat{\mu}^{(k,t)}\notin\mathcal{T}_n}}.
\end{align}

Therefore, we obtain:
\begin{align}\label{eq:variance:score}
    &\mathbb{P}\left(\frac{1}{n}\sum\limits_{i=1}^{n}\left|\hat{\psi}_{i}(\theta^{0})-\psi_{i}^{0}\right|^{2}>\varepsilon\right) = \nonumber \\
    = &\mathbb{P}\left(\sum\limits_{k=1}^{K}\frac{|A_k|}{n}\frac{1}{|A_k|}\sum\limits_{i\in A_k}\left|\hat{\psi}_{i}(\theta^{0})-\psi_{i}^{0}\right|^{2}>\varepsilon\right) \leq \nonumber \\
    \leq &\underbrace{\mathbb{P}\left(\max\limits_{k\in\{1,\ldots,K\}}\frac{1}{|A_k|}\sum\limits_{i\in A_k}\left|\hat{\psi}_{i}(\theta^{0})-\psi_{i}^{0}\right|^{2}>\varepsilon\right)}_{\substack{\text{Since the weighted average cannot}\\ \text{exceed the maximum}}} \leq \nonumber \\
    \leq &\underbrace{\sum\limits_{k=1}^{K}\mathbb{P}\left(\frac{1}{|A_k|}\sum\limits_{i\in A_k}\left|\hat{\psi}_{i}(\theta^{0})-\psi_{i}^{0}\right|^{2}>\varepsilon\right)}_{\text{Union bound}} \leq \nonumber \\
    \leq &\underbrace{K\left(\frac{(r_N')^{2}}{\varepsilon}+2\Delta_n\right)}_{\substack{\text{Inequality}\\ (\ref{eq:variance:fold-score})}} \xrightarrow[\substack{\text{Since }K\text{ is fixed,}\\ r_N'\to0\text{ and }\Delta_n\to0}]{n\to\infty} 0 \implies \nonumber \\
    \implies &\frac{1}{n}\sum\limits_{i=1}^{n}\left|\hat{\psi}_{i}(\theta^{0})-\psi_{i}^{0}\right|^{2}\xrightarrow{p}0.
\end{align}

Moreover, we obtain:
\begin{align}\label{eq:variance:estimated-score}
    &\left(\frac{1}{n}\sum\limits_{i=1}^{n}\left|\hat{\psi}_{i}(\hat{\theta})-\psi_{i}^{0}\right|^{2}\right)^{1/2} \leq \nonumber \\
    \leq &\underbrace{\left[\begin{aligned}
        &\left(\frac{1}{n}\sum\limits_{i=1}^{n}\left|\hat{\psi}_{i}(\hat{\theta})-\hat{\psi}_{i}(\theta^{0})\right|^{2}\right)^{1/2} + \\
        &+ \left(\frac{1}{n}\sum\limits_{i=1}^{n}\left|\hat{\psi}_{i}(\theta^{0})-\psi_{i}^{0}\right|^{2}\right)^{1/2}
    \end{aligned}\right]}_{\text{Minkowski's inequality}} = \nonumber \\
    = &\underbrace{\left(\frac{1}{n}\sum\limits_{i=1}^{n}\left|-\tilde{Z}_{i}\left(\hat{\theta}-\theta^{0}\right)\right|^{2}\right)^{1/2}}_{\substack{\text{Since the score is linear in }\theta\\ \text{and }\psi^{a}(\cdot)=-\tilde{Z}_{i}}} + \nonumber \\
    + &\left(\frac{1}{n}\sum\limits_{i=1}^{n}\left|\hat{\psi}_{i}(\theta^{0})-\psi_{i}^{0}\right|^{2}\right)^{1/2} = \nonumber \\
    = &\left|\hat{\theta}-\theta^{0}\right|\underbrace{\left(\frac{1}{n}\sum\limits_{i=1}^{n}\tilde{Z}_{i}^{2}\right)^{1/2}}_{\substack{\text{Bounded above by }1\\ \text{since }\tilde{Z}_{i}\in\{0,1\}}} + \left(\frac{1}{n}\sum\limits_{i=1}^{n}\left|\hat{\psi}_{i}(\theta^{0})-\psi_{i}^{0}\right|^{2}\right)^{1/2} \leq \nonumber \\
    \leq &\underbrace{\left|\hat{\theta}-\theta^{0}\right|}_{\substack{\xrightarrow{p}0\\ \text{Consistency implied by}\\ (\ref{eq:normality:distribution})}} + \underbrace{\left(\frac{1}{n}\sum\limits_{i=1}^{n}\left|\hat{\psi}_{i}(\theta^{0})-\psi_{i}^{0}\right|^{2}\right)^{1/2}}_{\substack{\xrightarrow{p}0\\ \text{Equation (\ref{eq:variance:score})}}}\xrightarrow{p}0.
\end{align}

Finally, we obtain:
\begin{align}\label{eq:variance:second-moment}
    &\left|\left(\frac{1}{n}\sum\limits_{i=1}^{n}\hat{\psi}_{i}(\hat{\theta})^{2}\right)^{1/2}-\left(\frac{1}{n}\sum\limits_{i=1}^{n}\left(\psi_{i}^{0}\right)^{2}\right)^{1/2}\right| \leq \nonumber \\
    \leq &\underbrace{\left(\frac{1}{n}\sum\limits_{i=1}^{n}\left|\hat{\psi}_{i}(\hat{\theta})-\psi_{i}^{0}\right|^{2}\right)^{1/2}}_{\text{Reverse triangle inequality}}\xrightarrow[\text{Equation (\ref{eq:variance:estimated-score})}]{p}0 \implies \nonumber \\
    \implies &\left(\frac{1}{n}\sum\limits_{i=1}^{n}\hat{\psi}_{i}(\hat{\theta})^{2}\right)^{1/2}\xrightarrow{p}\underbrace{\mathbb{E}\left(\left(\psi_{i}^{0}\right)^{2}\right)^{1/2}}_{\substack{\text{Law of large numbers}\\ \text{and Slutsky's theorem}}} \implies \nonumber \\
    \implies &\frac{1}{n}\sum\limits_{i=1}^{n}\hat{\psi}_{i}(\hat{\theta})^{2}\xrightarrow[\text{Continuous mapping theorem}]{p}\mathbb{E}\left(\left(\psi_{i}^{0}\right)^{2}\right).
\end{align}

Recall that $\hat{\sigma}^2=n\widehat{\operatorname{As.Var}}(\hat{\theta})$, where $\widehat{\operatorname{As.Var}}(\hat{\theta})$ is the corresponding variance estimator defined in the main text. Therefore, we obtain:
\begin{align}\label{eq:variance:consistency}
    \hat{\sigma}^{2} &= \underbrace{\left(\frac{1}{n}\sum\limits_{i=1}^{n}\tilde{Z}_{i}\right)^{-2}}_{\substack{\xrightarrow{p}J_0^{-2}\\ \text{Law of large numbers}\\ \text{and continuous mapping theorem}}}\underbrace{\frac{1}{n}\sum\limits_{i=1}^{n}\hat{\psi}_{i}(\hat{\theta})^{2}}_{\substack{\xrightarrow{p}\mathbb{E}\left((\psi_i^0)^2\right)\\ \text{Equation (\ref{eq:variance:second-moment})}}} \xrightarrow{p} \nonumber \\
    &\xrightarrow{p} \underbrace{J_0^{-2}\mathbb{E}\left(\left(\psi_{i}^{0}\right)^{2}\right)}_{\substack{\sigma^{2}\geq c>0\\ \text{Equation (\ref{eq:normality:sigma})}}} \implies \underbrace{\frac{1}{\hat{\sigma}}\xrightarrow{p}\frac{1}{\sigma}}_{\text{Continuous mapping theorem}}.
\end{align}

Finally, we obtain:
\begin{align}
    \frac{\sqrt{n}\left(\hat{\theta}-\theta^0\right)}{\hat{\sigma}} &= \underbrace{\sqrt{n}\left(\hat{\theta}-\theta^0\right)}_{\substack{\xrightarrow{d}\mathcal{N}\left(0,\sigma^{2}\right)\\ \text{Equation (\ref{eq:normality:distribution})}}}
    \underbrace{\frac{1}{\hat{\sigma}}}_{\substack{\xrightarrow{p}1/\sigma\\ \text{Equation (\ref{eq:variance:consistency})}}} \xrightarrow[\text{Slutsky's theorem}]{d}\mathcal{N}\left(0,1\right).
\end{align}

\subsection{Extension to ATTS, ATE, and ATT}

For ATTS, ATE, and ATT, the bounds on $m_N$ and $r_N'$ can be established analogously to those for ATES, since all additional probability ratios are uniformly bounded by Part 4. Moreover, Parts 4 and 5 imply that their deviations from the corresponding true ratios have $L^2$ norms of order $O(\delta_n n^{-1/4})$, so the arguments based on Hölder's inequality remain applicable. Unlike the scores for ATES and ATTS, the scores for ATE and ATT involve conditional outcome expectations not multiplied by $\tilde{Z}_i$. To bound the corresponding terms in $m_N$, notice that Parts 1 and 4 imply
\begin{align}
    \mathbb{E}\left(\left|\text{E}_i^{0(\overline{d})}\right|^q\right)^{1/q}\leq C/c^{2/q}.
\end{align}
This inequality follows from Jensen's inequality, the law of iterated expectations, and the fact that $\mathbb{E}(D_i^{(\overline{d})}\tilde{Z}_i\mid X_i,P_i^0)=P_i^{0(\overline{d})}\tilde{P}_i^0\geq c^2$ almost surely. Together with Parts 2 and 3, Minkowski's inequality, and the identification formulas for ATE and ATT, this ensures that the corresponding conditional outcome expectations and causal parameters are bounded in the $L^q$ norm and absolute value, respectively, as required to establish $m_N=O(1)$.

To establish the bound on $\lambda_N'$, denote
\begin{align}
    a_i^{\epsilon,\overline{d}} &= \mu_Y^\epsilon(\overline{d},X_i,P_i^\epsilon)-\mu_Y^0(\overline{d},X_i,P_i^0),\\
    b_i^{\epsilon,\overline{d}} &= a_i^{\epsilon,\overline{d}}-\epsilon\triangle p_Z R_{\partial Y}^{\epsilon,\overline{d}},\\
    H_i^\epsilon &= D_i^{(d^*)}R_{PD}^\epsilon-D_i^{(d)},\\
    L_i^\epsilon &= 1-\frac{\tilde{P}_i^0}{\tilde{P}_i^\epsilon}.
\end{align}
By applying the law of iterated expectations as in the derivation for ATES, we obtain:
\begin{align}
    \mathbb{E}(\psi_{\mathrm{ATTS}}^\epsilon)
    &=\mathbb{E}\left(H_i^\epsilon B_i^{\epsilon,d^*}\right),\\
    \mathbb{E}(\psi_{\mathrm{ATE}}^\epsilon)
    &=\mathbb{E}\left(\overline{R}_D^{\epsilon,d}b_i^{\epsilon,d}-\overline{R}_D^{\epsilon,d^*}b_i^{\epsilon,d^*}\right)+\nonumber\\
    &+\mathbb{E}\left(L_i^\epsilon\left(R_D^{\epsilon,d}a_i^{\epsilon,d}-R_D^{\epsilon,d^*}a_i^{\epsilon,d^*}\right)\right),\\
    \mathbb{E}(\psi_{\mathrm{ATT}}^\epsilon)
    &=\mathbb{E}\left(H_i^\epsilon b_i^{\epsilon,d^*}\right)+\nonumber\\
    &+\mathbb{E}\left(L_i^\epsilon\left(D_i^{(d)}a_i^{\epsilon,d}-D_i^{(d^*)}R_{PD}^\epsilon a_i^{\epsilon,d^*}\right)\right).
\end{align}
By Parts 4--6 and arguments analogous to those used above, $H_i^\epsilon$, $R_D^{\epsilon,\overline{d}}$, and $R_{PD}^\epsilon$ and their first two derivatives with respect to $\epsilon$ are uniformly bounded, while the $L^2$ norms of their first two derivatives are of order $O(\delta_n n^{-1/4})$. Inequality~(\ref{eq:lambda:au7}) and the bounds on the first and second derivatives of $B_i^{\epsilon,\overline{d}}$ established in the derivations of inequalities~(\ref{eq:lambda:T2}) and (\ref{eq:lambda:T3}), respectively, imply:
\begin{align}
    \left\|B_i^{\epsilon,\overline{d}}\right\|_{L^2}+\left\|\frac{\partial B_i^{\epsilon,\overline{d}}}{\partial\epsilon}\right\|_{L^2}&=O(\delta_n n^{-1/4}),\\
    \left\|\frac{\partial^2 B_i^{\epsilon,\overline{d}}}{\partial\epsilon^2}\right\|_{L^1}&=O(\delta_n^2n^{-1/2}).
\end{align}
The same bounds apply to $b_i^{\epsilon,\overline{d}}$ by setting $\tilde{P}_i^\epsilon=\tilde{P}_i^0=1$ in the corresponding derivations. Moreover, by Parts 4--6 and the chain rule, $a_i^{\epsilon,\overline{d}}$, $L_i^\epsilon$, and their first two derivatives with respect to $\epsilon$ have $L^2$ norms of order $O(\delta_n n^{-1/4})$. Since the probability ratios and their first two derivatives are uniformly bounded, the same $L^2$ bounds apply to the expressions in parentheses multiplied by $L_i^\epsilon$ in the expected scores for ATE and ATT and to their first two derivatives. As in the proof for ATES, the uniform derivative and moment bounds justify interchanging differentiation and expectation twice by the dominated convergence theorem. Consequently, by the Leibniz rule and the Cauchy--Schwarz inequality, the second derivatives of all the expected scores above are of order $O(\delta_n^2n^{-1/2})$, uniformly in $\epsilon\in[0,1]$ and $\mu\in\mathcal{T}_n$.

Therefore, the bounds required by Assumption 3.2 of \citet{Chernozhukov} hold for all four scores. The verification of Assumption 3.1 and the arguments establishing asymptotic normality and consistency of the variance estimators proceed as for ATES, with $\psi^a(O_i,\mu)$ replaced by $-D_i^{(d)}\tilde{Z}_i$, $-1$, and $-D_i^{(d)}$ for ATTS, ATE, and ATT, respectively. In each case, $|\psi^a(O_i,\mu)|\leq1$ and $c\leq|J_0|\leq1$ by Part 2.

\section{Proof of Theorem 4}

For comparability with \citet{Chernozhukov}, here we use the same notation as in their original paper. When conditioning on the training observations, we also condition on the sample partition. Consider an arbitrary sequence of probability distributions $P_N\in\mathcal{P}_N$. Under the proposed modifications of Assumptions 3.1 and 3.2 of \citet{Chernozhukov}, the only bound in the proof of their Theorem 3.1 that requires a different derivation is the bound on $\mathcal{I}_{4,k}$ in Step 3 of the proof for the DML2 estimator.

Consider the event $\mathcal{E}_{N} = \{\hat{\eta}_{0,k}\in \mathcal{T}_{N}\text{ for every }k\in\{1,\ldots,K\}\}$ and notice that:
\begin{align}\label{eq:modified:event}
    P_N\left(\mathcal{E}_N\right)
    &= 1-P_N\left(\mathcal{E}_N^{c}\right) \geq 1-\underbrace{\sum\limits_{k=1}^{K}P_N\left(\hat{\eta}_{0,k}\notin\mathcal{T}_N\right)}_{\text{Union bound}} \geq 1-\underbrace{K\Delta_N}_{\substack{\text{Condition (a) of}\\ \text{Assumption 3.2}}}\xrightarrow[\substack{\text{Since }K\text{ is fixed}\\ \text{and }\Delta_N\to0}]{N\to\infty}1.
\end{align}

Further, on the event $\mathcal{E}_N$, we obtain:
\begin{align}\label{eq:modified:I4}
    \mathcal{I}_{4,k}
    &= \underbrace{\sqrt{n}\left\|\left[\begin{aligned}
        &\mathbb{E}_{P_N}\left[\psi\left(W;\theta_{0},\hat{\eta}_{0,k}\right)\mid \left(W_{i}\right)_{i\in I_{k}^{c}}\right] \\
        &-\mathbb{E}_{P_N}\left[\psi\left(W;\theta_{0},\eta_{0}\right)\right]
    \end{aligned}\right]\right\|}_{\substack{\text{Definition of }\mathcal{I}_{4,k}\text{ in Step 3}\\ \text{of the proof of \citet{Chernozhukov}}}} = \nonumber \\
    &=\underbrace{\sqrt{n}\left\|\mathbb{E}_{P_N}\left[\psi\left(W;\theta_{0},\hat{\eta}_{0,k}\right)\mid \left(W_{i}\right)_{i\in I_{k}^{c}}\right]\right\|}_{\substack{\text{Since }\mathbb{E}_{P_N}\left[\psi\left(W;\theta_{0},\eta_{0}\right)\right] = 0\\ \text{by condition (a) of Assumption 3.1}}} \leq \nonumber \\
    &\leq \underbrace{\sqrt{n}\sup_{\eta\in \mathcal{T}_{N}}\left\|\mathbb{E}_{P_N}\left[\psi\left(W;\theta_{0},\eta\right)\right]\right\|}_{\substack{\text{Since }\hat{\eta}_{0,k}\in\mathcal{T}_N\text{ is fixed conditional on}\\ (W_i)_{i\in I_k^c}\text{ and the sample partition,}\\ \text{which are independent of }W}} = \nonumber \\
    &= \sqrt{n}\lambda_{N}'' \leq \underbrace{\sqrt{n}\frac{\delta_N}{\sqrt{N}}}_{\substack{\text{Inequality (\ref{eq:NO_alternative})}\\ \text{from Theorem~\ref{theorem:modified}}}} = \frac{\delta_N}{\sqrt{K}} \leq \delta_N.
\end{align}
Therefore, we obtain:
\begin{align}\label{eq:modified:I3I4}
    \underbrace{\mathcal{I}_{4,k} = O_{P_N}\left(\sqrt{n}\lambda_{N}''\right)}_{\text{Inequalities (\ref{eq:modified:event}) and (\ref{eq:modified:I4})}}, \qquad
    \underbrace{\mathcal{I}_{3,k} = O_{P_N}\left(r_{N}'\right)}_{\substack{\text{Step 3 of the proof}\\ \text{of \citet{Chernozhukov}}}}.
\end{align}
By plugging these bounds into inequality (A.16) from the original proof, we obtain:
\begin{align}\label{eq:modified:fold-bound}
    &\left\|\mathbb{E}_{n,k}\left[\psi\left(W;\theta_0,\hat{\eta}_{0,k}\right)\right]-\frac{1}{n}\sum\limits_{i\in I_k}\psi\left(W_i;\theta_0,\eta_0\right)\right\| \leq \underbrace{\frac{\mathcal{I}_{3,k}+\mathcal{I}_{4,k}}{\sqrt{n}}}_{\substack{\text{Inequality (A.16)}\\ \text{of \citet{Chernozhukov}}}} = \nonumber \\
    = &\underbrace{O_{P_N}\left(r_N'/\sqrt{n}+\lambda_N''\right)}_{\text{Equation (\ref{eq:modified:I3I4})}} =  \underbrace{O_{P_N}\left(r_N'/\sqrt{N}+\lambda_N''\right)}_{\substack{\text{Since }n=N/K\\ \text{and }K\text{ is fixed}}}.
\end{align}
Therefore, the remainder $R_{N,2}$ defined in Step 1 of the original proof obeys:
\begin{align}\label{eq:modified:R2}
    \left\|R_{N,2}\right\|
    &\leq \frac{1}{K}\sum\limits_{k=1}^{K}\frac{\mathcal{I}_{3,k}+\mathcal{I}_{4,k}}{\sqrt{n}} = \underbrace{O_{P_N}\left(r_N'/\sqrt{N}+\lambda_N''\right)}_{\substack{\text{Inequality (\ref{eq:modified:fold-bound})}\\ \text{and the fact that }K\text{ is fixed}}}.
\end{align}
By replacing the bound in (A.6) of the original proof with (\ref{eq:modified:R2}) and following the arguments in (A.9)--(A.12), we obtain the conclusion of Theorem 3.1 of \citet{Chernozhukov} for the DML2 estimator with the following remainder:
\begin{align}\label{eq:modified:remainder}
    \rho_N
    &= \frac{1}{\sqrt{N}}+r_N+r_N'+\sqrt{N}\lambda_N'' \leq \nonumber \\
    &\leq \underbrace{\frac{1}{\sqrt{N}}}_{\leq\delta_N}
    +\underbrace{r_N+r_N'}_{\substack{\leq2\delta_N\\ \text{Condition (c) of}\\ \text{Assumption 3.2}}}
    +\underbrace{\delta_N}_{\substack{\text{Inequality}\\ (\ref{eq:NO_alternative})}} \leq 4\delta_N\xrightarrow{N\to\infty}0.
\end{align}
Similarly, since $K$ is fixed, inequality~(\ref{eq:modified:fold-bound}) implies
\begin{align}
    \max_{k\in\{1,\ldots,K\}}\left\|R_{N,2,k}\right\|=O_{P_N}\left(r_N'/\sqrt{N}+\lambda_N''\right).
\end{align}
By replacing the bound in (A.19) of the original proof with this bound, we obtain the same conclusion for the DML1 estimator with the remainder specified in (\ref{eq:modified:remainder}). Since the sequence $P_N\in\mathcal{P}_N$ has been chosen arbitrarily and all the bounds hold uniformly over $P\in\mathcal{P}_N$, the conclusions hold uniformly over $P\in\mathcal{P}_N$.

\section{Proof of Theorem 5}

The proof is very similar to the proof of Theorem~\ref{theorem:normality}. In particular, the bounds on $m_N$, $m_N'$, $r_N$, and $r_N'$ established in the proof of Theorem~\ref{theorem:normality} remain valid when the derivative correction terms are omitted. There is no need to establish the bound on $\lambda_N'$ or twice continuous Gateaux differentiability of the expected score, since the corresponding conditions are dropped in Theorem~\ref{theorem:modified}. The only notable difference is the need to establish the bound $\lambda_{N}^{''}$ required by Theorem~\ref{theorem:modified}. Below, we establish this bound for each estimator.

\subsection{The bound for the estimator of ATES}

Consider the score without the derivative terms:
\begin{align}
     \psi_{\text{ATES}}^{*}\left(O_{i}, \theta, \mu\right) &= \tilde{Z}_{i}\left(\text{E}_{i}^{(d)} - \text{E}_{i}^{(d^{*})}\right) -\tilde{Z}_{i} \theta + \nonumber \\
        &+ \tilde{Z}_{i}\left(\frac{D_{i}^{(d)}}{P_{i}^{(d)}}\left(Y_{i} - \text{E}_{i}^{(d)}\right) - \frac{D_{i}^{(d^{*})}}{P_{i}^{(d^{*})}}\left(Y_{i} - \text{E}_{i}^{(d^{*})}\right)\right).
\end{align}

Notice that:
\begin{align}\label{eq:normality2:ates:mean}
    \mathbb{E}\left(\psi_{\text{ATES}}^{*}\left(O_{i}, \theta^{0}, \mu\right)\right) &= \underbrace{\mathbb{E}\left(\tilde{Z}_{i}\left(\left[\text{E}_{i}^{(d)} - \text{E}_{i}^{0(d)}\right] - \left[\text{E}_{i}^{(d^{*})} - \text{E}_{i}^{0(d^{*})}\right]\right) + \tilde{Z}_{i}\left(\text{E}_{i}^{0(d)} - \text{E}_{i}^{0(d^{*})} - \theta^{0}\right)\right)}_{\text{Add and subtract }\text{E}_{i}^{0(d)}, \text{E}_{i}^{0(d^{*})}} + \nonumber \\
    &+ \underbrace{\mathbb{E}\left(\tilde{Z}_{i}\left(\frac{D_{i}^{(d)}}{P_{i}^{(d)}}\left(\text{E}_{i}^{0(d)} - \text{E}_{i}^{(d)}\right) - \frac{D_{i}^{(d^{*})}}{P_{i}^{(d^{*})}}\left(\text{E}_{i}^{0(d^{*})} - \text{E}_{i}^{(d^{*})}\right)\right)\right)}_{\substack{\text{LIE conditional on }(D_i,X_i,W_i^{(Z)},\tilde{Z}_i)\\ \text{using the conditional expectations of }Y_i\text{ for }\tilde{Z}_i=1}} = \nonumber \\
    &= \underbrace{\mathbb{E}\left(\tilde{Z}_{i}\left(\Delta\text{E}_{i}^{(d)} - \Delta\text{E}_{i}^{(d^{*})}\right)\right)}_{\substack{\text{Since Theorem~\ref{theorem:ates} implies}\\ \mathbb{E}\left(\tilde{Z}_{i}\left(\text{E}_{i}^{0(d)} - \text{E}_{i}^{0(d^{*})} - \theta^{0}\right)\right) = 0}} - \mathbb{E}\left(\tilde{Z}_{i}\left(\frac{D_{i}^{(d)}}{P_{i}^{(d)}}\Delta\text{E}_{i}^{(d)} - \frac{D_{i}^{(d^{*})}}{P_{i}^{(d^{*})}}\Delta\text{E}_{i}^{(d^{*})}\right)\right) = \nonumber \\
    &= \mathbb{E}\left(\tilde{Z}_{i}\left(\Delta\text{E}_{i}^{(d)}\underbrace{\left(1-\frac{D_i^{(d)}}{P_i^{(d)}}\right)}_{\Delta P_i^{(d)}}-\Delta\text{E}_{i}^{(d^{*})}\underbrace{\left(1-\frac{D_i^{(d^{*})}}{P_i^{(d^{*})}}\right)}_{\Delta P_i^{(d^{*})}}\right)\right) = \nonumber \\
    &= \underbrace{\mathbb{E}\left(\tilde{P}_{i}^{0}\left(\Delta\text{E}_{i}^{(d)}\Delta P_{i}^{(d)} - \Delta\text{E}_{i}^{(d^{*})}\Delta P_{i}^{(d^{*})}\right)\right)}_{\substack{\text{LIE conditional on }(D_i,X_i,W_i^{(Z)})\\ \text{since }\mathbb{E}\left(\tilde{Z}_i\mid D_i,X_i,W_i^{(Z)}\right)=\tilde{P}_i^0}}.
\end{align}
Therefore, we obtain:
\begin{align}
    \mathbb{E}\left(\psi_{\text{ATES}}^{*}\left(O_i,\theta^0,\mu^0\right)\right)
    &= \underbrace{\mathbb{E}\left(\tilde{P}_i^0\left(\underbrace{\Delta\text{E}_i^{(d)}}_{=0}\Delta P_i^{(d)}-\underbrace{\Delta\text{E}_i^{(d^{*})}}_{=0}\Delta P_i^{(d^{*})}\right)\right)}_{\substack{\text{Equation (\ref{eq:normality2:ates:mean})}\\ \text{evaluated at }\mu=\mu^0}}=0.
\end{align}

Finally, using the notation of Theorem~\ref{theorem:modified}, we obtain:
\begin{align}
    \lambda_N'' &= \sup_{\mu\in\mathcal{T}_n}\left|\mathbb{E}\left(\psi_{\text{ATES}}^{*}\left(O_i,\theta^0,\mu\right)\right)\right| = \nonumber \\
    &= \underbrace{\sup_{\mu\in\mathcal{T}_n}\left|\mathbb{E}\left(\tilde{P}_{i}^{0}\left(\Delta\text{E}_{i}^{(d)}\Delta P_{i}^{(d)} - \Delta\text{E}_{i}^{(d^{*})}\Delta P_{i}^{(d^{*})}\right)\right)\right|}_{\text{Equation (\ref{eq:normality2:ates:mean})}} \leq \underbrace{\delta_n/\sqrt{n}}_{\substack{\text{Assumption~\ref{assumption:regularity2}}}}.
\end{align}

\subsection{The bound for the estimator of ATTS}

Consider the score without the derivative terms:
\begin{align}
     \psi_{\text{ATTS}}^{*}\left(O_{i}, \theta, \mu\right) &= D_{i}^{(d)}\tilde{Z}_{i}\left(Y_{i} - \text{E}_{i}^{(d^{*})}\right) - D_{i}^{(d)}\tilde{Z}_{i} \theta - D_{i}^{(d^{*})}\tilde{Z}_{i}\frac{P_{i}^{(d)}}{P_{i}^{(d^{*})}}\left(Y_{i} - \text{E}_{i}^{(d^{*})}\right).
\end{align}

Notice that:
\begin{align}
    \mathbb{E}\left(\psi_{\text{ATTS}}^{*}\left(O_{i}, \theta^{0}, \mu\right)\right) &= \underbrace{\mathbb{E}\left(\tilde{Z}_{i}\left(Y_{i} - \text{E}_{i}^{0(d^{*})}\right)\left(D_{i}^{(d)} - D_{i}^{(d^{*})}\frac{P_{i}^{(d)}}{P_{i}^{(d^{*})}}\right)\right)}_{\text{Add and subtract }\text{E}_{i}^{0(d^{*})}} - \nonumber \\
    &-\underbrace{\mathbb{E}\left(\tilde{Z}_{i}\Delta\text{E}_{i}^{(d^{*})}\left(D_{i}^{(d)} - D_{i}^{(d^{*})}\frac{P_{i}^{(d)}}{P_{i}^{(d^{*})}}\right)\right)}_{\text{Since }Y_{i} - \text{E}_{i}^{(d^{*})} = \left(Y_{i} - \text{E}_{i}^{0(d^{*})}\right) - \Delta\text{E}_{i}^{(d^{*})}} - \theta^{0}\mathbb{E}\left(D_{i}^{(d)}\tilde{Z}_{i}\right) = \nonumber \\
    &= \underbrace{\mathbb{E}\left(\tilde{Z}_{i}\left[\begin{aligned}
        &D_{i}^{(d)}\left(\text{E}_{i}^{0(d)} - \text{E}_{i}^{0(d^{*})} - \theta^{0}\right) - \\
        &-D_{i}^{(d^{*})}\frac{P_{i}^{(d)}}{P_{i}^{(d^{*})}}\underbrace{\left(\text{E}_{i}^{0(d^{*})} - \text{E}_{i}^{0(d^{*})}\right)}_{=0}
    \end{aligned}\right]\right)}_{\substack{\text{LIE conditional on }(D_i,X_i,W_i^{(Z)},\tilde{Z}_i)\\ \text{using the conditional expectations of }Y_i\text{ for }\tilde{Z}_i=1}} - \nonumber \\
    &-\mathbb{E}\left(\tilde{Z}_{i}\Delta\text{E}_{i}^{(d^{*})}\left(D_{i}^{(d)} - D_{i}^{(d^{*})}\frac{P_{i}^{(d)}}{P_{i}^{(d^{*})}}\right)\right) = \nonumber \\
    &= \underbrace{\mathbb{E}\left(\tilde{Z}_{i}D_{i}^{(d)}\left(\text{E}_{i}^{0(d)} - \text{E}_{i}^{0(d^{*})} - \theta^{0}\right)\right)}_{\substack{=0\\ \text{By Theorem~\ref{theorem:ates}}}} - \nonumber \\
    &-\mathbb{E}\left(\tilde{Z}_{i}\Delta\text{E}_{i}^{(d^{*})}\left(D_{i}^{(d)} - D_{i}^{(d^{*})}\frac{P_{i}^{(d)}}{P_{i}^{(d^{*})}}\right)\right) = \nonumber \\
    &= \mathbb{E}\left(\tilde{Z}_{i}\Delta\text{E}_{i}^{(d^{*})}P_{i}^{(d)}\underbrace{\left(\frac{D_{i}^{(d^{*})}}{P_{i}^{(d^{*})}}-\frac{D_{i}^{(d)}}{P_{i}^{(d)}}\right)}_{\substack{=\Delta P_{i}^{(d)}-\Delta P_{i}^{(d^{*})}\\ \text{By definition}}}\right) = \nonumber \\
    &= \underbrace{\mathbb{E}\left(\tilde{P}_{i}^{0}\Delta\text{E}_{i}^{(d^{*})}P_{i}^{(d)}\left(\Delta P_{i}^{(d)} - \Delta P_{i}^{(d^{*})}\right)\right)}_{\substack{\text{LIE conditional on }(D_i,X_i,W_i^{(Z)})\\ \text{since }\mathbb{E}\left(\tilde{Z}_i\mid D_i,X_i,W_i^{(Z)}\right)=\tilde{P}_i^0}}.
\end{align}
Therefore, we obtain:
\begin{align}
    \mathbb{E}\left(\psi_{\text{ATTS}}^{*}\left(O_i,\theta^0,\mu^0\right)\right)
    &= \mathbb{E}\left(\tilde{P}_{i}^{0}\underbrace{\Delta\text{E}_{i}^{(d^{*})}}_{\substack{=0\\ \text{Since }\mu=\mu^0}}P_{i}^{(d)}\left(\Delta P_{i}^{(d)}-\Delta P_{i}^{(d^{*})}\right)\right)=0.
\end{align}
Moreover, using the notation of Theorem~\ref{theorem:modified}, we obtain:
\begin{align}
    \lambda_N'' &= \sup_{\mu\in\mathcal{T}_n}\left|\mathbb{E}\left(\psi_{\text{ATTS}}^{*}\left(O_i,\theta^0,\mu\right)\right)\right| = \nonumber \\
    &= \sup_{\mu\in\mathcal{T}_n}\left|\mathbb{E}\left(\tilde{P}_{i}^{0}\Delta\text{E}_{i}^{(d^{*})}P_{i}^{(d)}\left(\Delta P_{i}^{(d)} - \Delta P_{i}^{(d^{*})}\right)\right)\right| \leq  \underbrace{\delta_n/\sqrt{n}}_{\text{Assumption~\ref{assumption:regularity2}}}.
\end{align}

\subsection{The bound for the estimator of ATE}

Consider the score without the derivative terms:
\begin{align}
     \psi_{\text{ATE}}^{*}\left(O_{i}, \theta, \mu\right) &= \left(\text{E}_{i}^{(d)} - \text{E}_{i}^{(d^{*})}\right) - \theta + \frac{\tilde{Z}_{i}}{\tilde{P}_{i}}\left(\frac{D_{i}^{(d)}}{P_{i}^{(d)}}\left(Y_{i} - \text{E}_{i}^{(d)}\right) - \frac{D_{i}^{(d^{*})}}{P_{i}^{(d^{*})}}\left(Y_{i} - \text{E}_{i}^{(d^{*})}\right)\right).
\end{align}

Notice that:
\begin{align}
    \mathbb{E}\left(\psi_{\text{ATE}}^{*}\left(O_{i}, \theta^{0}, \mu\right)\right) &= \underbrace{\mathbb{E}\left(\left[\text{E}_{i}^{(d)} - \text{E}_{i}^{0(d)}\right] - \left[\text{E}_{i}^{(d^{*})} - \text{E}_{i}^{0(d^{*})}\right] + \left(\text{E}_{i}^{0(d)} - \text{E}_{i}^{0(d^{*})} - \theta^{0}\right)\right)}_{\text{Add and subtract }\text{E}_{i}^{0(d)}, \text{E}_{i}^{0(d^{*})}} + \nonumber \\
    &+ \underbrace{\mathbb{E}\left(\frac{\tilde{Z}_{i}}{\tilde{P}_{i}}\left(\frac{D_{i}^{(d)}}{P_{i}^{(d)}}\left(\text{E}_{i}^{0(d)} - \text{E}_{i}^{(d)}\right) - \frac{D_{i}^{(d^{*})}}{P_{i}^{(d^{*})}}\left(\text{E}_{i}^{0(d^{*})} - \text{E}_{i}^{(d^{*})}\right)\right)\right)}_{\substack{\text{LIE conditional on }(D_i,X_i,W_i^{(Z)},\tilde{Z}_i)\\ \text{using the conditional expectations of }Y_i\text{ for }\tilde{Z}_i=1}} = \nonumber \\
    &= \underbrace{\mathbb{E}\left(\Delta\text{E}_{i}^{(d)} - \Delta\text{E}_{i}^{(d^{*})}\right)}_{\substack{\text{Since Theorem~\ref{theorem:ate} implies}\\ \mathbb{E}\left(\text{E}_{i}^{0(d)} - \text{E}_{i}^{0(d^{*})} - \theta^{0}\right) = 0}} - \nonumber \\
    &-\mathbb{E}\left(\frac{\tilde{Z}_{i}}{\tilde{P}_{i}}\left(\frac{D_{i}^{(d)}}{P_{i}^{(d)}}\Delta\text{E}_{i}^{(d)} - \frac{D_{i}^{(d^{*})}}{P_{i}^{(d^{*})}}\Delta\text{E}_{i}^{(d^{*})}\right)\right) = \nonumber \\
    &= \mathbb{E}\left(\Delta\text{E}_{i}^{(d)} - \Delta\text{E}_{i}^{(d^{*})}\right) - \nonumber \\
    &-\underbrace{\mathbb{E}\left(\frac{\tilde{P}_{i}^{0}}{\tilde{P}_{i}}\left(\frac{D_{i}^{(d)}}{P_{i}^{(d)}}\Delta\text{E}_{i}^{(d)} - \frac{D_{i}^{(d^{*})}}{P_{i}^{(d^{*})}}\Delta\text{E}_{i}^{(d^{*})}\right)\right)}_{\substack{\text{LIE conditional on }(D_i,X_i,W_i^{(Z)})\\ \text{since }\mathbb{E}\left(\tilde{Z}_i\mid D_i,X_i,W_i^{(Z)}\right)=\tilde{P}_i^0}} = \nonumber \\
    &= \underbrace{\mathbb{E}\left(\Delta\text{E}_{i}^{(d)}\left(1 - \frac{\tilde{P}_{i}^{0}}{\tilde{P}_{i}}\frac{D_{i}^{(d)}}{P_{i}^{(d)}}\right) - \Delta\text{E}_{i}^{(d^{*})}\left(1 - \frac{\tilde{P}_{i}^{0}}{\tilde{P}_{i}}\frac{D_{i}^{(d^{*})}}{P_{i}^{(d^{*})}}\right)\right)}_{\text{Collecting the terms}}.
\end{align}
Therefore, we obtain:
\begin{align}
    \mathbb{E}\left(\psi_{\text{ATE}}^{*}\left(O_i,\theta^0,\mu^0\right)\right)
    &= \mathbb{E}\left(\underbrace{\Delta\text{E}_{i}^{(d)}}_{\substack{=0\\ \text{Since }\mu=\mu^0}}\left(1 - \frac{D_{i}^{(d)}}{P_{i}^{0(d)}}\right) - \underbrace{\Delta\text{E}_{i}^{(d^{*})}}_{\substack{=0\\ \text{Since }\mu=\mu^0}}\left(1 - \frac{D_{i}^{(d^{*})}}{P_{i}^{0(d^{*})}}\right)\right) = 0.
\end{align}

Moreover, using the notation of Theorem~\ref{theorem:modified}, we obtain:
\begin{align}
    \lambda_N'' &= \sup_{\mu\in\mathcal{T}_n}\left|\mathbb{E}\left(\psi_{\text{ATE}}^{*}\left(O_i,\theta^0,\mu\right)\right)\right| = \nonumber \\
    &= \sup_{\mu\in\mathcal{T}_n}\left|\mathbb{E}\left(\Delta\text{E}_{i}^{(d)}\left(1 - \frac{\tilde{P}_{i}^{0}}{\tilde{P}_{i}}\frac{D_{i}^{(d)}}{P_{i}^{(d)}}\right) - \Delta\text{E}_{i}^{(d^{*})}\left(1 - \frac{\tilde{P}_{i}^{0}}{\tilde{P}_{i}}\frac{D_{i}^{(d^{*})}}{P_{i}^{(d^{*})}}\right)\right)\right| \leq \underbrace{\delta_n/\sqrt{n}}_{\text{Assumption~\ref{assumption:regularity2}}}.
\end{align}

\subsection{The bound for the estimator of ATT}

Consider the score without the derivative terms:
\begin{align}
     \psi_{\text{ATT}}^{*}\left(O_{i}, \theta, \mu\right) &= D_{i}^{(d)}\left(\text{E}_{i}^{(d)} - \text{E}_{i}^{(d^{*})}\right) - D_{i}^{(d)} \theta + \nonumber \\
        &+ \frac{\tilde{Z}_{i}}{\tilde{P}_{i}}D_{i}^{(d)}\left(Y_{i} - \text{E}_{i}^{(d)}\right) - \frac{\tilde{Z}_{i}}{\tilde{P}_{i}}\frac{P_{i}^{(d)}}{P_{i}^{(d^{*})}}D_{i}^{(d^{*})}\left(Y_{i} - \text{E}_{i}^{(d^{*})}\right).
\end{align}

Notice that:
\begin{align}
    \mathbb{E}\left(\psi_{\text{ATT}}^{*}\left(O_{i}, \theta^{0}, \mu\right)\right) &= \underbrace{\mathbb{E}\left(D_{i}^{(d)}\left(\left[\text{E}_{i}^{(d)} - \text{E}_{i}^{0(d)}\right] - \left[\text{E}_{i}^{(d^{*})} - \text{E}_{i}^{0(d^{*})}\right]\right)\right)}_{\text{Add and subtract }\text{E}_{i}^{0(d)}, \text{E}_{i}^{0(d^{*})}} + \nonumber \\ 
    &+\mathbb{E}\left(D_{i}^{(d)}\left(\text{E}_{i}^{0(d)} - \text{E}_{i}^{0(d^{*})} - \theta^{0}\right)\right) + \nonumber \\
    &+ \underbrace{\mathbb{E}\left(\frac{\tilde{Z}_{i}}{\tilde{P}_{i}}D_{i}^{(d)}\left(\text{E}_{i}^{0(d)} - \text{E}_{i}^{(d)}\right) - \frac{\tilde{Z}_{i}}{\tilde{P}_{i}}\frac{P_{i}^{(d)}}{P_{i}^{(d^{*})}}D_{i}^{(d^{*})}\left(\text{E}_{i}^{0(d^{*})} - \text{E}_{i}^{(d^{*})}\right)\right)}_{\substack{\text{LIE conditional on }(D_i,X_i,W_i^{(Z)},\tilde{Z}_i)\\ \text{using the conditional expectations of }Y_i\text{ for }\tilde{Z}_i=1}} = \nonumber \\
    &= \underbrace{\mathbb{E}\left(D_{i}^{(d)}\left(\Delta\text{E}_{i}^{(d)} - \Delta\text{E}_{i}^{(d^{*})}\right)\right)}_{\substack{\text{Since Theorem~\ref{theorem:ate} implies}\\ \mathbb{E}\left(D_{i}^{(d)}\left(\text{E}_{i}^{0(d)} - \text{E}_{i}^{0(d^{*})} - \theta^{0}\right)\right) = 0}} - \nonumber \\
    &- \mathbb{E}\left(\frac{\tilde{Z}_{i}}{\tilde{P}_{i}}D_{i}^{(d)}\Delta\text{E}_{i}^{(d)} - \frac{\tilde{Z}_{i}}{\tilde{P}_{i}}\frac{P_{i}^{(d)}}{P_{i}^{(d^{*})}}D_{i}^{(d^{*})}\Delta\text{E}_{i}^{(d^{*})}\right) = \nonumber \\
    &= \mathbb{E}\left(D_{i}^{(d)}\Delta\text{E}_{i}^{(d)} - D_{i}^{(d)}\Delta\text{E}_{i}^{(d^{*})}\right) - \nonumber \\
    &-\underbrace{\mathbb{E}\left(\frac{\tilde{P}_{i}^{0}}{\tilde{P}_{i}}D_{i}^{(d)}\Delta\text{E}_{i}^{(d)} - \frac{\tilde{P}_{i}^{0}}{\tilde{P}_{i}}\frac{P_{i}^{(d)}}{P_{i}^{(d^{*})}}D_{i}^{(d^{*})}\Delta\text{E}_{i}^{(d^{*})}\right)}_{\substack{\text{LIE conditional on }(D_i,X_i,W_i^{(Z)})\\ \text{since }\mathbb{E}\left(\tilde{Z}_i\mid D_i,X_i,W_i^{(Z)}\right)=\tilde{P}_i^0}} = \nonumber \\
    &= \underbrace{\mathbb{E}\left(\Delta\text{E}_{i}^{(d)}D_{i}^{(d)}\left(1 - \frac{\tilde{P}_{i}^{0}}{\tilde{P}_{i}}\right) + \Delta\text{E}_{i}^{(d^{*})}\left(\frac{\tilde{P}_{i}^{0}}{\tilde{P}_{i}}\frac{P_{i}^{(d)}}{P_{i}^{(d^{*})}}D_{i}^{(d^{*})} - D_{i}^{(d)}\right)\right)}_{\text{Collecting the terms}}.
\end{align}
Therefore, we obtain:
\begin{align}
    \mathbb{E}\left(\psi_{\text{ATT}}^{*}\left(O_i,\theta^0,\mu^0\right)\right)
    &= \mathbb{E}\left(\Delta\text{E}_{i}^{(d)}D_{i}^{(d)}\underbrace{\left(1 - \frac{\tilde{P}_{i}^{0}}{\tilde{P}_{i}^{0}}\right)}_{=0} + \underbrace{\Delta\text{E}_{i}^{(d^{*})}}_{\substack{=0\\ \text{Since }\mu=\mu^0}}\left(\frac{P_{i}^{0(d)}}{P_{i}^{0(d^{*})}}D_{i}^{(d^{*})} - D_{i}^{(d)}\right)\right) = 0.
\end{align}
Moreover, using the notation of Theorem~\ref{theorem:modified}, we obtain:
\begin{align}
    \lambda_N'' &= \sup_{\mu\in\mathcal{T}_n}\left|\mathbb{E}\left(\psi_{\text{ATT}}^{*}\left(O_i,\theta^0,\mu\right)\right)\right| = \nonumber \\
    &= \sup_{\mu\in\mathcal{T}_n}\left|\mathbb{E}\left(\Delta\text{E}_{i}^{(d)}D_{i}^{(d)}\left(1 - \frac{\tilde{P}_{i}^{0}}{\tilde{P}_{i}}\right) + \Delta\text{E}_{i}^{(d^{*})}\left(\frac{\tilde{P}_{i}^{0}}{\tilde{P}_{i}}\frac{P_{i}^{(d)}}{P_{i}^{(d^{*})}}D_{i}^{(d^{*})} - D_{i}^{(d)}\right)\right)\right| \leq \nonumber \\
    &\leq \underbrace{\delta_n/\sqrt{n}}_{\text{Assumption~\ref{assumption:regularity2}}}.
\end{align}

By combining the established bounds on $\lambda_N''$ with the bounds on $m_N$, $m_N'$, $r_N$, and $r_N'$ discussed above, we obtain the conclusions of Theorem~\ref{theorem:modified} for all four derivative-free scores. The arguments establishing asymptotic normality of the estimators averaged over the two nested cross-fitting branches and consistency of the corresponding variance estimators proceed as in the proof of Theorem~\ref{theorem:normality}.

\subsection{Alternative primitive conditions}

In this subsection, we show that the bounds on the expected scores imposed in Assumption~\ref{assumption:regularity2} may be replaced by two primitive conditions. First, suppose that:
\begin{align}\label{eq:primitive1}
    \sup_{\mu\in\mathcal{T}_{n}}\mathbb{E}\left(\left\|P_{i} - \mathbb{E}\left(P_{i}\mid X_{i},P_{i}^{0}\right)\right\|^{2}\right)^{1/2} &\leq \delta_{n}/\sqrt{n}.
\end{align}
This condition requires faster-than-parametric convergence only for the component of the estimation error in the generated control variables that cannot be explained by $X_i$ and $P_i^0$. In particular, it holds if $P_i$ is a measurable function of $X_i$ and $P_i^0$ for every $\mu\in\mathcal{T}_n$ or if the generated control variables converge at a faster-than-parametric rate in the $L^2$ norm uniformly over $\mu\in\mathcal{T}_n$.

Second, suppose that there exists a finite positive constant $L$ such that, for every $\overline{d}\in\{d, d^{*}\}$, $\mu\in\mathcal{T}_n$, $x\in\text{supp}(X_i)$, and $p_{1},p_{2}\in[0,1]^{2n_{Z}}$, the following inequality holds:
\begin{align}\label{eq:primitive21}
    \left|\mu_{Y}\left(\overline{d}, x, p_{2}\right) - \mu_{Y}\left(\overline{d}, x, p_{1}\right)\right| &\leq L\left\|p_{2} - p_{1}\right\|.
\end{align}

Below, we show that inequalities (\ref{eq:primitive1}) and (\ref{eq:primitive21}), along with the remaining conditions of Theorem~\ref{theorem:normality2}, imply the required bounds on $\lambda_N''$, possibly after rescaling $\delta_n$ by a sufficiently large constant.

\subsubsection{Preliminary bounds related to ATES}

First, consider the following decompositions:
\begin{align}
    \Delta E_{i}^{(\overline{d})} &= \Delta E_{1i}^{(\overline{d})} + \Delta E_{2i}^{(\overline{d})}, \\
    \Delta P_{i}^{(\overline{d})} &= \Delta P_{1i}^{(\overline{d})} + \Delta P_{2i}^{(\overline{d})},
\end{align}
where:
\begin{align}
    \Delta E_{1i}^{(\overline{d})} &= \mu_{Y}\left(\overline{d}, X_{i}, \mathbb{E}\left(P_i\mid X_i,P_i^0\right)\right) - \mu_{Y}^{0}\left(\overline{d}, X_{i}, P_{i}^{0}\right), \\
    \Delta E_{2i}^{(\overline{d})} &= \mu_{Y}\left(\overline{d}, X_{i}, P_{i}\right) - \mu_{Y}\left(\overline{d}, X_{i}, \mathbb{E}\left(P_i\mid X_i,P_i^0\right)\right), \\
    \Delta P_{1i}^{(\overline{d})} &= 1 - \frac{D_{i}^{(\overline{d})}}{P_{i}^{0(\overline{d})}}, \\
    \Delta P_{2i}^{(\overline{d})} &= \frac{D_{i}^{(\overline{d})}\left(P_{i}^{(\overline{d})} - P_{i}^{0(\overline{d})}\right)}{P_{i}^{(\overline{d})}P_{i}^{0(\overline{d})}}.
\end{align}
Notice that, for every fixed $\mu\in\mathcal{T}_n$, $\Delta E_{1i}^{(\overline{d})}$ is a function of $X_i$ and $P_i^0$. Moreover, since $P_i\in[0,1]^{2n_Z}$, we have $\mathbb{E}\left(P_i\mid X_i,P_i^0\right)\in[0,1]^{2n_Z}$ almost surely.

We start by establishing several useful bounds for the elements of the decomposition. The first useful bound is as follows:
\begin{align}\label{eq:uprimitive2}
    \mathbb{E}\left(\left\|\Delta E_{2i}^{(\overline{d})}\right\|^{2}\right)^{1/2} &\leq \underbrace{L\mathbb{E}\left(\left\|P_{i} - \mathbb{E}\left(P_i\mid X_i,P_i^0\right)\right\|^{2}\right)^{1/2}}_{\substack{\text{Lipschitz continuity}\\ \text{by (\ref{eq:primitive21})}}} \leq \underbrace{L \delta_{n}/\sqrt{n}}_{\text{By (\ref{eq:primitive1})}}.
\end{align}

The second useful bound is as follows:
\begin{align}\label{eq:uprimitive1}
    \mathbb{E}\left(\left\|\Delta E_{1i}^{(\overline{d})}\right\|^{2}\right)^{1/2} &\leq \underbrace{\mathbb{E}\left(\left\|\Delta E_{i}^{(\overline{d})}\right\|^{2}\right)^{1/2} + \mathbb{E}\left(\left\|\Delta E_{2i}^{(\overline{d})}\right\|^{2}\right)^{1/2}}_{\text{Minkowski's inequality}}\leq \nonumber \\ 
    &\leq \underbrace{\delta_{n}n^{-1/4}}_{\substack{\text{Part 5 of Assumption}\\ \ref{assumption:regularity}}} + \underbrace{L \delta_{n}/\sqrt{n}}_{\text{By (\ref{eq:uprimitive2})}} \leq \underbrace{\left(1 + L\right)\delta_{n}n^{-1/4}}_{\text{Since }n^{-1/2}\leq n^{-1/4}}.
\end{align}

The third useful bound follows by considering cases $D_{i}^{(\overline{d})} = 1$, $D_{i}^{(\overline{d})} = 0$ and using the inequality $P_{i}^{0(\overline{d})}\geq c$ from Part 4 of Assumption~\ref{assumption:regularity}:
\begin{align}
    \left|\Delta P_{1i}^{(\overline{d})}\right|\leq \frac{1}{c}. \label{eq:uprimitive3}
\end{align}

Notice that Lemma~\ref{lemma:independence} implies $D_i\perp\tilde{Z}_i\mid(X_i,P_i^0)$. Moreover, Part 4 of Assumption~\ref{assumption:regularity} implies that $\mathbb{P}(\tilde{Z}_i=1\mid X_i,P_i^0)=\tilde{P}_i^0\geq c$ almost surely. Therefore, we have
\begin{align}\label{eq:fourth_supplementary}
    \mathbb{E}\left(D_i^{(\overline{d})}\mid X_i,P_i^0\right)
    = \mathbb{P}\left(D_i=\overline{d}\mid X_i,P_i^0,\tilde{Z}_i=1\right)
    = P_i^{0(\overline{d})}
\end{align}
almost surely.

The fourth useful bound is as follows:
\begin{align}
    \mathbb{E}\left(\Delta P_{1i}^{(\overline{d})}\mid X_{i}, P_{i}^{0}\right) = \underbrace{1 - \frac{1}{P_{i}^{0(\overline{d})}}\mathbb{E}\left(D_{i}^{(\overline{d})}\mid X_{i}, P_{i}^{0}\right)}_{\substack{\text{Since }P_{i}^{0(\overline{d})}\text{ is a function}\\ \text{of }X_{i}, P_{i}^{0}}} = \underbrace{1 - \frac{1}{P_{i}^{0(\overline{d})}}P_{i}^{0(\overline{d})}}_{\text{Equality (\ref{eq:fourth_supplementary})}} = 1 - 1 = 0. \label{eq:uprimitive4}
\end{align}

The fifth useful bound is as follows:
\begin{align}\label{eq:uprimitive5}
    \mathbb{E}\left(\left\|\Delta P_{2i}^{(\overline{d})}\right\|^{2}\right)^{1/2} &\leq \underbrace{\mathbb{E}\left(\left\|\frac{\left(P_{i}^{(\overline{d})} - P_{i}^{0(\overline{d})}\right)}{P_{i}^{(\overline{d})}P_{i}^{0(\overline{d})}}\right\|^{2}\right)^{1/2}}_{\text{Since }D_{i}^{(\overline{d})}\in\{0, 1\}} \leq \nonumber \\ 
    &\leq \underbrace{\frac{1}{c^{2}}\mathbb{E}\left(\left\|P_{i}^{(\overline{d})} - P_{i}^{0(\overline{d})}\right\|^{2}\right)^{1/2}}_{\text{Part 4 of Assumption~\ref{assumption:regularity}}} \leq \nonumber \\
    &\leq \underbrace{\frac{1}{c^{2}}\delta_{n}n^{-1/4}}_{\text{Part 5 of Assumption~\ref{assumption:regularity}}}.
\end{align}

Notice that relations~(\ref{eq:uprimitive1})--(\ref{eq:uprimitive5}) hold uniformly over $\mu\in\mathcal{T}_n$.

\subsubsection{The bound for ATES}

Consider the following decomposition:
\begin{align}
    \tilde{P}_{i}^{0}\Delta E_{i}^{(\overline{d})}\Delta P_{i}^{(\overline{d})} = \tilde{P}_{i}^{0}\left(\Delta E_{1i}^{(\overline{d})}\Delta P_{1i}^{(\overline{d})} + \Delta E_{1i}^{(\overline{d})}\Delta P_{2i}^{(\overline{d})} + \Delta E_{2i}^{(\overline{d})}\Delta P_{1i}^{(\overline{d})} + \Delta E_{2i}^{(\overline{d})}\Delta P_{2i}^{(\overline{d})}\right).
\end{align}

Below, we bound the expectations of each term of the last equality. First, we obtain:
\begin{align}
    \mathbb{E}\left(\tilde{P}_{i}^{0}\Delta E_{1i}^{(\overline{d})}\Delta P_{1i}^{(\overline{d})}\right) &= \underbrace{\mathbb{E}\left(\mathbb{P}\left(\tilde{Z}_{i} = 1\mid X_{i}, P_{i}^{0}\right)\Delta E_{1i}^{(\overline{d})}\Delta P_{1i}^{(\overline{d})}\right)}_{\text{Lemma~\ref{lemma:useful}}} = \nonumber \\
    &= \underbrace{\mathbb{E}\left(\mathbb{P}\left(\tilde{Z}_{i} = 1\mid X_{i}, P_{i}^{0}\right)\Delta E_{1i}^{(\overline{d})}\mathbb{E}\left(\Delta P_{1i}^{(\overline{d})}\mid X_{i}, P_{i}^{0}\right)\right)}_{\substack{\text{LIE conditional on }(X_i,P_i^0)\\ \text{since }\Delta E_{1i}^{(\overline{d})}\text{ is a function of }X_i,P_i^0}} = \nonumber \\
    &= \underbrace{\mathbb{E}\left(\mathbb{P}\left(\tilde{Z}_{i} = 1\mid X_{i}, P_{i}^{0}\right)\Delta E_{1i}^{(\overline{d})}\times 0\right)}_{\text{By (\ref{eq:uprimitive4})}} = 0.
\end{align}

Second, we obtain:
\begin{align}
    \left|\mathbb{E}\left(\tilde{P}_{i}^{0}\Delta E_{1i}^{(\overline{d})}\Delta P_{2i}^{(\overline{d})}\right)\right| &\leq \mathbb{E}\left(\left|\tilde{P}_{i}^{0}\Delta E_{1i}^{(\overline{d})}\Delta P_{2i}^{(\overline{d})}\right|\right) \leq \nonumber \\
    &\leq \underbrace{\mathbb{E}\left(\left|\Delta E_{1i}^{(\overline{d})}\Delta P_{2i}^{(\overline{d})}\right|\right)}_{\text{Since }\left|\tilde{P}_{i}^{0}\right|\leq 1} \leq \nonumber \\
    &\leq \underbrace{\mathbb{E}\left(\left\|\Delta E_{1i}^{(\overline{d})}\right\|^{2}\right)^{1/2}\mathbb{E}\left(\left\|\Delta P_{2i}^{(\overline{d})}\right\|^{2}\right)^{1/2}}_{\text{Cauchy-Schwarz inequality}} \leq \nonumber \\
    &\leq \underbrace{\left(1 + L\right)\delta_{n}n^{-1/4}}_{\text{By (\ref{eq:uprimitive1})}}\times \underbrace{\frac{1}{c^{2}}\delta_{n}n^{-1/4}}_{\text{By (\ref{eq:uprimitive5})}} = \nonumber \\
    &= \frac{1}{c^{2}}\left(1 + L\right)\delta_{n}^{2}/\sqrt{n}.
\end{align}

Third, we obtain:
\begin{align}
    \left|\mathbb{E}\left(\tilde{P}_{i}^{0}\Delta E_{2i}^{(\overline{d})}\Delta P_{1i}^{(\overline{d})}\right)\right| &\leq \underbrace{\mathbb{E}\left(\left|\Delta E_{2i}^{(\overline{d})}\Delta P_{1i}^{(\overline{d})}\right|\right)}_{\text{Since }\left|\tilde{P}_{i}^{0}\right|\leq 1} \leq \nonumber \\
    &\leq \underbrace{\frac{1}{c}\mathbb{E}\left(\left|\Delta E_{2i}^{(\overline{d})}\right|\right)}_{\text{By (\ref{eq:uprimitive3})}} \leq \nonumber \\
    &\leq \underbrace{\frac{1}{c}\mathbb{E}\left(\left\|\Delta E_{2i}^{(\overline{d})}\right\|^{2}\right)^{1/2}}_{\text{Lyapunov's inequality}} \leq \underbrace{\frac{L}{c}\delta_{n}/\sqrt{n}}_{\text{By (\ref{eq:uprimitive2})}}.
\end{align}

Fourth, we obtain:
\begin{align}
    \left|\mathbb{E}\left(\tilde{P}_{i}^{0}\Delta E_{2i}^{(\overline{d})}\Delta P_{2i}^{(\overline{d})}\right)\right| &\leq \underbrace{\mathbb{E}\left(\left|\Delta E_{2i}^{(\overline{d})}\Delta P_{2i}^{(\overline{d})}\right|\right)}_{\text{Since }\left|\tilde{P}_{i}^{0}\right|\leq 1} \leq \nonumber \\
    &\leq \underbrace{\mathbb{E}\left(\left\|\Delta E_{2i}^{(\overline{d})}\right\|^{2}\right)^{1/2}\mathbb{E}\left(\left\|\Delta P_{2i}^{(\overline{d})}\right\|^{2}\right)^{1/2}}_{\text{Cauchy-Schwarz inequality}} \leq \nonumber \\
    &\leq \underbrace{L\delta_{n}/\sqrt{n}}_{\text{By (\ref{eq:uprimitive2})}}\times \underbrace{\frac{1}{c^{2}}\delta_{n}n^{-1/4}}_{\text{By (\ref{eq:uprimitive5})}} = \nonumber \\
    &= \frac{L}{c^{2}}\delta_{n}^{2}n^{-3/4}\leq \underbrace{\frac{L}{c^{2}}\delta_{n}^{2}/\sqrt{n}}_{\text{Since }n^{-3/4}\leq n^{-1/2}}.
\end{align}

By using the triangle inequality and the established bounds, we obtain:
\begin{align}\label{eq:primitive:ATES:component}
    \left|\mathbb{E}\left(\tilde{P}_{i}^{0}\Delta E_{i}^{(\overline{d})}\Delta P_{i}^{(\overline{d})}\right)\right| &\leq 0 + \frac{1}{c^{2}}\left(1 + L\right)\delta_{n}^{2}/\sqrt{n} + \frac{L}{c}\delta_{n}/\sqrt{n} + \frac{L}{c^{2}}\delta_{n}^{2}/\sqrt{n} = \nonumber \\
    &= \left(\frac{1 + 2L}{c^{2}}\delta_{n}^{2} + \frac{L}{c}\delta_{n}\right)/\sqrt{n}.
\end{align}

Finally, since the established bounds hold uniformly over $\mu\in\mathcal{T}_{n}$, we obtain:
\begin{align}
    &\sup_{\mu\in \mathcal{T}_{n}}\left|\mathbb{E}\left(\tilde{P}_{i}^{0}\left(\Delta\text{E}_{i}^{(d)}\Delta P_{i}^{(d)} - \Delta\text{E}_{i}^{(d^{*})}\Delta P_{i}^{(d^{*})}\right)\right)\right| \leq \nonumber \\
    &\leq \underbrace{\left[\begin{aligned}
        &\sup_{\mu\in \mathcal{T}_{n}}\left|\mathbb{E}\left(\tilde{P}_{i}^{0}\Delta E_{i}^{(d)}\Delta P_{i}^{(d)}\right)\right| + \\
        &+\sup_{\mu\in \mathcal{T}_{n}}\left|\mathbb{E}\left(\tilde{P}_{i}^{0}\Delta E_{i}^{(d^{*})}\Delta P_{i}^{(d^{*})}\right)\right|
    \end{aligned}\right]}_{\text{Triangle inequality}} \leq \nonumber \\
    &\leq \underbrace{2\left(\frac{1 + 2L}{c^{2}}\delta_{n}^{2} + \frac{L}{c}\delta_{n}\right)}_{\substack{\text{By (\ref{eq:primitive:ATES:component})}\\ \xrightarrow{n\to\infty}0}}\frac{1}{\sqrt{n}}.
\end{align}

\subsubsection{Preliminary bounds related to ATTS}

First, consider the following decomposition:
\begin{align}
    P_{i}^{(d)}\left(\Delta P_{i}^{(d)} - \Delta P_{i}^{(d^{*})}\right) &= \Delta Q_{1i} + \Delta Q_{2i},
\end{align}
where:
\begin{align}
    \Delta Q_{1i} &= \frac{P_{i}^{0(d)}}{P_{i}^{0(d^{*})}}D_{i}^{(d^{*})} - D_{i}^{(d)} = P_{i}^{0(d)}\left(\Delta P_{1i}^{(d)} - \Delta P_{1i}^{(d^{*})}\right), \\
    \Delta Q_{2i} &= \left(\frac{P_{i}^{(d)}}{P_{i}^{(d^{*})}} - \frac{P_{i}^{0(d)}}{P_{i}^{0(d^{*})}}\right)D_{i}^{(d^{*})}.
\end{align}

We start by establishing several useful bounds for the elements of the decomposition. The first useful bound follows by considering cases $D_{i} = d$, $D_{i} = d^{*}$, $D_{i}\notin \{d, d^{*}\}$ and using the inequality $P_{i}^{0(\overline{d})}\geq c$ from Part 4 of Assumption~\ref{assumption:regularity}:
\begin{align}
    \left|\Delta Q_{1i}\right|\leq \frac{1}{c}. \label{eq:uprimitive1B}
\end{align}

The second useful inequality is as follows:
\begin{align}
    \mathbb{E}\left(\Delta Q_{1i}\mid X_{i}, P_{i}^{0}\right) 
    &= \underbrace{P_{i}^{0(d)}\mathbb{E}\left(\Delta P_{1i}^{(d)} - \Delta P_{1i}^{(d^{*})}\mid X_{i}, P_{i}^{0}\right)}_{\substack{\text{Since }P_{i}^{0(d)}\text{ is a function}\\ \text{of }X_i,P_i^0}} = \nonumber \\
    &= P_{i}^{0(d)}\left(\underbrace{\mathbb{E}\left(\Delta P_{1i}^{(d)}\mid X_{i}, P_{i}^{0}\right)}_{\substack{=0\\ \text{By (\ref{eq:uprimitive4})}}} - \underbrace{\mathbb{E}\left(\Delta P_{1i}^{(d^{*})}\mid X_{i}, P_{i}^{0}\right)}_{\substack{=0\\ \text{By (\ref{eq:uprimitive4})}}}\right) = 0. \label{eq:uprimitive2B}
\end{align}

The third useful inequality is as follows:
\begin{align}
    \mathbb{E}\left(\left\|\Delta Q_{2i}\right\|^{2}\right)^{1/2} &\leq \underbrace{\mathbb{E}\left(\left\|\frac{P_{i}^{(d)}}{P_{i}^{(d^{*})}} - \frac{P_{i}^{0(d)}}{P_{i}^{0(d^{*})}}\right\|^{2}\right)^{1/2}}_{\text{Since }D_{i}^{(d^{*})}\in\{0, 1\}}\leq \nonumber \\
    &\leq \underbrace{\frac{1}{c^{2}}\mathbb{E}\left(\left\| P_{i}^{(d)}P_{i}^{0(d^{*})} - P_{i}^{0(d)}P_{i}^{(d^{*})} \right\|^{2}\right)^{1/2}}_{\substack{\text{Since }P_{i}^{(d^{*})}P_{i}^{0(d^{*})}\geq c^{2}\\ \text{by Part 4 of Assumption~\ref{assumption:regularity}}}} = \nonumber \\
    &= \frac{1}{c^{2}}\mathbb{E}\left(\left\| P_{i}^{(d)}\left(P_{i}^{0(d^{*})} - P_{i}^{(d^{*})}\right) + P_{i}^{(d^{*})}\left(P_{i}^{(d)} - P_{i}^{0(d)}\right) \right\|^{2}\right)^{1/2} \leq \nonumber \\
    &\leq \underbrace{\frac{1}{c^{2}}\mathbb{E}\left(\left\|\left|P_{i}^{0(d^{*})} - P_{i}^{(d^{*})}\right| + \left|P_{i}^{(d)} - P_{i}^{0(d)}\right|\right\|^{2}\right)^{1/2}}_{\substack{\text{Triangle inequality}\\ \text{and }P_{i}^{(d)},P_{i}^{(d^{*})}\in[0,1]}} \leq \nonumber \\
    &\leq \underbrace{\frac{1}{c^{2}}\mathbb{E}\left(\left\|P_{i}^{(d^{*})} - P_{i}^{0(d^{*})}\right\|^{2}\right)^{1/2} + \frac{1}{c^{2}}\mathbb{E}\left(\left\|P_{i}^{(d)} - P_{i}^{0(d)}\right\|^{2}\right)^{1/2}}_{\text{Minkowski's inequality}} \leq \nonumber \\
    &\leq \underbrace{\frac{1}{c^{2}}\delta_{n}n^{-1/4} + \frac{1}{c^{2}}\delta_{n}n^{-1/4}}_{\substack{\text{Part 5 of Assumption}\\ \ref{assumption:regularity}}} = \frac{2}{c^{2}}\delta_{n}n^{-1/4}. \label{eq:uprimitive3B}
\end{align}

Notice that inequalities (\ref{eq:uprimitive1B})--(\ref{eq:uprimitive3B}) hold uniformly over $\mu\in\mathcal{T}_n$.

\subsubsection{The bound for ATTS}

Consider the following decomposition:
\begin{align}
    \tilde{P}_{i}^{0}\Delta\text{E}_{i}^{(d^{*})}P_{i}^{(d)}\left(\Delta P_{i}^{(d)} - \Delta P_{i}^{(d^{*})}\right) &= \tilde{P}_{i}^{0}\Delta\text{E}_{i}^{(d^{*})}\left(\Delta Q_{1i} + \Delta Q_{2i}\right) = \nonumber \\
    &= \tilde{P}_{i}^{0}\Delta\text{E}_{1i}^{(d^{*})}\Delta Q_{1i} + \tilde{P}_{i}^{0}\Delta\text{E}_{1i}^{(d^{*})}\Delta Q_{2i} + \nonumber \\
    &+ \tilde{P}_{i}^{0}\Delta\text{E}_{2i}^{(d^{*})}\Delta Q_{1i} + \tilde{P}_{i}^{0}\Delta\text{E}_{2i}^{(d^{*})}\Delta Q_{2i}.
\end{align}

Below, we bound the expectations of each term of the last equality. First, we obtain:
\begin{align}\label{eq:primitive:ATTS:term1}
    \mathbb{E}\left(\tilde{P}_{i}^{0}\Delta\text{E}_{1i}^{(d^{*})}\Delta Q_{1i}\right)
    &= \underbrace{\mathbb{E}\left(\mathbb{P}\left(\tilde{Z}_{i}=1\mid X_i,P_i^0\right)\Delta\text{E}_{1i}^{(d^{*})}\Delta Q_{1i}\right)}_{\text{Lemma~\ref{lemma:useful}}} = \nonumber \\
    &= \underbrace{\mathbb{E}\left(\mathbb{P}\left(\tilde{Z}_{i}=1\mid X_i,P_i^0\right)\Delta\text{E}_{1i}^{(d^{*})}\mathbb{E}\left(\Delta Q_{1i}\mid X_i,P_i^0\right)\right)}_{\substack{\text{LIE conditional on }(X_i,P_i^0)\\ \text{since }\Delta\text{E}_{1i}^{(d^{*})}\text{ is a function of }X_i,P_i^0}} = \nonumber \\
    &= \mathbb{E}\left(\mathbb{P}\left(\tilde{Z}_{i}=1\mid X_i,P_i^0\right)\Delta\text{E}_{1i}^{(d^{*})}\times\underbrace{0}_{\text{By (\ref{eq:uprimitive2B})}}\right) = 0.
\end{align}

Second, we obtain:
\begin{align}\label{eq:primitive:ATTS:term2}
    \left|\mathbb{E}\left(\tilde{P}_{i}^{0}\Delta\text{E}_{1i}^{(d^{*})}\Delta Q_{2i}\right)\right|
    &\leq \underbrace{\mathbb{E}\left(\left|\Delta\text{E}_{1i}^{(d^{*})}\Delta Q_{2i}\right|\right)}_{\text{Since }\left|\tilde{P}_{i}^{0}\right|\leq1} \leq \nonumber \\
    &\leq \underbrace{\mathbb{E}\left(\left\|\Delta\text{E}_{1i}^{(d^{*})}\right\|^{2}\right)^{1/2}\mathbb{E}\left(\left\|\Delta Q_{2i}\right\|^{2}\right)^{1/2}}_{\text{Cauchy-Schwarz inequality}} \leq \nonumber \\
    &\leq \underbrace{\left(1+L\right)\delta_n n^{-1/4}}_{\text{By (\ref{eq:uprimitive1})}}\times\underbrace{\frac{2}{c^{2}}\delta_n n^{-1/4}}_{\text{By (\ref{eq:uprimitive3B})}} = \nonumber \\
    &= \frac{2}{c^{2}}\left(1+L\right)\delta_n^2/\sqrt{n}.
\end{align}

Third, we obtain:
\begin{align}\label{eq:primitive:ATTS:term3}
    \left|\mathbb{E}\left(\tilde{P}_{i}^{0}\Delta\text{E}_{2i}^{(d^{*})}\Delta Q_{1i}\right)\right|
    &\leq \underbrace{\mathbb{E}\left(\left|\Delta\text{E}_{2i}^{(d^{*})}\Delta Q_{1i}\right|\right)}_{\text{Since }\left|\tilde{P}_{i}^{0}\right|\leq1} \leq \nonumber \\
    &\leq \underbrace{\frac{1}{c}\mathbb{E}\left(\left|\Delta\text{E}_{2i}^{(d^{*})}\right|\right)}_{\text{By (\ref{eq:uprimitive1B})}} \leq \nonumber \\
    &\leq \underbrace{\frac{1}{c}\mathbb{E}\left(\left\|\Delta\text{E}_{2i}^{(d^{*})}\right\|^{2}\right)^{1/2}}_{\text{Lyapunov's inequality}} \leq \underbrace{\frac{L}{c}\delta_n/\sqrt{n}}_{\text{By (\ref{eq:uprimitive2})}}.
\end{align}

Fourth, we obtain:
\begin{align}\label{eq:primitive:ATTS:term4}
    \left|\mathbb{E}\left(\tilde{P}_{i}^{0}\Delta\text{E}_{2i}^{(d^{*})}\Delta Q_{2i}\right)\right|
    &\leq \underbrace{\mathbb{E}\left(\left|\Delta\text{E}_{2i}^{(d^{*})}\Delta Q_{2i}\right|\right)}_{\text{Since }\left|\tilde{P}_{i}^{0}\right|\leq1} \leq \nonumber \\
    &\leq \underbrace{\mathbb{E}\left(\left\|\Delta\text{E}_{2i}^{(d^{*})}\right\|^{2}\right)^{1/2}\mathbb{E}\left(\left\|\Delta Q_{2i}\right\|^{2}\right)^{1/2}}_{\text{Cauchy-Schwarz inequality}} \leq \nonumber \\
    &\leq \underbrace{L\delta_n/\sqrt{n}}_{\text{By (\ref{eq:uprimitive2})}}\times\underbrace{\frac{2}{c^{2}}\delta_n n^{-1/4}}_{\text{By (\ref{eq:uprimitive3B})}} = \nonumber \\
    &= \frac{2L}{c^{2}}\delta_n^2 n^{-3/4} \leq \underbrace{\frac{2L}{c^{2}}\delta_n^2/\sqrt{n}}_{\text{Since }n^{-3/4}\leq n^{-1/2}}.
\end{align}

Finally, since the established bounds hold uniformly over $\mu\in\mathcal{T}_{n}$, we obtain:
\begin{align}
    &\sup_{\mu\in \mathcal{T}_{n}}\left|\mathbb{E}\left(\tilde{P}_{i}^{0}\Delta\text{E}_{i}^{(d^{*})}P_{i}^{(d)}\left(\Delta P_{i}^{(d)} - \Delta P_{i}^{(d^{*})}\right)\right)\right|\leq \nonumber \\
    &\leq \underbrace{0}_{\text{By (\ref{eq:primitive:ATTS:term1})}} + \underbrace{\frac{2}{c^{2}}\left(1+L\right)\delta_n^2/\sqrt{n}}_{\text{By (\ref{eq:primitive:ATTS:term2})}} + \underbrace{\frac{L}{c}\delta_n/\sqrt{n}}_{\text{By (\ref{eq:primitive:ATTS:term3})}} + \underbrace{\frac{2L}{c^{2}}\delta_n^2/\sqrt{n}}_{\text{By (\ref{eq:primitive:ATTS:term4})}} = \nonumber \\
    &= \underbrace{\left(\frac{L}{c}\delta_n+\frac{2(1+2L)}{c^{2}}\delta_n^2\right)}_{\xrightarrow{n\to\infty}0}\frac{1}{\sqrt{n}}.
\end{align}

\subsubsection{Preliminary bound related to ATE}

First, consider the following decomposition:
\begin{align}
    \Delta\text{E}_{i}^{(\overline{d})}\left(1 - \frac{\tilde{P}_{i}^{0}}{\tilde{P}_{i}}\frac{D_{i}^{(\overline{d})}}{P_{i}^{(\overline{d})}}\right) = \Delta\text{E}_{i}^{(\overline{d})}\Delta R_{i}^{(\overline{d})},
\end{align}
where:
\begin{align}
    \Delta R_{i}^{(\overline{d})} &= 1 - \frac{\tilde{P}_{i}^{0}}{\tilde{P}_{i}}\frac{D_{i}^{(\overline{d})}}{P_{i}^{(\overline{d})}} = \Delta P_{1i}^{(\overline{d})} + \Delta R_{2i}^{(\overline{d})}, \\
    \Delta R_{2i}^{(\overline{d})} &= D_{i}^{(\overline{d})}\left(\frac{1}{P_{i}^{0(\overline{d})}} - \frac{\tilde{P}_{i}^{0}}{\tilde{P}_{i}P_{i}^{(\overline{d})}}\right).
\end{align}

We start by establishing the following useful bound:
\begin{align}
    \mathbb{E}\left(\left\|\Delta R_{2i}^{(\overline{d})}\right\|^{2}\right)^{1/2} &= \mathbb{E}\left(\left\|D_{i}^{(\overline{d})}\left(\frac{1}{P_{i}^{0(\overline{d})}} - \frac{1}{P_{i}^{(\overline{d})}} + \frac{1}{P_{i}^{(\overline{d})}} - \frac{\tilde{P}_{i}^{0}}{\tilde{P}_{i}P_{i}^{(\overline{d})}}\right)\right\|^{2}\right)^{1/2} = \nonumber \\
    &= \mathbb{E}\left(\left\|\underbrace{\frac{D_{i}^{(\overline{d})}\left(P_{i}^{(\overline{d})}-P_{i}^{0(\overline{d})}\right)}{P_{i}^{(\overline{d})}P_{i}^{0(\overline{d})}}}_{\Delta P_{2i}^{(\overline{d})}} + \frac{D_{i}^{(\overline{d})}\left(\tilde{P}_{i}-\tilde{P}_{i}^{0}\right)}{\tilde{P}_{i}P_{i}^{(\overline{d})}}\right\|^{2}\right)^{1/2} \leq \nonumber \\
    &\leq \underbrace{\mathbb{E}\left(\left\|\Delta P_{2i}^{(\overline{d})}\right\|^{2}\right)^{1/2} + \mathbb{E}\left(\left\|\frac{D_{i}^{(\overline{d})}\left(\tilde{P}_{i}-\tilde{P}_{i}^{0}\right)}{\tilde{P}_{i}P_{i}^{(\overline{d})}}\right\|^{2}\right)^{1/2}}_{\text{Minkowski's inequality}} \leq \nonumber \\
    &\leq \mathbb{E}\left(\left\|\Delta P_{2i}^{(\overline{d})}\right\|^{2}\right)^{1/2} + \underbrace{\frac{1}{c^{2}}\mathbb{E}\left(\left\|\tilde{P}_{i}-\tilde{P}_{i}^{0}\right\|^{2}\right)^{1/2}}_{\substack{\text{Since }D_{i}^{(\overline{d})}\in\{0,1\}\\ \text{and by Part 4 of Assumption~\ref{assumption:regularity}}}} \leq \nonumber \\
    &\leq \underbrace{\frac{1}{c^{2}}\delta_{n}n^{-1/4}}_{\text{By (\ref{eq:uprimitive5})}} + \underbrace{\frac{1}{c^{2}}\delta_{n}n^{-1/4}}_{\substack{\text{Part 5 of Assumption}\\ \ref{assumption:regularity}}} = \frac{2}{c^{2}}\delta_{n}n^{-1/4}. \label{eq:uprimitive1C}
\end{align}

Notice that inequality (\ref{eq:uprimitive1C}) holds uniformly over $\mu\in\mathcal{T}_n$.

\subsubsection{The bound for ATE}

Consider the following decomposition:
\begin{align}
    \Delta\text{E}_{i}^{(\overline{d})}\left(1 - \frac{\tilde{P}_{i}^{0}}{\tilde{P}_{i}}\frac{D_{i}^{(\overline{d})}}{P_{i}^{(\overline{d})}}\right) &= \Delta\text{E}_{i}^{(\overline{d})}\left(\Delta P_{1i}^{(\overline{d})} + \Delta R_{2i}^{(\overline{d})}\right) = \nonumber \\
    &= \Delta\text{E}_{1i}^{(\overline{d})}\Delta P_{1i}^{(\overline{d})} + \Delta\text{E}_{1i}^{(\overline{d})}\Delta R_{2i}^{(\overline{d})} + \nonumber \\
    &+ \Delta\text{E}_{2i}^{(\overline{d})}\Delta P_{1i}^{(\overline{d})} + \Delta\text{E}_{2i}^{(\overline{d})}\Delta R_{2i}^{(\overline{d})}.
\end{align}

Below, we bound the expectations of each term of the last equality. First, we obtain:
\begin{align}\label{eq:primitive:ATE:term1}
    \mathbb{E}\left(\Delta\text{E}_{1i}^{(\overline{d})}\Delta P_{1i}^{(\overline{d})}\right)
    &= \underbrace{\mathbb{E}\left(\Delta\text{E}_{1i}^{(\overline{d})}\mathbb{E}\left(\Delta P_{1i}^{(\overline{d})}\mid X_i,P_i^0\right)\right)}_{\substack{\text{LIE conditional on }(X_i,P_i^0)\\ \text{since }\Delta\text{E}_{1i}^{(\overline{d})}\text{ is a function of }X_i,P_i^0}} = \nonumber \\
    &= \mathbb{E}\left(\Delta\text{E}_{1i}^{(\overline{d})}\times\underbrace{0}_{\text{By (\ref{eq:uprimitive4})}}\right) = 0.
\end{align}

Second, we obtain:
\begin{align}\label{eq:primitive:ATE:term2}
    \left|\mathbb{E}\left(\Delta\text{E}_{1i}^{(\overline{d})}\Delta R_{2i}^{(\overline{d})}\right)\right|
    &\leq \underbrace{\mathbb{E}\left(\left\|\Delta\text{E}_{1i}^{(\overline{d})}\right\|^{2}\right)^{1/2}\mathbb{E}\left(\left\|\Delta R_{2i}^{(\overline{d})}\right\|^{2}\right)^{1/2}}_{\text{Cauchy-Schwarz inequality}} \leq \nonumber \\
    &\leq \underbrace{\left(1+L\right)\delta_n n^{-1/4}}_{\text{By (\ref{eq:uprimitive1})}}\times\underbrace{\frac{2}{c^{2}}\delta_n n^{-1/4}}_{\text{By (\ref{eq:uprimitive1C})}} = \nonumber \\
    &= \frac{2}{c^{2}}\left(1+L\right)\delta_n^2/\sqrt{n}.
\end{align}

Third, we obtain:
\begin{align}\label{eq:primitive:ATE:term3}
    \left|\mathbb{E}\left(\Delta\text{E}_{2i}^{(\overline{d})}\Delta P_{1i}^{(\overline{d})}\right)\right|
    &\leq \mathbb{E}\left(\left|\Delta\text{E}_{2i}^{(\overline{d})}\Delta P_{1i}^{(\overline{d})}\right|\right) \leq \nonumber \\
    &\leq \underbrace{\frac{1}{c}\mathbb{E}\left(\left|\Delta\text{E}_{2i}^{(\overline{d})}\right|\right)}_{\text{By (\ref{eq:uprimitive3})}} \leq \nonumber \\
    &\leq \underbrace{\frac{1}{c}\mathbb{E}\left(\left\|\Delta\text{E}_{2i}^{(\overline{d})}\right\|^{2}\right)^{1/2}}_{\text{Lyapunov's inequality}} \leq \underbrace{\frac{L}{c}\delta_n/\sqrt{n}}_{\text{By (\ref{eq:uprimitive2})}}.
\end{align}

Fourth, we obtain:
\begin{align}\label{eq:primitive:ATE:term4}
    \left|\mathbb{E}\left(\Delta\text{E}_{2i}^{(\overline{d})}\Delta R_{2i}^{(\overline{d})}\right)\right|
    &\leq \underbrace{\mathbb{E}\left(\left\|\Delta\text{E}_{2i}^{(\overline{d})}\right\|^{2}\right)^{1/2}\mathbb{E}\left(\left\|\Delta R_{2i}^{(\overline{d})}\right\|^{2}\right)^{1/2}}_{\text{Cauchy-Schwarz inequality}} \leq \nonumber \\
    &\leq \underbrace{L\delta_n/\sqrt{n}}_{\text{By (\ref{eq:uprimitive2})}}\times\underbrace{\frac{2}{c^{2}}\delta_n n^{-1/4}}_{\text{By (\ref{eq:uprimitive1C})}} = \nonumber \\
    &= \frac{2L}{c^{2}}\delta_n^2 n^{-3/4} \leq \underbrace{\frac{2L}{c^{2}}\delta_n^2/\sqrt{n}}_{\text{Since }n^{-3/4}\leq n^{-1/2}}.
\end{align}

By using the triangle inequality and the established bounds, we obtain:
\begin{align}\label{eq:primitive:ATE:component}
    &\left|\mathbb{E}\left(\Delta\text{E}_{i}^{(\overline{d})}\left(1 - \frac{\tilde{P}_{i}^{0}}{\tilde{P}_{i}}\frac{D_{i}^{(\overline{d})}}{P_{i}^{(\overline{d})}}\right)\right)\right| \leq \nonumber \\
    &\leq \underbrace{0}_{\text{By (\ref{eq:primitive:ATE:term1})}} + \underbrace{\frac{2}{c^{2}}\left(1+L\right)\delta_n^2/\sqrt{n}}_{\text{By (\ref{eq:primitive:ATE:term2})}} + \nonumber \\
    &+ \underbrace{\frac{L}{c}\delta_n/\sqrt{n}}_{\text{By (\ref{eq:primitive:ATE:term3})}} + \underbrace{\frac{2L}{c^{2}}\delta_n^2/\sqrt{n}}_{\text{By (\ref{eq:primitive:ATE:term4})}} = \nonumber \\
    &= \left(\frac{L}{c}\delta_n + \frac{2(1+2L)}{c^{2}}\delta_n^2\right)/\sqrt{n}.
\end{align}

Finally, since the established bounds hold uniformly over $\mu\in\mathcal{T}_{n}$, we obtain:
\begin{align}
    &\sup_{\mu\in \mathcal{T}_{n}}\left|\mathbb{E}\left(\Delta\text{E}_{i}^{(d)}\left(1 - \frac{\tilde{P}_{i}^{0}}{\tilde{P}_{i}}\frac{D_{i}^{(d)}}{P_{i}^{(d)}}\right) - \Delta\text{E}_{i}^{(d^{*})}\left(1 - \frac{\tilde{P}_{i}^{0}}{\tilde{P}_{i}}\frac{D_{i}^{(d^{*})}}{P_{i}^{(d^{*})}}\right)\right)\right| \leq \nonumber \\
    &\leq \underbrace{\left[\begin{aligned}
        &\sup_{\mu\in\mathcal{T}_{n}}\left|\mathbb{E}\left(\Delta\text{E}_{i}^{(d)}\left(1-\frac{\tilde{P}_{i}^{0}}{\tilde{P}_{i}}\frac{D_{i}^{(d)}}{P_{i}^{(d)}}\right)\right)\right| + \\
        &+\sup_{\mu\in\mathcal{T}_{n}}\left|\mathbb{E}\left(\Delta\text{E}_{i}^{(d^{*})}\left(1-\frac{\tilde{P}_{i}^{0}}{\tilde{P}_{i}}\frac{D_{i}^{(d^{*})}}{P_{i}^{(d^{*})}}\right)\right)\right|
    \end{aligned}\right]}_{\text{Triangle inequality}} \leq \nonumber \\
    &\leq \underbrace{2\left(\frac{L}{c}\delta_n+\frac{2(1+2L)}{c^{2}}\delta_n^2\right)}_{\substack{\text{By (\ref{eq:primitive:ATE:component})}\\ \xrightarrow{n\to\infty}0}}\frac{1}{\sqrt{n}}.
\end{align}

\subsubsection{Preliminary bounds related to ATT}

First, consider the following decomposition:
\begin{align}
    &\Delta\text{E}_{i}^{(d)}D_{i}^{(d)}\left(1 - \frac{\tilde{P}_{i}^{0}}{\tilde{P}_{i}}\right) + \Delta\text{E}_{i}^{(d^{*})}\left(\frac{\tilde{P}_{i}^{0}}{\tilde{P}_{i}}\frac{P_{i}^{(d)}}{P_{i}^{(d^{*})}}D_{i}^{(d^{*})} - D_{i}^{(d)}\right) = \nonumber \\
    =&\Delta\text{E}_{i}^{(d^{*})}\Delta V_{1i} + \Delta\text{E}_{i}^{(d^{*})}\Delta V_{2i} + \Delta\text{E}_{i}^{(d)}\Delta V_{3i},
\end{align}
where:
\begin{align}
    \Delta V_{1i} &= \frac{P_i^{0(d)}}{P_i^{0(d^*)}}D_i^{(d^*)} - D_i^{(d)},\\
    \Delta V_{2i} &= D_i^{(d^*)}\left(\frac{\tilde{P}_i^0}{\tilde{P}_i}\frac{P_i^{(d)}}{P_i^{(d^*)}} - \frac{P_i^{0(d)}}{P_i^{0(d^*)}}\right),\\
    \Delta V_{3i} &= D_i^{(d)}\left(1 - \frac{\tilde{P}_i^0}{\tilde{P}_i}\right).
\end{align}

We start by establishing several useful bounds for the elements of the decomposition. First, we obtain:
\begin{align}
    \left|\Delta V_{1i}\right| &= \underbrace{\left|\Delta Q_{1i}\right|}_{\substack{\text{Since }\Delta V_{1i}=\Delta Q_{1i}\\ \text{by definition}}} \leq \underbrace{\frac{1}{c}}_{\text{By (\ref{eq:uprimitive1B})}}. \label{eq:uprimitive1D}
\end{align}

Second, we obtain:
\begin{align}
    \mathbb{E}\left(\Delta V_{1i}\mid X_{i},P_{i}^{0}\right) &= \underbrace{\mathbb{E}\left(\Delta Q_{1i}\mid X_{i},P_{i}^{0}\right)}_{\text{Since }\Delta V_{1i}=\Delta Q_{1i}} = \underbrace{0}_{\text{By (\ref{eq:uprimitive2B})}}. \label{eq:uprimitive2D}
\end{align}

Third, we obtain:
\begin{align}
    \mathbb{E}\left(\left\|\Delta V_{2i}\right\|^2\right)^{1/2}
    &= \mathbb{E}\left(\left\|D_i^{(d^*)}\left(\frac{P_i^{(d)}}{P_i^{(d^*)}} - \frac{P_i^{0(d)}}{P_i^{0(d^*)}} + \frac{P_i^{(d)}}{P_i^{(d^*)}}\frac{\tilde{P}_i^0-\tilde{P}_i}{\tilde{P}_i}\right)\right\|^2\right)^{1/2} = \nonumber \\
    &= \mathbb{E}\left(\left\|\underbrace{D_i^{(d^*)}\left(\frac{P_i^{(d)}}{P_i^{(d^*)}} - \frac{P_i^{0(d)}}{P_i^{0(d^*)}}\right)}_{\Delta Q_{2i}} + \frac{D_i^{(d^*)}P_i^{(d)}\left(\tilde{P}_i^0-\tilde{P}_i\right)}{\tilde{P}_iP_i^{(d^*)}}\right\|^2\right)^{1/2} \leq \nonumber \\
    &\leq \underbrace{\mathbb{E}\left(\left\|\Delta Q_{2i}\right\|^2\right)^{1/2} + \mathbb{E}\left(\left\|\frac{D_i^{(d^*)}P_i^{(d)}\left(\tilde{P}_i^0-\tilde{P}_i\right)}{\tilde{P}_iP_i^{(d^*)}}\right\|^2\right)^{1/2}}_{\text{Minkowski's inequality}} \leq \nonumber \\
    &\leq \mathbb{E}\left(\left\|\Delta Q_{2i}\right\|^2\right)^{1/2} + \underbrace{\frac{1}{c^2}\mathbb{E}\left(\left\|\tilde{P}_i^0-\tilde{P}_i\right\|^2\right)^{1/2}}_{\substack{\text{Since }D_i^{(d^*)}P_i^{(d)}\in[0,1]\\ \text{and by Part 4 of Assumption~\ref{assumption:regularity}}}} \leq \nonumber \\
    &\leq \underbrace{\frac{2}{c^2}\delta_n n^{-1/4}}_{\text{By (\ref{eq:uprimitive3B})}} + \underbrace{\frac{1}{c^2}\delta_n n^{-1/4}}_{\substack{\text{Part 5 of Assumption}\\ \ref{assumption:regularity}}} = \frac{3}{c^2}\delta_n n^{-1/4}. \label{eq:uprimitive3D}
\end{align}

Fourth, we obtain:
\begin{align}
    \mathbb{E}\left(\left\|\Delta V_{3i}\right\|^2\right)^{1/2}
    &= \mathbb{E}\left(\left\|\frac{D_i^{(d)}\left(\tilde{P}_i-\tilde{P}_i^0\right)}{\tilde{P}_i}\right\|^2\right)^{1/2} \leq \nonumber \\
    &\leq \underbrace{\frac{1}{c}\mathbb{E}\left(\left\|\tilde{P}_i-\tilde{P}_i^0\right\|^2\right)^{1/2}}_{\substack{\text{Since }D_i^{(d)}\in\{0,1\}\\ \text{and by Part 4 of Assumption~\ref{assumption:regularity}}}} \leq \nonumber \\
    &\leq \underbrace{\frac{1}{c}\delta_n n^{-1/4}}_{\text{Part 5 of Assumption~\ref{assumption:regularity}}}. \label{eq:uprimitive4D}
\end{align}

Notice that the bounds (\ref{eq:uprimitive1D})--(\ref{eq:uprimitive4D}) hold uniformly over $\mu\in\mathcal{T}_n$.

\subsubsection{The bound for ATT}

Consider the following decomposition:
\begin{align}
    &\Delta\text{E}_{i}^{(d)}D_{i}^{(d)}\left(1 - \frac{\tilde{P}_{i}^{0}}{\tilde{P}_{i}}\right) + \Delta\text{E}_{i}^{(d^{*})}\left(\frac{\tilde{P}_{i}^{0}}{\tilde{P}_{i}}\frac{P_{i}^{(d)}}{P_{i}^{(d^{*})}}D_{i}^{(d^{*})} - D_{i}^{(d)}\right) = \nonumber \\
    =&\Delta E_{1i}^{(d)}\Delta V_{3i} + \Delta E_{2i}^{(d)}\Delta V_{3i} + \nonumber \\
    +&\Delta E_{1i}^{(d^*)}\Delta V_{1i} + \Delta E_{2i}^{(d^*)}\Delta V_{1i} + \nonumber \\
    +&\Delta E_{1i}^{(d^*)}\Delta V_{2i} + \Delta E_{2i}^{(d^*)}\Delta V_{2i}.
\end{align}

Below, we bound the expectations of each term of the last equality. First, we obtain:
\begin{align}\label{eq:primitive:ATT:term1}
    \mathbb{E}\left(\Delta\text{E}_{1i}^{(d^{*})}\Delta V_{1i}\right)
    &= \underbrace{\mathbb{E}\left(\Delta\text{E}_{1i}^{(d^{*})}\mathbb{E}\left(\Delta V_{1i}\mid X_i,P_i^0\right)\right)}_{\substack{\text{LIE conditional on }(X_i,P_i^0)\\ \text{since }\Delta\text{E}_{1i}^{(d^{*})}\text{ is a function of }X_i,P_i^0}} = \nonumber \\
    &= \mathbb{E}\left(\Delta\text{E}_{1i}^{(d^{*})}\times\underbrace{0}_{\text{By (\ref{eq:uprimitive2D})}}\right) = 0.
\end{align}

Second, we obtain:
\begin{align}\label{eq:primitive:ATT:term2}
    \left|\mathbb{E}\left(\Delta\text{E}_{2i}^{(d^{*})}\Delta V_{1i}\right)\right|
    &\leq \mathbb{E}\left(\left|\Delta\text{E}_{2i}^{(d^{*})}\Delta V_{1i}\right|\right) \leq \nonumber \\
    &\leq \underbrace{\frac{1}{c}\mathbb{E}\left(\left|\Delta\text{E}_{2i}^{(d^{*})}\right|\right)}_{\text{By (\ref{eq:uprimitive1D})}} \leq \nonumber \\
    &\leq \underbrace{\frac{1}{c}\mathbb{E}\left(\left\|\Delta\text{E}_{2i}^{(d^{*})}\right\|^2\right)^{1/2}}_{\text{Lyapunov's inequality}} \leq \underbrace{\frac{L}{c}\delta_n/\sqrt{n}}_{\text{By (\ref{eq:uprimitive2})}}.
\end{align}

Third, we obtain:
\begin{align}\label{eq:primitive:ATT:term3}
    \left|\mathbb{E}\left(\Delta\text{E}_{1i}^{(d)}\Delta V_{3i}\right)\right|
    &\leq \underbrace{\mathbb{E}\left(\left\|\Delta\text{E}_{1i}^{(d)}\right\|^2\right)^{1/2}\mathbb{E}\left(\left\|\Delta V_{3i}\right\|^2\right)^{1/2}}_{\text{Cauchy-Schwarz inequality}} \leq \nonumber \\
    &\leq \underbrace{(1+L)\delta_n n^{-1/4}}_{\text{By (\ref{eq:uprimitive1})}}\times\underbrace{\frac{1}{c}\delta_n n^{-1/4}}_{\text{By (\ref{eq:uprimitive4D})}} = \nonumber \\
    &= \frac{1+L}{c}\delta_n^2/\sqrt{n}.
\end{align}

Fourth, we obtain:
\begin{align}\label{eq:primitive:ATT:term4}
    \left|\mathbb{E}\left(\Delta\text{E}_{2i}^{(d)}\Delta V_{3i}\right)\right|
    &\leq \underbrace{\mathbb{E}\left(\left\|\Delta\text{E}_{2i}^{(d)}\right\|^2\right)^{1/2}\mathbb{E}\left(\left\|\Delta V_{3i}\right\|^2\right)^{1/2}}_{\text{Cauchy-Schwarz inequality}} \leq \nonumber \\
    &\leq \underbrace{L\delta_n/\sqrt{n}}_{\text{By (\ref{eq:uprimitive2})}}\times\underbrace{\frac{1}{c}\delta_n n^{-1/4}}_{\text{By (\ref{eq:uprimitive4D})}} = \nonumber \\
    &= \frac{L}{c}\delta_n^2 n^{-3/4} \leq \underbrace{\frac{L}{c}\delta_n^2/\sqrt{n}}_{\text{Since }n^{-3/4}\leq n^{-1/2}}.
\end{align}

Fifth, we obtain:
\begin{align}\label{eq:primitive:ATT:term5}
    \left|\mathbb{E}\left(\Delta\text{E}_{1i}^{(d^{*})}\Delta V_{2i}\right)\right|
    &\leq \underbrace{\mathbb{E}\left(\left\|\Delta\text{E}_{1i}^{(d^{*})}\right\|^2\right)^{1/2}\mathbb{E}\left(\left\|\Delta V_{2i}\right\|^2\right)^{1/2}}_{\text{Cauchy-Schwarz inequality}} \leq \nonumber \\
    &\leq \underbrace{(1+L)\delta_n n^{-1/4}}_{\text{By (\ref{eq:uprimitive1})}}\times\underbrace{\frac{3}{c^2}\delta_n n^{-1/4}}_{\text{By (\ref{eq:uprimitive3D})}} = \nonumber \\
    &= \frac{3(1+L)}{c^2}\delta_n^2/\sqrt{n}.
\end{align}

Sixth, we obtain:
\begin{align}\label{eq:primitive:ATT:term6}
    \left|\mathbb{E}\left(\Delta\text{E}_{2i}^{(d^{*})}\Delta V_{2i}\right)\right|
    &\leq \underbrace{\mathbb{E}\left(\left\|\Delta\text{E}_{2i}^{(d^{*})}\right\|^2\right)^{1/2}\mathbb{E}\left(\left\|\Delta V_{2i}\right\|^2\right)^{1/2}}_{\text{Cauchy-Schwarz inequality}} \leq \nonumber \\
    &\leq \underbrace{L\delta_n/\sqrt{n}}_{\text{By (\ref{eq:uprimitive2})}}\times\underbrace{\frac{3}{c^2}\delta_n n^{-1/4}}_{\text{By (\ref{eq:uprimitive3D})}} = \nonumber \\
    &= \frac{3L}{c^2}\delta_n^2 n^{-3/4} \leq \underbrace{\frac{3L}{c^2}\delta_n^2/\sqrt{n}}_{\text{Since }n^{-3/4}\leq n^{-1/2}}.
\end{align}

Finally, since the established bounds hold uniformly over $\mu\in\mathcal{T}_{n}$, by using these bounds and the triangle inequality, we obtain:
\begin{align}
    &\sup_{\mu\in \mathcal{T}_{n}}\left|\mathbb{E}\left(\left[\begin{aligned}
        &\Delta\text{E}_{i}^{(d)}D_{i}^{(d)}\left(1 - \frac{\tilde{P}_{i}^{0}}{\tilde{P}_{i}}\right) + \\
        &+\Delta\text{E}_{i}^{(d^{*})}\left(\frac{\tilde{P}_{i}^{0}}{\tilde{P}_{i}}\frac{P_{i}^{(d)}}{P_{i}^{(d^{*})}}D_{i}^{(d^{*})} - D_{i}^{(d)}\right)
    \end{aligned}\right]\right)\right| \leq \nonumber \\
    \leq &\underbrace{0}_{\text{By (\ref{eq:primitive:ATT:term1})}} + \underbrace{\frac{L}{c}\delta_n/\sqrt{n}}_{\text{By (\ref{eq:primitive:ATT:term2})}} + \underbrace{\frac{1+L}{c}\delta_n^2/\sqrt{n}}_{\text{By (\ref{eq:primitive:ATT:term3})}} + \nonumber \\
    +&\underbrace{\frac{L}{c}\delta_n^2/\sqrt{n}}_{\text{By (\ref{eq:primitive:ATT:term4})}} + \underbrace{\frac{3(1+L)}{c^2}\delta_n^2/\sqrt{n}}_{\text{By (\ref{eq:primitive:ATT:term5})}} + \underbrace{\frac{3L}{c^2}\delta_n^2/\sqrt{n}}_{\text{By (\ref{eq:primitive:ATT:term6})}} = \nonumber \\
    =&\underbrace{\left(\frac{L}{c}\delta_n + \frac{1+2L}{c}\left(1+\frac{3}{c}\right)\delta_n^2\right)}_{\xrightarrow{n\to\infty}0}\frac{1}{\sqrt{n}}.
\end{align}

\subsubsection{Conclusion}

Therefore, inequalities~(\ref{eq:primitive1}) and (\ref{eq:primitive21}), together with the remaining regularity conditions, imply that $\lambda_N''=O((\delta_n+\delta_n^2)/\sqrt{n})=O(\delta_n/\sqrt{n})$ for all four derivative-free scores. By combining this result with the previously established bounds on $r_N$ and $r_N'$, we conclude that there exists a sufficiently large constant $C_1\geq1$ such that $\overline{\delta}_n=C_1\max(\delta_n^{(q-2)/q},\delta_n,n^{-1/2})$ satisfies all the rate conditions required by Theorem~\ref{theorem:modified}. Since $\overline{\delta}_n\to0$, the conclusions of Theorem~\ref{theorem:normality2} hold under these alternative primitive conditions.

\clearpage
\normalsize
\bibliography{bibliography}
\end{document}